\documentclass[sigconf,edbt]{acmart-edbt2023}

\usepackage[frozencache=true,cachedir=minted-cache]{minted}

\def\BibTeX{{\rm B\kern-.05em{\sc i\kern-.025em b}\kern-.08em
    T\kern-.1667em\lower.7ex\hbox{E}\kern-.125emX}}

\usepackage{booktabs} % For formal tables

\usepackage{listings}
\usepackage{soul}
\usepackage{subcaption}
\usepackage{color}
\definecolor{lightgray}{rgb}{0.75,0.75,0.75}
\usepackage{colortbl}

\newcolumntype{L}[1]{>{\raggedright\let\newline\\\arraybackslash\hspace{0pt}}m{#1}}
\newcolumntype{C}[1]{>{\centering\let\newline\\\arraybackslash\hspace{0pt}}m{#1}}
\newcolumntype{R}[1]{>{\raggedleft\let\newline\\\arraybackslash\hspace{0pt}}m{#1}}

\usepackage[ruled,linesnumbered]{algorithm2e}

\newenvironment{listingAlgorithm}[1][htb]
  {% Update algorithm name
   \begin{algorithm}[#1]%
  }{\end{algorithm}}
  
\newenvironment{listingAlgorithm*}[1][htb]
  {% Update algorithm name
   \begin{algorithm*}[#1]%
  }{\end{algorithm*}}

\newif\ifArxivVersion

\ArxivVersiontrue

\setcopyright{rightsretained}

\acmDOI{}

\acmISBN{978-3-89318-088-2}

\acmConference[EDBT 2023]{26th International Conference on Extending Database Technology (EDBT)}{28th March-31st March, 2023}{Ioannina, Greece}
\acmYear{2023}

\begin{document}
\title{Integration of Skyline Queries into Spark SQL}
% \titlenote{Produces the permission block, and copyright information}
% \subtitle{Extended Abstract}
% \subtitlenote{The full version of the author's guide is available as
%   \texttt{acmart.pdf} document}

\author{Lukas Grasmann}
\email{lukas.grasmann@tuwien.ac.at}
\affiliation{%
  \institution{ TU Wien}
%   \streetaddress{Favoritenstraße 9-11}
   \city{Vienna}
   \country{Austria}
%  \postcode{1040}
}

\author{Reinhard Pichler}
\email{reinhard.pichler@tuwien.ac.at}
\affiliation{%
  \institution{ TU Wien}
%  \streetaddress{Favoritenstraße 9-11}
   \city{Vienna}
   \country{Austria}
%  \postcode{1040}
}

\author{Alexander Selzer}
\email{alexander.selzer@tuwien.ac.at}
\affiliation{%
  \institution{ TU Wien}
%  \streetaddress{Favoritenstraße 9-11}
 \city{Vienna}
  \country{Austria}
%  \postcode{1040}
}

% \renewcommand{\shortauthors}{B. Trovato et al.}
\renewcommand{\shortauthors}{Grasmann, Pichler, and Selzer}

\begin{abstract}
Skyline queries are frequently used in data analytics and multi-criteria decision support applications to filter relevant information from big amounts of data. Apache Spark is a popular framework for processing big, distributed data. The framework even provides a convenient SQL-like interface via the Spark SQL module.
However, skyline queries are not natively supported and require tedious rewriting to fit the SQL standard or Spark's SQL-like language. 

The goal of our work is to fill this gap. We thus provide a full-fledged integration of the skyline operator into Spark SQL. This allows for a simple and easy to use syntax to input skyline queries. Moreover, our empirical results show that this integrated solution by far outperforms a solution based on rewriting into standard SQL.
\end{abstract}

%
% % The code below should be generated by the tool at
% % http://dl.acm.org/ccs.cfm
% % Please copy and paste the code instead of the example below. 

% \keywords{ACM proceedings, \LaTeX, text tagging}


\maketitle
\section{Introduction}\label{section:introduction}

Skyline queries are an important tool in data analytics and decision making to find \textit{interesting} points in a multi-dimensional dataset. Given a set $P$ of data points, we can compute the skyline (also called Pareto front) using the \textit{skyline operator}~\cite{TheSkylineOperator}. Intuitively, a data point $p \in P$ belongs to the skyline if it is not \textit{dominated} by any other point. For a given set of dimensions, a data point $q$ \textit{dominates} $p$ (denoted $q \prec p$) if $q$ is better in at least one dimension while being at least as good in every other dimension. 
More formally, the skyline $S$ is obtained as $S = \{p \in P \mid \not\exists q \in P : q \prec p\}$.

\begin{figure}[htp]
    \includegraphics{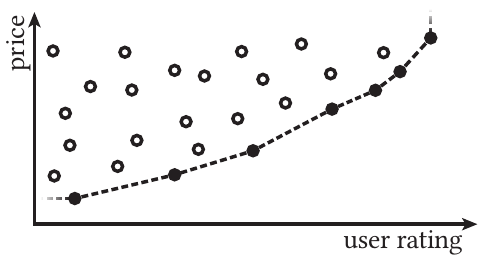}
    \caption{Example skyline for hotels% with dimensions price and user rating
    }
    \label{fig:skyline_example}
    % \Description[Example skyline for hotels with skyline dimensions price and star rating]{A skyline of hotels with regards to the skyline dimensions price and star rating contains all non-dominated hotels for which no strictly better hotel exists.}
\end{figure}

A common example for skyline queries in the literature can be found in Figure~\ref{fig:skyline_example}. Suppose that we want to find the perfect hotel for our holiday stay. Each hotel has multiple properties that express its quality such as the price per night or the distance to the beach among others. In this example, we limit ourselves to the price per night (y-axis) and the user rating (x-axis) for simplicity. Such attributes relevant to the skyline are called \textit{skyline dimensions}. A hotel dominates another hotel if it is strictly better in at least one skyline dimension and not worse in all others. For the price per night, minimization is desirable while we want the rating to be as high as possible (maximization).
Figure~\ref{fig:skyline_example} depicts the skyline of the hotels according to these two dimensions. In this simple (two-dimensional) example, it is immediately visible that for each hotel not part of the skyline there exists a ``better'' (dominating) alternative. In practice, things get more complex as skyline queries usually handle a larger number of skyline dimensions.

Skyline queries have a wide range of applications, including recommender systems to propose suitable items to the user~\cite{TheSkylineOperator} and location-based systems to find ``the best'' routes~\cite{RouteSkylineQueries, InRouteSkylineQuerying}.
Skyline queries have also been used to improve the quality of (web) services~\cite{SkylineServicesQoSWebServiceComposition} and they have applications in privacy~\cite{PrivacySkyline}, authentication~\cite{AuthenticationLocationBasedSkyline}, and DNA searching~\cite{DynamicSkylineQueriesMetricSpaces}.
Many further use cases for skyline queries are given in these surveys:~\cite{SurveyOfSkylineQueryProcessing, SurveyOfSkylineQueryProcessingHighlyDistributedEnvironments, SkylineQueriesFrontBack}. 
Skyline queries are particularly useful when dealing with big datasets and filtering the relevant data items. 
Indeed, in data science and data analytics, we are typically confronted with big data, which is stored in an optimized and distributed manner. We thus need a processing engine that is capable of efficiently operating on such data.

Apache Spark~\cite{ApacheSparkUnifiedEngine} is a unified framework that is specifically designed for distributed processing of potentially big data and has gained huge popularity in recent years.
It works for a wide range of applications, and it can interoperate with a great variety of data sources. Conceptually, it is similar to the MapReduce framework, where instead of map-combine-reduce nodes, there are individual nodes in an execution plan for each processing step. Spark also provides a multitude of modular extensions 
such as MLlib, PySpark, SparkR, and others, built on top of its core functionality. Another important extension is Spark SQL, which provides relational query functionality and comes with the powerful \textit{Catalyst} optimizer.

Hence, Apache Spark would be the perfect fit for executing skyline queries over big, distributed data. Despite this, to date, the skyline operator has not been integrated into Spark SQL. Nevertheless, it is possible to formulate skyline queries using ``plain'' SQL. Listing~\ref{alg:skyline_hotel_query_modified_sql} shows 
the ``plain'' skyline query for our hotel example.

\begin{listingAlgorithm}% [t]
SELECT price, user\_rating FROM hotels AS o WHERE NOT EXISTS(\\
	\Indp SELECT * FROM hotels AS i WHERE \\
		\Indp	$i$.price $\leq$ $o$.price \\
				AND $i$.user\_rating $\geq$ $o$.user\_rating \\
				AND ( \\
				\Indp	$i$.price $<$ $o$.price \\
						OR $i$.user\_rating $<$ $o$.user\_rating \\
				\Indm) \\
\Indm\Indm);
\caption{Hotel skyline query in plain SQL~\cite{TheSkylineOperator, ExtendingPostgreSQLSkylineOperator}}
\label{alg:skyline_hotel_query_plain_sql}
\end{listingAlgorithm}

There are several drawbacks to such a formulation of skyline queries in plain SQL. First of all, as more skyline dimensions are used, the SQL code gets more and more cumbersome. Hence, such an approach is error-prone, and readability and maintainability will inevitably be lost. Moreover, most SQL engines (this also applies to Spark SQL) are not optimized for this kind of nested \texttt{SELECT}s. Hence, also performance is 
of course 
an issue. Therefore, a better solution is called for. This is precisely the goal of our work:

\begin{itemize}
    \item We aim at an integration of skyline queries into Spark SQL with a simple, easy to use 
    % (and easy to read and maintain)
    syntax, e.g., the skyline query for our hotel example should look as depicted in Listing~\ref{alg:skyline_hotel_query_modified_sql}.
    \item We want to make use of existing query optimizations provided by the Catalyst optimizer of Spark SQL and extend the optimizer to also cover skyline queries.
    \item At the same time, we want to make sure that our integration of the skyline operator has no negative effect on the optimization and execution of other queries.
    \item Maintainability of the new code is an issue. We aim at a modular implementation, making use of existing functionality and making future enhancements as easy as possible.
    \item Spark and Spark SQL support many different data sources. The integration of skyline queries should work independently of the 
    data source that is being used.
\end{itemize}

\begin{listingAlgorithm}% [t]
    SELECT price, user\_rating FROM hotels SKYLINE OF price MIN, user\_rating MAX;
    \caption{Hotel skyline query in extended SQL~\cite{TheSkylineOperator, ExtendingPostgreSQLSkylineOperator}}
    \label{alg:skyline_hotel_query_modified_sql}
\end{listingAlgorithm}

\paragraph{Our contribution}
The main result of this work is a full-fledged integration of skyline queries. This includes the following: 

\begin{itemize}
    \item Virtually every single component involved in the query optimization and execution process of Spark SQL (see~Figure~\ref{fig:spark_sql_query_processing_execution}) had to be extended: the parser, analyzer, optimizer, etc.
    \item While a substantial portion of new code had to be provided, at the same time, we took care to make use of existing code as much as possible and to keep the extensions modular for easy maintenance and future enhancements.
    \item For the optimization of skyline queries, the specifics of distributed query processing had to be taken into account. Apart from incorporating new rules into the Catalyst optimizer of Spark SQL, we have extended the skyline query syntax compared with previous proposals~\cite{TheSkylineOperator, ExtendingPostgreSQLSkylineOperator} to allow the user to control further skyline-specific optimization.
    \item We have carried out an extensive empirical evaluation which shows that the integrated version of skyline queries by far outperforms the rewritten version in the style of Listing~\ref{alg:skyline_hotel_query_plain_sql}. 
\end{itemize}

\paragraph{Structure of the paper}
In Section~\ref{section:related_work}, we discuss relevant related work. The syntax and semantics of skyline queries are introduced in Section~\ref{section:skyline_queries}. A brief introduction to Apache Spark and Spark SQL is given in Section~\ref{section:apache_spark}. In Section~\ref{section:integration}, we describe the main ideas of our integration of the skyline operator into Spark SQL. We report on the results of our empirical evaluation in Section~\ref{section:performance_evaluation}. In Section~\ref{section:conclusion}, we conclude and identify some lines of future work.

The source code of our implementation is available here: \url{https://github.com/Lukas-Grasmann/skyline-queries-spark}.
Further material, comprised of the binaries of our implementation, all input data, and queries from our empirical evaluation is provided at~\cite{Zen:data/Skyline}.

\section{Related Work}\label{section:related_work}

Since the original publication of the skyline operator by Börzsönyi, Kossmann, and Stocker~\cite{TheSkylineOperator}, various  algorithms and approaches to skyline queries have been proposed. In this section, we give an overview of the relevant related work. We limit ourselves to, broadly, the ``original'' definition of skyline queries and exclude related query types like reverse skyline queries~\cite{EfficientComputationReverseSkylineQueries}.

The main cost factor of skyline computation is the time spent on dominance testing, which in turn depends on how many dominance tests between tuples need to be performed. A simple, straightforward algorithm runs in quadratic time w.r.t.\ the size of the data. It checks for each pair of tuples whether one of them is dominated and, therefore, not part of the resulting skyline. This basic idea of pairwise dominance tests is realized in the Block-Nested-Loop skyline algorithm~\cite{TheSkylineOperator}, which performs well especially when the resulting output is not too big, and the input dataset fits in memory. 

In the most recent algorithms, there are two main approaches that are used to either decrease the total number of dominance checks or to increase the parallelism of these checks.

\begin{enumerate}
    \item Using a score function on the skyline dimensions and sorting the tuples w.r.t. to this 
    score function~\cite{TheSkylineOperator, SDI}. These approaches may require the entire dataset to be available on a single node, which 
    is not ideal in a distributed environment.
    \item Partitioning (and distributing) the skyline computation by first independently computing the skyline of partitions (\textit{local skylines}) before proceeding to the computation of the final skyline (\textit{global skyline})~\cite{EfficientParallelProcessingMethodSkylineMapReduce, ParallelDistributedProcessingConstrainedSkyline, EfficientParallelSkylineQueryProcessingHighDimensionalData, OptimizingSkylineQueryProcessingIncompleteData}. This does not necessarily decrease the total number of dominance checks but speeds up the algorithm 
    through parallelism.
\end{enumerate}

Most of these approaches are centered around special data structures and indexes, which are provided by the respective database system. This makes them less suitable for Spark, which currently has no internal support for building and maintaining indexes. Such algorithms include \texttt{Index}~\cite{SkylineBitmapIndex} which uses a bitmap index to compute the skyline. Many algorithms of this class also make use of presorting, i.e., sorting the tuples from the dataset according to a monotone scoring function. Such scoring can, by definition, rule out certain dominance relationships between tuples such that fewer comparisons become necessary. This includes the algorithms \texttt{SFS}~\cite{SFS_1, SFS_2}, \texttt{LESS}~\cite{LESS_1, LESS_2}, SALSA~\cite{SALSA_1, SALSA_2}, and SDI~\cite{SDI}.

A simple partition-based algorithm is the Divide-and-Conquer algorithm presented in the original skyline paper~\cite{TheSkylineOperator}. Its working principle is to recursively divide the data into partitions until the units are small enough such that the skyline can be computed efficiently. The resulting (local) skylines are then merged step by step until the entire recursion stack 
has been processed. Related approaches include the nearest-neighbor search~\cite{NearestNeighbor} as well as other similar algorithms~\cite{BBS_1, BBS_2, LS, OSPS, WorstCaseIOEfficientSkylineAlgorithms}.

The algorithm \texttt{BSkyTree}~\cite{BSkyTree_1, BSkyTree_2} supports both a sorting-based and a partition-based variant. These make use of an index tree-structure to increase performance. Alternatively, there also exist tree structures based on quad-trees~\cite{ParallelComputationSkylineReverseSkyline}. Here, the data is partitioned dynamically and gradually into smaller and smaller ``rectangular'' chunks until the skylines can be computed efficiently.

There are also newer algorithms specifically targeted towards 
the MapReduce framework~\cite{EfficientParallelProcessingMethodSkylineMapReduce, ApproachingSkylineZOrder, EfficientParallelSkylineQueryProcessingHighDimensionalData}.
MapReduce approaches use \textit{mappers} to generate partitions by assigning keys to each tuple while the \textit{reducers} are responsible for computing the actual skylines 
for all tuples with the same key. Each intermediate skyline for a partition is called a \textit{local skyline}. This approach usually employs two or more rounds of map and reduce operations (\textit{stages}) where each stage first (re-)partitions the data and then computes skylines. The final result is then derived by the last stage which generates only one partition and computes the \textit{global skyline}.

The success of MapReduce approaches relies on keeping the intermediate results as small as possible and maximizing parallelism~\cite{EfficientParallelProcessingMethodSkylineMapReduce}. To achieve this, multiple different partitioning schemes exist such as \textit{random}, \textit{grid-based}, and \textit{angle-based} partitioning~\cite{EfficientParallelProcessingMethodSkylineMapReduce}. Keeping the last local skylines before the global skyline computation small is especially desirable since the global skyline computation cannot be fully parallelized. Indeed, this is usually the bottleneck which slows the skyline computation down~\cite{EfficientParallelProcessingMethodSkylineMapReduce}. This approach is not suitable for our purposes since it requires data from different partitions to be passed along with meta-information about the partitions. Spark does not currently support such functionality and adding it would be beyond the scope of this work.

There are efforts to eliminate multiple data points at once and
make the global skyline computation step distributed by partitioning the data. In the case of \textit{grid-based-partitioning} it is possible to eliminate entire \textit{cells} of data if they are dominated by another (non-empty) cell~\cite{EfficientParallelSkylineQueryProcessingHighDimensionalData}. It is then also possible to cut down the number of data points which each tuple must be compared against in the global skyline since some cells can never dominate each other according to the basic properties of the skyline (if a tuple is better in at least one dimension then it cannot be dominated)~\cite{EfficientParallelProcessingMethodSkylineMapReduce, EfficientParallelSkylineQueryProcessingHighDimensionalData}.

There have been two past efforts to implement skyline computation in Spark. The first one is the system \textit{SkySpark}~\cite{SkySpark} which uses Spark as the data source and realizes the 
skyline computation on the level of RDD operations (see Section~\ref{section:apache_spark} for details on 
Spark and RDDs).
A similar effort was made in the context of the thesis by Ioanna Papanikolaou~\cite{DistributedAlgorithmsApacheSpark},
which uses the map-reduce functionality of RDDs in Spark to compute the skyline. Both systems thus realize skyline computation as standalone Spark programs rather than as part of Spark SQL,
which would make, for instance, a direct performance comparison with our fully integrated solution difficult (even though it might still provide some interesting insights). However, the most important issue is that these implementations are {\em not ready to use}:
SkySpark~\cite{SkySpark}, with the last source code update in 2016, implements skyline computation for an old version of Spark. The code has been deprecated and archived by the author with a note that the code does not meet his standards anymore.
For the system reported in~\cite{DistributedAlgorithmsApacheSpark}, there is no runnable code provided since it is only available as code snippets in the PDF document.

As will be detailed in Section~\ref{section:integration}, we have chosen the Block-Nested-Loop skyline algorithm~\cite{TheSkylineOperator} for our integration into Spark SQL due to its simplicity.
By the modular structure of our implementation, replacing this algorithm by more sophisticated ones in the future 
should not be too difficult. 
We will come back to suggestions for future work in Section~\ref{section:conclusion}. 
Partitioning also plays an important role in our distributed approach (see Section~\ref{section:integration} for limits of distributed processing in the case of incomplete data though). 
However, to avoid unnecessary communication cost, we refrain from overriding Spark's partitioning mechanism.

\section{Skyline Queries}\label{section:skyline_queries}

In this section, we take a deeper look into the syntax and semantics of skyline queries.
Skyline queries can be expressed in a simple extension of SELECT-FROM-WHERE queries in SQL. 
This syntax
was proposed together with the original skyline operator in~\cite{TheSkylineOperator} and can be found in Listing~\ref{lst:skyline_query_sql_syntax}. The $m$ skyline dimensions out of $n$ total dimensions are denoted as $d_1$ through $d_m$ (where $m \leq n$). The \texttt{DISTINCT} keyword defines that we only return a single tuple if there are tuples with the same values in the skyline dimensions. If there are multiple possibilities, the exact tuple that is returned is chosen arbitrarily. The keyword \texttt{COMPLETE} is an extension compared to~\cite{TheSkylineOperator} that we have introduced. It allows the user to inform the system that the dataset is complete in the sense that no \texttt{null} occurs in the skyline dimensions. 
The system can, therefore, safely choose the \textit{complete} skyline algorithm, 
which is more efficient than the \textit{incomplete} one.
This will be discussed in more detail in Section~\ref{section:integration}.

\begin{listingAlgorithm}
  SELECT ... FROM ... WHERE ... GROUP BY ... HAVING ...\\
  SKYLINE OF [DISTINCT][COMPLETE]\\
  $d_1$ [MIN | MAX | DIFF], ..., $d_m$ [MIN | MAX | DIFF]\\
  ORDER BY ...
\caption{Syntax of skyline queries in SQL~\cite{TheSkylineOperator}}
\label{lst:skyline_query_sql_syntax}
\end{listingAlgorithm}

In Section~\ref{section:introduction}, we have already introduced the dominance relationship and the skyline itself in a semi-formal way. Below, we provide formal definitions of both concepts.

\begin{definition}[Dominance relationship between tuples~\cite{TheSkylineOperator, EfficientParallelProcessingMethodSkylineMapReduce}]
Given a set of tuples $R$, two $n$-ary tuples $r, s \in R$ and a set of skyline dimensions $D$ which always consists of four (potentially empty) disjoint subsets $D_{\mathit{min}}, D_{\mathit{max}}, D_{\mathit{diff}}, D_{\mathit{extra}}$, we define $r_i$ and $s_i$ as the value of the tuple $r$ and $s$ in dimension $d_i \in D$ respectively. The subsets of the dimensions correspond to the skyline dimensions ($D_{\mathit{min}}, D_{\mathit{max}}, D_{\mathit{diff}}$) and ``extra'' (non-skyline) dimensions ($D_{\mathit{extra}}$). Then $r$ \textit{dominates} $s$ ($r \prec s$) if and only if:
\begin{align*}
    \bigg( \bigwedge_{d_i \in D_{\mathit{min}}} r_i \leq s_i \bigg) \wedge \bigg( \bigwedge_{d_i \in D_{\mathit{max}}} r_i \geq s_i \bigg) \wedge \bigg( \bigwedge_{d_i \in D_{\mathit{diff}}} r_i = s_i \bigg) \\
    \wedge \Bigg( \bigg( \bigvee_{d_i \in D_{\mathit{min}}} r_i < s_i \bigg) \vee \bigg( \bigvee_{d_i \in D_{\mathit{max}}} r_i > s_i \bigg) \Bigg)
\end{align*}
\end{definition}

\noindent
In words, the above definition means that  a tuple $r \in R$ \textit{dominates} another tuple $s \in R$ if and only if:

\begin{itemize}
    \item The values in all \texttt{DIFF} dimensions are equal \textbf{and}
    \item $r$ is \textit{at least as good} in all \texttt{MIN/MAX} skyline dimensions \textbf{and}
    \item $r$ is \textit{strictly better} in at least one \texttt{MIN/MAX} skyline dimension
\end{itemize}

The dominance relationships are \textit{transitive}, i.e., if $a$ dominates $b$ and $b$ dominates $c$ then $a$ also dominates $c$.

Given this formal definition of the dominance relationship, we can now also formally define the skyline itself.

\begin{definition}[Skyline~\cite{TheSkylineOperator}]
Given a set of skyline dimensions $D$, let $R$ be a set of tuples. The \textit{skyline} $R$ (denoted $SKY(R)$) 
is a set of tuples defined as follows:
$
    SKY(R) := \{r \in R \mid \nexists s \in R : s \prec r\}.
$
\end{definition}

Note that the above definitions apply to {\em complete} datasets. 
For {\em incomplete} datasets (if \texttt{null} may occur in some skyline dimension), we need to slightly modify the definition of dominance in that the comparison of two tuples is always restricted to those dimensions where both are not \texttt{null}. Hence, the above definition  for the complete case is modified as follows~\cite{OptimizingSkylineQueryProcessingIncompleteData}:

\begin{itemize}
    \item The values in all \texttt{DIFF} dimensions where $r$ and $s$ are not \texttt{null} are equal \textbf{and}
    \item $r$ is \textit{at least as good} in all \texttt{MIN/MAX} skyline dimensions where $r$ and $s$ are not \texttt{null} \textbf{and}
    \item $r$ is \textit{strictly better} in at least one \texttt{MIN/MAX} skyline dimension where $r$ and $s$ are not \texttt{null}
\end{itemize}

Given the modified dominance relationship, we now run into the problem that for incomplete datasets the \textit{transitivity} property of skyline dominance is lost and there may be \textit{cyclic dominance relationships}. 
For instance, assume a dataset similar to the one given in~\cite{OptimizingSkylineQueryProcessingIncompleteData} consisting of three tuples $a = (1, *, 10)$, $b = (3, 2, *), c = (*, 5, 3)$ where $*$ is a placeholder for a missing value. If all three dimensions are skyline dimensions, then, given the definition of dominance in incomplete datasets, it follows that 
$a \prec b$ since $1 < 3$ for the first dimension. Similarly, we have $b \prec c$ since $2 < 5$ on the second dimension. Lastly, we note that $a \not\prec c$ since $10 \nless 3$ but $c \prec a$ since $3 < 10$ for the third dimension. Under the assumption of transitivity, it would follow from $a \prec b$ and $b \prec c$ that $a \prec c$ holds, which is not the case. 
In other words, for incomplete data, the transitivity property gets lost~\cite{OptimizingSkylineQueryProcessingIncompleteData}. Since $a \prec b$, $b \prec c$, and $c \prec a$, the dominance relationship forms a cycle and can therefore be referred to as a \textit{cyclic dominance relationship}~\cite{OptimizingSkylineQueryProcessingIncompleteData}.
Hence, when computing the skyline of a potentially incomplete dataset, prematurely deleting dominated tuples may lead to erroneous results. Indeed, this is 
%exactly
the trap into which the skyline algorithm proposed in~\cite{OptimizingSkylineQueryProcessingIncompleteData} fell.
For details, see 
\ifArxivVersion
Appendix~\ref{appendix:algorithm_incomplete_datasets}.
\else
\cite{unsereLangversion}. 
\fi

As already mentioned in Section~\ref{section:introduction}, skyline queries can be formulated in plain SQL without the specialized skyline syntax from Listing~\ref{lst:skyline_query_sql_syntax}.
A general schema of rewriting skyline queries in plain SQL is given in Listing~\ref{lst:skyline_query_plain_sql}. 
A first version of this rewriting was informally described in~\cite{TheSkylineOperator}. The full details and the corresponding syntax used here were introduced in~\cite{ExtendingPostgreSQLSkylineOperator}:
First, the outer query selects all tuples from some relation (which may itself be the result of a complex
SELECT-statement) and then we use a subquery with \texttt{WHERE NOT EXISTS} to eliminate all dominated tuples. 

\begin{listingAlgorithm}[htp]
    SELECT column\_list FROM rel AS o WHERE condition(s) AND NOT EXISTS( \label{alg:skyline_query_plain_sql:outerquery} \\
    	\Indp SELECT * FROM rel AS i WHERE condition(s) \label{alg:skyline_query_plain_sql:innerquerystart} \\
    		\Indp	AND $i.a_1 \leq o.a_1$ AND $\dots$ AND $i.a_j \leq o.a_j$ \\
    				AND $i.a_{j+1} \geq o.a_{j+1}$ AND $\dots$ AND $i.a_k \geq o.a_k$ \\
    				AND $i.a_{k+1} = o.a_{k+1}$ AND $\dots$ AND $i.a_m = o.a_m$ \label{alg:skyline_query_plain_sql:diff} \\
    				AND ( \\
    				\Indp	$i.a_1 < o.a_1$ OR $\dots$ OR $i.a_j < o.a_j$ \label{alg:skyline_query_plain_sql:strictlybetterstart} \\
    						OR $i.a_{j+1} > o.a_{j+1}$ OR $\dots$ OR  $i.a_k > o.a_k$ \label{alg:skyline_query_plain_sql:strictlybetterend} \\
    				\Indm) \label{alg:skyline_query_plain_sql:innerqueryend} \\
    \Indm\Indm);
    \caption{Translated skyline query in plain SQL~\cite{TheSkylineOperator, ExtendingPostgreSQLSkylineOperator}}
    \label{lst:skyline_query_plain_sql}
\end{listingAlgorithm}

\section{Apache Spark}\label{section:apache_spark}

\begin{figure*}
    \centering
    \includegraphics[width=17cm]{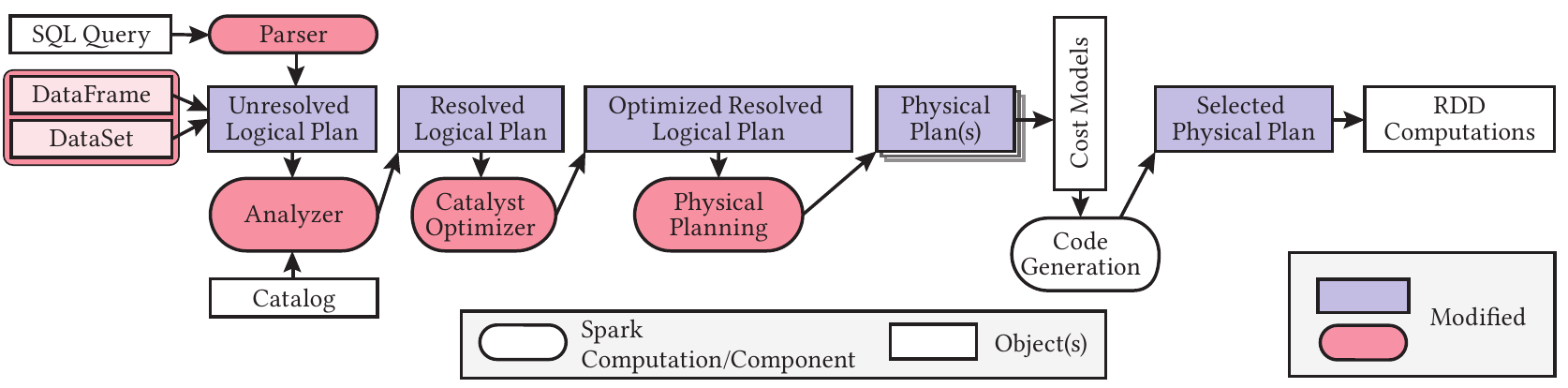}
    \caption{Apache Spark SQL Query Processing and Execution (adapted from~\cite{SparkClusterComputing})}
    \label{fig:spark_sql_query_processing_execution}
\end{figure*}

Apache Spark is a unified and distributed framework for processing large volumes of data. The framework can handle many different data sources. Through various specialized modules on top of its core module, it supports a wide range of processing workloads.
The central data structure of the Spark Core are \textit{resilient distributed datasets} (RDDs). An RDD is a collection of elements distributed over the nodes of the cluster. Working with Spark comes down to defining transformations and actions on RDDs, while 
data distribution, parallel execution, synchronization, fault-tolerance, etc. are taken care of by the system and are largely hidden from the user. 

The focus of our work lies on the Spark SQL module, which extends the Spark Core by providing an SQL interface. Queries can be formulated either via query strings or using an API. This API is based on specialized data structures called \textit{DataFrame} and \textit{DataSet}. They are (meanwhile) closely related in that every DataFrame is also a DataSet, with the only difference being that DataSets are strongly typed while DataFrames are not.
The processing and execution of input queries follows a well-defined schema with multiple steps that are depicted in Figure~\ref{fig:spark_sql_query_processing_execution}.

First, the queries are formulated either via SQL query strings or via APIs. In the case of query strings, it is necessary to parse them first. Both methods produce a \textit{logical} execution plan that contains the information which operations the plan
consists of. For instance, filtering is represented as a specific node in the plan and can be used for both \texttt{WHERE} and \texttt{HAVING} clauses in the query. 
The references to tables or columns are not yet assigned to actual objects in the database in this step. To solve this, the \texttt{Analyzer} takes each identifier and translates it using the \texttt{Catalog}. The result returned by the \texttt{Analyzer} is the \textit{resolved} logical plan where all placeholders have been replaced by actual references to objects in the database.

The \textit{Catalyst Optimizer} is a crucial part for the success of Spark SQL. It is a powerful rule-based optimizer that can be extended to include new rules for specific (newly introduced) nodes. Optimizations are applied to the resolved logical plan.

To actually execute the query, the logical execution plan must first be translated into a \textit{physical} execution plan, which contains information about how the various operations are to be realized. For example, Spark provides different join implementations, which are each represented by different nodes in the physical plan and one of them is selected during the physical planning phase. There may be more than one physical plan generated during this step. Based on the physical plan(s), one specific plan must be selected, and \textit{Code Generation} is carried out according to the chosen physical plan. In Spark, the generated code is a computation based on RDDs.

It will turn out that, for a full-fledged integration of skyline queries into Spark SQL, virtually all of the components of query processing and execution mentioned above have to be extended or modified. This will be the topic of the next section. 

\section{Integrating Skylines into Apache Spark}\label{section:integration}

Below, we give an overview of the required modifications and extensions of each component of the query processing flow depicted in Figure~\ref{fig:spark_sql_query_processing_execution}. An important aspect of integrating skyline queries 
%%%
into Spark SQL 
is the algorithm selection for the actual computation of the skyline. Here, we will discuss in some detail the challenges arising from potentially incomplete data and our solution.

\subsection{Skyline Query Syntax in Spark SQL}
\label{sect:SkylineQuerySyntax}

The parser is the first step of processing a Spark SQL query. It uses the ANTLR parser grammar generator to generate a lexer and subsequent parser which take query strings as input to obtain nodes in a logical plan. We modify both parts to extend Spark's SQL-like syntax such that it also includes skyline queries.

\begin{listingAlgorithm}
\begin{minted}[fontsize=\small,linenos,breaklines]{antlr}
skylineClause
  : SKYLINE
    skylineDistinct=DISTINCT?
    skylineComplete=COMPLETE?
    skylineItems+=skylineItem (',' skylineItems+=skylineItem)*
  ;

skylineItem
  : skylineItemExpression=expression skylineMinMaxDiff=(MIN | MAX | DIFF)
  ;
\end{minted}
\caption{ANTLR grammar for skyline queries}
\label{lst:skyline_antlr_grammar}
\end{listingAlgorithm}

The grammar corresponding to the syntax introduced in Listing~\ref{lst:skyline_query_sql_syntax} is shown in Listing~\ref{lst:skyline_antlr_grammar}. Given a ``regular'' \texttt{SELECT} statement, note that a skyline always comes after the \texttt{HAVING} clause (if any) but before any \texttt{ORDER BY} clause.
Each skyline clause consists of the \texttt{SKYLINE OF} keyword (line~2), an optional \texttt{DISTINCT} (line~3) and \texttt{COMPLETE} (line~4), as well as an arbitrary number of skyline dimensions (line~5 and 8~-~9) where the ``type'' \texttt{MIN}/\texttt{MAX}/\texttt{DIFF} is given separately for each dimension (line~9). The ordering of skyline dimensions in this syntax has no bearing on the outcome of the query (except for potential ordering). It may, however, have a slight effect on the performance of dominance checks since it determines the order in which the skyline dimensions of 
two tuples are compared.

% The parser is the first step of processing a Spark SQL query. It uses the ANTLR parser grammar generator to generate a lexer and subsequent parser which take query strings as input to obtain nodes in a logical plan. We modify both parts to extend Spark's SQL-like syntax such that it also includes skyline queries.
% The details of the ANTLR grammar of skyline
% queries are given in 
% \ifArxivVersion
% Appendix~\ref{appendix:ANTLR}.
% \else
% the appendix, 
% \fi

\subsection{Extending the Logical Plan}
\label{sect:ExtendingLogicalPlan}

Each node in the logical plan stores the details necessary to choose an algorithm for the specific operator and 
to derive a physical plan. In the case of the skyline operator, we use a single node with a single child in the logical plan. This node contains information such as the skyline dimensions \texttt{SkylineDimension} as well as other vital information like whether the skyline should be \texttt{DISTINCT} or not. The child node of the skyline operator node provides the input data for the skyline computation.

Each \texttt{SkylineDimension} extends the default Spark \texttt{Expression} such that it stores both the reference to the database dimension and the type (i.e., \texttt{MIN}/\texttt{MAX}/\texttt{DIFF}). The database dimension is usually a column but can also be a more complex \texttt{Expression} 
(e.g., an aggregate) in Spark. It is stored as the \texttt{child} of the \texttt{SkylineDimension} which allows us to make use of the generic functionality responsible for resolving \texttt{Expression}s in the analyzer.

\subsection{Extending the Analyzer}
\label{sect:ExtendingAnalyzer}

The analyzer of Spark already offers a wide range of rules which can be used to resolve the expressions. Since the skyline dimensions used by our skyline operator as well as their respective children are also expressions, they will be resolved automatically by the existing rules in most cases. We have to ensure that all common queries also work when the skyline operator is used. To achieve this, we have to extend the existing rules from the \texttt{Analyzer} to also incorporate the skyline operator. Such rules mainly pertain to the propagation of aggregates across the (unresolved) logical execution plan of Spark.

First, we ensure that we are also able to compute skylines which include dimensions not present in the final projection (i.e., the attributes specified in the \texttt{SELECT} clause). To achieve this, we expand the function \texttt{ResolveMissingReferences} by adding another case for the \texttt{SkylineOperator}. The code is shown in Listing~\ref{alg:skyline_analyzer_resolve_missing_references}.
The rule is applied if and only if the \texttt{SkylineOperator} has not yet been resolved, has missing input attributes, and the child is already fully resolved (line~1~-~3). Then, we resolve the expressions and add the missing attributes (line~4~-~5) which are then used to generate a new set of skyline dimensions (line~6~-~7). If no output was added, we simply replace the skyline dimensions (line~8~-~9). Otherwise, we create a new skyline with the newly generated child (line~10~-~11) and add a projection to eliminate redundant attributes (line~12).

\begin{listingAlgorithm}
\begin{minted}[fontsize=\small,linenos]{scala}
case s @ SkylineOperator(_, _, skylineItems, child)
  if (!s.resolved || s.missingInput.nonEmpty)
     && child.resolved =>
    val (exprs, newChild) =
      resolveExprsAndAddMissingAttrs(skylineItems, child)
    val dimensions =
      exprs.map(_.asInstanceOf[SkylineDimension])
    if (child.output == newChild.output) {
      s.copy(skylineItems = dimensions)
    } else {
      val newSkyline = s.copy(dimensions, newChild)
      Project(child.output, newSkyline)
    }
\end{minted}
  \caption{Analyzer extension to allow dimensions not present in the projection in the skyline operator}
  \label{alg:skyline_analyzer_resolve_missing_references}
\end{listingAlgorithm}

Next, we need to take care that aggregate attributes are also propagated to the skyline properly. The code to accomplish this is given in Listing~\ref{alg:skyline_analyzer_resolve_aggregate_functions}. We will only explain it briefly since it was modified from existing Spark code for similar nodes.

To do this, we match the skyline operator which is a parent or ancestor node of an \texttt{Aggregate} (line~3). Then we try to resolve the skyline dimensions using the output of the aggregate as a basis (line~5~-~6). This will also introduce missing aggregates in the \texttt{Aggregate node} which is necessary, for example, if the output of the query only contains the \texttt{sum} but the skyline is based on the \texttt{count} (line~7~-~13). Using the output of the resolution, we (re-)construct the skyline operator (line~14~-~17). Additionally, we introduce analogous rules for other similar cases like plans where there is a \texttt{Filter} node introduced by a \texttt{HAVING} clause between the \texttt{Aggregate} and the skyline operator node among others.

\begin{listingAlgorithm}
\begin{minted}[fontsize=\small,linenos]{scala}
case SkylineOperator(
    distinct, complete, skylineItems,
    agg: Aggregate
) if agg.resolved =>
  val maybeResolved = skylineItems.map(_.child)
    .map(resolveExpressionByPlanOutput(_, agg))
  resolveOperatorWithAggregate(
    maybeResolved, agg,
    (newExprs, newChild) => {
      val newSkylineItems = skylineItems.zip(newExprs)
          .map {
        case (skylineItems, expr) =>
          skylineItems.copy(child = expr)
      }
      SkylineOperator(
        distinct, complete, newSkylineItems, newChild
      )
    }
  )
\end{minted}
  \caption{Analyzer extension to propagate aggregate attributes to skylines}
  \label{alg:skyline_analyzer_resolve_aggregate_functions}
\end{listingAlgorithm}

Note that the correct handling of \texttt{Filter} and \texttt{Aggregate} nodes is non-trivial in general -- not only in combination with the skyline operator.  
Indeed, when working on the integration of skyline queries into Spark SQL, we noticed that aggregates are in some cases not resolved correctly by the default rules of Spark.
This is, for example, the case when \texttt{Sort} nodes are resolved in combination with \texttt{Filter}s and \texttt{Aggregate}s.
Such cases may be introduced by a \texttt{HAVING} clause in the query. For a more detailed description of this erroneous behavior of Spark SQL together with a proposal how to fix this bug, see
\ifArxivVersion
Appendix~\ref{appendix:aggregate_resolution_error}.
\else
\cite{unsereLangversion}.
\fi

\subsection{Additional Optimizations}

Given a resolved logical plan, the \textit{Catalyst Optimizer} aims at 
improving the logical query plan by applying rule-based optimizations (see Figure~\ref{fig:spark_sql_query_processing_execution}). 
The default optimizations of Spark also apply to skyline queries -- especially to the ones where the data itself is retrieved by a more complex query. Skyline queries integrated into Apache Spark thus benefit from existing optimizations, which can be further improved by adding the following specialized rules.

If a skyline query contains a single \texttt{MIN} or \texttt{MAX} dimension, then this is equivalent to selecting the tuples with the lowest or highest values for the skyline dimension respectively.
In other words, the Pareto optimum in a single dimension is simply the optimum.
There are two possibilities to rewrite the skyline. We can either first sort the values according to the skyline dimension and then only select the top results or find the lowest or highest value in a scalar subquery (i.e., a subquery that yields a single value) and then select the tuples which have the desired value. Given that, for a relation with $n$ tuples, sorting and selecting exhibits a worst-case runtime of $\mathcal{O}(n \cdot \log (n))$ while the scalar subquery and selection can, when optimized, be done in $\mathcal{O}(n)$ time, we opt for the latter.

A more sophisticated optimization is taken from~\cite{TheSkylineOperator}, where~\cite{SayingEnoughAlready} is referenced for the correctness of this transformation. It is applicable if the skyline appears in combination with a \texttt{Join} node in the tree. If the output of the join serves as the input to the skyline operator, then we can check whether it is possible to move the skyline into one of the ``sides'' of the join. This is applicable if the join is \textit{non-reductive}~\cite{TheSkylineOperator} which is defined in~\cite{SayingEnoughAlready}. Intuitively, non-reductiveness in our case means that due to the constraints in the database, it can be inferred that for every tuple in the first table joined there must exist at least one ``partner'' in the other join table if the tuple is part of the skyline. 
The main benefit of this optimization is that computing the skyline before the join is likely to reduce the input size for both the skyline operator and the (now subsequent) join operation,
which typically reduces the total query execution time.

\subsection{Algorithm Selection}
\label{sect:AlgorithmSelection}

Given the problems that arise with skyline queries on incomplete datasets as mentioned in Section~\ref{section:skyline_queries}, we have to take extra care which algorithm is used for which dataset. The decision is mainly governed by whether we can assume the input dataset to be complete or not. If the dataset is (potentially) incomplete, we have to select an algorithm that is capable of handling incomplete data since otherwise, the computation may not yield the correct results.

As will be explained in Section \ref{section:skyline_incomplete_data}
(and it will also be observed in the empirical evaluation in 
Section \ref{section:performance_evaluation}), 
skyline algorithms that are able to handle incomplete datasets 
are lacking in performance compared to their complete counterparts. Given that Spark can handle multiple different data sources and cannot always detect the \texttt{nullability} of a column, it is desirable to provide an override that decides whether a complete algorithm is used regardless of the detected input columns. We do this by introducing the optional \texttt{COMPLETE} keyword in the skyline query syntax. This gives the user the possibility to enforce the use of the complete algorithms also on datasets that can technically be incomplete but are known to be complete. The correctness of the algorithm only depends on whether \texttt{null} values actually 
%%% do appear in the data or not. 
appear in the data.

We implement the algorithm selection using the nodes in the physical execution plan. Which algorithm is selected depends on which physical nodes are chosen during translation of the (resolved and optimized) logical execution plan. In a minimally viable implementation, it is possible to implement the skyline operator in a single physical node. This has, however, the disadvantage that 
Spark's potential of parallelism would be lost for the skyline computation. 
We therefore split the computation into two steps, represented by two separate nodes in the physical plan. The pseudocode of the algorithm selection is shown in 
Listing~\ref{lst:skyline_node_selection}.

\begin{listingAlgorithm}
\SetKwInOut{Input}{input}
\SetKwInOut{Output}{output}

\Input{resolved logical plan with skyline operator node \texttt{sky} as root}
\Output{optimized (resolved) logical plan}

$\mathit{skylineNullable} \leftarrow \exists d \in D_{\mathit{SKY}} : isnullable(d)$

\uIf{\texttt{COMPLETE} is set OR $\neg \mathit{skylineNullable}$}{
    \texttt{local\_skyline} $\leftarrow$ \texttt{local\_node()} \\
    \texttt{global\_skyline} $\leftarrow$ \texttt{complete\_global\_node()}
}
\uElse{
    \texttt{local\_skyline} $\leftarrow$ \texttt{local\_node()} \\
    \texttt{global\_skyline} $\leftarrow$ \texttt{incomplete\_global\_node()}
}

\Return \texttt{global\_skyline}
  \caption{Selection of appropriate skyline nodes in physical execution plan}
  \label{lst:skyline_node_selection}
\end{listingAlgorithm}

The main decision to be made (line~1,~2) is if we may use the 
complete algorithm. 
This is the case when the \texttt{SKYLINE} clause contains the \texttt{COMPLETE} keyword or when the skyline dimensions are recognized as not \texttt{nullable} by the system. 
Note that the local skylines use the same basic nodes both in the complete and incomplete cases, while the nodes for the global skyline differ.

The first node in the physical plan corresponds to the \textit{local skyline} computation, which can be done in a distributed manner (line~3 and 6). This means that there may exist multiple instances of the node which can exist distributed across the cluster on which Spark runs. For the actual distribution of the data, we can use multiple different schemes provided by Spark or leave it as-is from the child in the physical plan. When left at standard distribution (\texttt{UnspecifiedDistribution}), Spark will usually try to distribute the data equally given the number of available executors. For example, if there are 10 executors available for 10.000.000 tuples in the original dataset, then each executor will receive roughly 1 million tuples each. For the algorithms that can handle incomplete datasets, 
we have to use a specialized distribution scheme, which will be discussed in more detail in Section~\ref{section:skyline_incomplete_data}.

The output of the local skyline computation 
is used as input for calculating the global skyline (line~4 and 7). This means that, in the physical execution plan, we create a global skyline node whose only child is the local skyline.
In contrast to the local skyline, we may not freely choose the distribution but must instead ensure that all tuples from the local skyline are handled by the same executor for the global skyline computation. This is enforced by choosing the \texttt{AllTuples} distribution provided by Spark.

To keep our solution modular, we encapsulate the \textit{dominance check} in a new utility that we introduce in Spark. It takes as input the values and types of the skyline dimensions of two tuples and checks if one tuple dominates the other. 
For every skyline dimension, we match the data type to avoid costly casting and (in the case of casting to floating types) potential loss of accuracy.

With our approach to algorithm selection in the physical plan and the modular structure of the code, it is easy to incorporate further skyline algorithms if so desired in the future. Choosing a different algorithm mainly entails replacing the local and/or global skyline computation nodes in the physical plan by new ones.

\subsection{Skylines in Complete Datasets}\label{section:skyline_complete_data}

For complete datasets, we adapt the Block-Nested-Loop skyline algorithm~\cite{TheSkylineOperator}.
The main idea is to keep a window of tuples in which the skyline of all tuples processed up to this point is stored. 
We iterate through the entire dataset, and, for each tuple, 
we check which dominance relationships exist with the tuples in the current window. If tuple $t$ is dominated by a tuple in the window, then $t$ is eliminated. Here, it is not necessary to check $t$ against the remaining tuples since it cannot dominate any tuples in the window due to transitivity. 
If tuple $t$ dominates one or more tuples in the window, then the dominated tuples are eliminated. In this case, $t$ is inserted into the window since, by transitivity, $t$ cannot be dominated by other tuples in the window. Tuple $t$ is also inserted into the window if it is found incomparable with all tuples in the current window. 

We can use the same algorithm for both the local and the global skyline computation. The only necessary difference is the distribution of the data, where we let Spark handle the distribution for the local skyline while we force the \texttt{AllTuples} distribution for the global skyline computation. 
This has the additional advantage that partitioning done in prior processing steps can be kept, which increases the locality of the data for the local skyline computation and 
may thus help to further improve the overall performance.

The Block-Nested-Loop approach is most efficient if the window fits into main memory. 
Note that also Spark, in general, performs best if large portions of the data fit into the main memory available. Especially in cloud-based platforms, there is typically sufficient RAM available (in the order of magnitude of terabytes or even petabytes).
At any rate, if RAM does not suffice, Spark will swap data to disk like any other program -- with the usual expected performance loss.
%This makes the Block-Nested-Loop approach a natural first choice for skyline computation in Spark.

\subsection{Skylines in Incomplete Datasets}\label{section:skyline_incomplete_data}

When computing the local skyline for an incomplete dataset, we must take care that no potential cyclic dominance relationships are lost such that tuples may appear in the result even though they are dominated by another tuple. To combat this, we use a special form of a \textit{bitmap}-based skyline algorithm~\cite{OptimizingSkylineQueryProcessingIncompleteData}.

Given a set $P$ of tuples, we can assign each tuple $p \in P$ a bitmap $b$ (an index in binary format) such that each bit in $b$ corresponds to a skyline dimension. If a tuple $p$ has a \texttt{null} value in a skyline dimension, then the corresponding bit in bitmap $b$ is set to $1$; otherwise, it is set to $0$. 
Then, subsequently, the data is partitioned according to the bitmap indexes such that all tuples with a specific bitmap $b$ are assigned to the same subset (partition) $P_b$ of $P$. We can then calculate the local skyline $SKY(P_b)$ for each set of tuples $P_b$ without losing the transitivity property or running into issues with cyclic dominance relationships.

In Spark, this sort of partitioning is done via the integrated distribution of the nodes. We craft an expression for the distribution which uses the predefined \texttt{IsNull()} method to achieve this effect.

For the global skyline computation, we cannot use the standard Block-Nested-Loop approach since cyclic dominance relationships may occur and the transitivity of the skyline operator is not guaranteed. We therefore opt for the less efficient approach where we compare all tuples against each other. Even if a tuple $t$ is dominated, we may not immediately delete it since it may be the only tuple that dominates another tuple $t'$. 
In such a case, by prematurely deleting $t$, tuple $t'$ would be erroneously added to the skyline. This is a subtle point which was, for instance, overlooked in the algorithm proposed in~\cite{OptimizingSkylineQueryProcessingIncompleteData}. 
In
\ifArxivVersion
Appendix~\ref{appendix:algorithm_incomplete_datasets}.
\else
\cite{unsereLangversion}, 
\fi
we illustrate the error resulting from premature deletion
of dominated tuples in detail. 

The following lemma guarantees that our skyline computation in case of a potentially incomplete dataset yields the correct result: 

\begin{lemma}
Let a dataset $P$ be partitioned according to the \texttt{null} values and let $S_{local}$ denote the resulting union of local skylines. Then it holds for every tuple $p \in P$ not part of the global skyline $S_{global}$ that either $p \not\in S_{local}$ or there exists $q \in S_{local}$ with $q \prec p$.
\end{lemma}

\begin{proof}
Let $p \in P$ be a tuple that is not part of the global skyline $S_{global}$, 
i.e., there exists some tuple $q \in P$ with $q \prec p$.
If both $p$ and $q$ belong to the same partition during local skyline computation, then this dominance relation will be detected in this step and $p$ will be deleted. Hence, in this case, $p \not\in S_{local}$.

Now suppose that $p$ and $q$ belong to different partitions. If $q \in S_{local}$, then $q$ is the desired tuple in $S_{local}$ with $q \prec p$.

It remains to consider the case that $q \not\in S_{local}$. This means that $q$ gets deleted during the local skyline computation. In other words, there exists a tuple $r$ in the same partition as $q$ with $r \prec q$ such that $r \in S_{local}$. The latter property is guaranteed by the fact that, inside each partition, all tuples have \texttt{null}s at the same positions and there can be no cyclic dominance relationships. 

By the assumption $q \prec p$ and the definition of skylines in incomplete datasets, $q$ is at least as good as $p$ in all non-missing skyline dimensions and strictly better in at least one. 
Since $r \prec q$ and both tuples have the same set of missing dimensions due to being in the same partition, it follows that also $r$ is at least as good as $p$ in all non-missing skyline dimensions and strictly better in at least one. That is, $r$ is the desired tuple in $S_{local}$ with $r \prec p$.
\end{proof}

Selecting an algorithm which can handle incomplete datasets yields the correct result also for a complete dataset (while the converse is, of course, not true). There is, however, a major drawback to always using an incomplete algorithm. Since the partitioning in the incomplete algorithm is done according to which values of a tuple are \texttt{null},
only limited partitioning is possible. 
The worst case occurs when the algorithm for incomplete datasets is applied to a complete dataset. In this case, since there are no \texttt{null} values, there is only a single partition which all tuples belong to. This means that any potential of parallelism gets lost. Hence, in the case of a complete dataset, the algorithm for incomplete data performs even worse than a strictly non-distributed approach which immediately proceeds to the global skyline
computation by a single executor. 

\subsection{Extending the User Interfaces}
\label{sect:extending-userinterfaces}

In addition to our extension of the SQL string-interface of Spark SQL, we integrate the skyline queries also into the basic DataFrame API provided in Java and Scala by adding new API functions.
Here, the data about the skyline dimensions can either be passed via pairs of skyline dimension and associated dimension type or by using the \textit{columnar} definition of Spark directly. 
For the latter purpose, we define the functions \texttt{smin()}, \texttt{smax()}, and \texttt{sdiff()}, which each take a single argument that provides the 
skyline dimension in Spark columnar format as input.
The API calls bypass the parsing step and 
directly create a new skyline operator node in the logical plan.

We also integrate skyline queries into Python and R via PySpark and SparkR, respectively -- two of the most popular languages for analytical and statistical applications. Rather than re-implementing skyline queries
in these languages, we build an intermediate layer that calls 
the Scala-implementation of the DataFrame API.

In Python, this method relies on the external, open-source library Py4J (\url{https://www.py4j.org/}). Its usage requires the skyline dimensions and skyline types to first be translated into Java objects, which can then be passed on using Py4J.
A slight complication arises from the fact that Python is a weakly typed language, which imposes restrictions on the method signatures.
Skyline dimensions and their types are therefore passed as separate lists of strings. The first dimension is then matched to the first type of skyline dimension and so forth. Passing the skyline dimensions via \texttt{Column}s (i.e., \texttt{Expression}s) works similarly to the Scala/Java APIs.

In R, the integration is simpler since R allows for tuples of data to be entered more easily. Hence, in addition to the column-based input analogous to the Python interface, the R interface also accepts a list of pairs of skyline dimension plus type. 

\subsection{Ensuring Correctness}
\label{sect:correctness}

As far as the impact of the skyline handling on the overall behavior of Spark SQL is concerned, we recall from Sections~\ref{sect:ExtendingLogicalPlan} and \ref{sect:ExtendingAnalyzer} that, in the logical plan, the skyline operator gives rise to a single node with a single input (from the child node) and a single output (to the parent node). 
When translating the logical node to a physical plan, this still holds since, even if there are multiple physical nodes in the plan, there is only a single input and output.
 
Therefore, the main concern of ensuring correctness of our integration of skyline queries into Spark SQL is the handling of skyline queries itself. 
There are no side effects whatsoever of the skyline integration on the rest of query processing in Spark SQL. This also applies to potential effects of the skyline integration on the performance of Spark SQL commands not using the skyline feature at all.
In this case, the only difference in the query processing flow is an additional clause in the parser. The additional cost is negligible.
 
We have intensively tested the skyline handling to provide evidence for its correctness.
Unit tests are provided as part of our implementation in the GitHub repository.
Additionally, for a significant portion of the experiments reported in Section~\ref{section:performance_evaluation}, we have verified that our integrated skyline computation yields the same result as the equivalent ``plain'' SQL query 
in the style of Listing~\ref{lst:skyline_query_plain_sql}.

\section{Empirical Evaluation}\label{section:performance_evaluation}

In this section, we take a closer look at how our integration performs based on a series of benchmarks executed on a cluster. We will first provide information about the experimental setup as well as the input data and the queries which are executed on them. Following that, we will provide an overview of the outcomes of the benchmark. To round off the section, we will give a brief summary of the performance measurements and draw some conclusions.

\subsection{Experimental Setup}
\label{sect:experimental-setup}

We have implemented the integration of skyline queries into Apache Spark,
and provide our implementation as open-source code at
\url{https://github.com/Lukas-Grasmann/skyline-queries-spark}.
Our implementation uses the same languages as Apache Spark, which include Java, Scala, Python, and R. The bulk of the main functionality was written in Scala, which is also the language in which the core functionalities of Spark are implemented.
The experiments can be easily reproduced by using the provided open-source software. 
The binaries, benchmark data, queries, and additional scripts can be found at~\cite{Zen:data/Skyline}.

All tests were run on a cluster consisting of 2 namenodes and 18 datanodes. 
The latter are used to execute the actual program. Every node consists of 2 Xeon E5-2650 v4 CPUs by Intel that provide 24 cores each, which equals 48 cores per node and up to 864 cores across all worker nodes in total. Each node provides up to 256GB of RAM and 4 hard disks with a total capacity of 4TB each.

The resource management of the cluster is entirely based on Cloudera and uses YARN to deploy the applications. It provides the possibility to access data stored in Hive, which we found convenient for storing and maintaining the test data in our experiments. And it 
is also a common way how Apache Spark is used in practice.

For fine-tuning the parameters, we use the command line arguments provided by Spark to deploy the applications to YARN. In these tests, we tell Spark how many executors are to be spawned. The actual resource assignment is then left to Spark to provide conditions as close to the real-world setting as possible.

\subsection{Test Data and Queries}
\label{sect:testdata-and-queries}

For our tests with a {\bf real-world dataset}, 
we use the freely available subset of the Inside Airbnb dataset as input, which contains accommodations from Airbnb over a certain timespan. 
% For the sake of brevity, we will simply refer to this dataset as the ``Airbnb'' dataset.
The time span chosen is 30 days and contains approximately 1 million tuples. The tuples were downloaded from the Inside Airbnb website~\cite{InsideAirbnb} and then subsequently merged while eliminating duplicates and string-based columns. For the complete variant of the dataset, we have eliminated all tuples containing a \texttt{null} 
value in at least one skyline dimension. Content-wise, this dataset is similar to the hotel example provided in Section~\ref{section:introduction}. The use cases considered here are also similar and encompass finding the ``best'' listings according to the chosen dimensions. The relevant key (identifying) dimension and the 6 skyline dimensions can be found in Table~\ref{tab:skyline_dimensions_airbnb_dataset}. From this relation, we constructed skyline queries with 1 dimension, 2 dimensions, \dots, 6 dimensions by selecting the dimensions 
in the same order as they appear in Table~\ref{tab:skyline_dimensions_airbnb_dataset}, 
i.e., the one-dimensional skyline query only uses the first skyline dimension (price) in the table,
the two-dimensional query uses the first two dimensions (price, accommodates), etc.

\begin{table}[htp]
% \begin{table}[b]
\centering
\begin{tabular}{|l|l|p{3.75cm}|}
	\hline
	\textbf{Dimension} & \textbf{Type} & \textbf{Description} \\
	\hline
	id & \texttt{KEY} & identification number \\
	\arrayrulecolor{lightgray}\hline
	price & \texttt{MIN} & price for renting \\
	\arrayrulecolor{lightgray}\hline
	accommodates & \texttt{MAX} & (max) number of accommodated people \\
	\arrayrulecolor{lightgray}\hline
	bedrooms & \texttt{MAX} & number of bedrooms \\
	\arrayrulecolor{lightgray}\hline
	beds & \texttt{MAX} & number of beds \\
	\arrayrulecolor{lightgray}\hline
	number\_of\_reviews & \texttt{MAX} & number of reviews \\
	\arrayrulecolor{lightgray}\hline
	review\_scores\_rating & \texttt{MAX} & total review score ratings (all categories) \\
	\arrayrulecolor{black}\hline
\end{tabular}
\caption{Skyline dimensions in the Inside Airbnb dataset}
\label{tab:skyline_dimensions_airbnb_dataset}
\end{table}

We also carried out tests on a {\bf synthetic dataset}, namely the \texttt{store\_sales} table from the benchmark DSB~\cite{DSB}. 
The skyline dimensions used in the benchmarks can be found in Table~\ref{tab:skyline_dimensions_store_sales_dataset}. 
The table has 2 identifying key dimensions and, again, 6 skyline dimensions. 
The selection of skyline dimensions to derive 6 skyline queries (with 
the number of skyline dimensions ranging from 1 to 6) is done exactly as described above for the Airbnb test case.

As far as the test data is concerned, we randomly generated data such that the total size of the table is approximately 15,000,000 tuples (note that data generation in DSB works by indicating the data volume in bytes not tuples; in our case, we generated 1.5 GB of data). 
For the tests with varying data size, we simply select
the first tuples from the table until the desired size of the dataset is reached. 
For the complete dataset, we only select the tuples which are not \texttt{null} in all six potential skyline dimensions. 

\begin{table}[H]
% \begin{table}[t]
\centering
\begin{tabular}{|l|l|p{3.75cm}|}
	\hline
	\textbf{Dimension} & \textbf{Type} & \textbf{Description} \\
	\hline
	ss\_item\_sk & \texttt{KEY} & stock item identifier \\
	\arrayrulecolor{lightgray}\hline
	ss\_ticket\_number & \texttt{KEY} & ticket number identifier \\
	\arrayrulecolor{lightgray}\hline
	ss\_quantity & \texttt{MAX} & quantity purchased in sale \\
	\arrayrulecolor{lightgray}\hline
	ss\_wholesale\_cost & \texttt{MIN} & wholesale cost \\
	\arrayrulecolor{lightgray}\hline
	ss\_list\_price & \texttt{MIN} & list price \\
	\arrayrulecolor{lightgray}\hline
	ss\_sales\_price & \texttt{MIN} & sales price \\
	\arrayrulecolor{lightgray}\hline
	ss\_ext\_discount\_amt & \texttt{MAX} & total discount given \\
	\arrayrulecolor{lightgray}\hline
	ss\_ext\_sales\_price & \texttt{MIN} & sum of sales price \\
	\arrayrulecolor{black}\hline
\end{tabular}
\caption{Skyline dimensions in the store\_sales dataset}
\label{tab:skyline_dimensions_store_sales_dataset}
\end{table}

For both datasets, there exists a complete as well as an incomplete variant. The difference between them is that for the complete dataset all tuples that contain \texttt{null} values in at least one skyline dimension have been removed. For the real-world dataset, this means that the incomplete dataset is bigger than the complete variant while for the synthetic dataset, both variants have the same size.

\subsection{Tested Algorithms}
\label{sect:tests-algorithms}

In total, we run our tests on up to four skyline algorithms: 

\begin{enumerate}
    \item The distributed algorithm for complete datasets (described in Section \ref{section:skyline_complete_data}), which splits the skyline computation into local and global skyline computation; strictly speaking, only the local part is distributed (whence, parallelized).
    \item The non-distributed complete algorithm, which completely gives up on parallelism, skips the local skyline computation and immediately proceeds to the global skyline computation from the previous algorithm.
    \item The distributed algorithm for incomplete datasets described in Section \ref{section:skyline_incomplete_data}. Recall, however, that here, distribution is based on the occurrence of \texttt{null}s in skyline dimensions, which severely restricts the potential of parallelism. 
    \item As the principal reference algorithm for our skyline algorithms, we run our tests also with the rewriting of skyline queries into plain SQL as described in Listing~\ref{lst:skyline_query_plain_sql}.
\end{enumerate}

In our performance charts, we refer to these 4 algorithms as ``distributed complete'', ``non-distributed complete'', ``distributed incomplete'', and ``reference''.
In all tests with complete datasets, we compare all four algorithms against each other. 
For incomplete datasets, the complete algorithms are not applicable; we thus only compare the remaining two algorithms in these cases. 

\subsection{Experimental Results}
\label{sect:experimental-results}

In our experimental evaluation, we want to find out how the following parameters affect the execution time:

\begin{itemize}
    \item number of skyline dimensions;
    \item number of input tuples in the dataset;
    \item number of executors used by Spark.
\end{itemize}

For the number of skyline dimensions, we have crafted queries with 1~--~6 dimensions. The number of executors used by Spark is chosen from 1, 2, 3, 5, 10. For the size of the dataset, we distinguish between the real-world (Airbnb) and synthetic dataset (DSB): all tests with the real-world dataset are carried out with all tuples contained in the data, i.e., ca.~1,200,000 tuples if \texttt{null}s are allowed and ca.~820,000 tuples after removing all tuples with a \texttt{null} in one of the skyline dimensions.
In contrast, the synthetic dataset is randomly generated and we choose subsets of size $10^6$, $2 \cdot 10^6$, $5 \cdot 10^6$, and $10^7$ both, for the complete and the incomplete dataset.
%In particular, the elimination of tuples with \texttt{null}s can of course 
%be compensated by generating further tuples. 

\begin{figure*}[p]
    \begin{subfigure}{.5\linewidth}
      \centering
      \includegraphics[width=\linewidth]{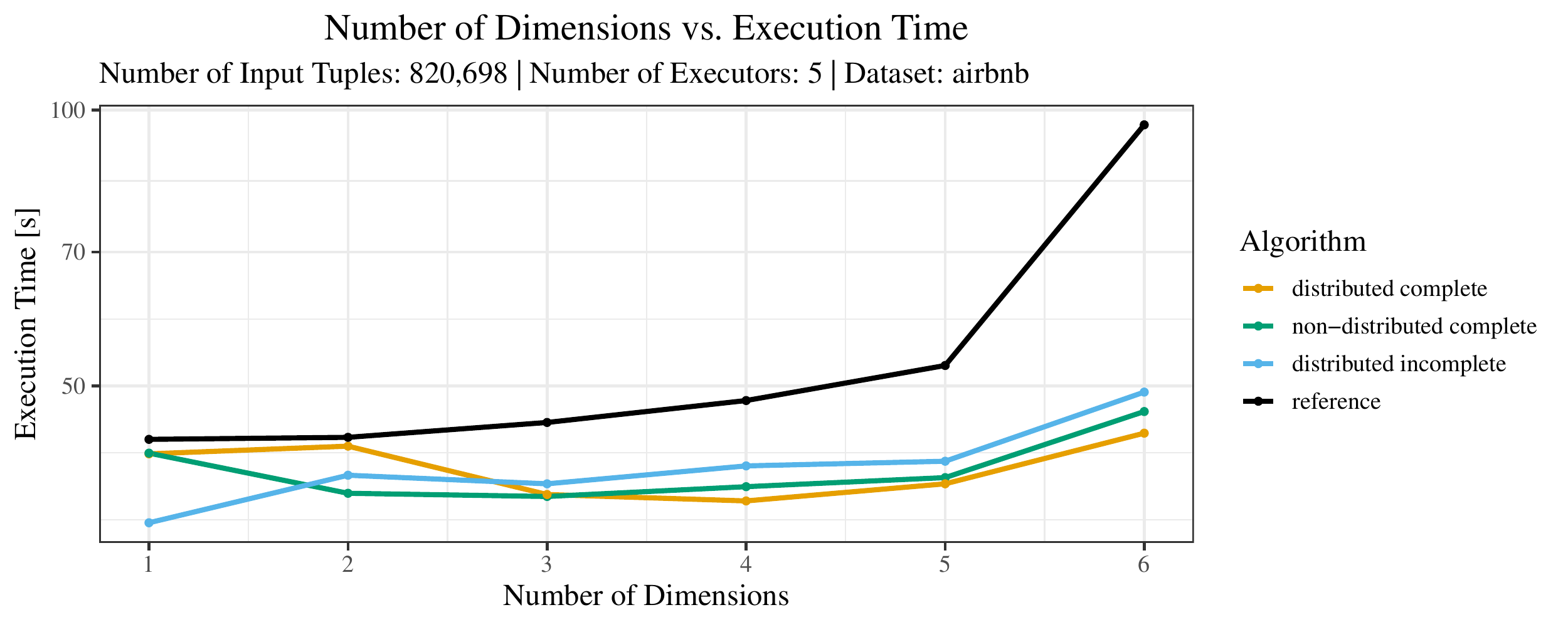}
    \end{subfigure}%
    \begin{subfigure}{.5\linewidth}
      \centering
      \includegraphics[width=\linewidth]{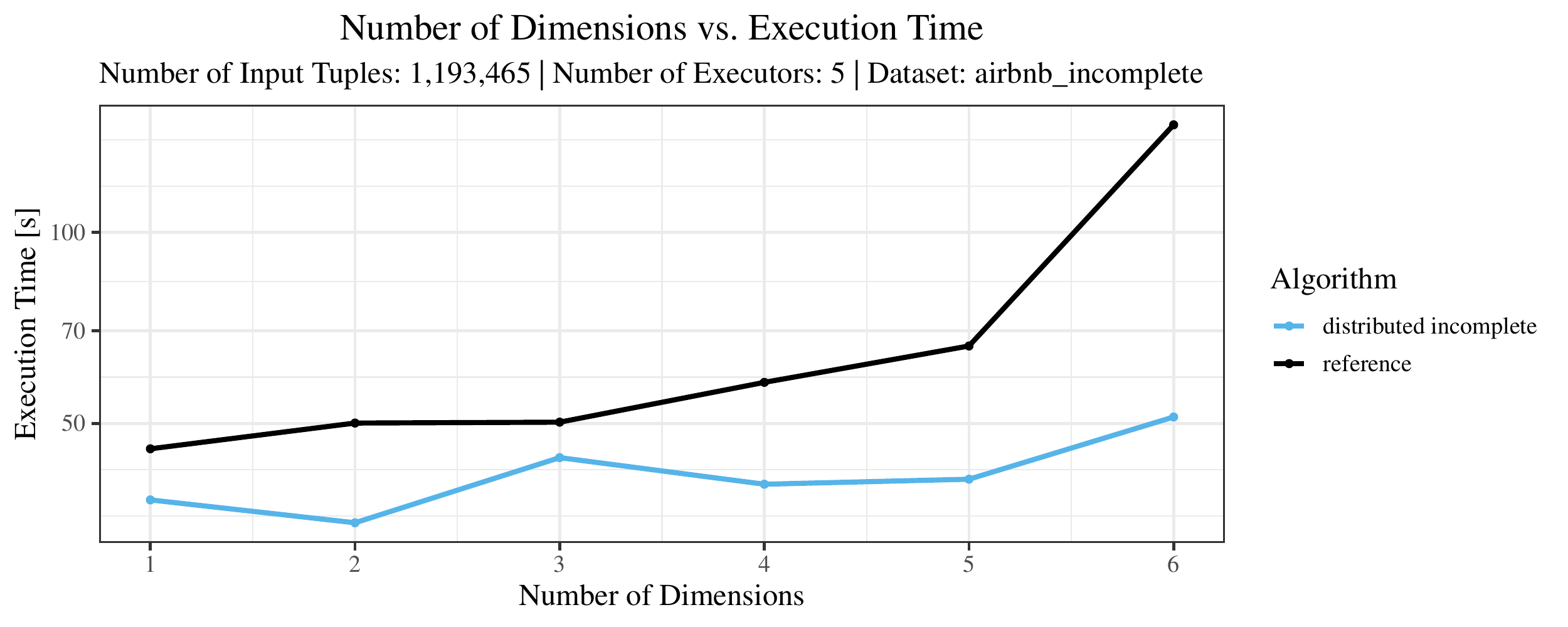}
    \end{subfigure}
    \caption{Number of dimensions vs. execution time on the Inside Airbnb dataset}
    \label{fig:cluster_dimensions_vs_time_complete_airbnb}
\end{figure*}

We have carried out tests with virtually all combinations of the above mentioned value ranges for the three parameters under investigation, resulting in a big collection of test cases: 
complete vs.\ incomplete data, 1 -- 6 skyline dimensions, ``all'' tuples (for the Airbnb data) vs. varying between $10^6$, $2 \cdot 10^6$, $5 \cdot 10^6$, and $10^7$ tuples (for the DSB data), and between 1 and 10 executors. 
In all these cases, we ran 4 algorithms (in the case of complete data) or 2 algorithms (for incomplete data) as explained in Section~\ref{sect:tests-algorithms}.
Below we report on a representative subset of our experiments. The runtime measurements are shown in Figures~\ref{fig:cluster_dimensions_vs_time_complete_airbnb} -- \ref{fig:cluster_nodes_vs_time_store_sales}. Further performance charts are provided 
\ifArxivVersion
in Appendix~\ref{appendix:additional_benchmarks}.
\else
in~\cite{unsereLangversion}. 
\fi

In our experiments, we have defined a timeout of $3600$ seconds. 
Timeouts are ``visualized'' in our performance charts by missing data points. 
Actually, timeouts did occur 
%%% in our experiments 
quite frequently -- especially with incomplete data and here, in particular, with the ``reference'' algorithm. 
Therefore, in Figures~\ref{fig:cluster_dimensions_vs_time_complete_airbnb} -- \ref{fig:cluster_nodes_vs_time_store_sales}, for the case of incomplete data (always the plot on the right-hand side), we only show the results for datasets smaller than \ $10^7$ tuples. In contrast, for complete data (shown in the plots on the left-hand side) we typically scale up to the max. size of $10^7$ tuples.

\begin{figure*}[p]
    \begin{subfigure}{.5\linewidth}
      \centering
      \includegraphics[width=\linewidth]{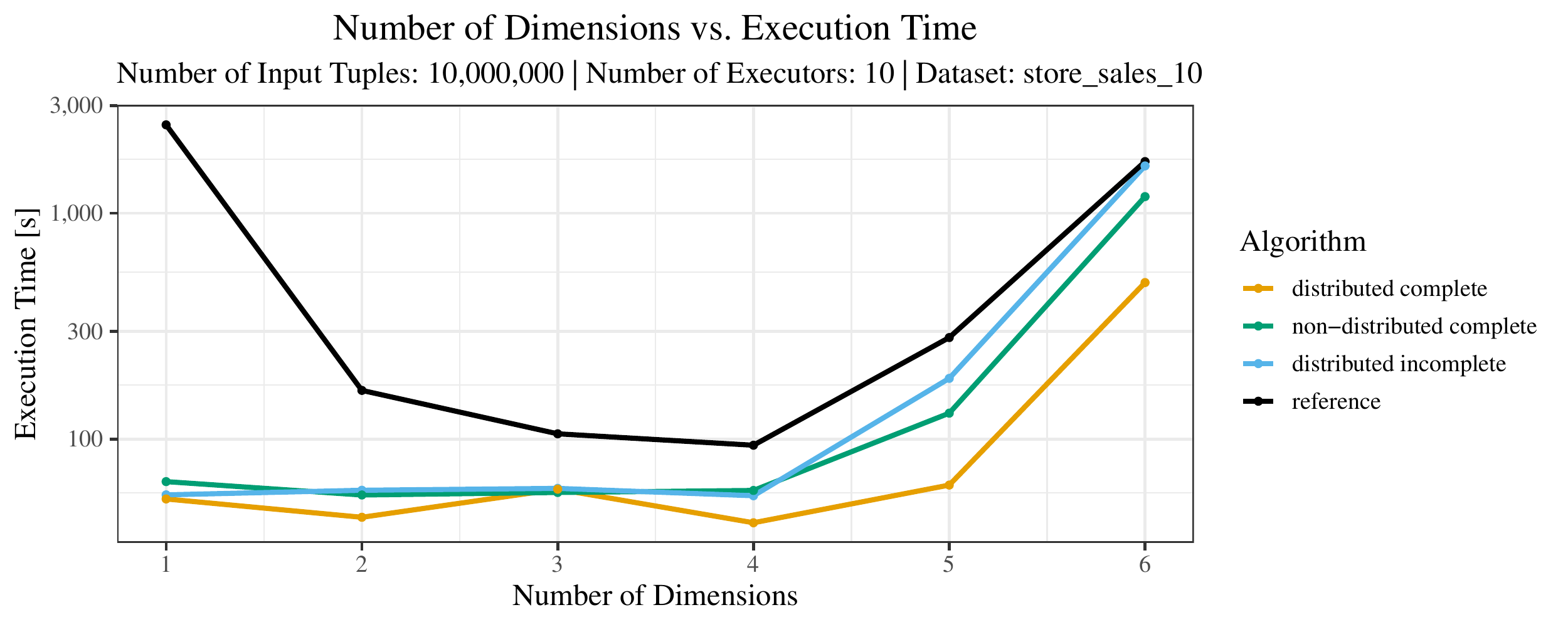}
    \end{subfigure}%
    \begin{subfigure}{.5\linewidth}
      \centering
      \includegraphics[width=\linewidth]{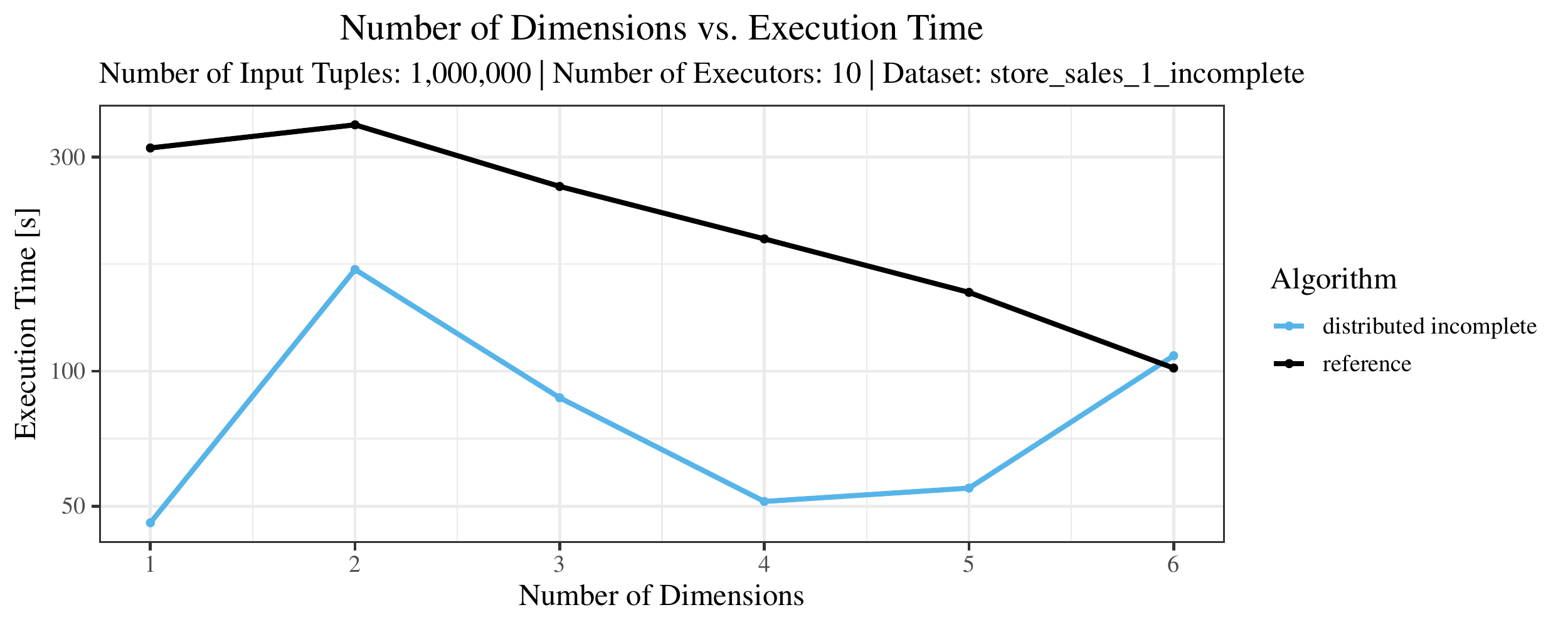}
    \end{subfigure}
    \caption{Number of dimensions vs. execution time on the store\_sales dataset}
    \label{fig:cluster_dimensions_vs_time_complete_store_sales}
\end{figure*}

\begin{figure*}[p]
    \begin{subfigure}{.5\linewidth}
      \centering
      \includegraphics[width=\linewidth]{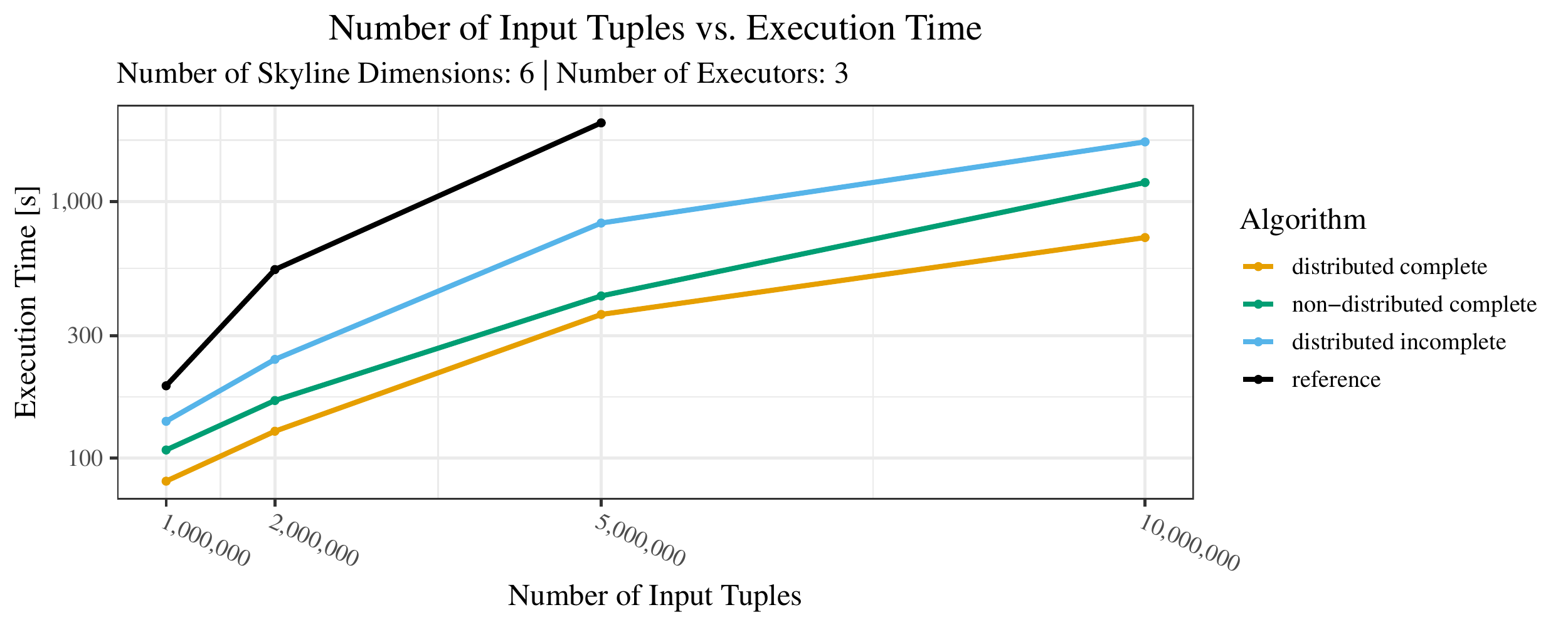}
    \end{subfigure}%
    \begin{subfigure}{.5\linewidth}
      \centering
      \includegraphics[width=\linewidth]{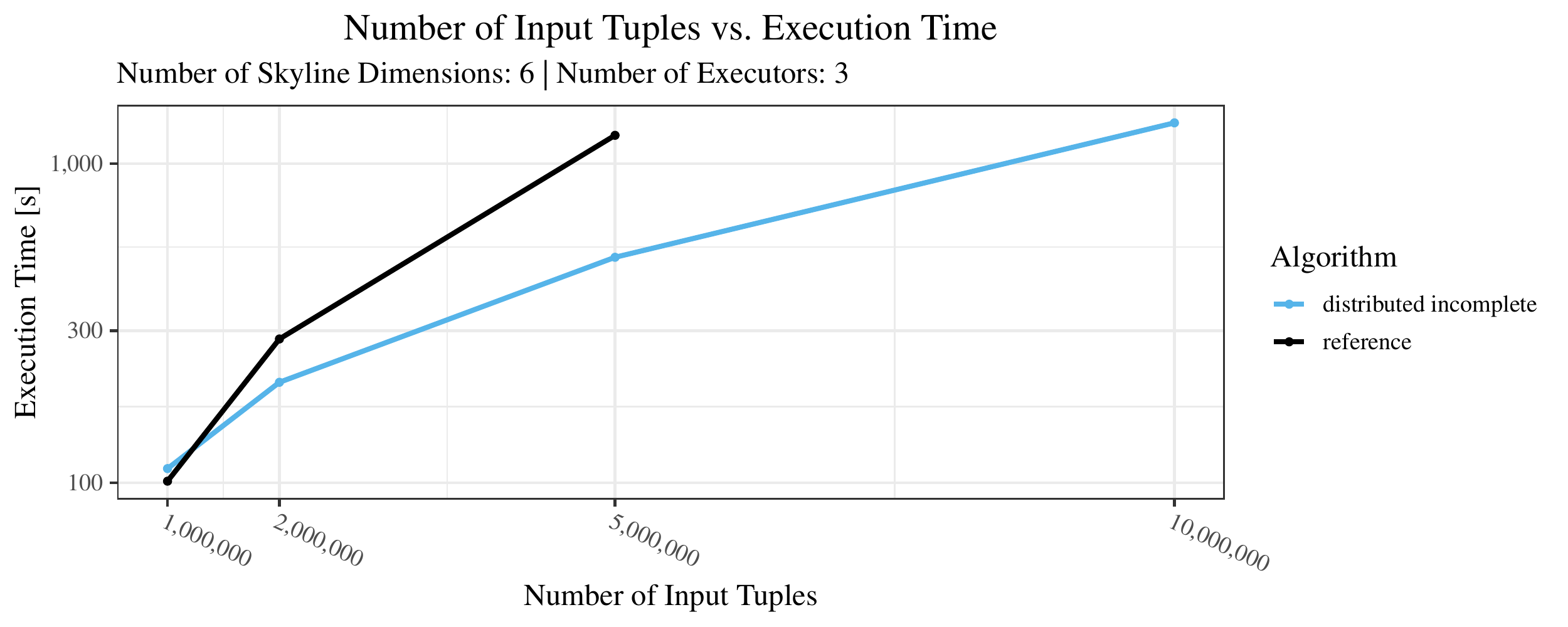}
    \end{subfigure}
    \caption{Number of input tuples vs. execution time on the store\_sales dataset}
    \label{fig:cluster_size_vs_time}
\end{figure*}

\begin{figure*}[p]
    \begin{subfigure}{.5\linewidth}
      \centering
      \includegraphics[width=\linewidth]{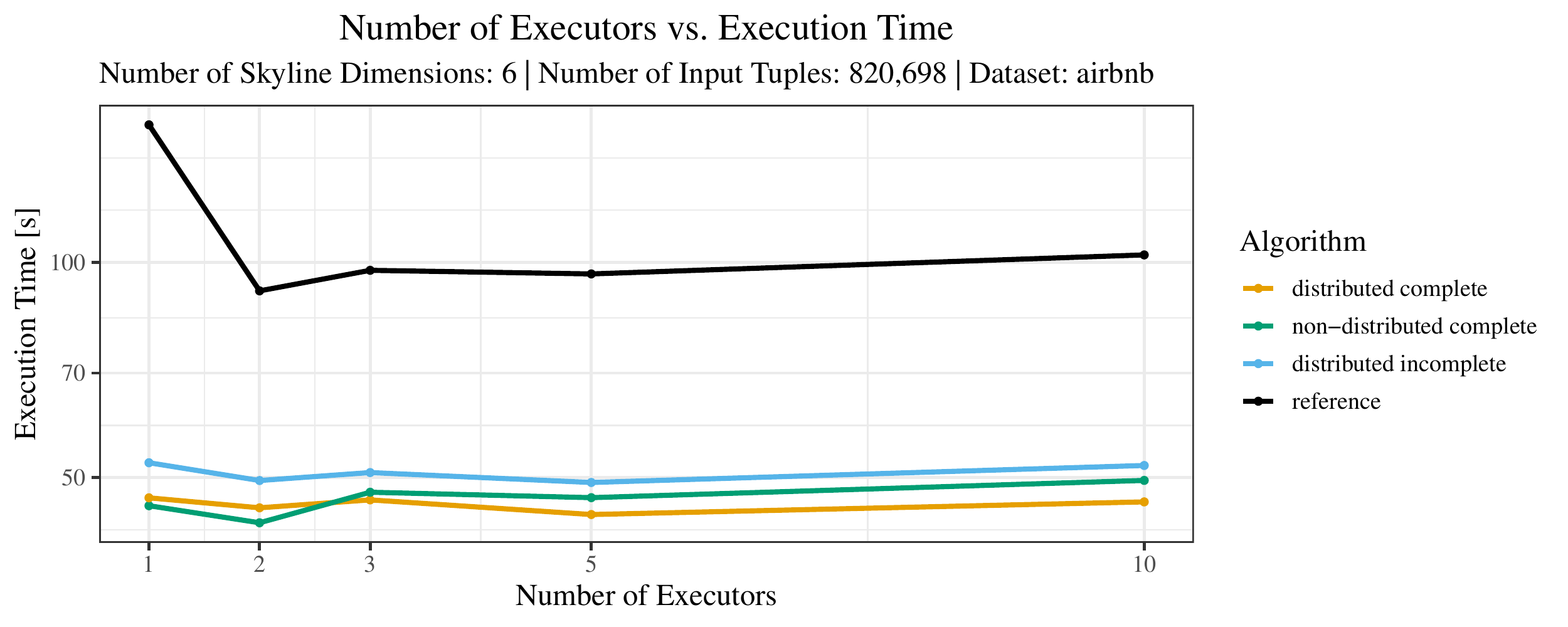}
    \end{subfigure}%
    \begin{subfigure}{.5\linewidth}
      \centering
      \includegraphics[width=\linewidth]{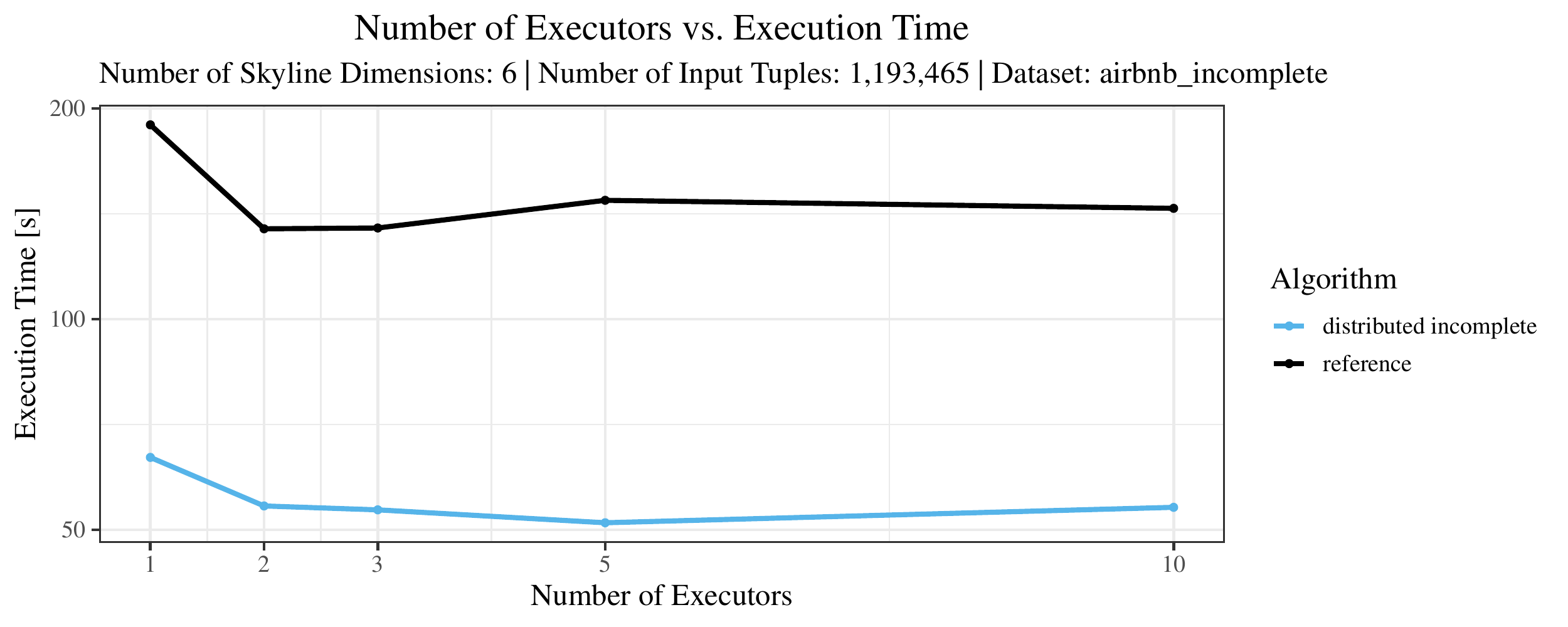}
    \end{subfigure}
    \caption{Number of executors vs. execution time on the Inside Airbnb dataset}
    \label{fig:cluster_nodes_vs_time_airbnb}
\end{figure*}

\begin{figure*}[p]
    \begin{subfigure}{.5\linewidth}
      \centering
      \includegraphics[width=\linewidth]{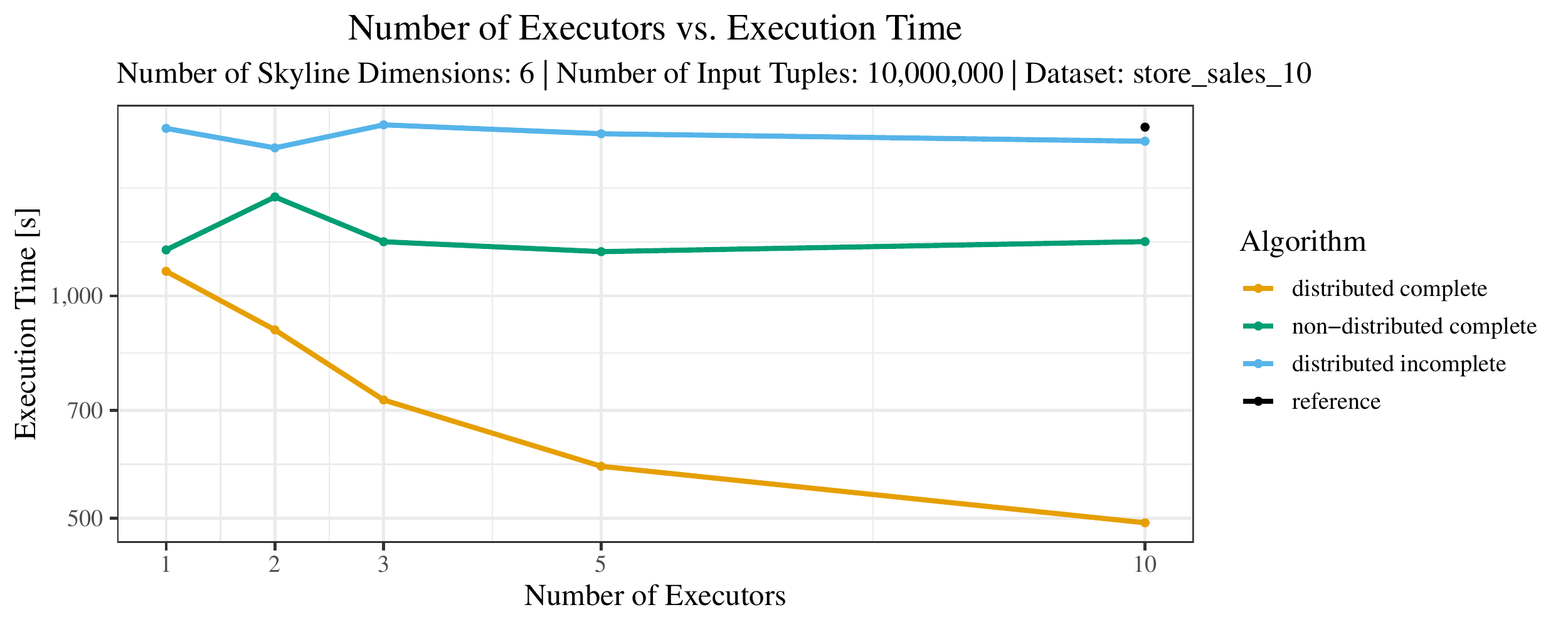}
    \end{subfigure}%
    \begin{subfigure}{.5\linewidth}
      \centering
      \includegraphics[width=\linewidth]{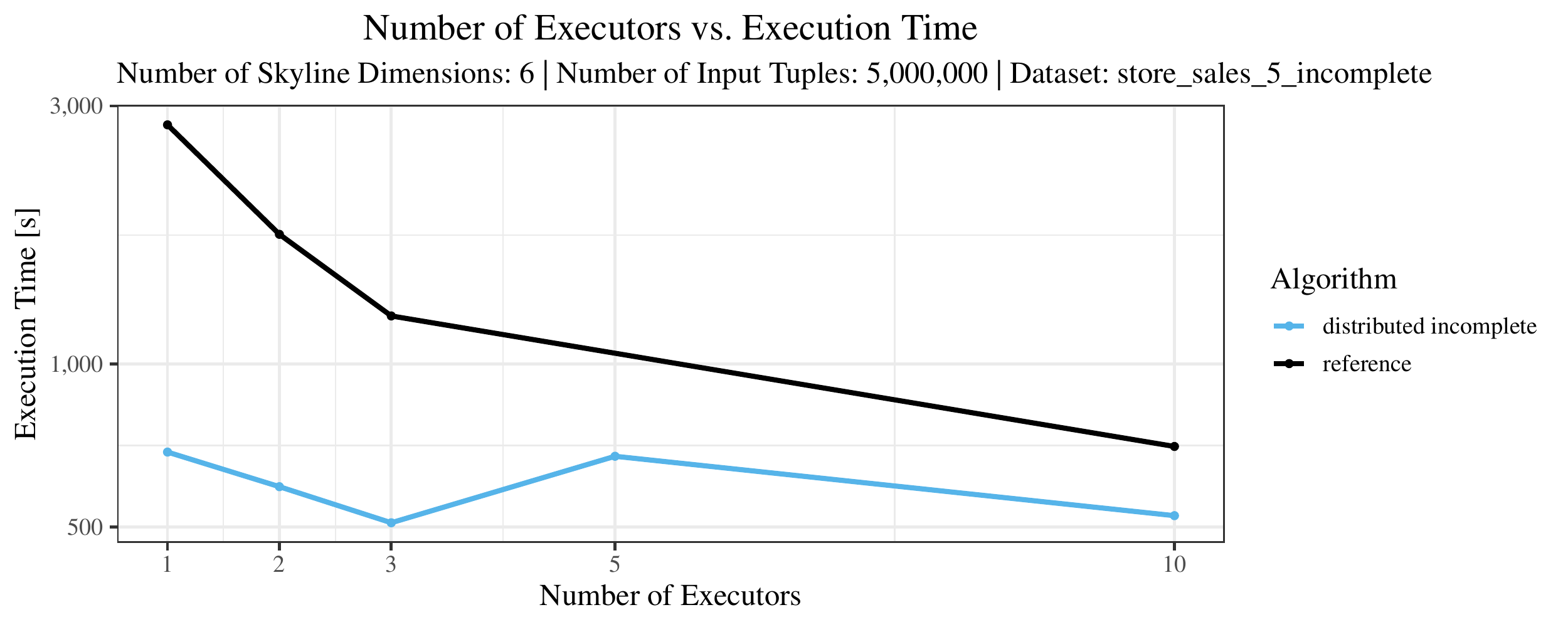}
    \end{subfigure}
    \caption{Number of executors vs. execution time on the store\_sales dataset}
    \label{fig:cluster_nodes_vs_time_store_sales}
\end{figure*}

The plots for the impact of the \textbf{number of skyline dimensions} on the execution time can be found in Figure~\ref{fig:cluster_dimensions_vs_time_complete_airbnb} for the real-world dataset and Figure~\ref{fig:cluster_dimensions_vs_time_complete_store_sales} for the synthetic dataset. 
Of course, if the number of dimension increases, the dominance checks get slightly more costly. But the most important effect on the execution time is via the size of the skyline (in particular, of the intermediate skyline in the window).

Increasing the number of dimensions can have two opposing effects. First, adding a dimension can make tuples incomparable, where previously one tuple dominated the other. In this case, the size of the skyline increases due to the additional skyline dimension. 
Second, however, it may also happen that two tuples had identical values in the previously considered dimensions and the additional dimension introduces a dominance relation between these tuples. In this case, the size of the skyline decreases.

For the real-world data (Figure~\ref{fig:cluster_dimensions_vs_time_complete_airbnb}), the execution time tends to increase with the number of skyline dimensions. This is, in particular, the case for the reference algorithm. In other words, here we see the first possible effect of increasing the number of dimensions. 
In contrast, for the synthetic dataset, we also see the second possible effect. This effect is best visible for the reference algorithm in the left plot in Figure~\ref{fig:cluster_dimensions_vs_time_complete_store_sales}. Apparently, there are many tuples with the maximal value in the first dimension (=~\texttt{ss\_quantity}), which become distinguishable by the second dimension (=~\texttt{ss\_wholesale\_cost}). 
For the right plot in Figure~\ref{fig:cluster_dimensions_vs_time_complete_store_sales}, it should be noted that the tests were carried out with a 10 times smaller dataset to avoid timeouts, which makes the test results less robust. 
The peak in the case of two dimensions is probably due to ``unfavorable'' partitioning of tuples causing some of the local skylines to get overly big.

At any rate, when comparing the algorithms, we note that the specialized algorithms are faster in almost all cases and scale significantly better than the ``reference'' algorithm (in particular, in case of the real-word data).
For complete data, the ``distributed complete'' algorithm performs best. 
The discrepancy between the best specialized algorithm and the reference algorithm is over 50\% in most cases and can get up to ca.~80\% (complete synthetic data, 5 dimensions) and even over 95\% (complete synthetic data, 1 dimension).

The impact of the \textbf{size of the dataset} on the execution time is shown in Figure~\ref{fig:cluster_size_vs_time}. 
Recall that 
we consider the size of the real-world dataset as fixed. Hence, the experiments with varying data size were only done with the synthetic data.
Of course, the execution time increases with the size of the dataset, since a larger dataset also increases the number of dominance checks needed. 
The exact increase depends on the used algorithm; it is particularly dramatic for the ``reference'' algorithm, which even reaches the timeout threshold as bigger datasets are considered. 
Again, the ``distributed complete'' algorithm, when applicable, performs best and, in almost all cases, all specialized skyline algorithms outperform the ``reference'' algorithm. 
The discrepancy between the best specialized algorithm and the reference algorithm reaches ca. 60\% for incomplete data and over 80\% for complete data (for $5\cdot 10^6$ tuples).
For the biggest datasets ($10^7$ tuples), the reference algorithm even times out.

The impact of the \textbf{number of executors} on the performance is shown in Figure~\ref{fig:cluster_nodes_vs_time_airbnb} for the real-world dataset and in Figure~\ref{fig:cluster_nodes_vs_time_store_sales} for the synthetic dataset. 
Ideally, we want the algorithms to be able to use an arbitrary number of executors productively. Here, we use the term ``executors'' since it is the parameter by which we can instruct Spark how many instances of the code it should run in parallel. 

We observe that the benefit of additional executors tapers off after a certain number of executors has been reached. This point is reached sooner, the smaller the dataset is. This behavior, especially in the case of the distributed algorithms, can be explained as follows: As we increase the parallelism of the local skyline computation, the portion of data contained in each partition becomes smaller; this means that more and more dominance relationships get lost and fewer tuples are eliminated from the local skylines. Hence, more work is left to the global skyline computation, which allows for little to no parallelism and, thus, becomes the bottleneck of the entire computation. In other words, there is a sweet spot in terms of the amount of parallelism relative to the size of the input data.

Note that the ``reference'' algorithm is also able to make (limited) use of parallelism. Indeed, the execution plan generated by Spark from the plain SQL query is still somewhat distributed albeit not as much as the truly distributed and specialized approaches. As in the previous experiments, the ``reference'' algorithm never outperforms any of the specialized algorithms -- not even the non-distributed or quasi-non-distributed (incomplete algorithm on complete dataset) approaches.
The discrepancy between the best specialized algorithm and the reference algorithm is constantly above 50\% and may even go up to 90\% (complete synthetic data, 10 executors).

\subsection{Further Measurements and Experiments}
\label{sect:further}

Our main concern with the experiments reported in Section \ref{sect:experimental-results} was to measure the execution time of our integrated skyline algorithms depending on several factors (number of dimensions/tuples/executors). 
Apart from the execution time, we made further measurements in our experiments. In particular, we were interested in the \textbf{memory consumption} of the various algorithms in the various settings. In short, the results were unsurprising: the number of dimensions does not influence the memory consumption while the number of executors and, even more so, the number of input tuples does. 
In the case of the executors this is mainly due to the fact that each executor loads its entire execution environment, including the Java execution framework, into main memory. 
In the case of the number of tuples, of course, more memory is needed if more tuples are loaded. Importantly, we did not observe any significant difference in memory consumption between the four algorithms studied here. 
For the sake of completeness, we provide plots with the memory consumption in various scenarios
\ifArxivVersion
in Appendix~\ref{appendix:additional_benchmarks}.
\else
in~\cite{unsereLangversion}.
\fi

The test setup described in Section \ref{sect:testdata-and-queries} was intentionally minimalistic in the sense that we applied skyline queries to base tables of the database rather than to the result of (possibly complex) SQL queries. This allowed us to focus on the behavior of our new skyline algorithms vs.~the reference queries without having to worry about side-effects of the overall query evaluation. But, of course, it is also interesting to test the behavior of skyline queries on top of complex queries. To this end, we have carried out experiments on the MusicBrainz database~\cite{Musicbrainz} with more complex queries including joins and aggregates.
Due to lack of space, the detailed results are given 
\ifArxivVersion
in Appendix~\ref{appendix:complexQueries}.
\else
in~\cite{unsereLangversion}. 
\fi
In a nutshell, the results are quite similar to the ones reported in Section \ref{sect:experimental-results}: 

\begin{itemize}
    \item The execution time of the reference solution is almost always above the time of the specialized algorithms. The only cases where the reference solution performs best are the easiest ones with execution times below 50 seconds. For the harder cases it is always slower and -- in contrast to the specialized algorithms -- causes several timeouts.
    \item The memory consumption is comparable for all 4 algorithms. There are, however, some peaks in case of the reference algorithms that do not occur with the specialized algorithm. In general, the behavior of the reference algorithm seems to be less stable than the other algorithms. For instance, with 4, 5, or 6 dimensions, the execution time jumps from below 50 seconds for 3 executors to over 1000 seconds for 5 executors and then slightly decreases again for 10 executors.
    \item It is also notable, that skyline queries rewritten to plain SQL are much less readable and less intuitive in the context of the entire query.
\end{itemize}

\subsection{Summary of Results}
\label{sect:summary}

From our experimental evaluation, we conclude that, in almost all cases, our specialized algorithms outperform the ``reference'' algorithm and they also provide better scalability.
Parallelization (if applicable) has proved profitable -- but only up to a certain point, that depends on the size of the data.
Moreover, it has become clear that using the incomplete distributed algorithm on a complete dataset is not advisable and may perform worse than the non-distributed algorithm. As such, the introduction of additional syntax to select an appropriate algorithm through the keyword \texttt{COMPLETE} may help to significantly boost the performance.

\section{Conclusion and Future Work}
\label{section:conclusion}

In this work, we have presented the extension of Spark SQL by the skyline operator. It provides users with a simple, easy to use syntax for formulating skyline queries. To the best of our knowledge, this is the first distributed data processing system with native support of skyline queries.
By our extension of the various components of Spark SQL query execution and, in particular, of the Catalyst optimizer, we have obtained an implementation that clearly outperforms the formulation of skyline queries in plain SQL.
Nevertheless, there is still ample space for future enhancements:

So far, we have implemented only the Block-Nested-Loop skyline algorithm and variants thereof. It would be interesting to implement additional algorithms based on other paradigms like ordering~\cite{SFS_1, SFS_2, LESS_1, LESS_2, SALSA_1, SALSA_2, SDI} or index structures~\cite{SkylineBitmapIndex, BSkyTree_1, BSkyTree_2}
and to evaluate their strengths and weaknesses in the Apache Spark context. Also, for the partitioning scheme, further options such as angle-based partitioning~\cite{EfficientParallelProcessingMethodSkylineMapReduce, AngleBasedPartitioning} are worth trying. Further specialized algorithms, that require a deeper modification of Spark (such as Z-order partitioning~\cite{EfficientParallelSkylineQueryProcessingHighDimensionalData}, which requires the computation of a Z-address for each tuple~\cite{Zsearch}) are more long-term projects.

For our skyline algorithm that can cope with (potentially) incomplete datasets, we have chosen a straightforward extension of the Block-Nested-Loop algorithm.
% , which partitions the data according to the occurrence of NULLs in the skyline dimensions. 
As mentioned in Section~\ref{section:skyline_incomplete_data}, this may severely limit the potential of parallelism. Moreover, in the worst case, the algorithm may thus have to compare each tuple against any other tuple. Clearly, 
a more sophisticated algorithm for potentially incomplete datasets would be highly desirable.

We have included {\em rule-based optimizations} in our integration of skyline queries. 
The support of {\em cost-based optimization} by Spark SQL is only somewhat rudimentary as of now. Fully integrating skyline queries into a future cost-based optimizer will be an important, highly non-trivial research and development project. However, as soon as further skyline algorithms are implemented, a light-weight form of cost-based optimization should be implemented that selects the best-suited skyline algorithm for a particular query.

Finally, from the user perspective, the integration into different Spark modules such as structured streaming would be desirable.

To conclude, the goal of this work was a full integration of skyline queries into Spark SQL. 
The favorable experimental results with simple skyline algorithms (Block-Nested-Loops and variants thereof) demonstrate that an integrated solution is clearly superior to a rewriting of the query on SQL level. To leverage the full potential of the integrated solution, the implementation of further (more sophisticated) skyline algorithms is the most important task for future work. It should be noted that, in our implementation, we have paid particular attention to a modular structure of our new software so that the implementation of further algorithms is possible without having to worry about Spark SQL as a whole. Moreover, our code is provided under Apache Spark’s own open-source license, and we explicitly invite other groups to join in this effort of enabling convenient and efficient skyline queries in Spark~SQL.

\begin{acks}
This work was supported by the Austrian Science
Fund (FWF): P30930 and 
by the Vienna Science and Technology Fund
(WWTF) grant VRG18-013.
\end{acks}

% custom push to new page
\clearpage
\newpage

\bibliographystyle{ACM-Reference-Format}
\bibliography{references}

%% Please note that all the content must fit within the page limits, including any appendices.
%\appendix
% ...
\ifArxivVersion
\appendix
\section{Problems with Incomplete Datasets}\label{appendix:algorithm_incomplete_datasets}

Recall from Section \ref{section:skyline_queries} that, in the case of incomplete data, 
cyclic dominance relationships may arise. Therefore, 
care is required with deleting dominated tuples, 
since premature deletion of dominated tuples may prevent us from identifying other
tuples as being dominated.
Below, we illustrate this problem by revisiting the 
skyline algorithm from~\cite{OptimizingSkylineQueryProcessingIncompleteData},
which behaves incorrectly in the presence 
of cyclic dominance relationships. 

The algorithm in~\cite{OptimizingSkylineQueryProcessingIncompleteData},
divides the dataset into $n$ clusters (= partitions) according to the occurrence
of \texttt{null}s in skyline dimensions and computes the local skyline in each cluster. 
It then arranges these clusters, i.e., $C_1$ through $C_n$,
in some order and computes the global skyline by 
visiting the clusters in this order and
carrying out dominance checks for the data points in the local skylines of all clusters.
For each data point, the following two steps are applied:

\begin{itemize}
	\item For the current data point $p$ of the iteration, we check for every not yet deleted data point $q$ in all subsequent clusters if one dominates the other. For instance, if $C_i$ is the current cluster, we check for all clusters $C_j$ with $j > i$. Checking in the same cluster is not necessary since dominated data points have already been removed when computing the local skyline. If dominance is detected there are two scenarios:
	\begin{itemize}
		\item If $p \prec q$, then $q$ is immediately eliminated.
		\item If $q \prec p$, then a \textit{domination flag} is set for $p$.
	\end{itemize}
	\item If after checking against all data points $q$, the \textit{domination flag} is set for $p$, then $p$ is subsequently also eliminated from the candidate list, since it is dominated by another point.
\end{itemize}

\bigskip
\noindent
\textbf{Counterexample.}
We now argue that the approach 
of~\cite{OptimizingSkylineQueryProcessingIncompleteData}
to compute the global skyline 
as described above is incorrect. 
To show this, we give a counterexample with simple data where applying the algorithm does not yield a correct skyline.
We thus revisit the example that was used in Section~\ref{section:skyline_queries}
to illustrate the existence of dominance cycles in the case of incomplete data:
Suppose that we have 3 dominance dimension and consider the following 3
data points: $a = (1, *, 10)$, $b = (3, 2, *)$, and $c = (*, 5, 3)$. Here,
``$*$'' indicates a \texttt{null} value.

As already discussed, we know that $a \prec b$, $b \prec c$ and $c \prec a$ in this example under the assumption that all dimensions are minimized in the skyline. From the definition of the dominance relation and the skyline
for incomplete data, it follows that the skyline of these values should be empty as every single tuple is dominated by another tuple. We now ``execute'' the above algorithm to check whether we get the same result.

\begin{itemize}
    \item For clustering, the tuples correspond to the bitmaps \texttt{101}, \texttt{110}, and \texttt{011} respectively. It follows that every tuple is in its own cluster according to the algorithm.
%     \item For the preprocessing step, none of the items are removed since we do not set a threshold or condition.
    \item Clearly, each tuple is contained in its local skyline since 
    each cluster contains only a single element.
\end{itemize}

This leaves the global skyline computation as the only step where tuples are eliminated. We now assume w.l.o.g. that $a$ belongs to $C_1$, $b$ belongs to $C_2$, and $c$ belongs to $C_3$ as per the algorithm introduced 
in~\cite{OptimizingSkylineQueryProcessingIncompleteData}. It follows that the algorithm starts by comparing $a$ from $C_1$ to $b$ which is the only element in $C_2$. Since $a \prec b$, we remove $b$ from $C_2$ after which $C_2$ is empty. Next, we go to the cluster $C_3$ and compare $a$ to $c$. Since $c \prec a$, we set the domination flag for $a$. No further clusters and tuples remain. Therefore, $a$ is deleted from $C_1$ since the domination flag is set. For the next iteration, we check the cluster $C_2$. 
Since $C_2$ is now empty, no comparisons remain for this cluster. 
We move on to the last cluster $C_3$. Since there are no further clusters, we note that $C_3$ still contains $c$ after these steps. Hence, 
the algorithm returns tuple $c$ as part of the global skyline,
which contradicts our expected result of an empty skyline.

\medskip
\noindent
\textbf{Correct Skyline Computation.}
As was detailed in Section~\ref{section:skyline_incomplete_data},
we take over from~\cite{OptimizingSkylineQueryProcessingIncompleteData}
the idea of partitioning the data according to the occurrence of \texttt{null}s
for the {\em local} skyline computation of incomplete data. 
However, 
in our {\em global} skyline computation for incomplete data, we compare all pairs of tuples with each other and set a dominance flag 
if one tuple is dominated. 
Only after all pairs have been processed in this way, we actually 
delete the tuples for which the dominance flag is set.
In this way, we can guarantee that all dominance relationships are found even if they are cyclic.

\section{Spark SQL Error When Sorting on Aggregates With Filter}\label{appendix:aggregate_resolution_error}

While integrating the skyline queries into Apache Spark, we have noticed that aggregates are sometimes not resolved correctly by the default rules of Spark. This occurs, for example, when an \texttt{Aggregate} (introduced by a \texttt{GROUP BY} clause) is used in combination with a \texttt{Filter} (introduced by a \texttt{HAVING} clause) as input to the \texttt{Skyline} node. The aggregates are then not correctly resolved since they are either not introduced into the \texttt{Aggregate} node or lost during a projection after the \texttt{Filter}. 
As part of the implementation, we have taken preventive measures such that these queries can be resolved correctly for skyline queries. 
% These can be found in Section~\ref{section:integration}.

% Algorithm moved here such that it is correctly rendered in Overleaf

\begin{listingAlgorithm}
\begin{minted}[fontsize=\small,linenos,breaklines]{scala}
object PreventPrematureProjections extends Rule[LogicalPlan] {
  def apply(plan: LogicalPlan): LogicalPlan = plan.resolveOperatorsUp {
    case sort@Sort(_, _,
      project@Project(_,
        filter@Filter(_,
          aggregate: Aggregate
        )
      )
    ) if filter.resolved && aggregate.resolved =>
      val newSort = sort.copy(
        child = filter.copy(
          child = aggregate
        )
      )
  
      val newSortMaybeResolved = ResolveAggregateFunctions
        .apply(newSort)
        .asInstanceOf[Sort]
  
      if (!newSortMaybeResolved.equals(newSort)) {
        project.copy(
          child = newSortMaybeResolved
        )
      } else {
        sort
      }
  }
}
\end{minted}
\caption{Prevent premature Project node in plan}
\label{lst:prevent_premature_projection}
\end{listingAlgorithm}

Note, however, that this problem is not skyline-specific. 
A similar (erroneous) behavior in the \texttt{Sort} node leads to errors 
with queries in standard Apache Spark SQL. 
Below we discuss how this error can be fixed for standard Apache Spark SQL. The bugfix in the case of the skyline extension is similar. 
First, we eliminate \texttt{Project} nodes that may be introduced as part of the \texttt{Filter} caused by a \texttt{HAVING} clause in the query. To achieve this, we introduce a new analyzer rule which can be found in Listing~\ref{lst:prevent_premature_projection}.

First, we match the plan that was described above using the matching syntax of Scala (line~3~-~6). We then try to change the plan to not include the filter (line~10~-~14). Only if this allows us to resolve additional attributes (line~16 and 18), then we return the changed plan with the \texttt{Filter} reintroduced as the ``parent'' node (line~19~-~20). Otherwise, we return the old plan as-is (line~23).

% Algorithm moved here such that it is correctly rendered in Overleaf

\begin{listingAlgorithm}
\begin{minted}[fontsize=\small,linenos,breaklines]{scala}
case Sort(sortOrder, global, filter@Filter(_, agg: Aggregate)) =>
  val maybeResolved = sortOrder
    .map(_.child)
    .map(resolveExpressionByPlanOutput(_, agg))
  resolveOperatorWithAggregate(maybeResolved, agg, (newExprs, newChild) => {
    val newSortOrder = sortOrder.zip(newExprs).map {
      case (sortOrder, expr) => sortOrder.copy(child = expr)
    }
    Sort(newSortOrder, global, filter.copy(child = newChild))
})
\end{minted}
\caption{Extension of \texttt{ResolveAggregateFunctions}}
\label{lst:extend_aggregate_resolution}
\end{listingAlgorithm}

In addition to the new rule introduced above, we also extend the \texttt{ResolveAggregateFunctions} rule such that it can handle plans consisting of \texttt{Sort}, \texttt{Filter}, and \texttt{Aggregate}. This is necessary, since otherwise the aggregates from the \texttt{Aggregate} node may not be resolved correctly in the \texttt{Sort} node. To achieve this, we modify the existing source code from 
Spark to fit our needs. The modifications can be found in Listing~\ref{lst:extend_aggregate_resolution}.

This code first takes all ordering dimensions (line~2) and subsequently tries to resolve them (line~3~-~7). In this step, we insert the (potentially) resolved dimensions into a new list of sort orders (line~4~-~5) that is used to create a new \texttt{Sort} node that is now (more) resolved than before.

\section{Additional Benchmarks}\label{appendix:additional_benchmarks}

In this section, we provide additional benchmarks and performance data that were left out in Section~\ref{section:performance_evaluation} due to lack of space.
First, we look at the peak {\bf memory consumption} across all nodes in 
Figure~\ref{fig:cluster_memory_vs_executors_real_world} for the
real-world dataset and in Figures~\ref{fig:cluster_memory_vs_executors_synthetic} and
\ref{fig:cluster_memory_vs_datasize_synthetic} for the synthetic data set. 
As in Section~\ref{section:performance_evaluation}, 
the measurements for complete datasets are shown on the left-hand side, 
while the results for incomplete datasets are shown on the right-hand side.

In Figures~\ref{fig:cluster_memory_vs_executors_real_world} and \ref{fig:cluster_memory_vs_executors_synthetic}, 
we are interested in the impact of the number of executors on the 
memory 
consumption, while the plots in 
Figure~\ref{fig:cluster_memory_vs_datasize_synthetic} are about the 
relationship between data size and memory 
consumption. 
In Figures \ref{fig:cluster_memory_vs_executors_real_world} and \ref{fig:cluster_memory_vs_executors_synthetic}, 
we observe that 
the memory consumption mainly increases with the number of executors -- 
but with a considerable amount of jitter. The increase of memory consumption 
is the expected behavior as every single executor must include the entire execution environment of Spark to be able to execute the code generated by the nodes in the physical plan. However, there are some outliers such as, 
in Figure \ref{fig:cluster_memory_vs_executors_real_world}, 
for 5 executors in the case of the non-distributed complete algorithm (left plot) 
and the distributed incomplete algorithm (right plot). 
The problem here is probably that Spark's history server sometimes reports the memory consumption inaccurately. These inaccuracies occur since Spark uses periodical ``heartbeats'' to report the current memory consumption and its beat is then computed by selecting the highest value. If the timing of the heartbeats is off, then the peak memory consumption may not be captured correctly.
There is notably less jitter in 
Figures~\ref{fig:cluster_memory_vs_executors_synthetic} and
\ref{fig:cluster_memory_vs_datasize_synthetic}.
This is likely due to the fact that the execution times of the queries are higher and the memory consumption improves the longer each Spark application runs.

To sum up, the plots show that the improved runtime of our specialized 
algorithms compared with the reference algorithm is not achieved at the 
expense of significantly higher memory usage. On the contrary, in most cases, 
the memory consumption of our algorithms is even below that of the reference 
algorithm. In the remaining cases, the additional memory consumption (in particular, of the distributed complete algorithm) is within a reasonable 
margin. A comparison among our algorithms shows that the distributed complete algorithm has the highest memory requirements. This is most obvious as 
the data size increases. Here, the size of the ``window''
(see Section \ref{section:skyline_complete_data} for explanations) 
maintained in the 
skyline computation seems to be the predominant cost factor in terms of 
memory consumption.

In Figures~\ref{fig:cluster_dimensions_vs_time_complete_airbnb} and
\ref{fig:cluster_dimensions_vs_time_complete_store_sales}, we have already 
seen plots depicting the relationship between the {\bf number of dimensions} and the 
execution time. Further results -- with varying numbers of executors -- 
are shown in Figure \ref{fig:appendix_dimensions_vs_time_real_world} for the real-world dataset and 
in Figure~\ref{fig:appendix_dimensions_vs_time_synthetic} for the synthetic dataset.
In the case of the real-world dataset, the plots provide a similar picture as in 
Section~\ref{section:performance_evaluation}. In the case of the synthetic dataset, two observations 
in Figure~\ref{fig:appendix_dimensions_vs_time_synthetic} 
seem 
particularly noteworthy: since the tests with incomplete data were carried out on 
a bigger dataset than in Figure~\ref{fig:cluster_dimensions_vs_time_complete_store_sales},
timeouts (in particular, of the reference algorithm) occur much more frequently. In the case of the 
tests with complete datasets, the two opposing effects of increasing the number of dimensions
are very well visible: first, when moving from 1 dimension to 2 and further to 3 dimensions, 
apparently the size of the skyline shrinks and, therefore, in particular the reference 
algorithm requires significantly less time. As further dimensions are added, apparently the 
size of the skyline increases and, hence, in particular the reference 
algorithm starts requiring significantly more time again. However, what remains unchanged 
compared with 
Figures~\ref{fig:cluster_dimensions_vs_time_complete_airbnb} and
\ref{fig:cluster_dimensions_vs_time_complete_store_sales} in 
Section \ref{section:performance_evaluation} is, that in almost all cases, the time needed by our specialized algorithms is (significantly) below the time
needed by the reference algorithm.

In Figure~\ref{fig:appendix_size_vs_time_synthetic}, 
we follow up on Figure~\ref{fig:cluster_size_vs_time}, 
providing further details on the impact of the {\bf size of the dataset} 
on the execution time. In the plots in 
Figure~\ref{fig:appendix_size_vs_time_synthetic},
the number of executors may take the values 2,3,5, or 10.
We notice that, only when we increase the number of executors
to 5 or even 10, then the reference algorithm can cope with the biggest dataset of size $10^7$. In all tests with complete data, the distributed complete algorithm performs best
and also the other specialized algorithms outperform the reference algorithm. 
In the case of incomplete data and a higher number of executors, the performance of the
reference algorithm is comparable to the distributed incomplete algorithm.

In Figures~\ref{fig:cluster_nodes_vs_time_airbnb} and \ref{fig:cluster_nodes_vs_time_store_sales}, we analyzed the relationship between the {\bf number of executors} and the running time. 
We report on further experiments in this direction in Figure~\ref{fig:appendix_cluster_executors_vs_time_real_world} for the real-world dataset and in Figure~\ref{fig:appendix_cluster_executors_vs_time_synthetic} for the synthetic dataset. Again, the picture in Figures~\ref{fig:appendix_cluster_executors_vs_time_real_world} and \ref{fig:appendix_cluster_executors_vs_time_synthetic} is quite similar to the one in Figures~\ref{fig:cluster_nodes_vs_time_airbnb} and \ref{fig:cluster_nodes_vs_time_store_sales}. 
For instance, in almost all cases, all of the specialized algorithms outperform the reference algorithm and, for complete data, in almost all cases, the distributed complete algorithm performs best. Most noteworthy in the case of the tests with the real-world dataset (Figures~\ref{fig:appendix_cluster_executors_vs_time_real_world}) seems the observation that the distributed complete algorithm (which normally performs best) hardly profits from additional executors. We already discussed in Section~\ref{section:performance_evaluation} that, if the dataset is too small, then the distributed computation of the local skylines loses its effectiveness -- leaving more work for the global skyline computation. 
This is in sharp contrast to the setting in Figure~\ref{fig:cluster_nodes_vs_time_store_sales}, where we considered an almost 10 times bigger dataset and where additional executors were clearly helpful. 
This effect is still visible but weaker in Figure~\ref{fig:appendix_cluster_executors_vs_time_synthetic}, where we consider the complete dataset with $5 \cdot 10^6$ tuples.
In the case of the incomplete dataset of this size, the reference algorithm runs into several timeouts and, in all cases, the distributed incomplete algorithm performs better than the reference algorithm.

\begin{figure*}[p]
    \begin{subfigure}{.5\linewidth}
      \centering
      \includegraphics[width=\linewidth]{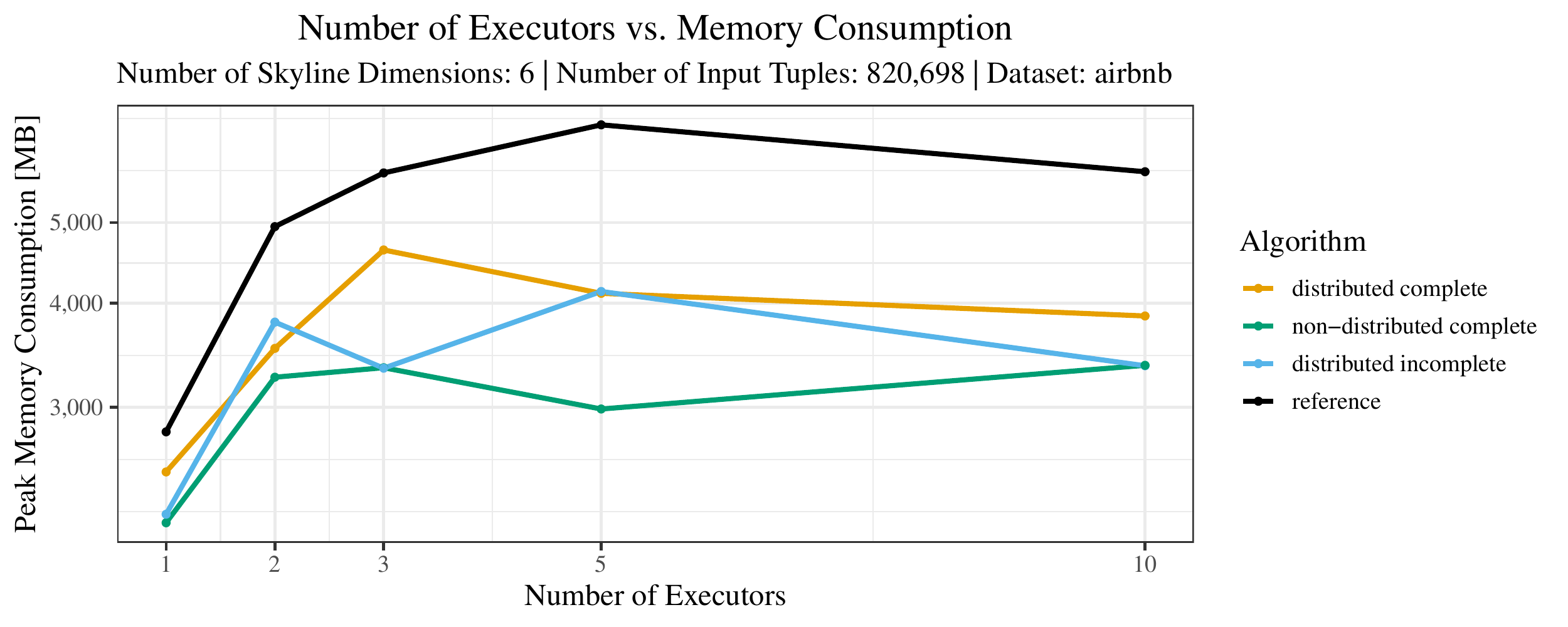}
    \end{subfigure}%
    \begin{subfigure}{.5\linewidth}
      \centering
      \includegraphics[width=\linewidth]{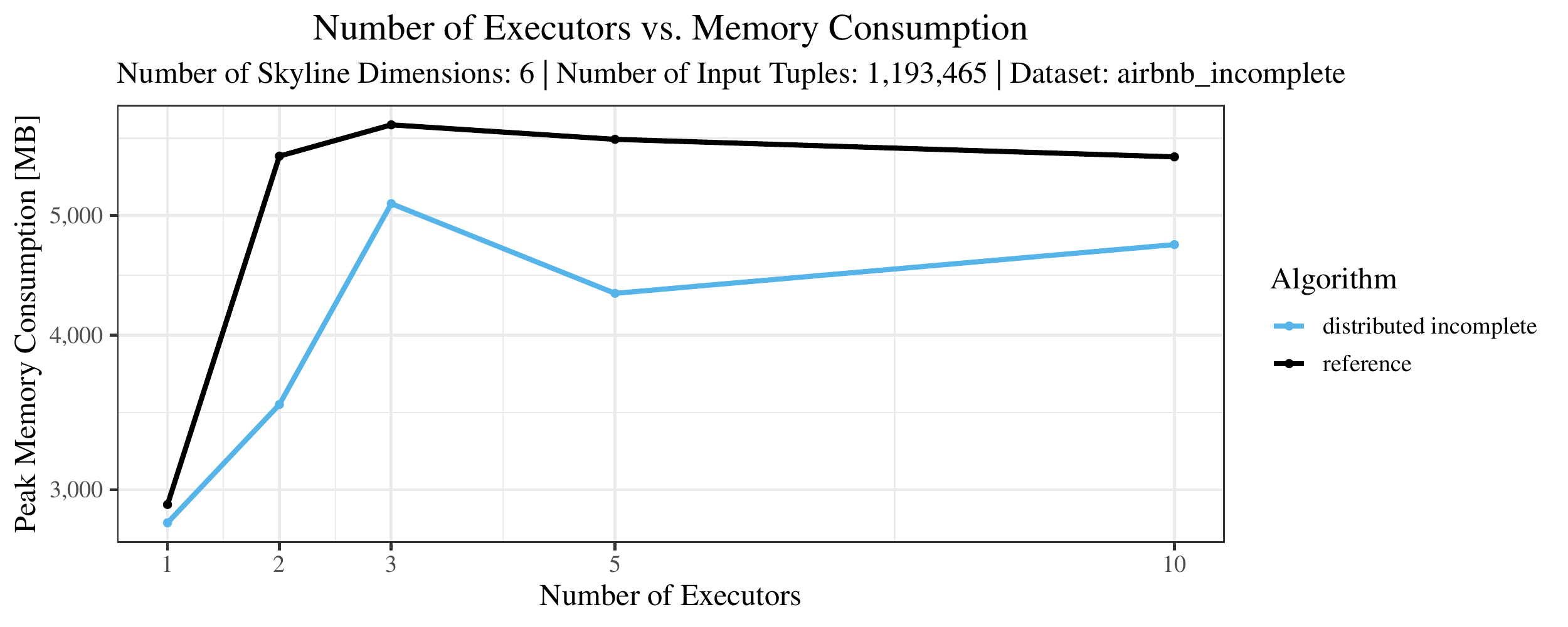}
    \end{subfigure}
    \caption{Number of executors vs. memory consumption on the Inside Airbnb dataset}
    \label{fig:cluster_memory_vs_executors_real_world}
\end{figure*}

\begin{figure*}[p]
    \begin{subfigure}{.5\linewidth}
      \centering
      \includegraphics[width=\linewidth]{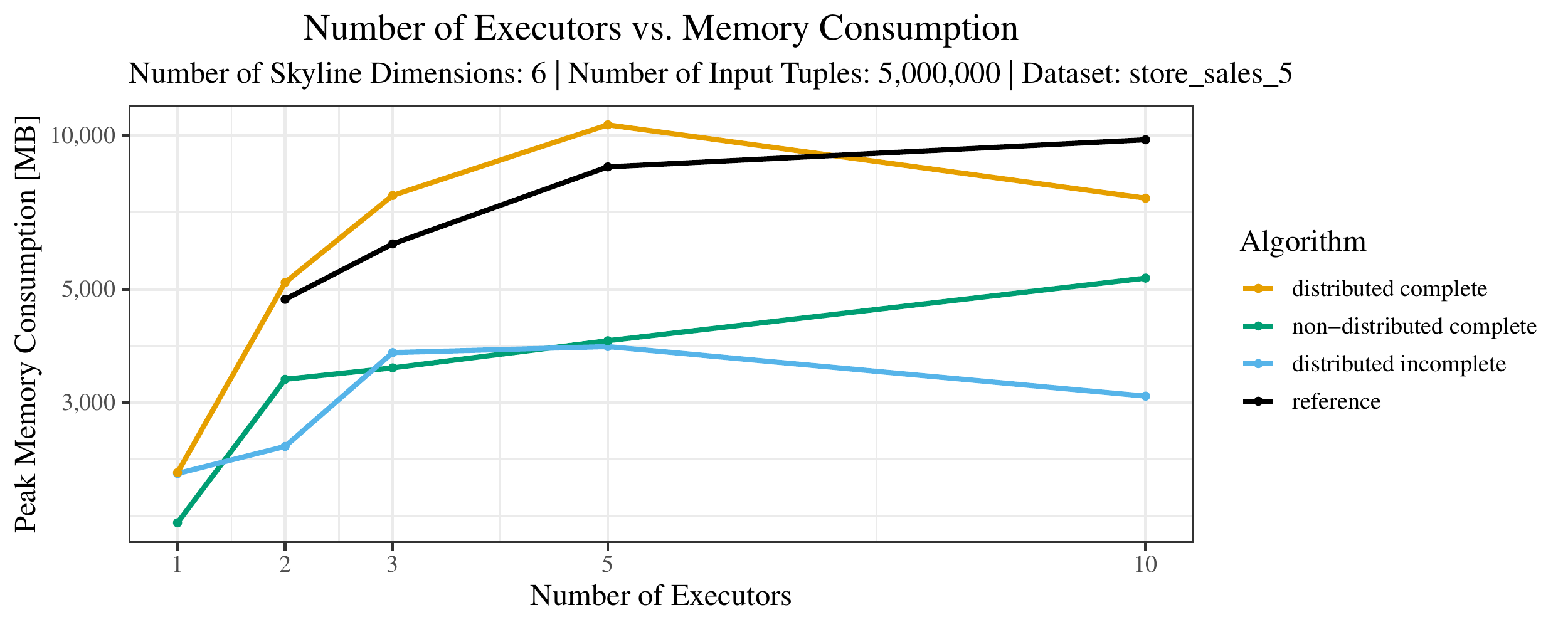}
    \end{subfigure}%
    \begin{subfigure}{.5\linewidth}
      \centering
      \includegraphics[width=\linewidth]{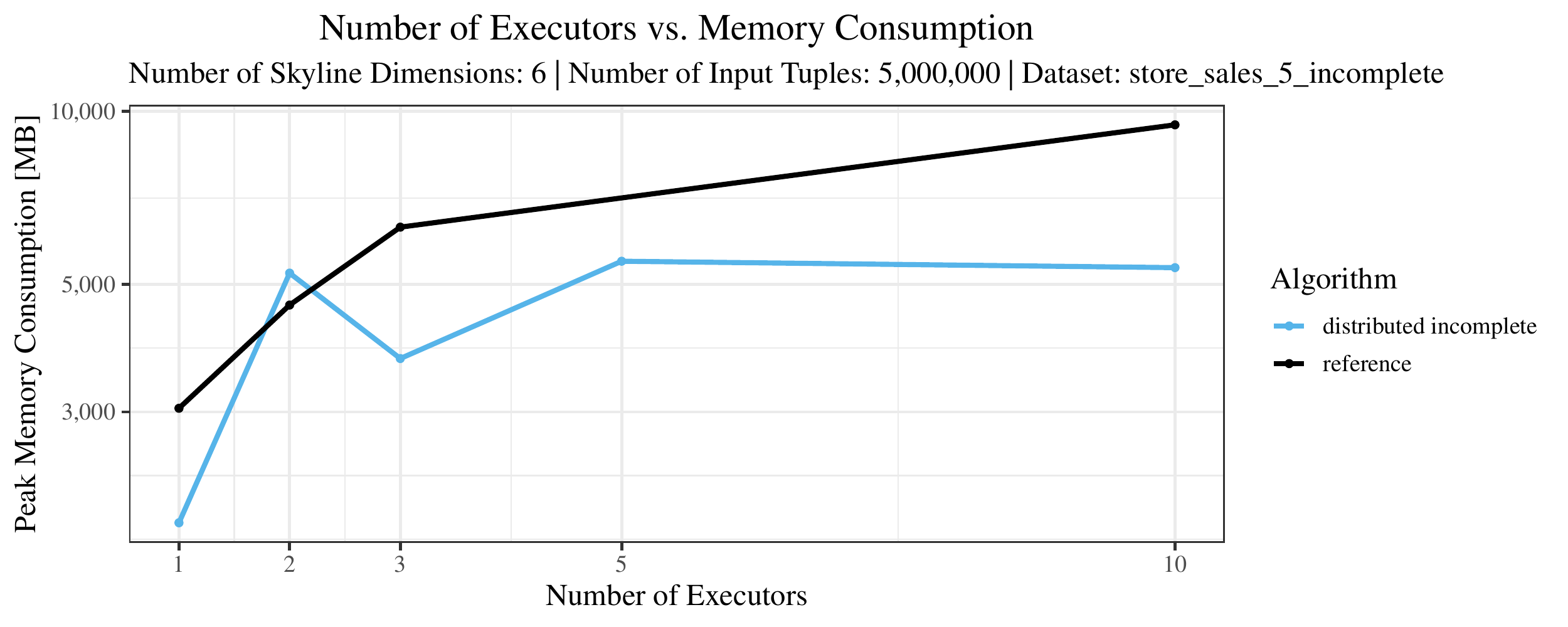}
    \end{subfigure}
    \caption{Number of executors vs. memory consumption on the store\_sales dataset}
    \label{fig:cluster_memory_vs_executors_synthetic}
\end{figure*}

\begin{figure*}[p]
    \begin{subfigure}{.5\linewidth}
      \centering
      \includegraphics[width=\linewidth]{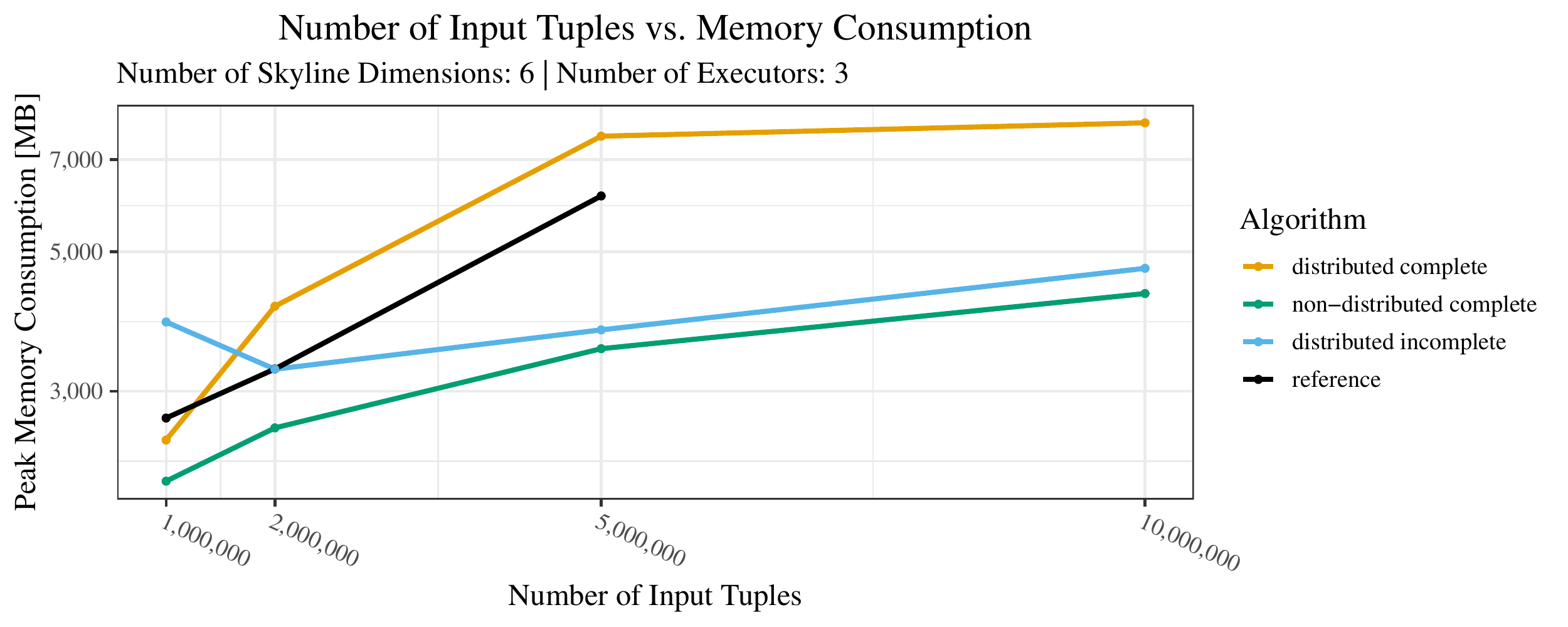}
    \end{subfigure}%
    \begin{subfigure}{.5\linewidth}
      \centering
      \includegraphics[width=\linewidth]{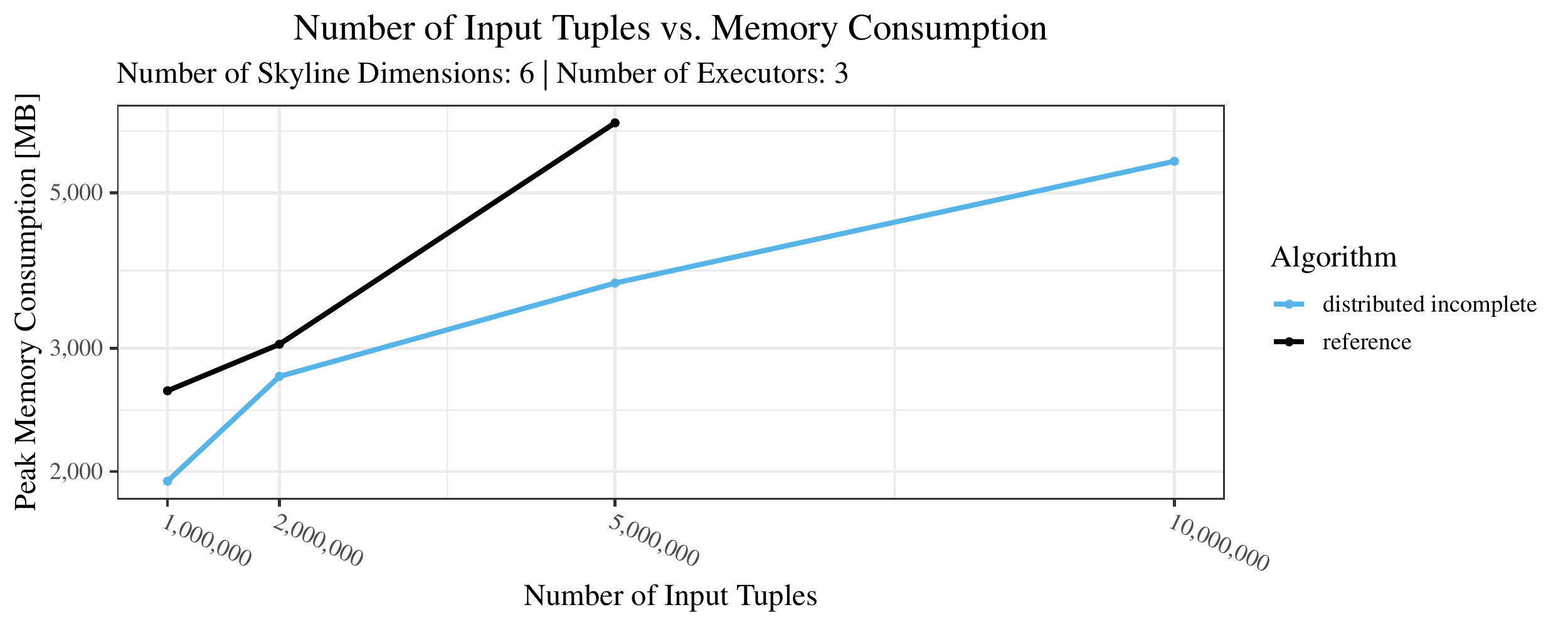}
    \end{subfigure}
    \begin{subfigure}{.5\linewidth}
      \centering
      \includegraphics[width=\linewidth]{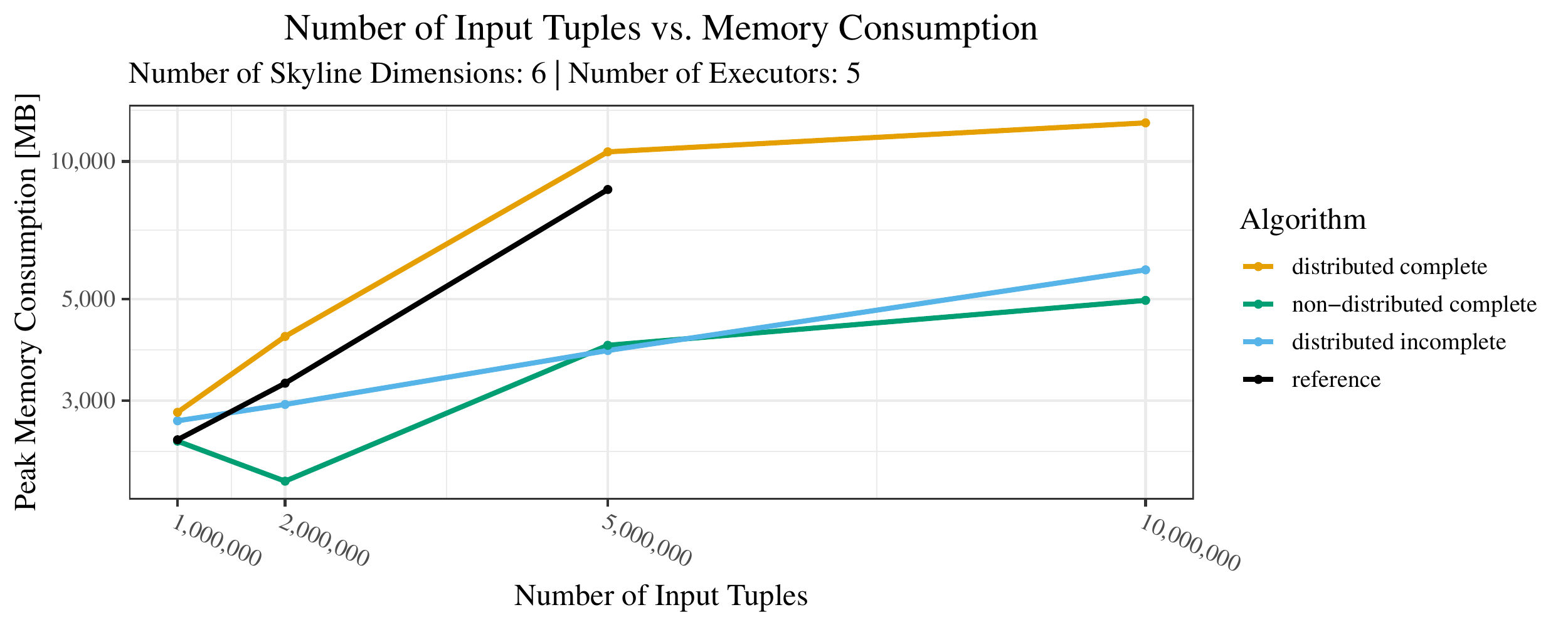}
    \end{subfigure}%
    \begin{subfigure}{.5\linewidth}
      \centering
      \includegraphics[width=\linewidth]{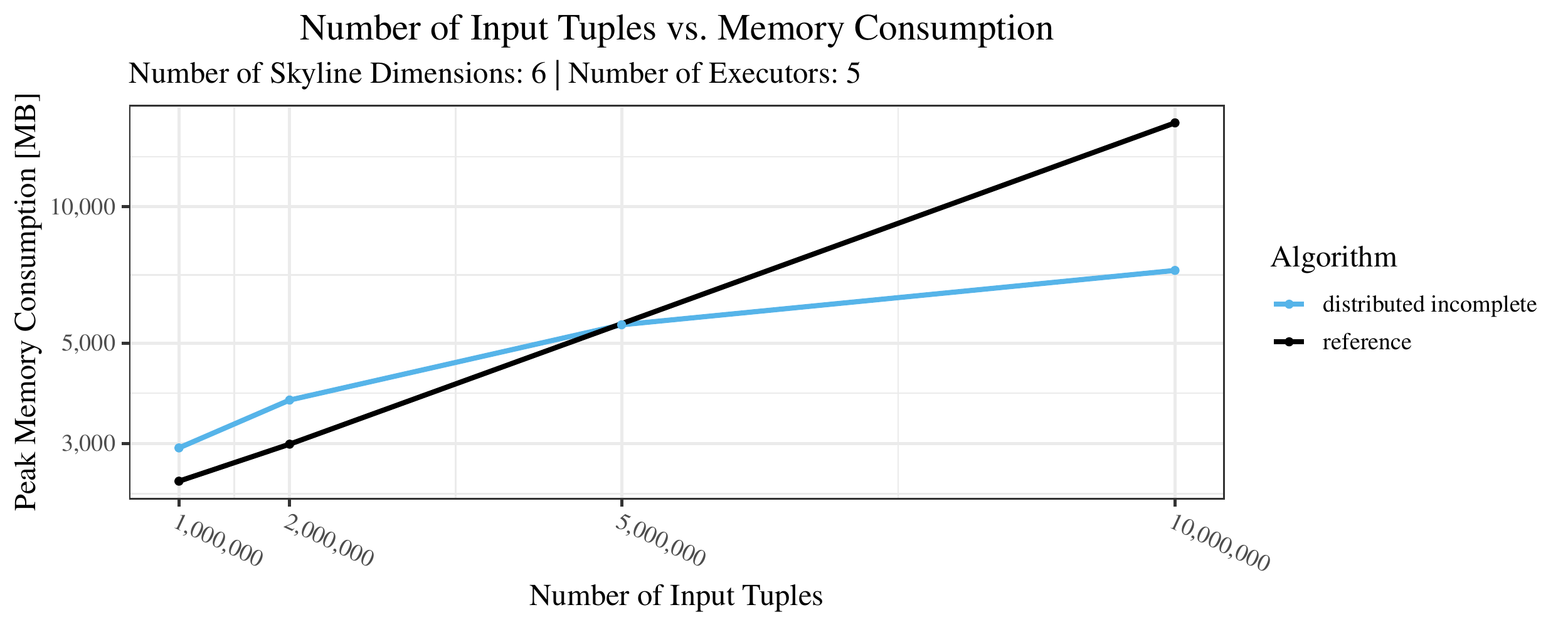}
    \end{subfigure}
    \begin{subfigure}{.5\linewidth}
      \centering
      \includegraphics[width=\linewidth]{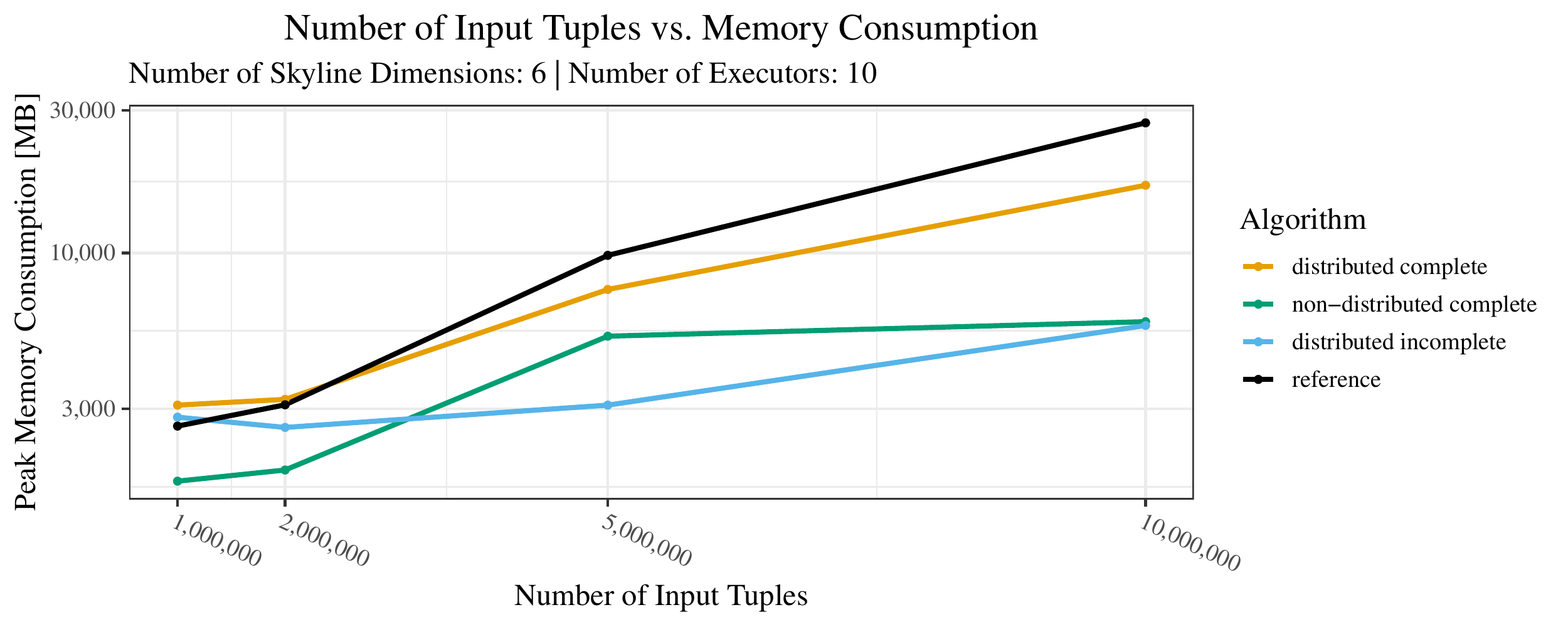}
    \end{subfigure}%
    \begin{subfigure}{.5\linewidth}
      \centering
      \includegraphics[width=\linewidth]{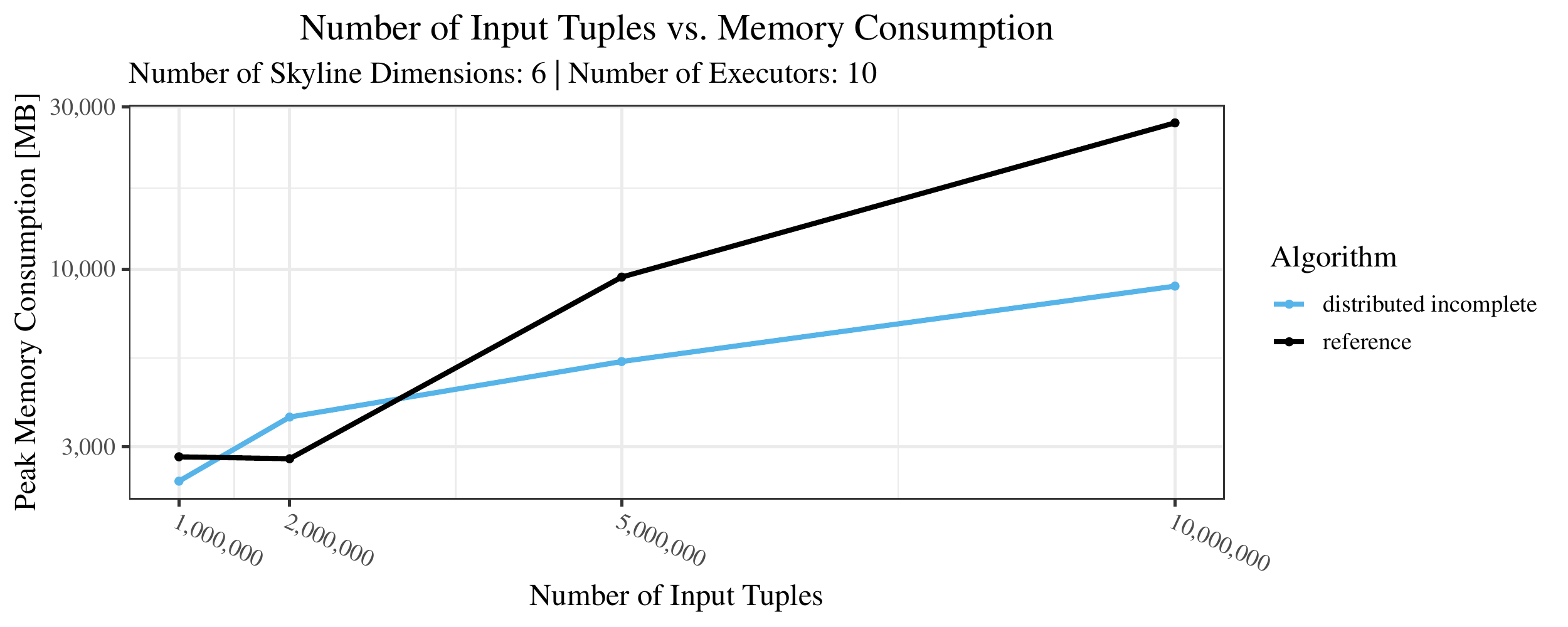}
    \end{subfigure}
    \caption{Number of input tuples vs. memory consumption on the store\_sales dataset}
    \label{fig:cluster_memory_vs_datasize_synthetic}
\end{figure*}

\begin{figure*}[p]
    \begin{subfigure}{.5\linewidth}
      \centering
      \includegraphics[width=\linewidth]{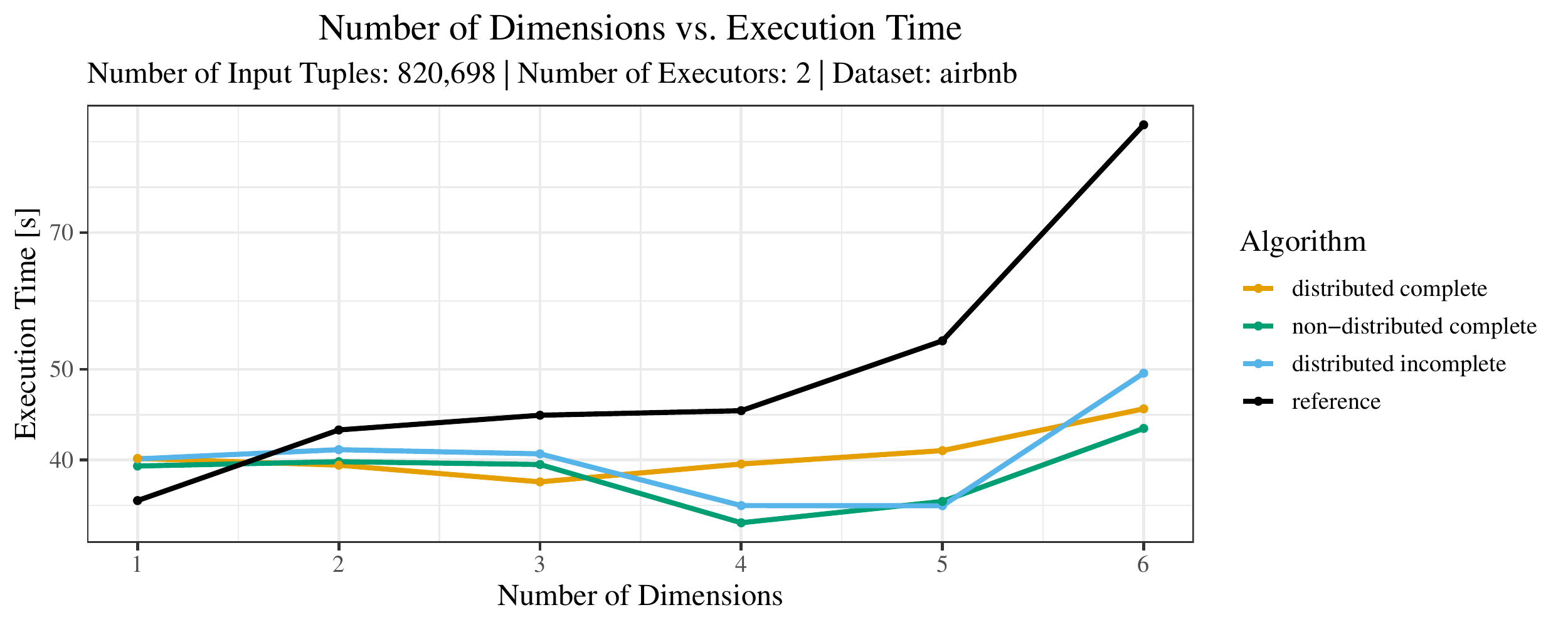}
    \end{subfigure}%
    \begin{subfigure}{.5\linewidth}
      \centering
      \includegraphics[width=\linewidth]{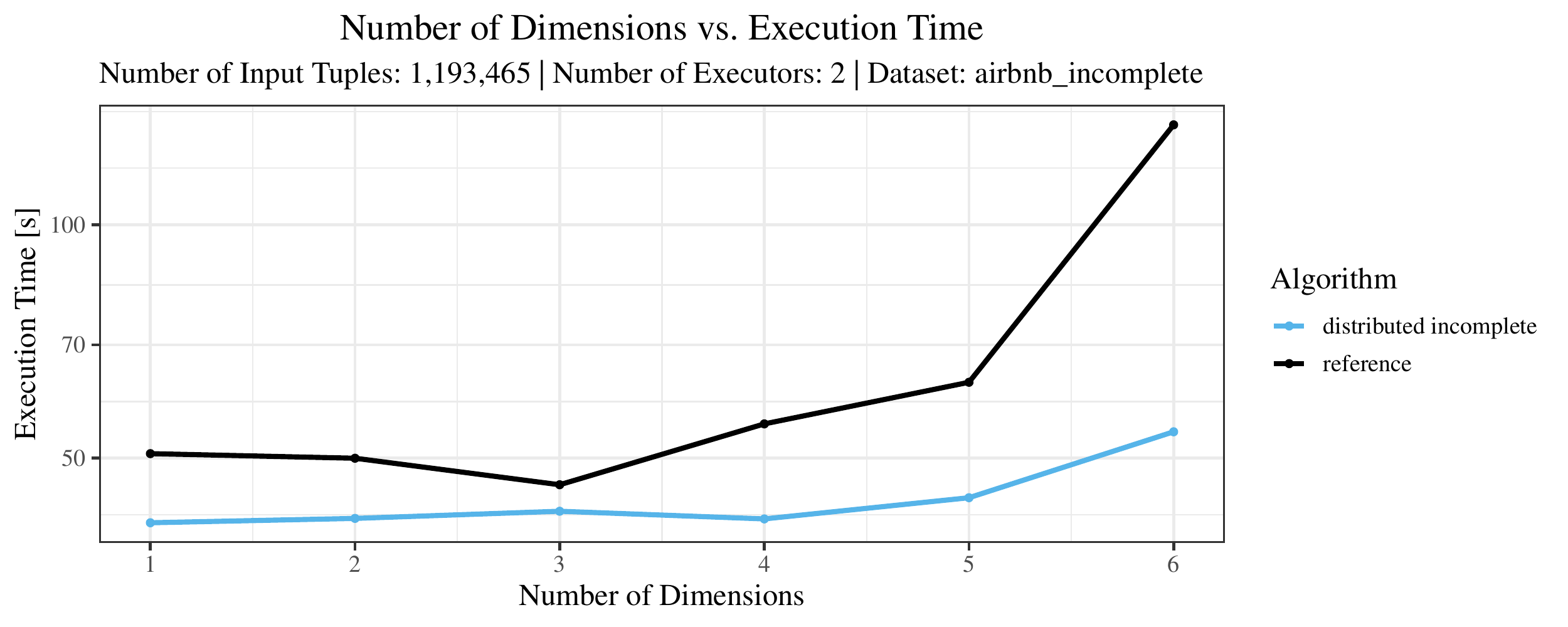}
    \end{subfigure}
    \begin{subfigure}{.5\linewidth}
      \centering
      \includegraphics[width=\linewidth]{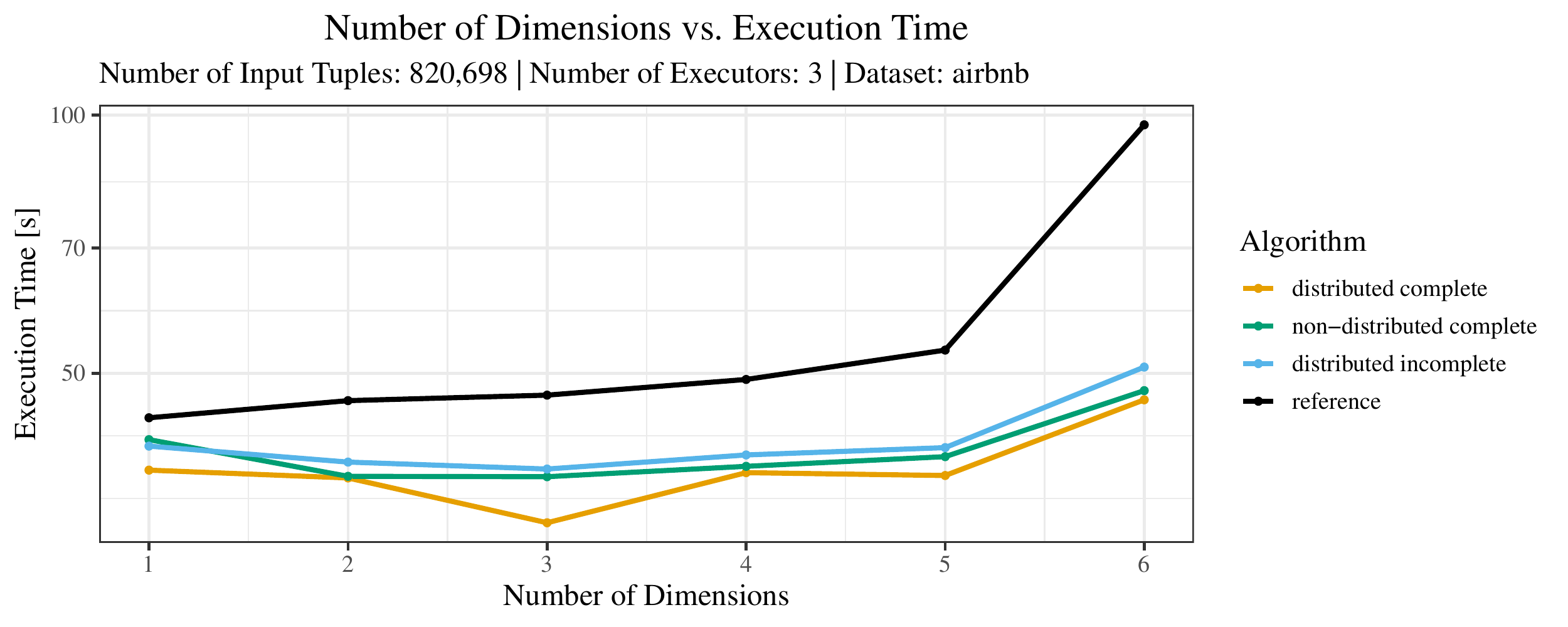}
    \end{subfigure}%
    \begin{subfigure}{.5\linewidth}
      \centering
      \includegraphics[width=\linewidth]{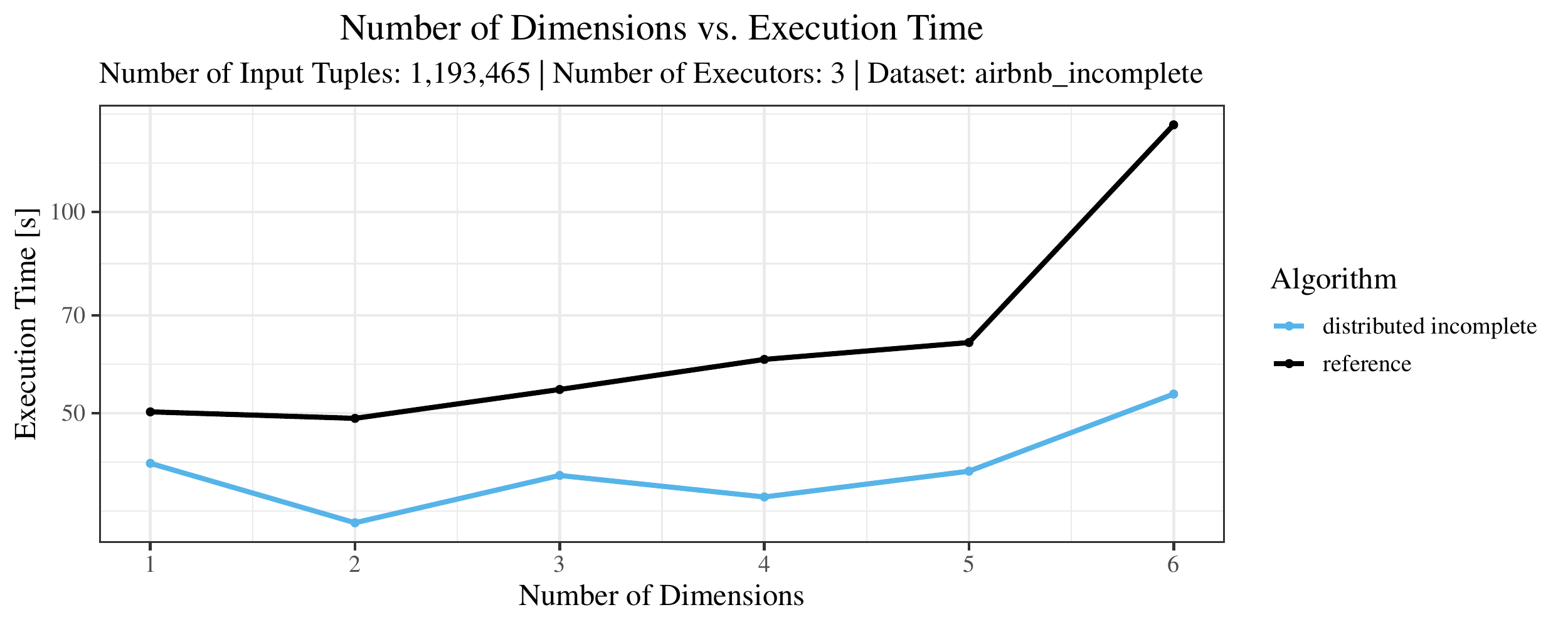}
    \end{subfigure}
    \begin{subfigure}{.5\linewidth}
      \centering
      \includegraphics[width=\linewidth]{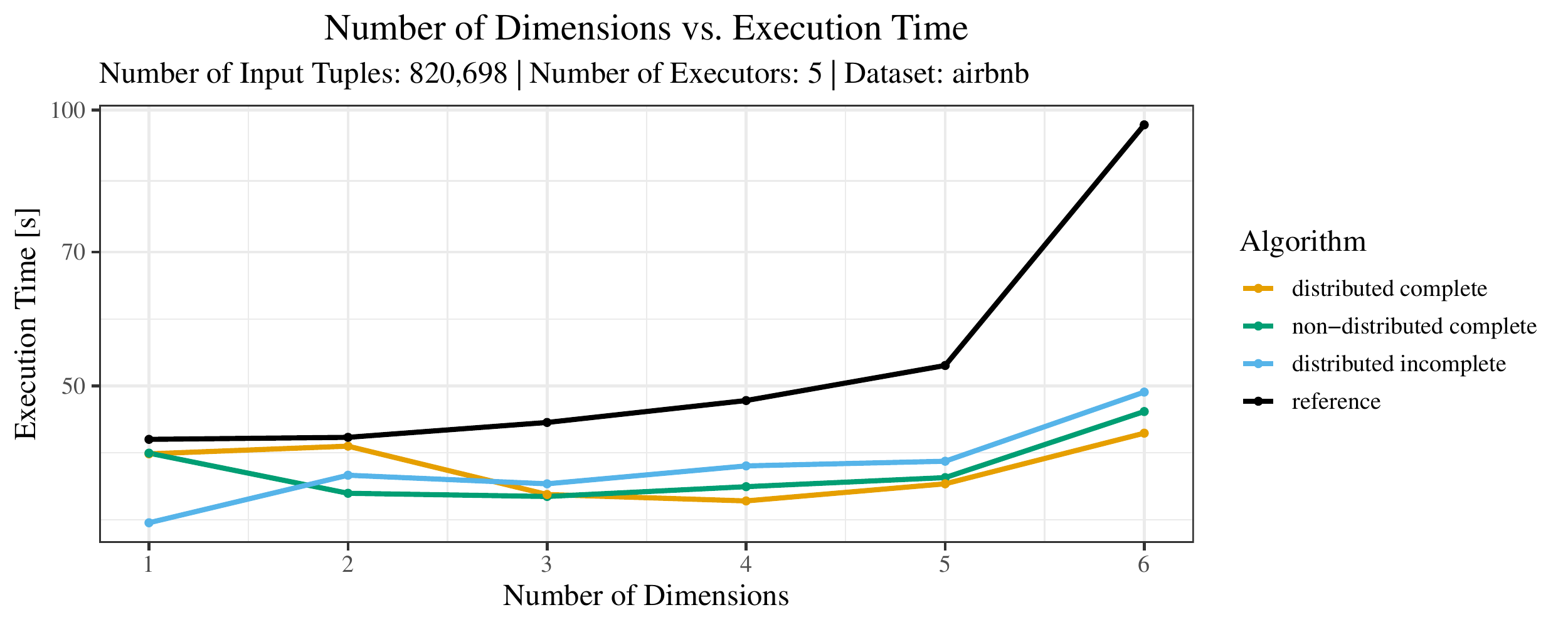}
    \end{subfigure}%
    \begin{subfigure}{.5\linewidth}
      \centering
      \includegraphics[width=\linewidth]{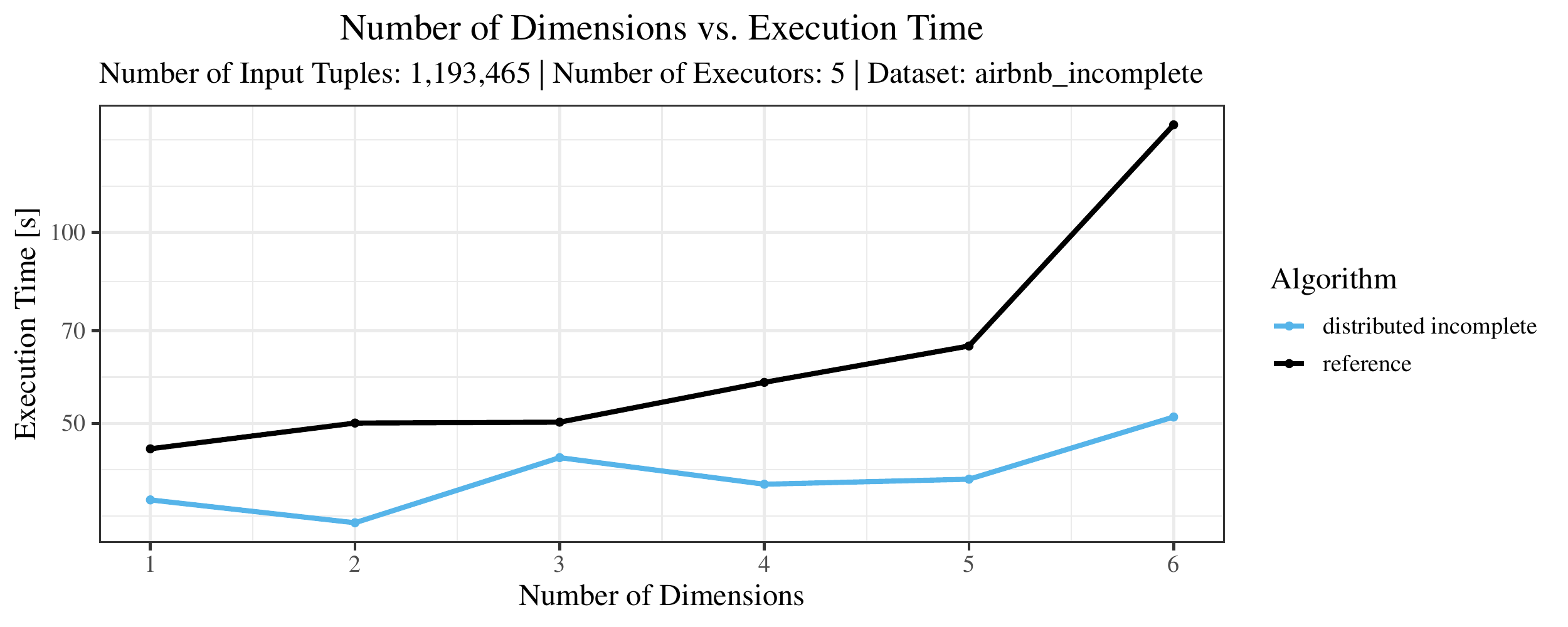}
    \end{subfigure}
    \begin{subfigure}{.5\linewidth}
      \centering
      \includegraphics[width=\linewidth]{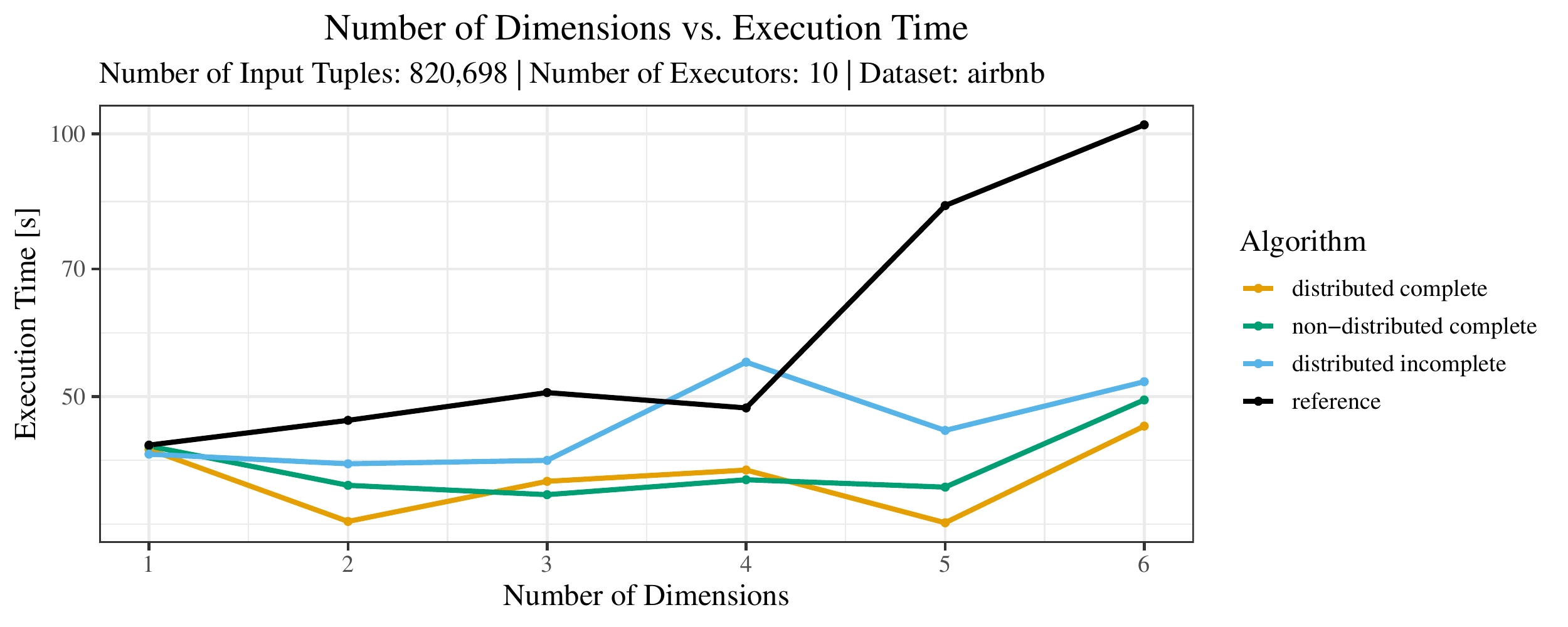}
    \end{subfigure}%
    \begin{subfigure}{.5\linewidth}
      \centering
      \includegraphics[width=\linewidth]{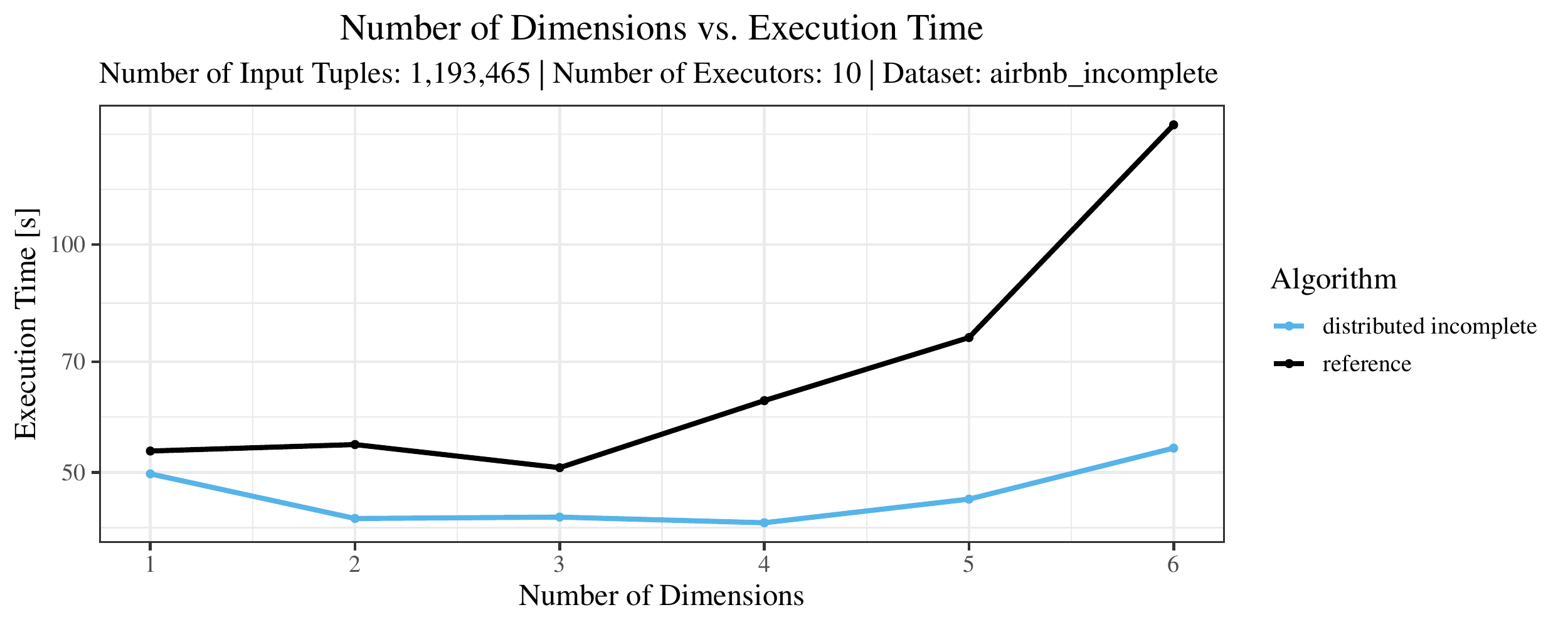}
    \end{subfigure}
    \caption{Number of dimensions vs. execution time on the Inside Airbnb dataset}
    \label{fig:appendix_dimensions_vs_time_real_world}
\end{figure*}

\begin{figure*}[p]
    \begin{subfigure}{.5\linewidth}
      \centering
      \includegraphics[width=\linewidth]{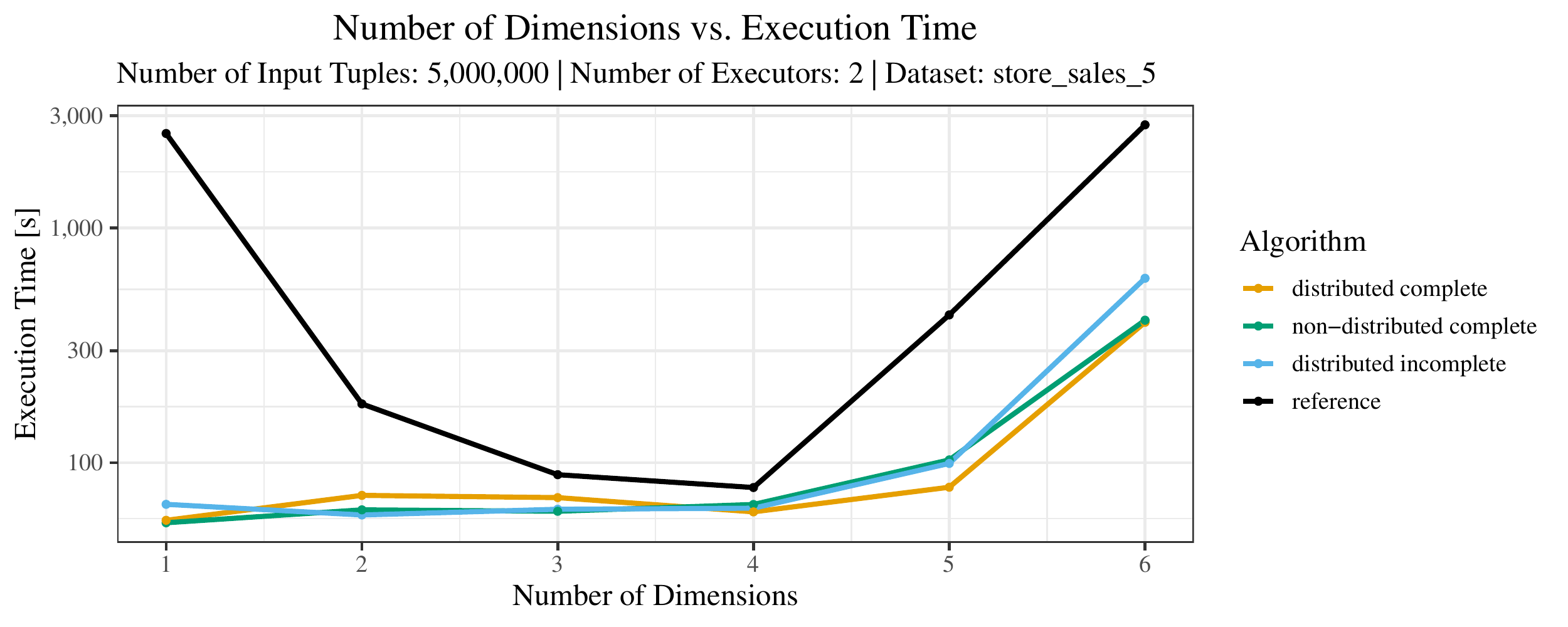}
    \end{subfigure}%
    \begin{subfigure}{.5\linewidth}
      \centering
      \includegraphics[width=\linewidth]{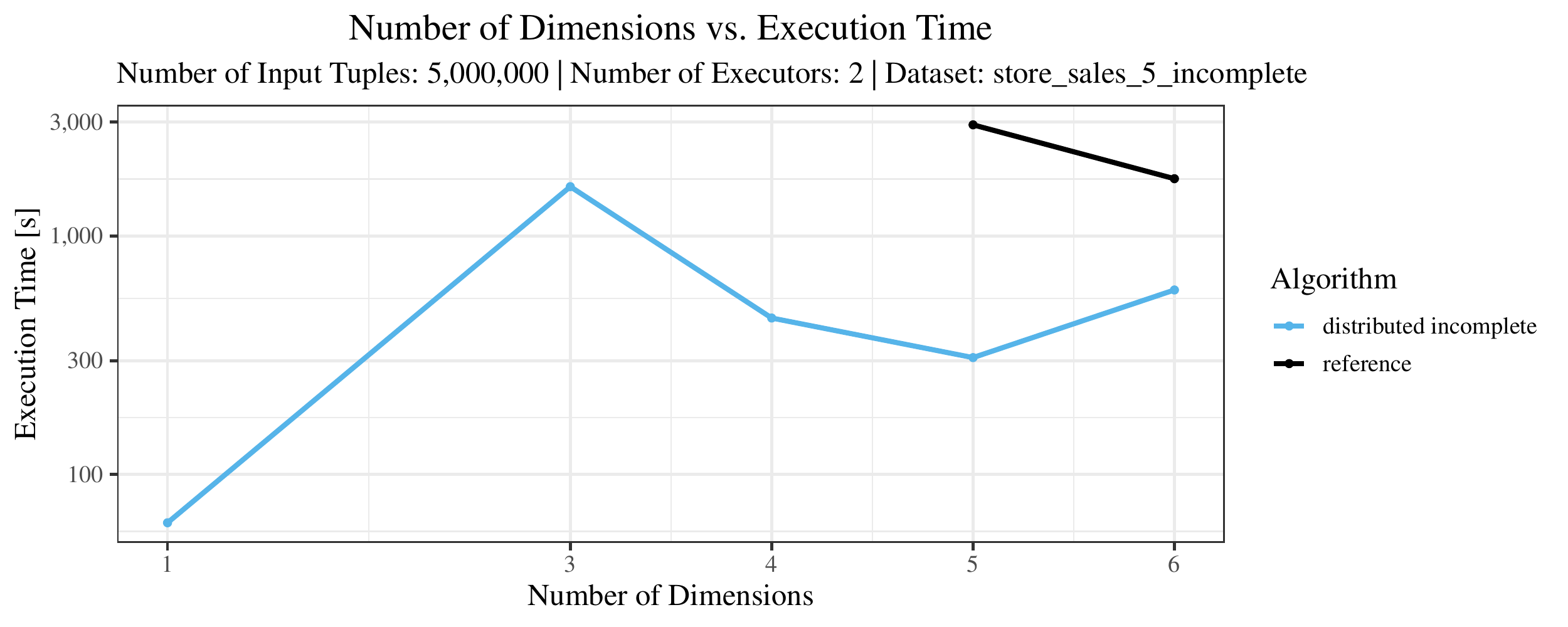}
    \end{subfigure}
    \begin{subfigure}{.5\linewidth}
      \centering
      \includegraphics[width=\linewidth]{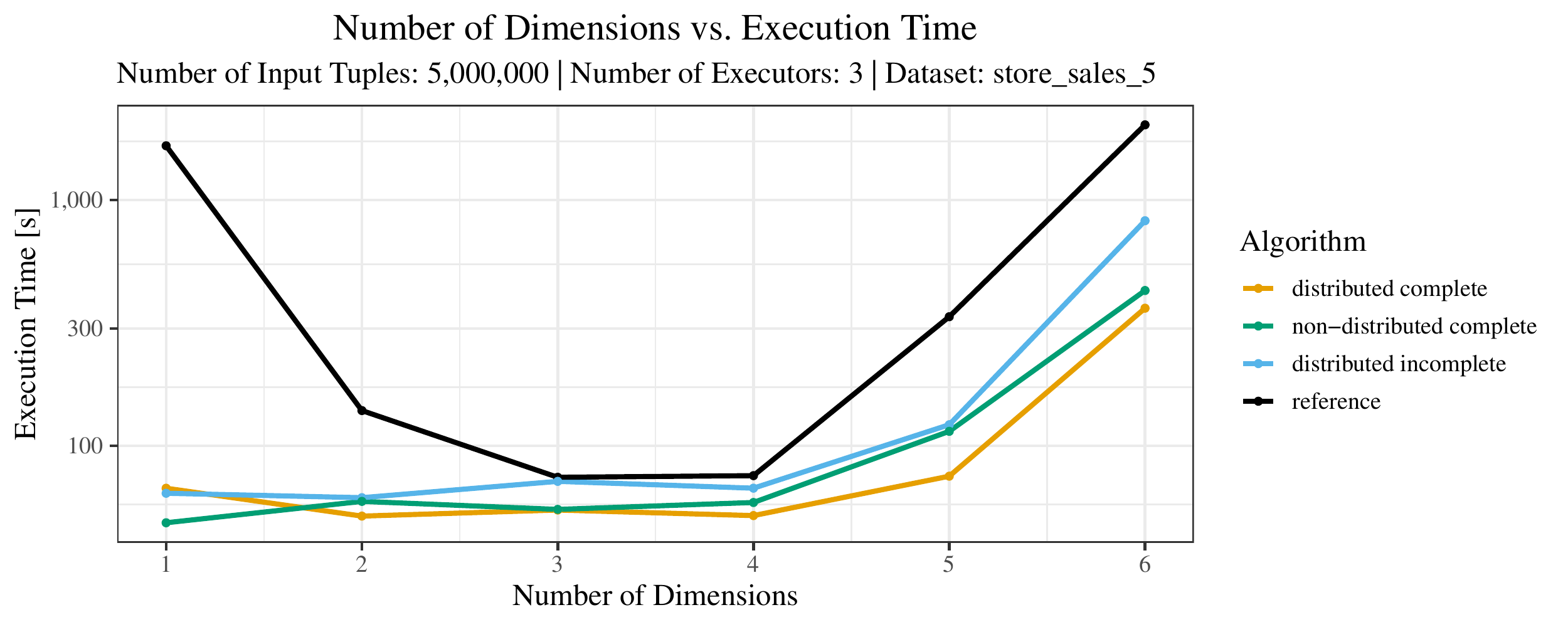}
    \end{subfigure}%
    \begin{subfigure}{.5\linewidth}
      \centering
      \includegraphics[width=\linewidth]{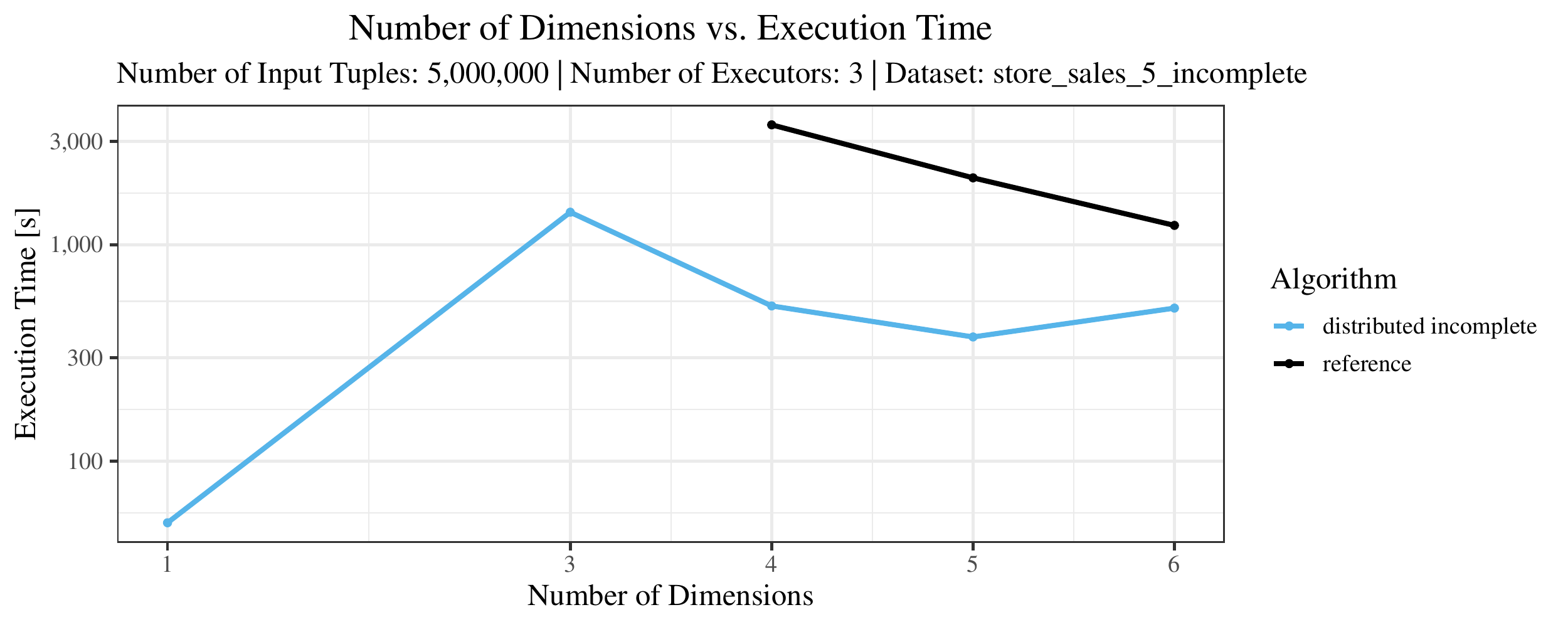}
    \end{subfigure}
    \begin{subfigure}{.5\linewidth}
      \centering
      \includegraphics[width=\linewidth]{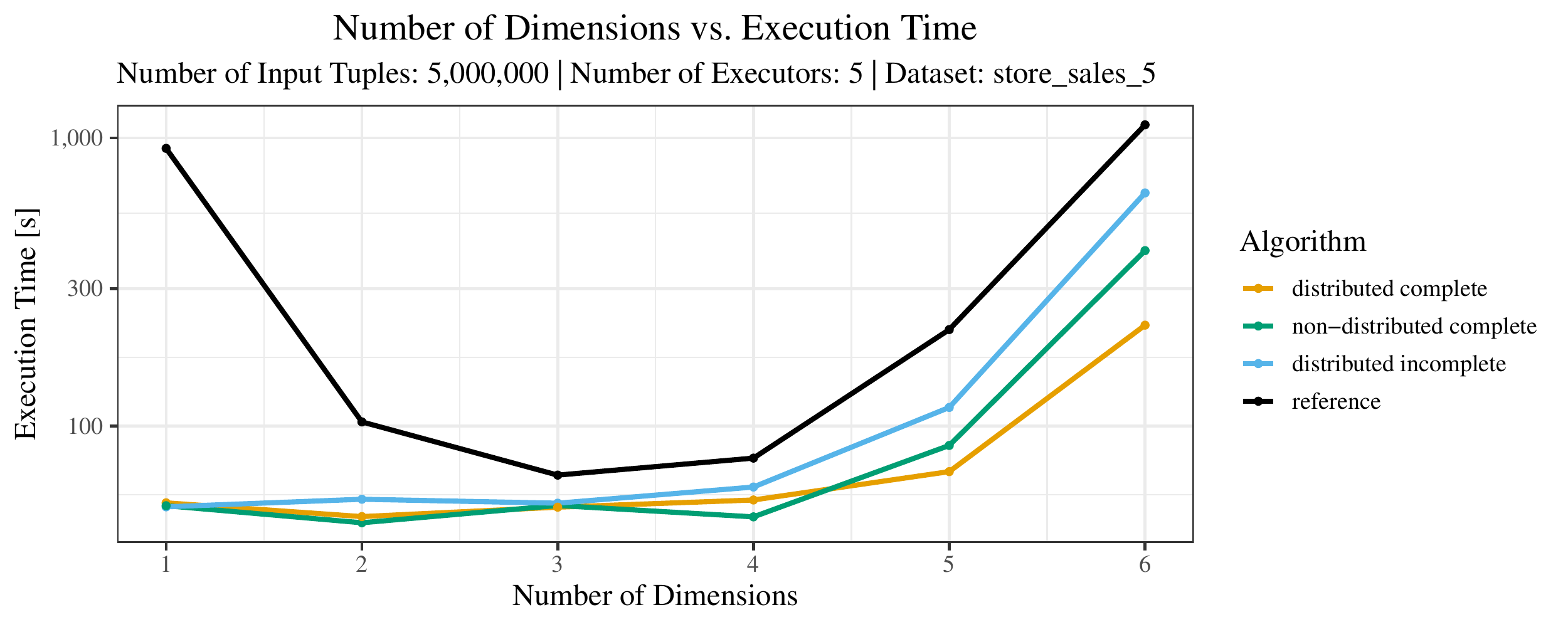}
    \end{subfigure}%
    \begin{subfigure}{.5\linewidth}
      \centering
      \includegraphics[width=\linewidth]{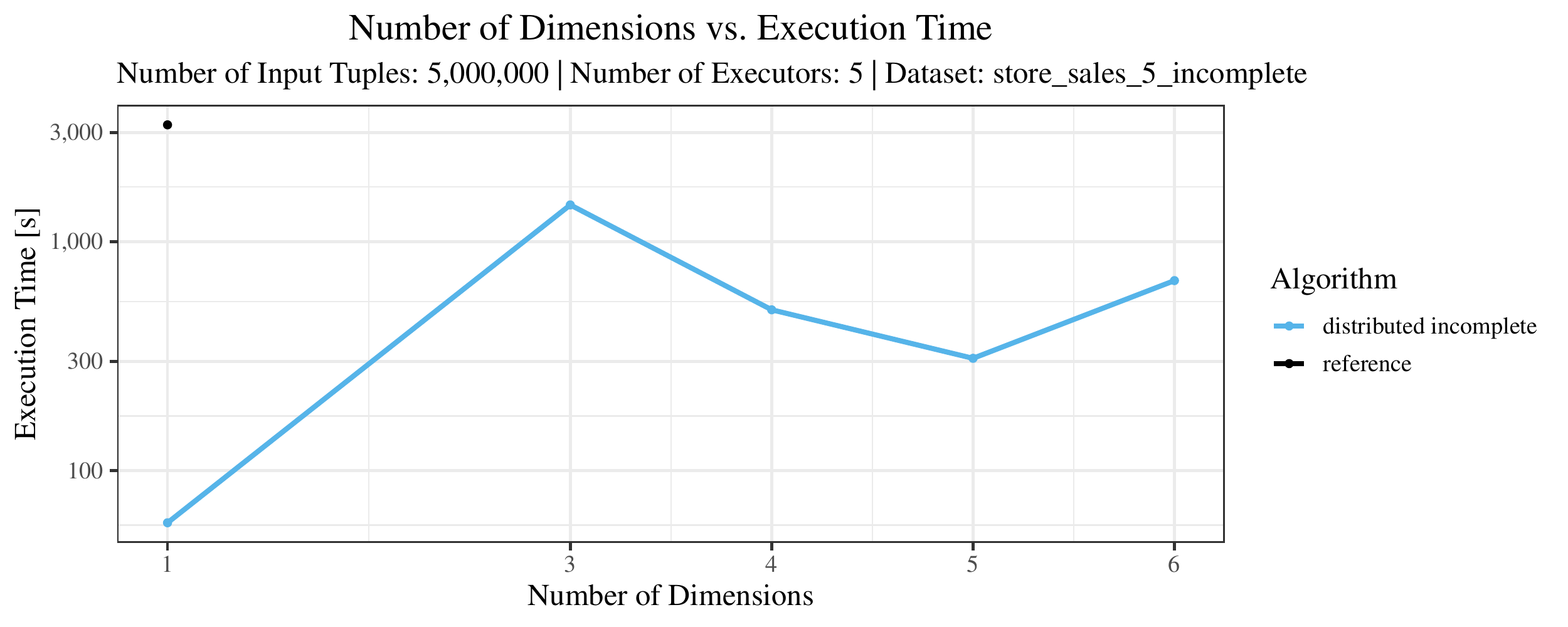}
    \end{subfigure}
    \begin{subfigure}{.5\linewidth}
      \centering
      \includegraphics[width=\linewidth]{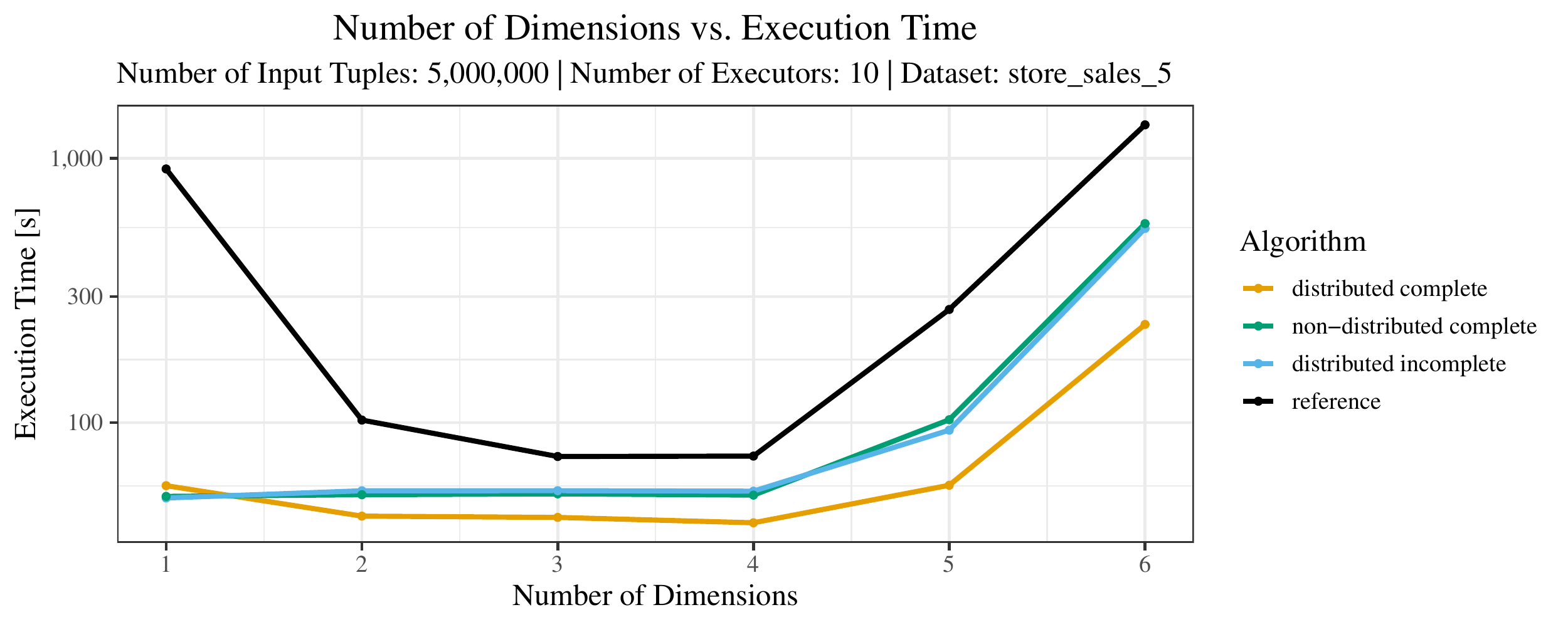}
    \end{subfigure}%
    \begin{subfigure}{.5\linewidth}
      \centering
      \includegraphics[width=\linewidth]{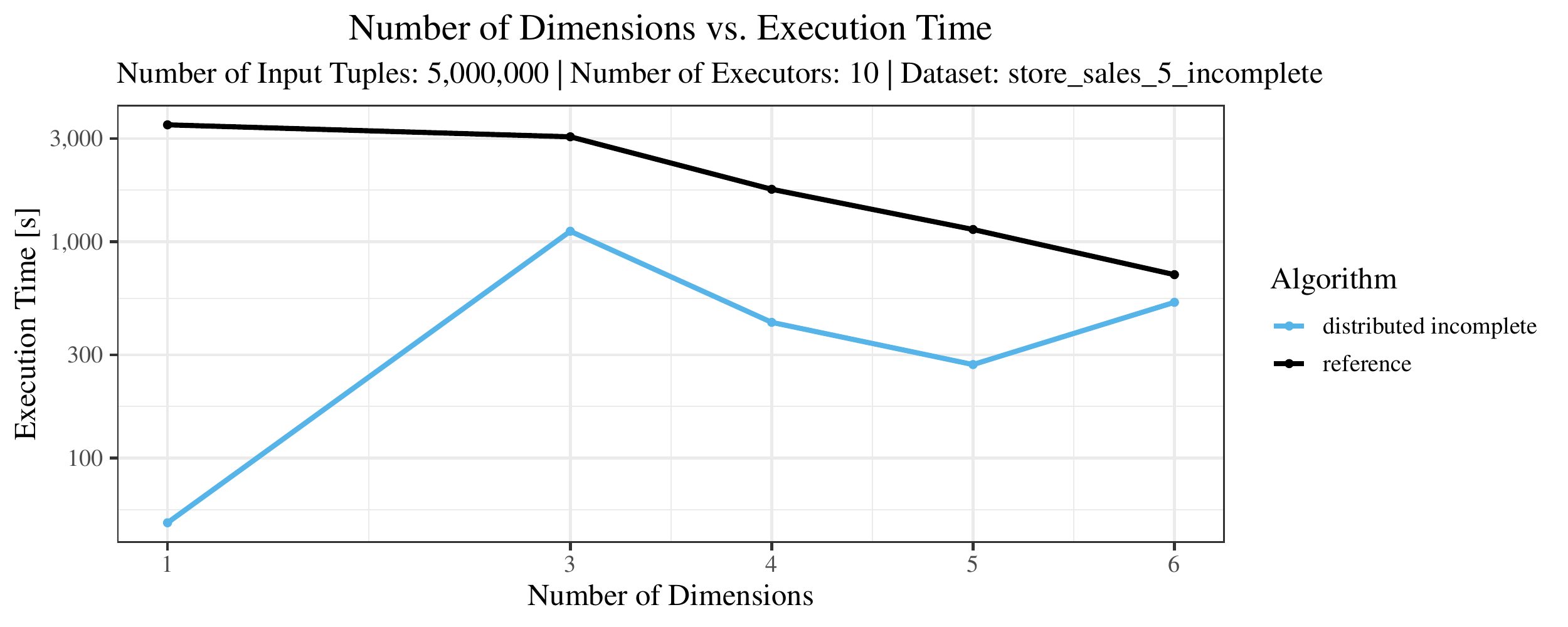}
    \end{subfigure}
    \caption{Number of dimensions vs. execution time on the store\_sales dataset}
    \label{fig:appendix_dimensions_vs_time_synthetic}
\end{figure*}

\begin{figure*}[p]
    \begin{subfigure}{.5\linewidth}
      \centering
      \includegraphics[width=\linewidth]{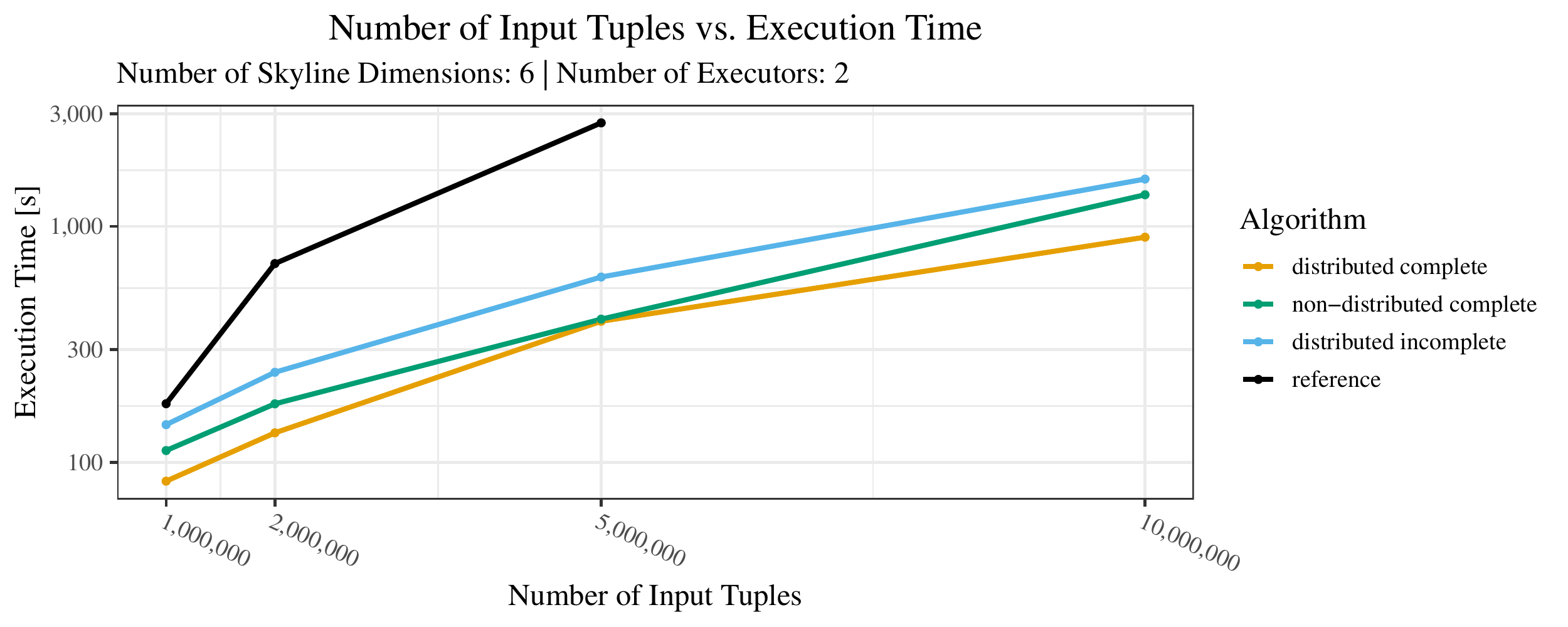}
    \end{subfigure}%
    \begin{subfigure}{.5\linewidth}
      \centering
      \includegraphics[width=\linewidth]{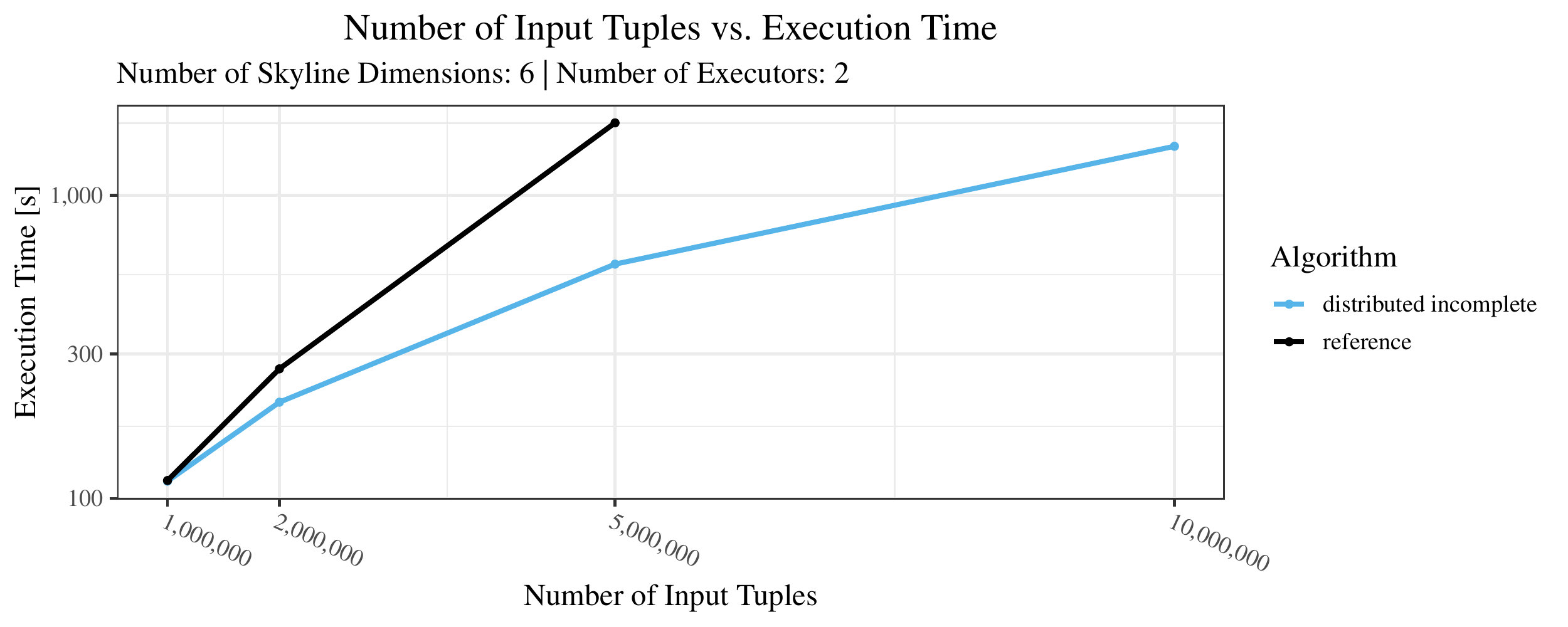}
    \end{subfigure}
    \begin{subfigure}{.5\linewidth}
      \centering
      \includegraphics[width=\linewidth]{Plots/benchmarks/appendix_cluster_synthetic/size_vs_time_complete-skyline-6d-3n.pdf}
    \end{subfigure}%
    \begin{subfigure}{.5\linewidth}
      \centering
      \includegraphics[width=\linewidth]{Plots/benchmarks/appendix_cluster_synthetic/size_vs_time_incomplete-skyline-6d-3n.pdf}
    \end{subfigure}
    \begin{subfigure}{.5\linewidth}
      \centering
      \includegraphics[width=\linewidth]{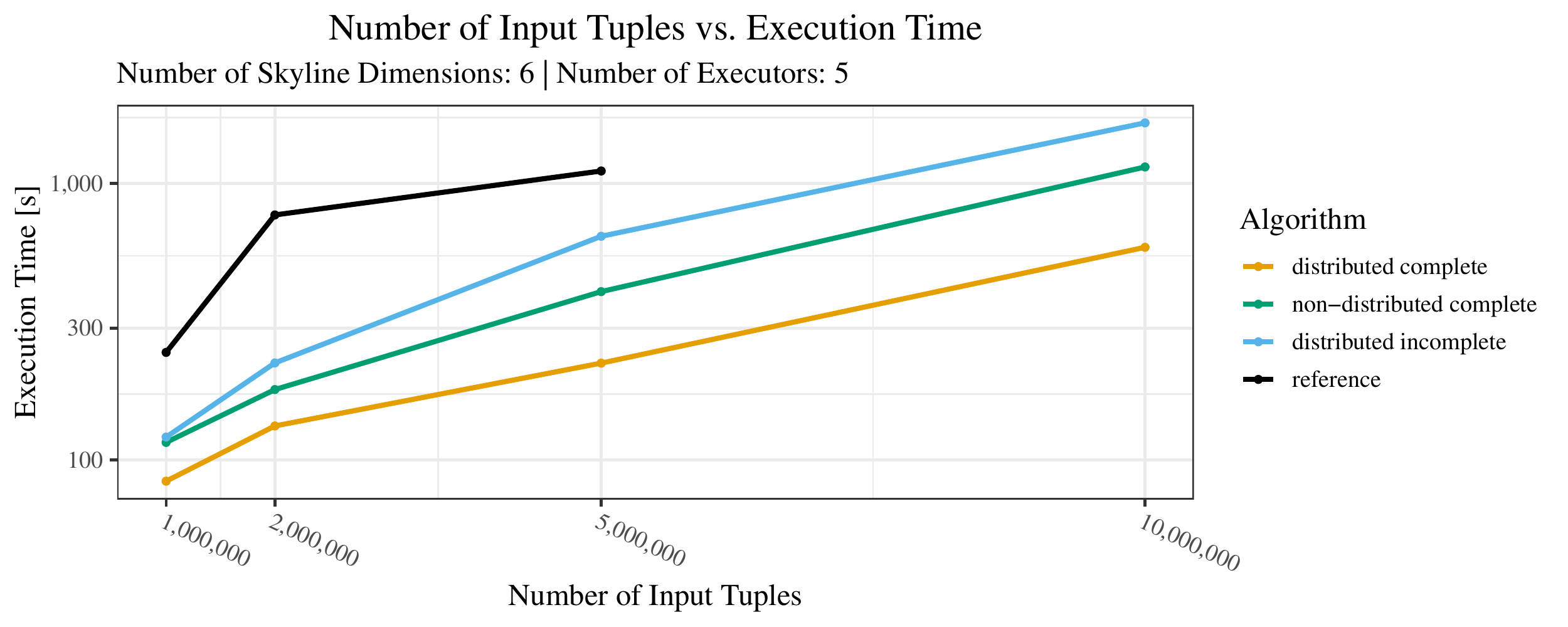}
    \end{subfigure}%
    \begin{subfigure}{.5\linewidth}
      \centering
      \includegraphics[width=\linewidth]{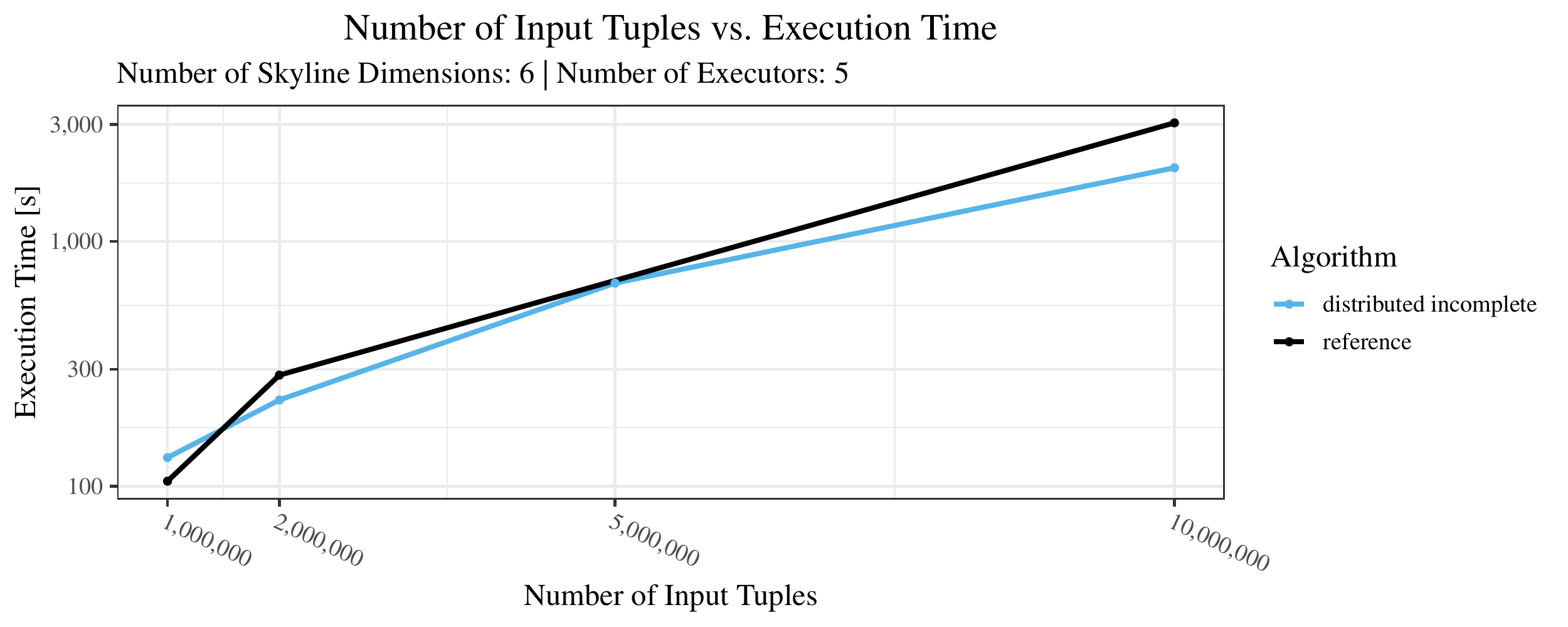}
    \end{subfigure}
    \begin{subfigure}{.5\linewidth}
      \centering
      \includegraphics[width=\linewidth]{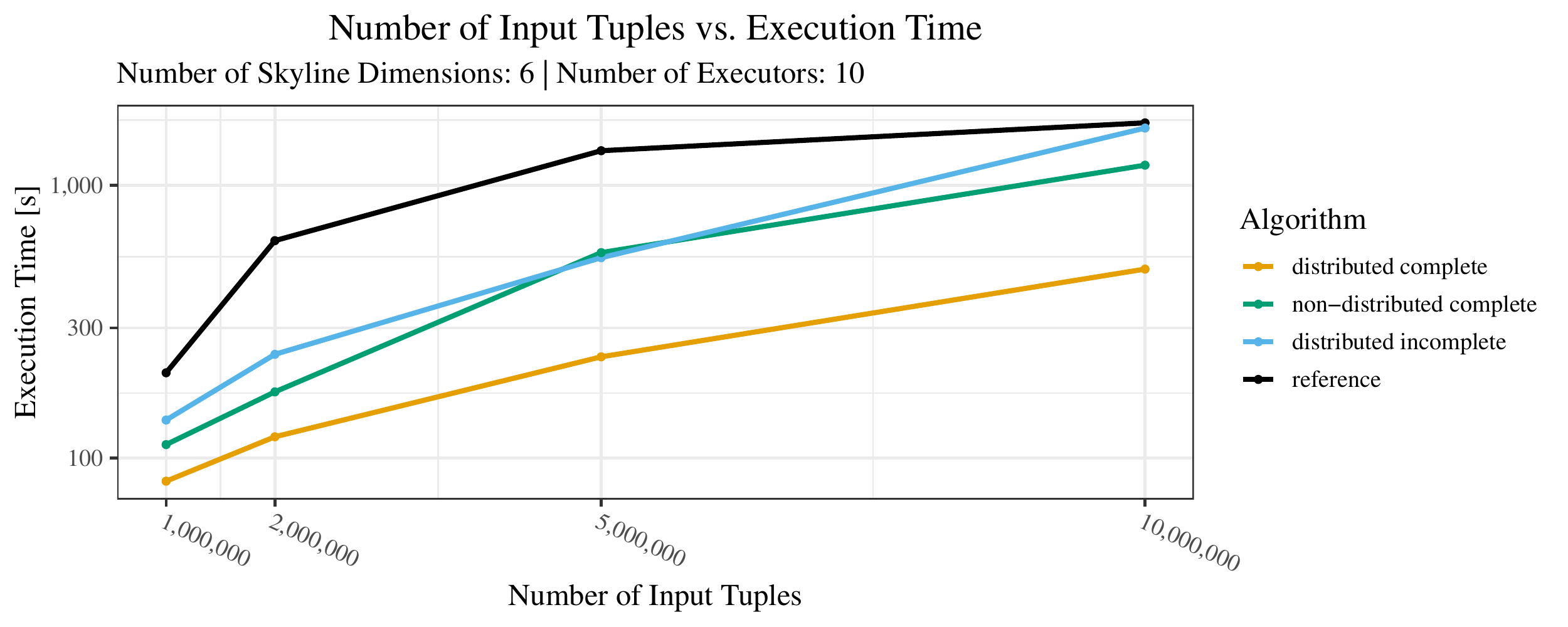}
    \end{subfigure}%
    \begin{subfigure}{.5\linewidth}
      \centering
      \includegraphics[width=\linewidth]{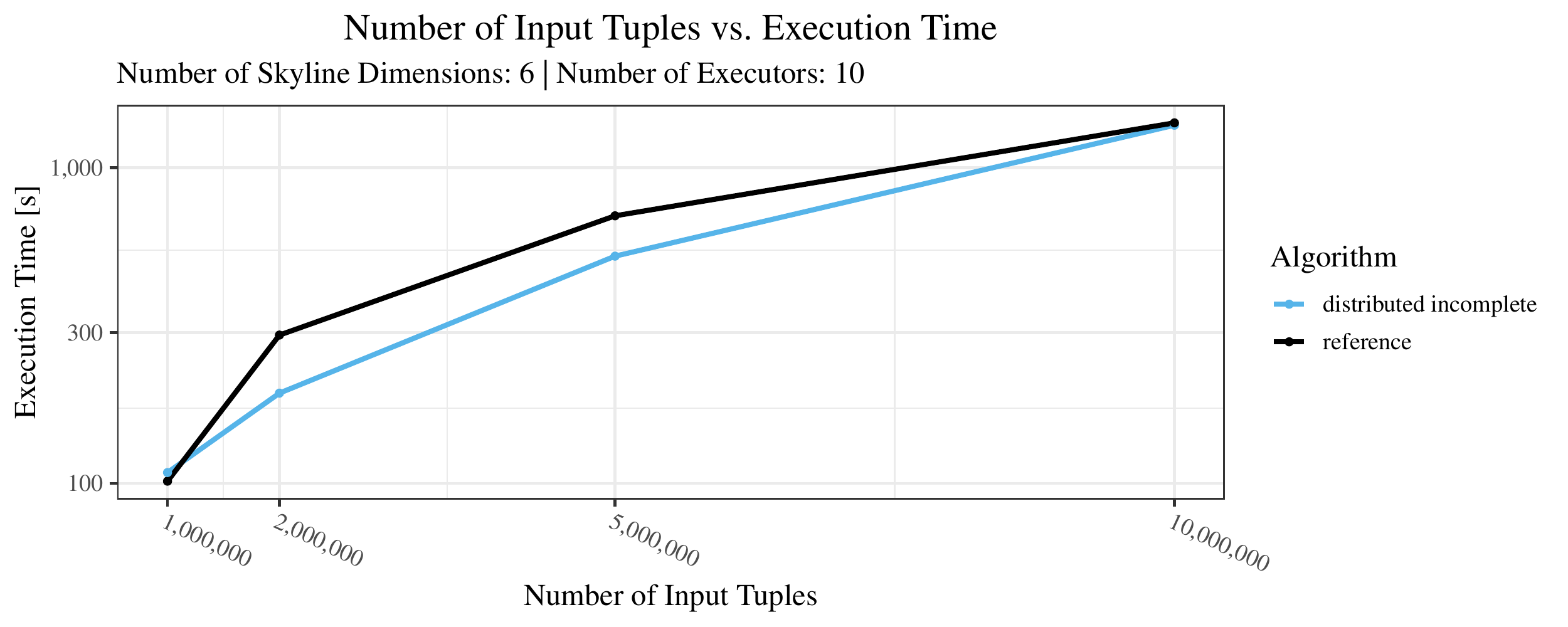}
    \end{subfigure}
    \caption{Number of input tuples vs. execution time  on the store\_sales dataset}
    \label{fig:appendix_size_vs_time_synthetic}
\end{figure*}

\begin{figure*}[p]
    \begin{subfigure}{.5\linewidth}
      \centering
      \includegraphics[width=\linewidth]{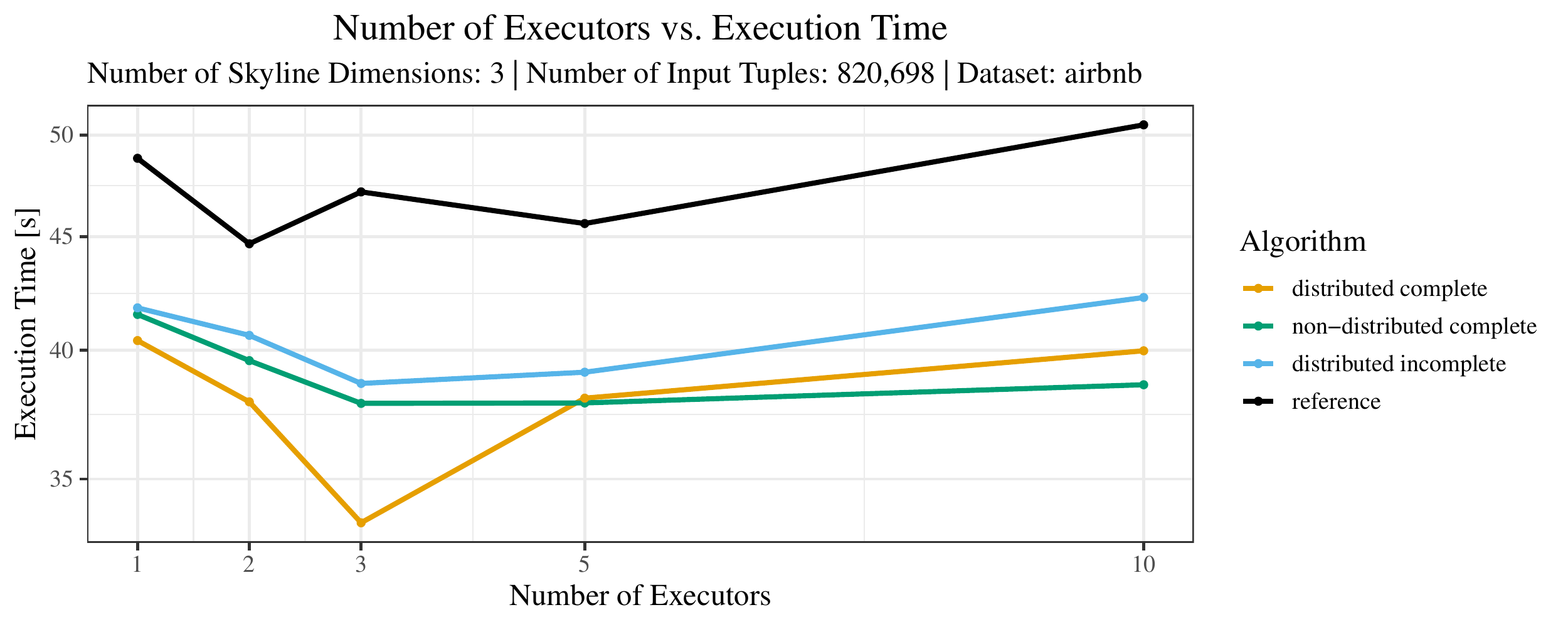}
    \end{subfigure}%
    \begin{subfigure}{.5\linewidth}
      \centering
      \includegraphics[width=\linewidth]{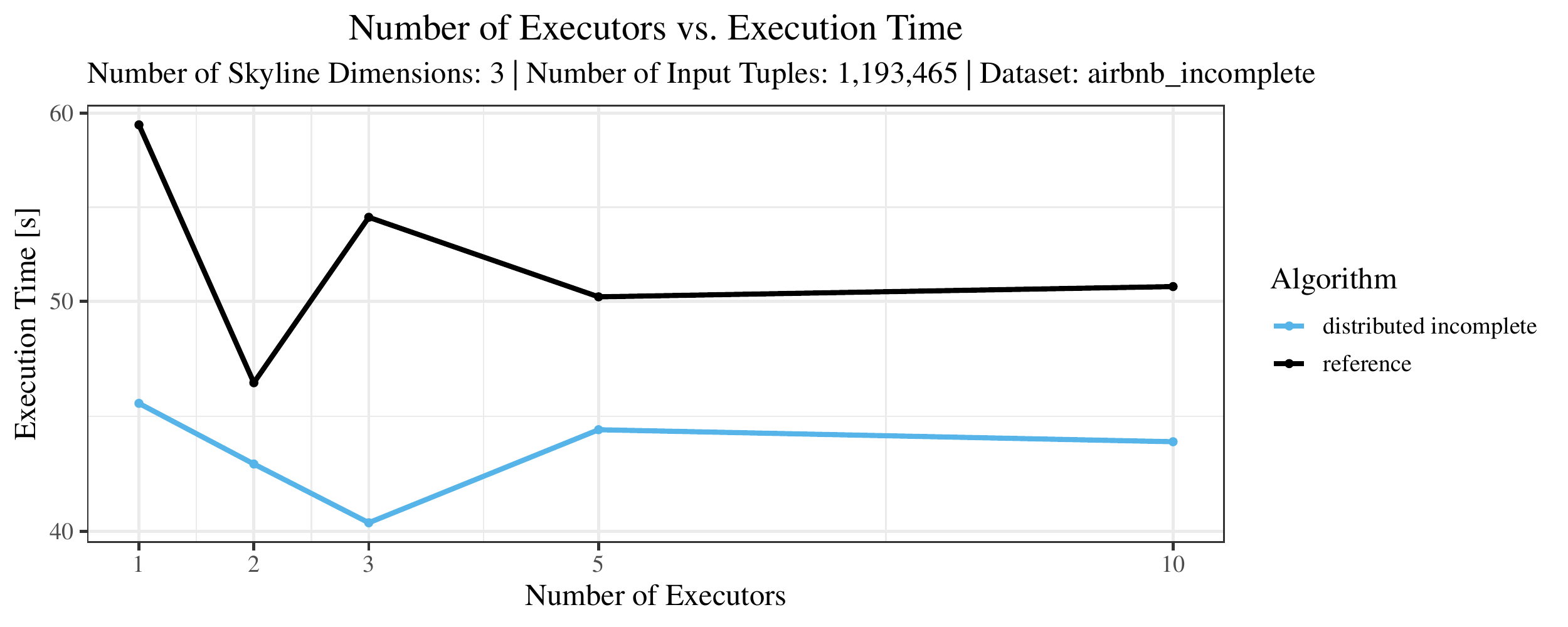}
    \end{subfigure}
    \begin{subfigure}{.5\linewidth}
      \centering
      \includegraphics[width=\linewidth]{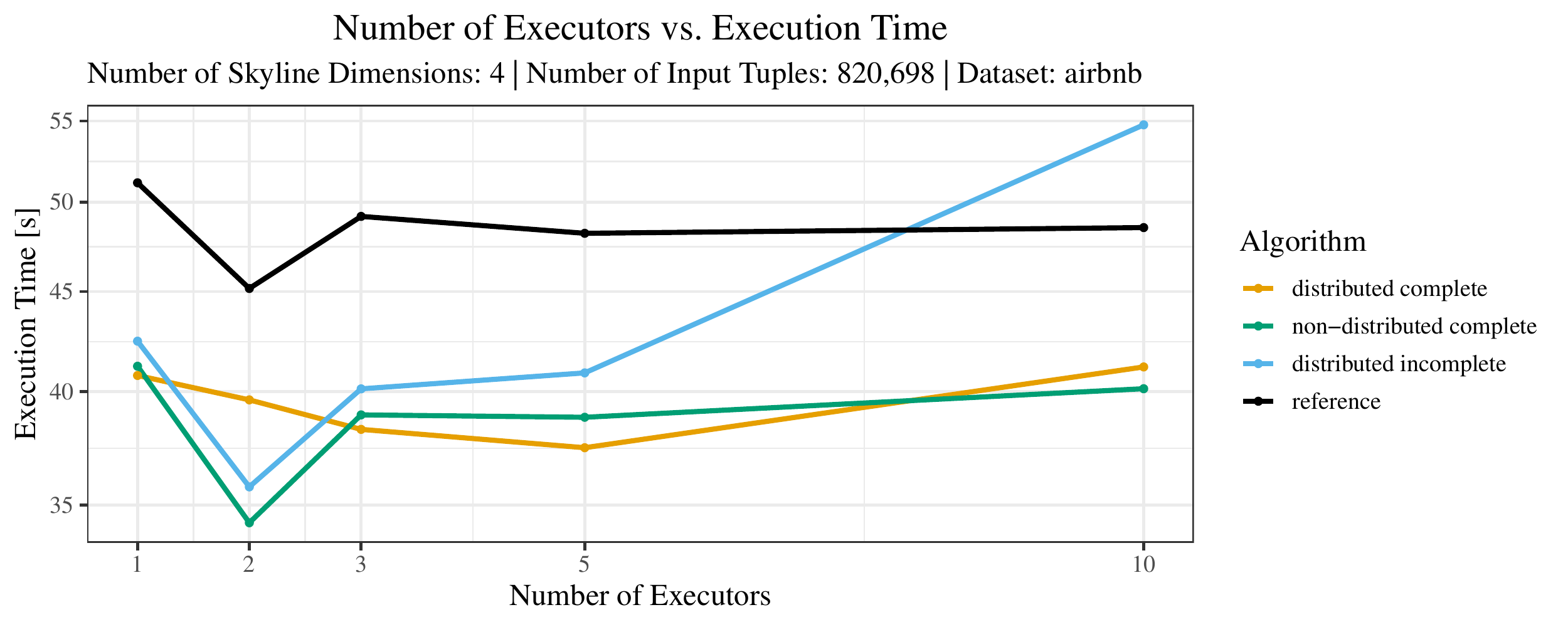}
    \end{subfigure}%
    \begin{subfigure}{.5\linewidth}
      \centering
      \includegraphics[width=\linewidth]{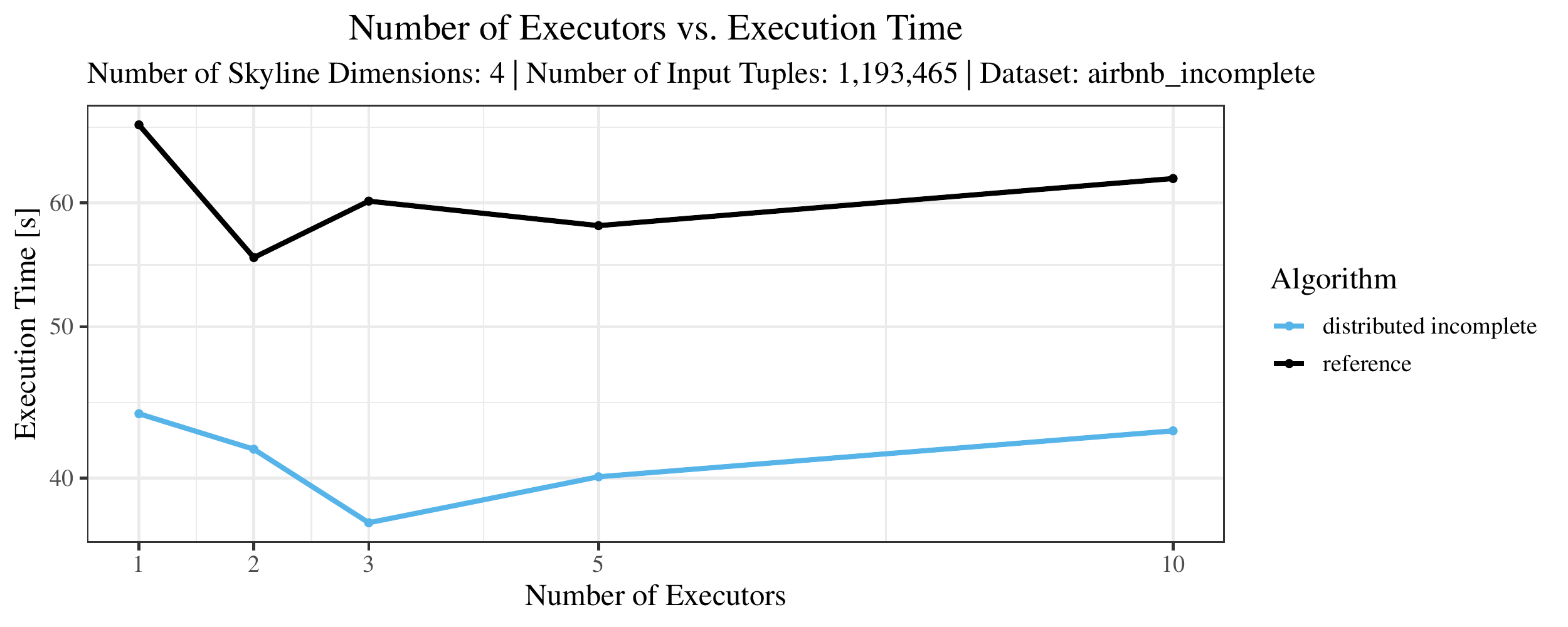}
    \end{subfigure}
    \begin{subfigure}{.5\linewidth}
      \centering
      \includegraphics[width=\linewidth]{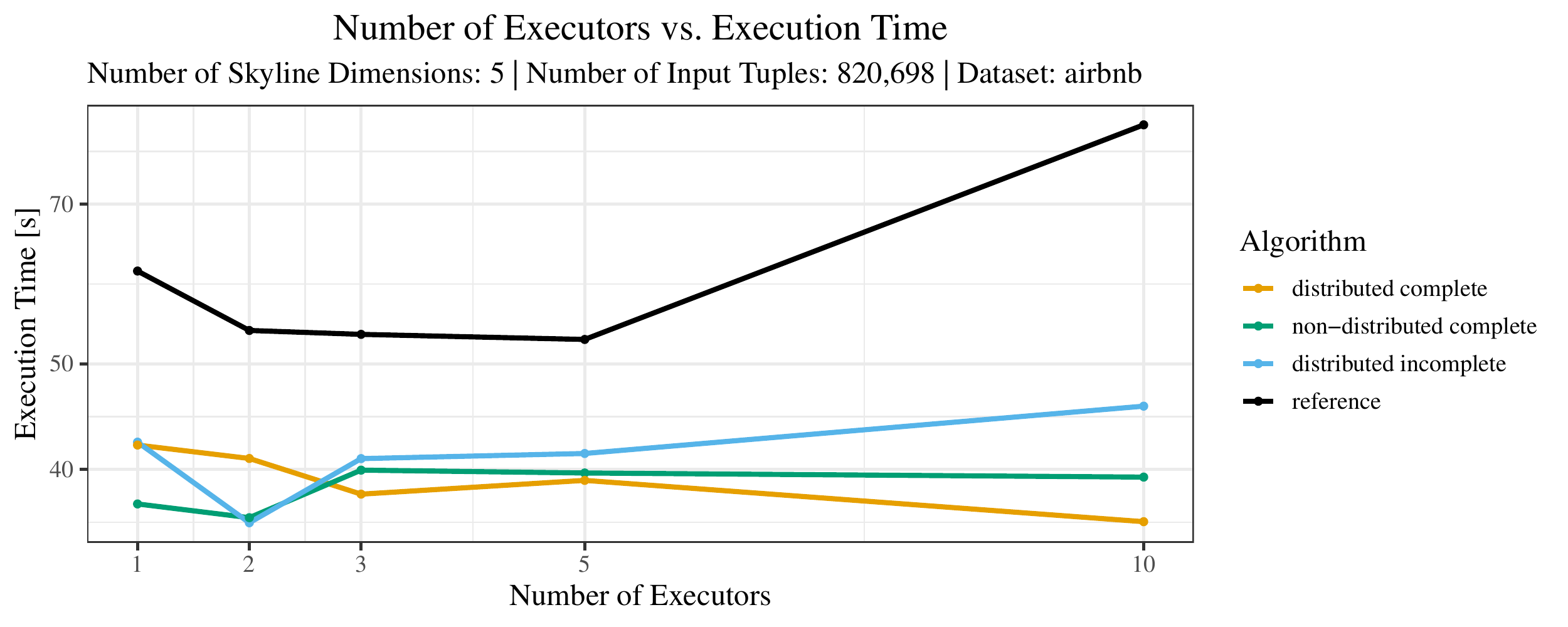}
    \end{subfigure}%
    \begin{subfigure}{.5\linewidth}
      \centering
      \includegraphics[width=\linewidth]{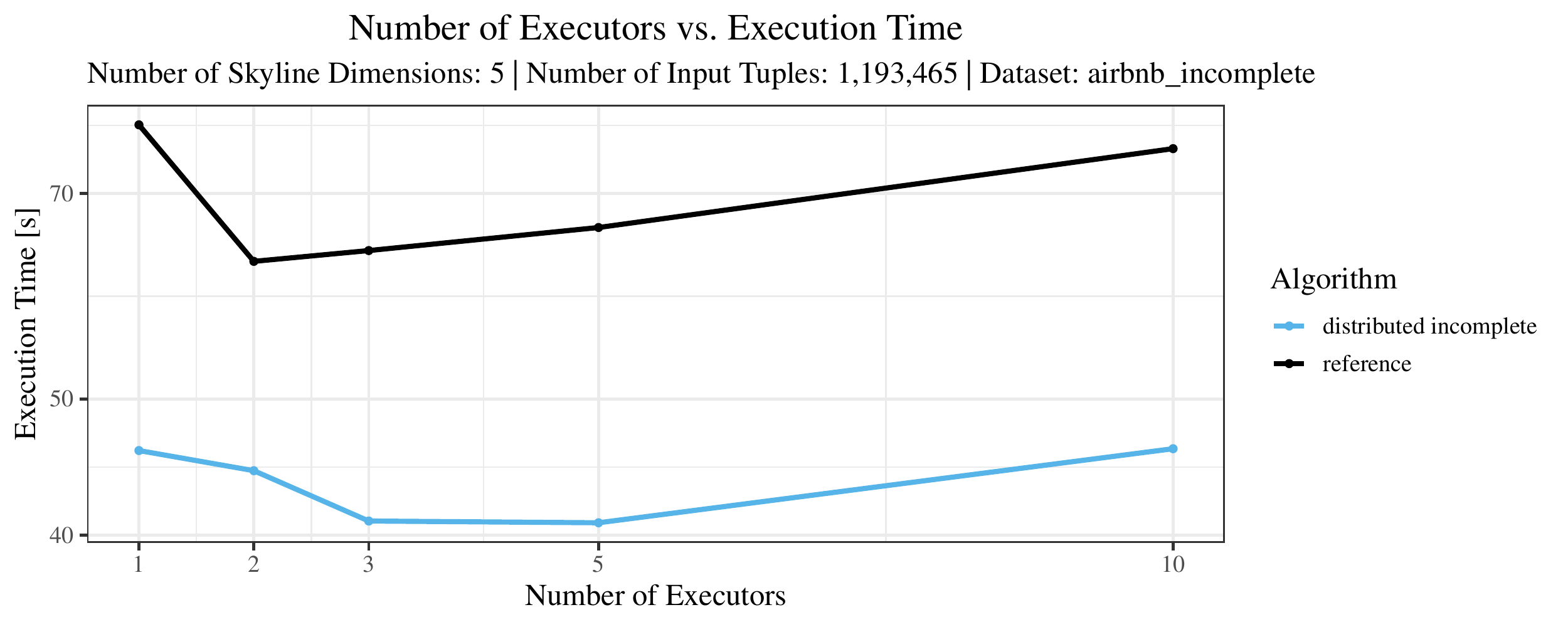}
    \end{subfigure}
    \begin{subfigure}{.5\linewidth}
      \centering
      \includegraphics[width=\linewidth]{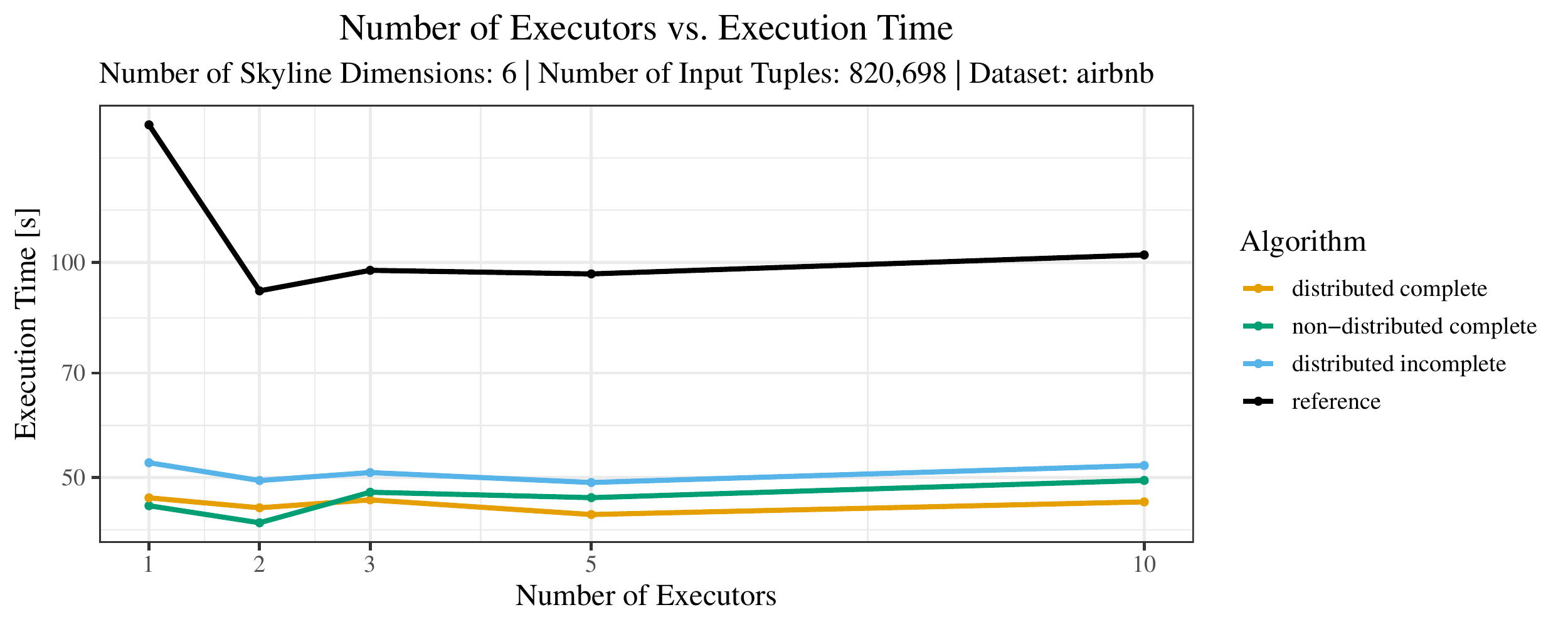}
    \end{subfigure}%
    \begin{subfigure}{.5\linewidth}
      \centering
      \includegraphics[width=\linewidth]{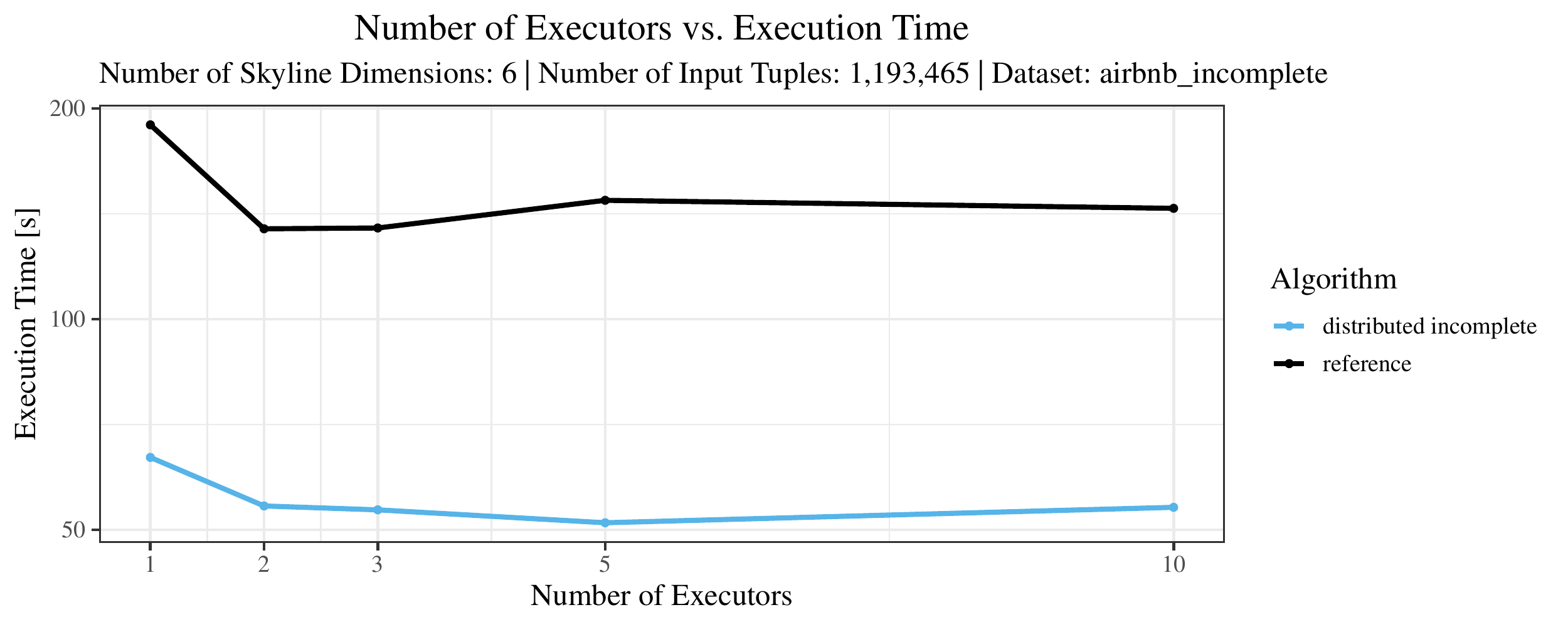}
    \end{subfigure}
    \caption{Number of executors vs. execution time on the Inside Airbnb dataset}
    \label{fig:appendix_cluster_executors_vs_time_real_world}
\end{figure*}

\begin{figure*}[p]
    \begin{subfigure}{.5\linewidth}
      \centering
      \includegraphics[width=\linewidth]{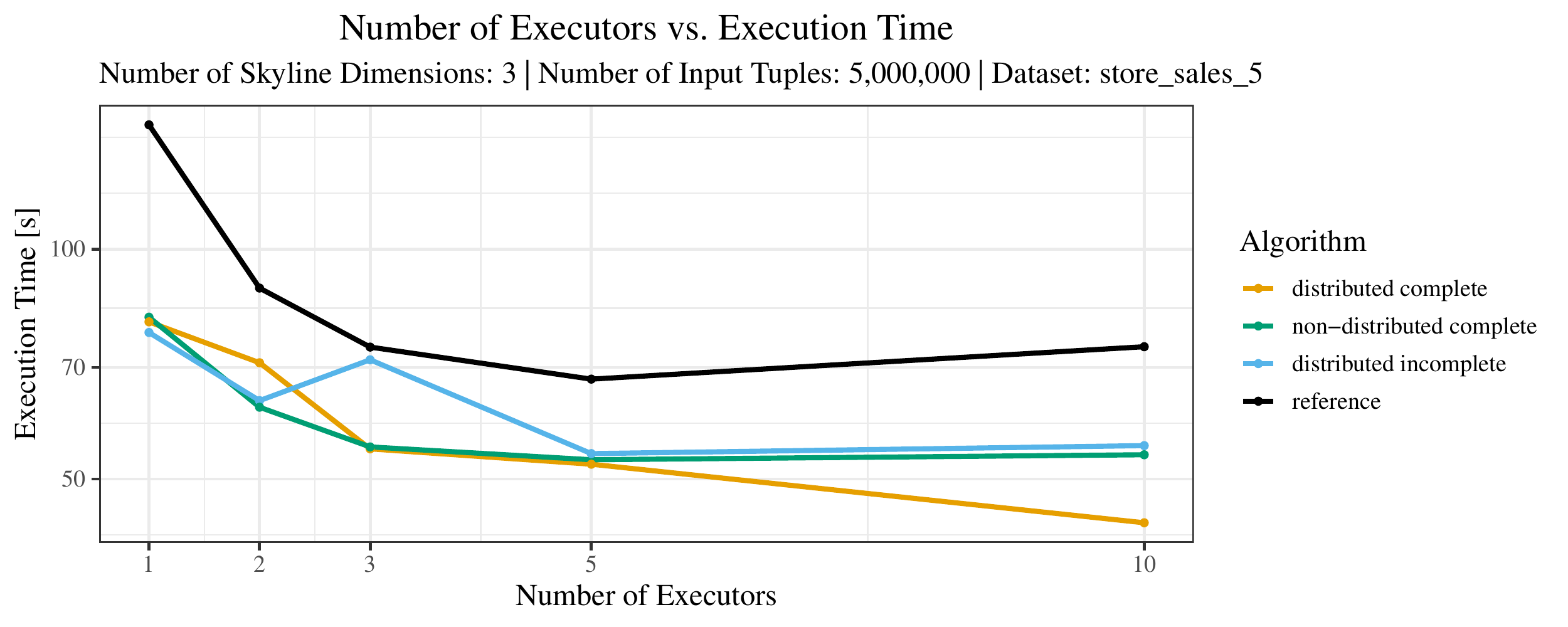}
    \end{subfigure}%
    \begin{subfigure}{.5\linewidth}
      \centering
      \includegraphics[width=\linewidth]{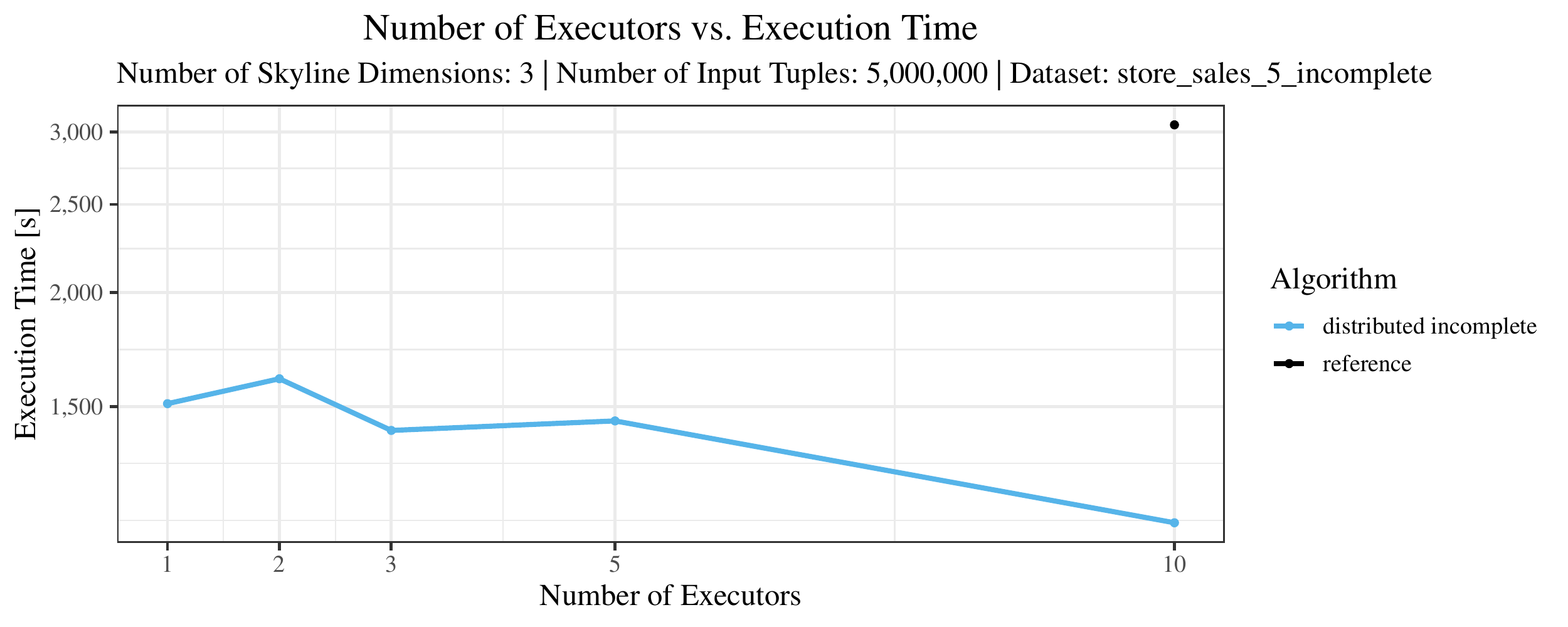}
    \end{subfigure}
    \begin{subfigure}{.5\linewidth}
      \centering
      \includegraphics[width=\linewidth]{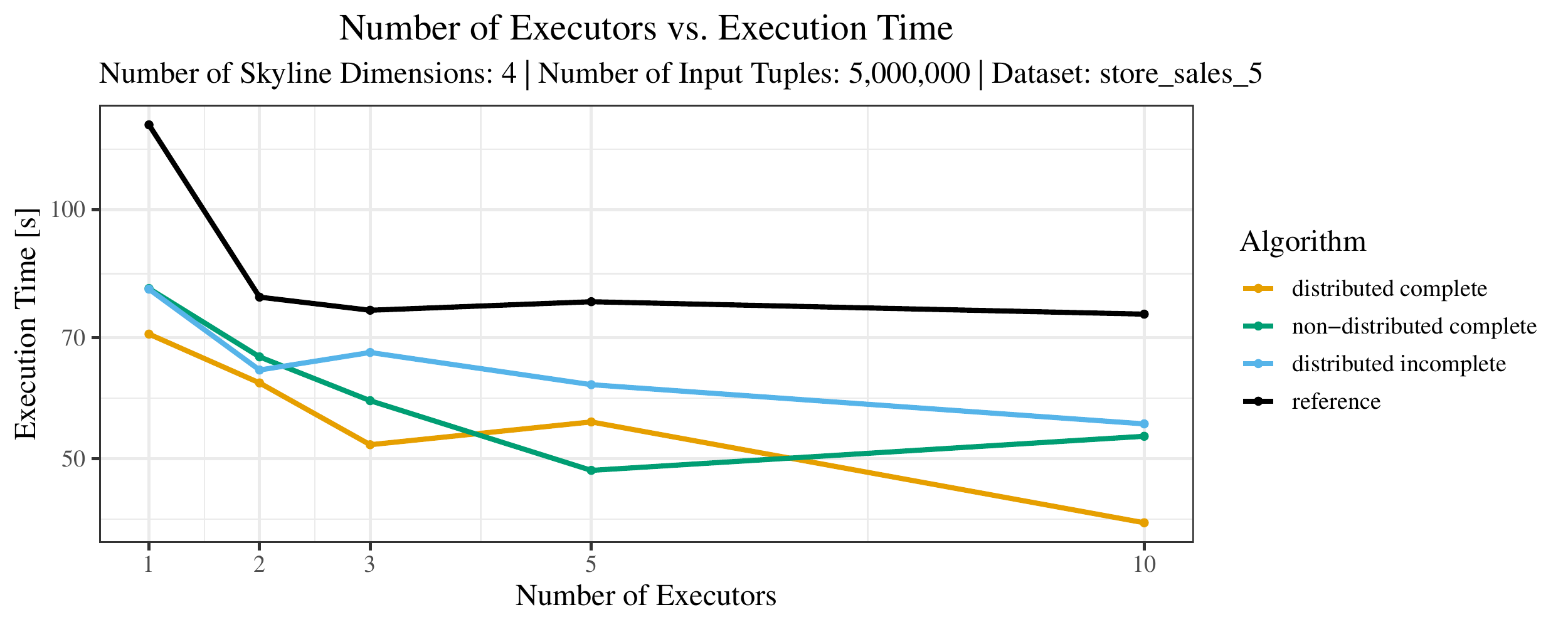}
    \end{subfigure}%
    \begin{subfigure}{.5\linewidth}
      \centering
      \includegraphics[width=\linewidth]{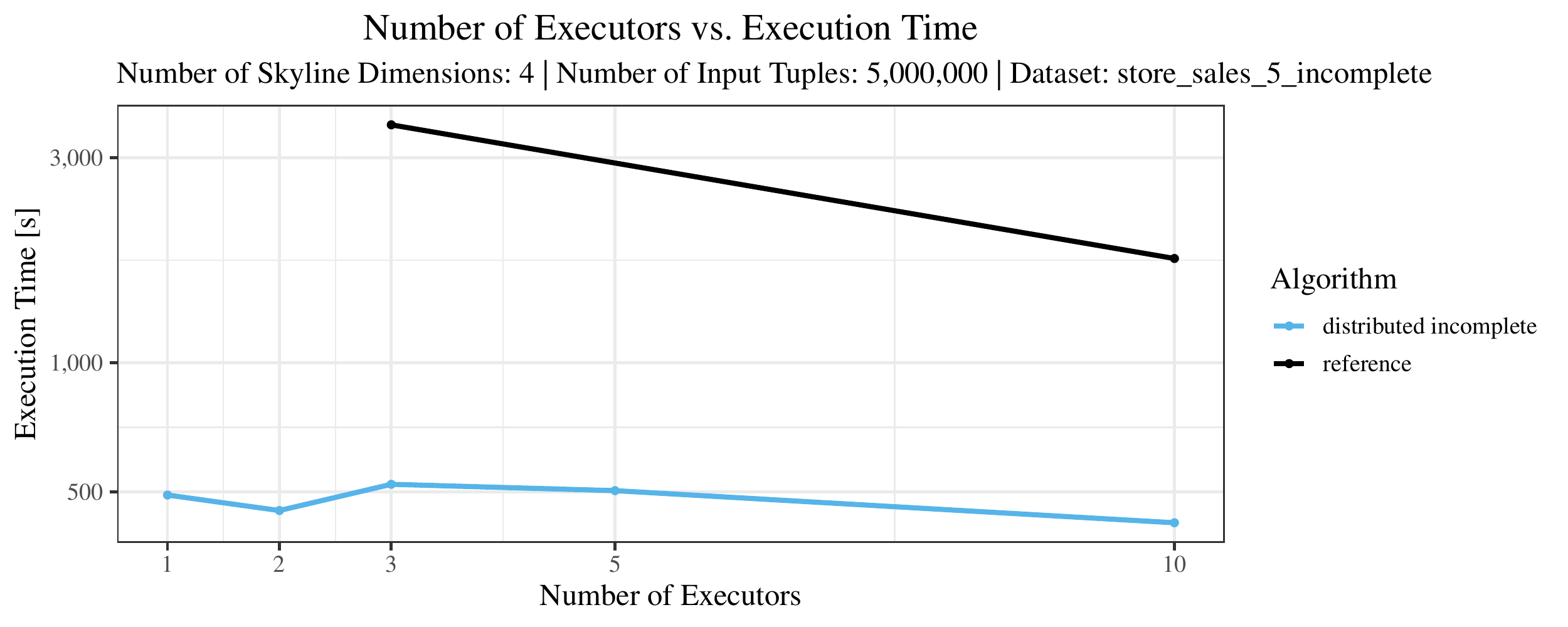}
    \end{subfigure}
    \begin{subfigure}{.5\linewidth}
      \centering
      \includegraphics[width=\linewidth]{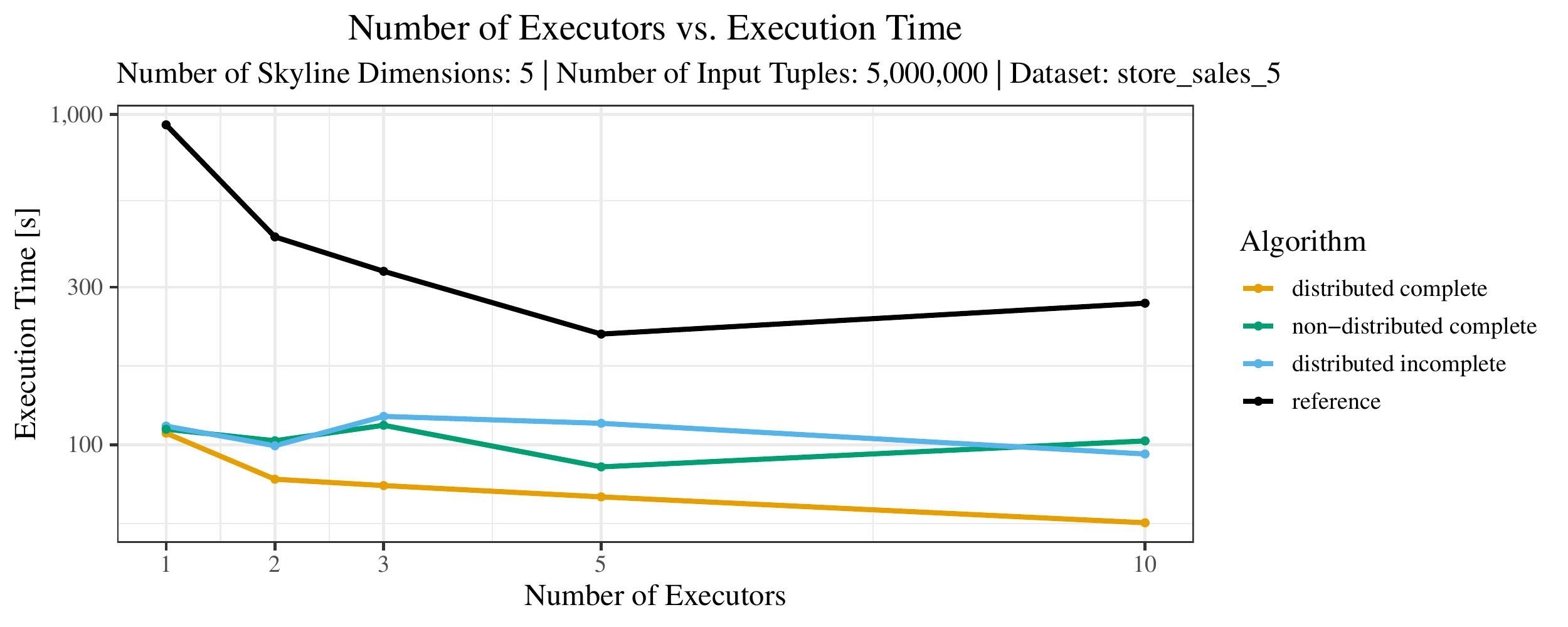}
    \end{subfigure}%
    \begin{subfigure}{.5\linewidth}
      \centering
      \includegraphics[width=\linewidth]{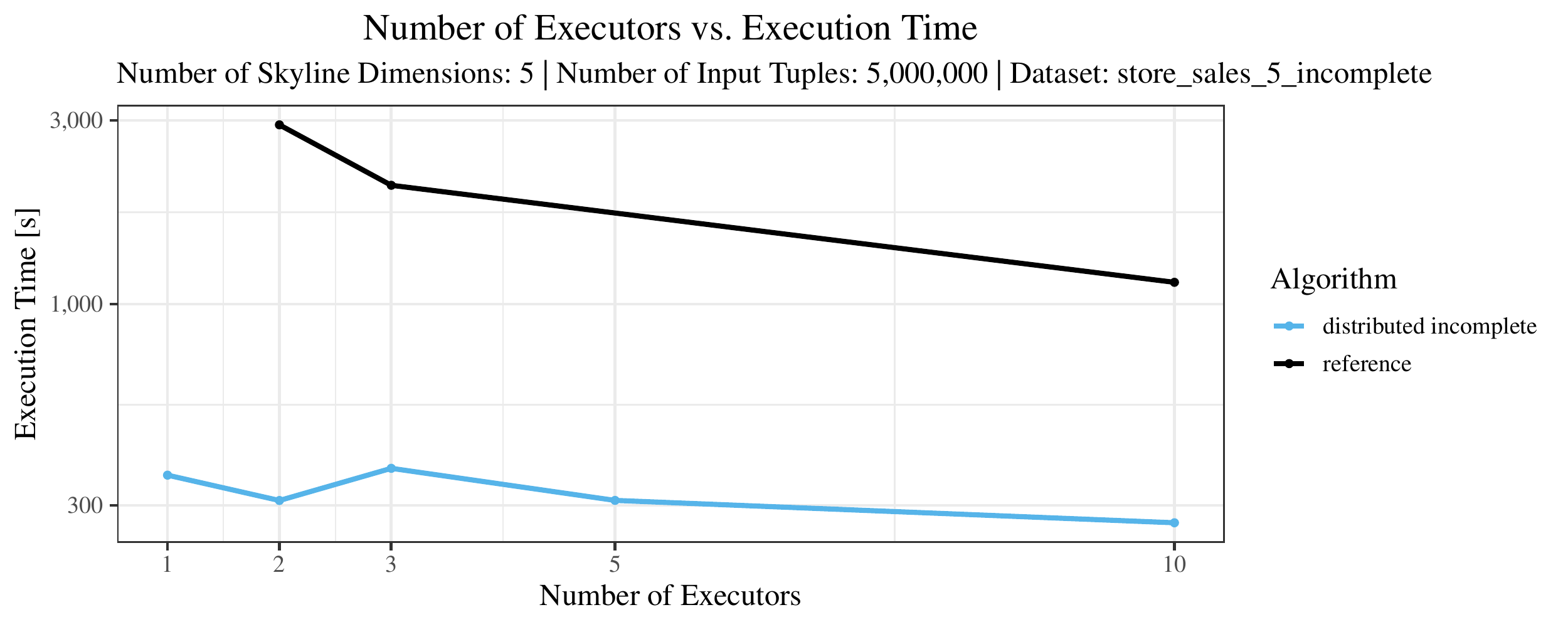}
    \end{subfigure}
    \begin{subfigure}{.5\linewidth}
      \centering
      \includegraphics[width=\linewidth]{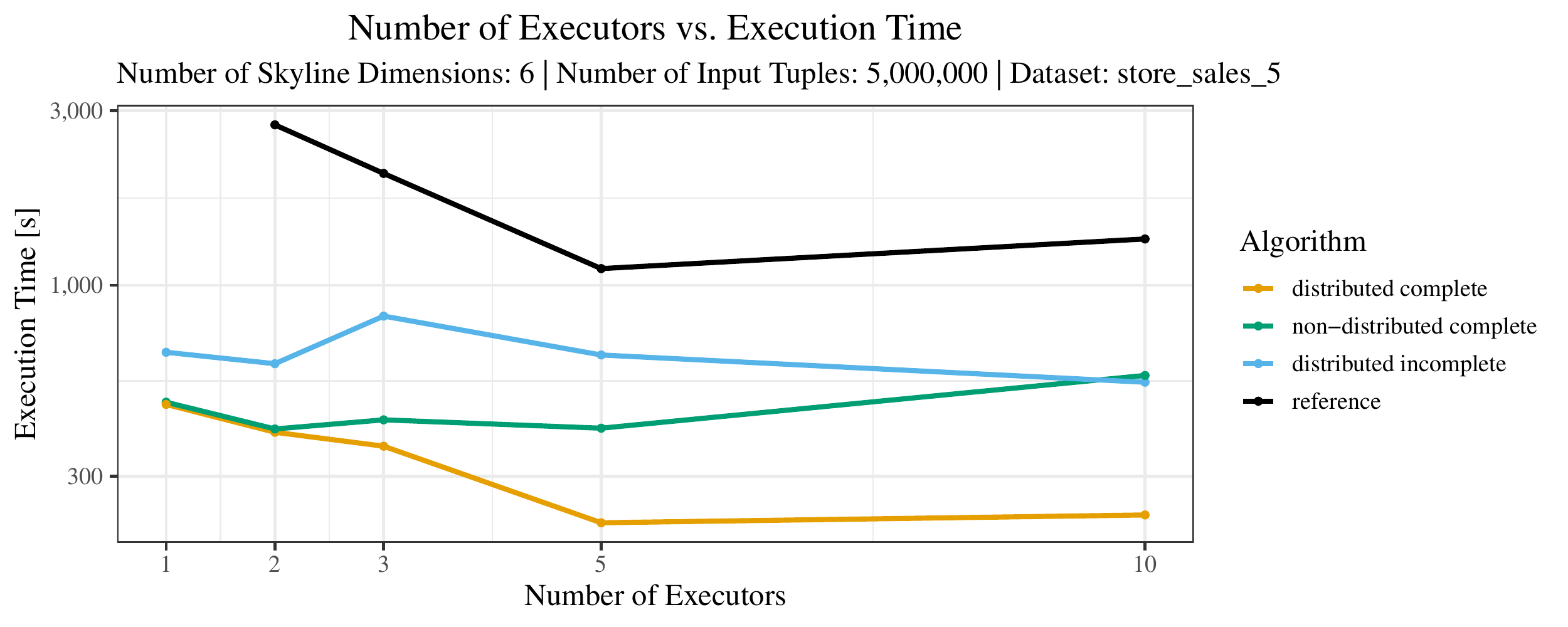}
    \end{subfigure}%
    \begin{subfigure}{.5\linewidth}
      \centering
      \includegraphics[width=\linewidth]{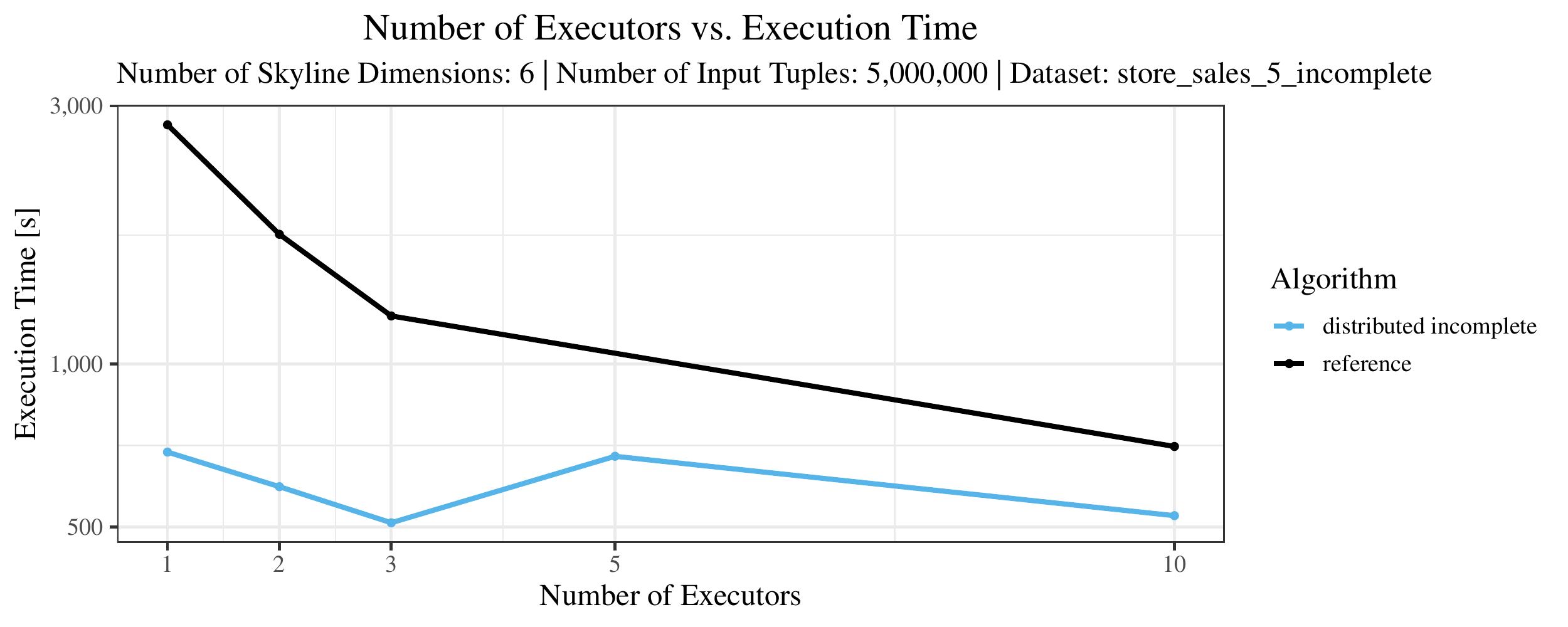}
    \end{subfigure}
    \caption{Number of executors vs. execution time on the store\_sales dataset}
    \label{fig:appendix_cluster_executors_vs_time_synthetic}
\end{figure*}

\clearpage
\newpage

\section{Tabulation of Execution Time}
\label{appendix:tabulation}

In this section, we revisit the experimental evaluation from 
Section~\ref{sect:experimental-results}. In Figures
\ref{fig:cluster_dimensions_vs_time_complete_airbnb} -- 
\ref{fig:cluster_nodes_vs_time_store_sales},
we presented plots of the execution times 
depending on the number of dimensions, tuples, and executors, respectively. 
We now present  the same results in table form to make the precise execution times easier to read. Since some queries do not finish in the allotted time frame, we use ``t.o.'' to mark such timeouts in the tables.

The tables also contain the execution times relative to the reference query. 
Here, we always denote the time consumed by the reference query as 100\%. 
For the other algorithms, a value below 100\% means a speed-up while a value above
100\% means a slow-down. If a timeout has occurred, no meaningful percentage value can be given. We use ``n.a.'' to denote that the calculation of a percentage value is not applicable. In fact, if the reference query has timed out, we use ``n.a.'' for the entire column since no comparison with the reference value is possible in this case even if the specialized algorithms do not time out. 
Note that we never have the opposite situation that a specialized algorithm times out but not the reference algorithm.

The results are shown in Tables \ref{table_dimension_vs_time-airbnb-skyline-820698t-5n}
--
\ref{table_nodes_vs_time-store_sales_5_incomplete-skyline-6d-5000000t}.
The correspondence with the plots in Figures
\ref{fig:cluster_dimensions_vs_time_complete_airbnb}
--
\ref{fig:cluster_nodes_vs_time_store_sales}
is as follows: 

\begin{itemize}
\item The values of the plots in 
Figure~\ref{fig:cluster_dimensions_vs_time_complete_airbnb}
(number of dimensions vs. execution time on the Inside Airbnb dataset)
are shown in 
Table~\ref{table_dimension_vs_time-airbnb-skyline-820698t-5n}
for the complete dataset and 
in 
Table~\ref{table_dimension_vs_time-airbnb_incomplete-skyline-1193465t-5n}
for the incomplete dataset.

\item The values of the plots in 
Figure~\ref{fig:cluster_dimensions_vs_time_complete_store_sales}
(number of dimensions vs. execution time on the store\_sales dataset)
are shown in 
Table~\ref{table_dimension_vs_time-store_sales_10-skyline-10000000t-10n}
for the complete dataset and in 
Table~\ref{table_dimension_vs_time-store_sales_1_incomplete-skyline-1000000t-10n}
for the incomplete dataset.

\item The values of the plots in 
Figure~\ref{fig:cluster_size_vs_time}
(number of input tuples vs. execution time on the store\_sales dataset)
are shown in 
Table~\ref{table_size_vs_time-store_sales_10-skyline-6d-3n}
for the complete dataset and in 
Table~\ref{table_size_vs_time-store_sales_incomplete-skyline-6d-3n}
for the incomplete dataset.

\item The values of the plots in 
Figure~\ref{fig:cluster_nodes_vs_time_airbnb}
(number of executors vs. execution time on the Inside Airbnb dataset)
are shown in 
Table~\ref{table_nodes_vs_time-airbnb-skyline-6d-820698t}
for the complete dataset and in 
Table~\ref{table_nodes_vs_time-airbnb_incomplete-skyline-6d-1193465t}
for the incomplete dataset.

\item The values of the plots in 
Figure~\ref{fig:cluster_nodes_vs_time_store_sales}
(number of executors vs. execution time on the store\_sales dataset)
are shown in 
Table~\ref{table_nodes_vs_time-store_sales_10-skyline-6d-10000000t}
for the complete dataset and in 
Table~\ref{table_nodes_vs_time-store_sales_5_incomplete-skyline-6d-5000000t}
for the incomplete dataset.
\end{itemize}

From these tables, it becomes very apparent that if the reference query can execute the skyline in a reasonable time, then the computation with our 
specialized algorithms displays a similar or even better behavior.
Actually, a notable fact with regards to the results discussed in 
Section~\ref{sect:experimental-results} can be derived from 
Table~\ref{table_dimension_vs_time-store_sales_1_incomplete-skyline-1000000t-10n}. Here, we encounter one of the rare cases where the reference query is better than our implementation. However, we also see that the increase is only $\sim 6.6$ seconds or $\sim 6.5\%$. While this is an interesting result, it is unlikely to ever have an impact on the performance or usability of our integration of the skyline operator.

\begin{table*}[ht]
\begin{tabular}{R{0.25cm}L{3.25cm}R{1.25cm}R{1.25cm}R{1.25cm}R{1.25cm}R{1.25cm}R{1.25cm}}
  \toprule
   & algorithm & 1 & 2 & 3 & 4 & 5 & 6 \\ 
  \midrule
  1 & reference & 100.00\% & 100.00\% & 100.00\% & 100.00\% & 100.00\% & 100.00\% \\ 
  2 & non-distributed complete & 96.60\% & 86.88\% & 83.03\% & 80.54\% & 75.45\% & 48.66\% \\ 
  3 & distributed complete & 96.42\% & 97.81\% & 83.44\% & 77.71\% & 74.28\% & 46.08\% \\ 
  4 & distributed incomplete & 81.09\% & 90.92\% & 85.72\% & 84.86\% & 78.62\% & 51.10\% \\ 
   \bottomrule
\end{tabular}
\begin{tabular}{R{0.25cm}L{3.25cm}R{1.25cm}R{1.25cm}R{1.25cm}R{1.25cm}R{1.25cm}R{1.25cm}}
  \toprule
 & algorithm & 1 & 2 & 3 & 4 & 5 & 6 \\ 
  \midrule
  1 & reference & 43.72 & 43.94 & 45.61 & 48.19 & 52.63 & 96.34 \\ 
  2 & non-distributed complete & 42.23 & 38.17 & 37.87 & 38.81 & 39.71 & 46.88 \\ 
  3 & distributed complete & 42.15 & 42.98 & 38.06 & 37.45 & 39.09 & 44.40 \\ 
  4 & distributed incomplete & 35.45 & 39.95 & 39.10 & 40.90 & 41.38 & 49.23 \\ 
   \bottomrule
\end{tabular}
\caption{Number of dimensions vs. execution time on complete Inside Airbnb dataset (Executors: 5, tuples: 820698)} 
\label{table_dimension_vs_time-airbnb-skyline-820698t-5n}
\end{table*}

\begin{table*}[ht]
\begin{tabular}{R{0.25cm}L{3.25cm}R{1.25cm}R{1.25cm}R{1.25cm}R{1.25cm}R{1.25cm}R{1.25cm}}
  \toprule
 & algorithm & 1 & 2 & 3 & 4 & 5 & 6 \\ 
  \midrule
1 & reference & 100.00\% & 100.00\% & 100.00\% & 100.00\% & 100.00\% & 100.00\% \\ 
  2 & distributed incomplete & 83.07\% & 69.62\% & 87.92\% & 69.08\% & 61.64\% & 34.61\% \\ 
   \bottomrule
\end{tabular}
\begin{tabular}{R{0.25cm}L{3.25cm}R{1.25cm}R{1.25cm}R{1.25cm}R{1.25cm}R{1.25cm}R{1.25cm}}
  \toprule
 & algorithm & 1 & 2 & 3 & 4 & 5 & 6 \\ 
  \midrule
1 & reference & 45.58 & 50.05 & 50.21 & 58.03 & 66.22 & 147.82 \\ 
  2 & distributed incomplete & 37.87 & 34.85 & 44.15 & 40.08 & 40.82 & 51.17 \\ 
   \bottomrule
\end{tabular}
\caption{Number of dimensions vs. execution time on incomplete Inside Airbnb dataset (Executors: 5, tuples: 1193465)} 
\label{table_dimension_vs_time-airbnb_incomplete-skyline-1193465t-5n}
\end{table*}

\begin{table*}[ht]
\centering
\begin{tabular}{R{0.25cm}L{3.25cm}R{1.25cm}R{1.25cm}R{1.25cm}R{1.25cm}R{1.25cm}R{1.25cm}}
  \toprule
 & algorithm & 1 & 2 & 3 & 4 & 5 & 6 \\ 
  \midrule
1 & reference & 100.00\% & 100.00\% & 100.00\% & 100.00\% & 100.00\% & 100.00\% \\ 
  2 & non-distributed complete & 2.63\% & 34.41\% & 54.98\% & 63.06\% & 46.26\% & 69.97\% \\ 
  3 & distributed complete & 2.20\% & 27.42\% & 56.83\% & 45.35\% & 22.22\% & 29.12\% \\ 
  4 & distributed incomplete & 2.30\% & 36.12\% & 57.33\% & 59.74\% & 65.95\% & 95.69\% \\ 
   \bottomrule
\end{tabular}
\begin{tabular}{R{0.25cm}L{3.25cm}R{1.25cm}R{1.25cm}R{1.25cm}R{1.25cm}R{1.25cm}R{1.25cm}}
  \toprule
 & algorithm & 1 & 2 & 3 & 4 & 5 & 6 \\ 
  \midrule
1 & reference & 2463.29 & 164.18 & 105.44 & 93.86 & 281.42 & 1693.31 \\ 
  2 & non-distributed complete & 64.77 & 56.50 & 57.97 & 59.18 & 130.17 & 1184.86 \\ 
  3 & distributed complete & 54.26 & 45.02 & 59.93 & 42.56 & 62.52 & 493.03 \\ 
  4 & distributed incomplete & 56.58 & 59.29 & 60.45 & 56.06 & 185.60 & 1620.40 \\ 
   \bottomrule
\end{tabular}
\caption{Number of dimensions vs. execution time on the complete store\_sales dataset (Executors: 10, tuples: 10000000)} 
\label{table_dimension_vs_time-store_sales_10-skyline-10000000t-10n}
\end{table*}

\begin{table*}[ht]
\centering
\begin{tabular}{R{0.25cm}L{3.25cm}R{1.25cm}R{1.25cm}R{1.25cm}R{1.25cm}R{1.25cm}R{1.25cm}}
  \toprule
 & algorithm & 1 & 2 & 3 & 4 & 5 & 6 \\ 
  \midrule
1 & reference & 100.00\% & 100.00\% & 100.00\% & 100.00\% & 100.00\% & 100.00\% \\ 
  2 & distributed incomplete & 14.60\% & 47.59\% & 33.80\% & 25.98\% & 36.61\% & 106.51\% \\ 
   \bottomrule
\end{tabular}
\begin{tabular}{R{0.25cm}L{3.25cm}R{1.25cm}R{1.25cm}R{1.25cm}R{1.25cm}R{1.25cm}R{1.25cm}}
  \toprule
 & algorithm & 1 & 2 & 3 & 4 & 5 & 6 \\ 
  \midrule
1 & reference & 314.47 & 354.12 & 258.10 & 197.19 & 149.84 & 101.62 \\ 
  2 & distributed incomplete & 45.92 & 168.53 & 87.23 & 51.23 & 54.86 & 108.23 \\ 
   \bottomrule
\end{tabular}
\caption{Number of dimensions vs. execution time on the incomplete store\_sales dataset (Executors: 10, tuples: 1000000)} 
\label{table_dimension_vs_time-store_sales_1_incomplete-skyline-1000000t-10n}
\end{table*}

\begin{table*}[ht]
\centering
\begin{tabular}{R{0.25cm}L{3.25cm}R{1.25cm}R{1.25cm}R{1.25cm}R{1.25cm}R{1.25cm}R{1.25cm}}
  \toprule
 & algorithm & 1000000 & 2000000 & 5000000 & 10000000 \\ 
  \midrule
1 & reference & 100.00\% & 100.00\% & 100.00\% & n.a. \\ 
  2 & non-distributed complete & 56.23\% & 30.91\% & 21.18\% & n.a. \\ 
  3 & distributed complete & 42.51\% & 23.46\% & 17.94\% & n.a. \\ 
  4 & distributed incomplete & 72.73\% & 44.69\% & 40.72\% & n.a. \\ 
   \bottomrule
\end{tabular}
\begin{tabular}{R{0.25cm}L{3.25cm}R{1.25cm}R{1.25cm}R{1.25cm}R{1.25cm}R{1.25cm}R{1.25cm}}
  \toprule
 & algorithm & 1000000 & 2000000 & 5000000 & 10000000 \\ 
  \midrule
1 & reference & 191.35 & 542.55 & 2022.67 & t.o. \\ 
  2 & non-distributed complete & 107.59 & 167.69 & 428.31 & 1184.43 \\ 
  3 & distributed complete & 81.35 & 127.31 & 362.83 & 723.11 \\ 
  4 & distributed incomplete & 139.16 & 242.47 & 823.56 & 1705.47 \\ 
   \bottomrule
\end{tabular}
\caption{Number of tuples vs. execution time on complete store\_sales dataset (executors: 3, dimensions: 6)} 
\label{table_size_vs_time-store_sales_10-skyline-6d-3n}
\end{table*}

\begin{table*}[ht]
\centering
\begin{tabular}{R{0.25cm}L{3.25cm}R{1.25cm}R{1.25cm}R{1.25cm}R{1.25cm}R{1.25cm}R{1.25cm}}
  \toprule
 & algorithm & 1000000 & 2000000 & 5000000 & 10000000 \\ 
  \midrule
1 & reference & 100.00\% & 100.00\% & 100.00\% & n.a. \\ 
  2 & distributed incomplete & 109.52\% & 73.15\% & 41.47\% & n.a. \\ 
   \bottomrule
\end{tabular}
\begin{tabular}{R{0.25cm}L{3.25cm}R{1.25cm}R{1.25cm}R{1.25cm}R{1.25cm}R{1.25cm}R{1.25cm}}
  \toprule
 & algorithm & 1000000 & 2000000 & 5000000 & 10000000 \\ 
  \midrule
1 & reference & 101.17 & 282.33 & 1227.49 & t.o. \\ 
  2 & distributed incomplete & 110.80 & 206.53 & 509.06 & 1342.98 \\ 
   \bottomrule
\end{tabular}
\caption{Number of tuples vs. execution time on incomplete store\_sales dataset (Executors: 3, dimensions: 6)} 
\label{table_size_vs_time-store_sales_incomplete-skyline-6d-3n}
\end{table*}

\begin{table*}[ht]
\centering
\begin{tabular}{R{0.25cm}L{3.25cm}R{1.25cm}R{1.25cm}R{1.25cm}R{1.25cm}R{1.25cm}R{1.25cm}}
  \toprule
 & algorithm & 1 & 2 & 3 & 5 & 10 \\ 
  \midrule
1 & reference & 100.00\% & 100.00\% & 100.00\% & 100.00\% & 100.00\% \\ 
  2 & non-distributed complete & 29.34\% & 47.40\% & 48.95\% & 48.66\% & 48.38\% \\ 
  3 & distributed complete & 30.09\% & 49.74\% & 47.76\% & 46.08\% & 45.15\% \\ 
  4 & distributed incomplete & 33.69\% & 54.30\% & 52.15\% & 51.10\% & 50.77\% \\ 
   \bottomrule
\end{tabular}
\begin{tabular}{R{0.25cm}L{3.25cm}R{1.25cm}R{1.25cm}R{1.25cm}R{1.25cm}R{1.25cm}R{1.25cm}}
  \toprule
 & algorithm & 1 & 2 & 3 & 5 & 10 \\ 
  \midrule
1 & reference & 155.69 & 91.23 & 97.47 & 96.34 & 102.42 \\ 
  2 & non-distributed complete & 45.68 & 43.24 & 47.71 & 46.88 & 49.55 \\ 
  3 & distributed complete & 46.85 & 45.37 & 46.55 & 44.40 & 46.24 \\ 
  4 & distributed incomplete & 52.46 & 49.54 & 50.82 & 49.23 & 52.00 \\ 
   \bottomrule
\end{tabular}
\caption{Number of executors vs. execution time on Inside Airbnb dataset (tuples: 820698, dimensions: 6)} 
\label{table_nodes_vs_time-airbnb-skyline-6d-820698t}
\end{table*}

\begin{table*}[ht]
\centering
\begin{tabular}{R{0.25cm}L{3.25cm}R{1.25cm}R{1.25cm}R{1.25cm}R{1.25cm}R{1.25cm}R{1.25cm}}
  \toprule
 & algorithm & 1 & 2 & 3 & 5 & 10 \\ 
  \midrule
1 & reference & 100.00\% & 100.00\% & 100.00\% & 100.00\% & 100.00\% \\ 
  2 & distributed incomplete & 33.48\% & 40.18\% & 39.56\% & 34.61\% & 37.39\% \\ 
   \bottomrule
\end{tabular}
\begin{tabular}{R{0.25cm}L{3.25cm}R{1.25cm}R{1.25cm}R{1.25cm}R{1.25cm}R{1.25cm}R{1.25cm}}
  \toprule
 & algorithm & 1 & 2 & 3 & 5 & 10 \\ 
  \midrule
1 & reference & 189.48 & 134.60 & 134.95 & 147.82 & 143.96 \\ 
  2 & distributed incomplete & 63.44 & 54.08 & 53.39 & 51.17 & 53.83 \\ 
   \bottomrule
\end{tabular}
\caption{Number of executors vs. execution time on incomplete Inside Airbnb dataset (tuples: 1193465, dimensions: 6)} 
\label{table_nodes_vs_time-airbnb_incomplete-skyline-6d-1193465t}
\end{table*}

\begin{table*}[ht]
\centering
\begin{tabular}{R{0.25cm}L{3.25cm}R{1.25cm}R{1.25cm}R{1.25cm}R{1.25cm}R{1.25cm}R{1.25cm}}
  \toprule
 & algorithm & 1 & 2 & 3 & 5 & 10 \\ 
  \midrule
1 & reference & n.a. & n.a. & n.a. & n.a. & 100.00\%  \\ 
  2 & non-distributed complete & n.a. & n.a. & n.a. & n.a. & 69.97\%\\ 
  3 & distributed complete & n.a. & n.a. & n.a. & n.a. & 29.12\% \\ 
  4 & distributed incomplete & n.a. & n.a. & n.a. & n.a. & 95.69\%  \\ 
   \bottomrule
\end{tabular}
\begin{tabular}{R{0.25cm}L{3.25cm}R{1.25cm}R{1.25cm}R{1.25cm}R{1.25cm}R{1.25cm}R{1.25cm}}
  \toprule
 & algorithm & 1 & 2 & 3 & 5 & 10 \\ 
  \midrule
1 & reference & t.o. & t.o. & t.o. & t.o. & 1693.31 \\ 
  2 & non-distributed complete  & 1154.56 & 1361.92 & 1184.43 & 1148.37 & 1184.86\\ 
  3 & distributed complete & 1080.17 & 899.84 & 723.11 & 587.94 & 493.03 \\ 
  4 & distributed incomplete & 1686.67 & 1587.29 & 1705.47 & 1658.78 & 1620.40\\ 
   \bottomrule
\end{tabular}
\caption{Number of executors vs. execution time on complete store\_sales dataset (tuples: 10000000, dimensions: 6)} 
\label{table_nodes_vs_time-store_sales_10-skyline-6d-10000000t}
\end{table*}

\begin{table*}[ht]
\centering
\begin{tabular}{R{0.25cm}L{3.25cm}R{1.25cm}R{1.25cm}R{1.25cm}R{1.25cm}R{1.25cm}R{1.25cm}}
  \toprule
 & algorithm & 1 & 2 & 3 & 5 & 10 \\ 
  \midrule
1 & reference & 100.00\% & 100.00\% & 100.00\% & n.a. & 100.00\% \\ 
  2 & distributed incomplete & 24.85\% & 34.17\% & 41.47\% & n.a. & 74.49\% \\ 
   \bottomrule
\end{tabular}
\begin{tabular}{R{0.25cm}L{3.25cm}R{1.25cm}R{1.25cm}R{1.25cm}R{1.25cm}R{1.25cm}R{1.25cm}}
  \toprule
 & algorithm & 1 & 2 & 3 & 5 & 10 \\ 
  \midrule
1 & reference & 2768.18 & 1737.20 & 1227.49 & t.o. & 704.39 \\ 
  2 & distributed incomplete & 687.76 & 593.55 & 509.06 & 675.54 & 524.68 \\ 
   \bottomrule
\end{tabular}
\caption{Number of executors vs. execution time on incomplete store\_sales dataset (tuples: 5000000, dimensions: 6)} 
\label{table_nodes_vs_time-store_sales_5_incomplete-skyline-6d-5000000t}
\end{table*}

\clearpage
\newpage

\section{Skylines of More Complex Queries}
\label{appendix:complexQueries}

We have also evaluated the performance of our skyline implementation using more complex queries. For this purpose, we have adapted a part of the MusicBrainz database \cite{Musicbrainz} to allow for 
more complex queries, which contain multiple joins and aggregates.

\subsection{Test Data and Queries}

We have selected tuples from Musicbrainz' \texttt{recording}s such that we have a dataset for complete and one for incomplete algorithms. Both datasets contain almost exactly 1.5 million \texttt{recording} tuples, which we will use as the size for both sets. Tuples were selected such that all recordings with ratings ($\sim 500,000$) were selected first and the rest ($\sim 1,000,000$) were sampled randomly. To limit the complexity of the joins, we also have limited the tables of the join ``partners'' in similar ways. For details on the generation of the dataset and the datasets themselves as \texttt{.csv} files, we refer to our collection of supplemental material under \cite{Zen:data/Skyline}.

As opposed to our more simple queries, there are now slight differences between the complete and incomplete queries due to the more complex nature of the queries. Nevertheless, they should still be comparable with regards to both execution time and memory consumption.

Our ``base'' queries are given in
Listing \ref{alg:skyline_complex_query_base_incomplete_sql} and
Listing \ref{alg:skyline_complex_query_base_complete_sql},
respectively. They can be easily extended to skyline queries by adding the skyline clause and potentially adding a subquery where applicable.
In Table \ref{tab:skyline_dimensions_musicbrainz_dataset},
we give the skyline dimensions as we have already done in 
Section~\ref{sect:experimental-setup}. The actual skyline queries are then constructed 
by using the base queries in combination with the skyline dimensions from this table.

\begin{listingAlgorithm}% [t]
  SELECT \\
    \Indp r.id, \\
    ifnull(r.length, 0) AS length, \\
    r.video, \\
    ifnull(rm.rating, 0) AS rating, \\
    ifnull(rm.rating\_count, 0) AS rating\_count, \\
    recording\_tracks.num\_tracks, \\
    recording\_tracks.min\_position \\
 \Indm FROM recording\_complete r LEFT OUTER JOIN ( \\
    \Indp SELECT \\
      \Indp ri.id AS id, \\
      count(ti.recording) AS num\_tracks, \\
      min(ti.position) AS min\_position \\
    \Indp FROM recording\_complete ri \\
    JOIN track ti ON (ti.recording = ri.id) \\
    GROUP BY ri.id \\
  \Indm ) recording\_tracks USING (id) \\
  JOIN recording\_meta rm USING (id) \\
\Indm
\caption{Base query (complete) of the more complex benchmarks on a subset of the MusicBrainz database}
\label{alg:skyline_complex_query_base_complete_sql}
\end{listingAlgorithm}

\begin{listingAlgorithm}% [t]
SELECT * FROM recording\_incomplete r \\
LEFT OUTER JOIN ( \\
  \Indp SELECT \\
    \Indp ri.id AS id, \\
    count(ti.recording) AS num\_tracks, \\
    min(ti.position) AS min\_position \\
  \Indm FROM recording\_incomplete ri \\
  JOIN track ti ON (ti.recording = ri.id) \\
  GROUP BY ri.id \\
\Indm ) recording\_tracks USING (id) \\
JOIN recording\_meta rm USING (id)
\caption{Base query (incomplete) of the more complex benchmarks on a subset of the MusicBrainz database}
\label{alg:skyline_complex_query_base_incomplete_sql}
\end{listingAlgorithm}

\begin{table}[H]
\centering
\begin{tabular}{|L{3cm}|L{1cm}|L{3.75cm}|}
	\hline
	\textbf{Dimension} & \textbf{Type} & \textbf{Description} \\
	\hline
	id & \texttt{KEY} & id of the recording \\
	\arrayrulecolor{lightgray}\hline
	rating & \texttt{MAX} & rating of the recording (cumulative) \\
	\arrayrulecolor{lightgray}\hline
	rating\_count & \texttt{MAX} & number of ratings \\
	\arrayrulecolor{lightgray}\hline
	length & \texttt{MIN} & length of the recording \\
	\arrayrulecolor{lightgray}\hline
	video & \texttt{MAX} & whether the recording has a video \\
	\arrayrulecolor{lightgray}\hline
	num\_tracks & \texttt{MAX} & how many tracks the recording is on \\
	\arrayrulecolor{lightgray}\hline
	min\_position & \texttt{MIN} & lowest track number of recording \\
	\arrayrulecolor{black}\hline
\end{tabular}
\caption{Skyline dimensions for recordings in the MusicBrainz dataset}
\label{tab:skyline_dimensions_musicbrainz_dataset}
\end{table}

Our queries can be described as finding the best and most often rated recordings which are the shortest and have an associated video. Additionally, we also seek recordings, which were on many tracks and had a very low position on the associated album (i.e., at the beginning of the album).

\medskip

The more complex queries considered in this section clearly demonstrate the advantage of the 
easy to use skyline syntax. While the queries in skyline syntax are relatively straightforward by only containing 
the base query and a skyline clause, the corresponding reference queries become quite extensive and unwieldy. As we can see in Listing \ref{alg:skyline_complex_query_example_complete_reference_sql}, the rewritten query is not 
easy to read anymore, even after heavy formatting to balance space restrictions and readability. In comparison, the specialized skyline syntax given in Listing \ref{alg:skyline_complex_query_example_complete_sql},  which expresses the same query, is significantly more concise and readable.

\begin{listingAlgorithm}[p]
SELECT * FROM (SELECT * FROM ( SELECT \\
      \Indp\Indp\Indp r.id,  ifnull(r.length, 0) AS length, \\
      r.video, ifnull(rm.rating, 0) AS rating, \\
      ifnull(rm.rating\_count, 0) AS rating\_count, \\
      recording\_tracks.num\_tracks, \\
      recording\_tracks.min\_position \\
    \Indm FROM recording\_complete r \\
    LEFT OUTER JOIN ( \\
      \Indp SELECT \\
        \Indp ri.id AS id, \\
        count(ti.recording) AS num\_tracks, \\
        min(ti.position) AS min\_position \\
      \Indm FROM recording\_complete ri \\
      JOIN track ti ON (ti.recording = ri.id) \\
      GROUP BY ri.id \\
    \Indm ) recording\_tracks USING (id) \\
    JOIN recording\_meta rm USING (id) ) \\
\Indm\Indm ) AS o WHERE NOT EXISTS( \\
  \Indp SELECT * FROM (SELECT * FROM ( \\
      \Indp SELECT \\
        \Indp r.id, \\
        ifnull(r.length, 0) AS length, r.video, \\
        ifnull(rm.rating, 0) AS rating, \\
        ifnull(rm.rating\_count, 0) AS rating\_count, \\
        recording\_tracks.num\_tracks, \\
        recording\_tracks.min\_position \\
      \Indm FROM recording\_complete r \\
      LEFT OUTER JOIN ( \\
        \Indp SELECT \\
          \Indp ri.id AS id, \\
          count(ti.recording) AS num\_tracks, \\
          min(ti.position) AS min\_position \\
        \Indm FROM recording\_complete ri \\
        JOIN track ti ON (ti.recording = ri.id) \\
        GROUP BY ri.id \\
      \Indm ) recording\_tracks USING (id) \\
      JOIN recording\_meta rm USING (id) \\
  \Indm\Indm ) ) AS i WHERE \\
    \Indp i.rating >= o.rating AND \\
    i.rating\_count >= o.rating\_count AND \\
    i.length <= o.length AND \\
    i.video >= o.video AND \\
    i.num\_tracks >= o.num\_tracks AND \\
    i.min\_position <= o.min\_position AND ( \\
      \Indp i.rating > o.rating OR \\
      i.rating\_count > o.rating\_count OR \\
      i.length < o.length OR \\
      i.video > o.video OR \\
      i.num\_tracks > o.num\_tracks OR \\
      i.min\_position < o.min\_position ) ) \\
\Indm\Indm\Indm
\caption{Reference query of the more complex benchmarks on a subset of the MusicBrainz database (6 dimensions)}
\label{alg:skyline_complex_query_example_complete_reference_sql}
\end{listingAlgorithm}

\begin{listingAlgorithm}% [t]
SELECT * FROM ( \\
  \Indp SELECT \\
    \Indp r.id, \\
    ifnull(r.length, 0) AS length, \\
    r.video, \\
    ifnull(rm.rating, 0) AS rating, \\
    ifnull(rm.rating\_count, 0) AS rating\_count, \\
    recording\_tracks.num\_tracks, \\
    recording\_tracks.min\_position \\
 \Indm FROM recording\_complete r LEFT OUTER JOIN ( \\
    \Indp SELECT \\
      \Indp ri.id AS id, \\
      count(ti.recording) AS num\_tracks, \\
      min(ti.position) AS min\_position \\
    \Indm FROM recording\_complete ri \\
    JOIN track ti ON (ti.recording = ri.id) \\
    GROUP BY ri.id \\
  \Indm ) recording\_tracks USING (id) \\
  JOIN recording\_meta rm USING (id) \\
\Indm ) SKYLINE OF COMPLETE \\
  \Indp rating MAX, \\
  rating\_count MAX,
  length MIN, \\
  video MAX, \\
  num\_tracks MAX, \\
  min\_position MIN \\
\caption{Skyline query (complete) of the more complex benchmarks on a subset of the MusicBrainz database}
\label{alg:skyline_complex_query_example_complete_sql}
\end{listingAlgorithm}

\subsection{Experimental Results}

In this section, we will take a closer look at the performance results and some conclusions we can draw from them. For this, we will again look at both the execution time and the memory consumption depending on the number of dimensions and executors. Since we are using a real-world dataset to do this, considering the number of tuples as a parameter is not applicable here (as was already the case with the Airbnb dataset in Section \ref{sect:experimental-results}).
The impact of the number of dimensions on the execution time and on the memory consumption is shown in Figure~\ref{fig:appendix_dimensions_vs_time_musicbrainz} and in Figure~\ref{fig:appendix_dimensions_vs_memory_musicbrainz_complete}, respectively. 
Likewise, the impact of the number of executors on the execution time and on the memory consumption is shown in Figure~\ref{fig:appendix_nodes_vs_time_musicbrainz} and in Figure~\ref{fig:appendix_nodes_vs_memory_musicbrainz}, respectively. 

First, we note that more complex queries result in the measurements to be both more oblique and more opaque. The way how joins and aggregates are handled by Spark has a huge influence on the query execution and affects both the execution time and the memory consumption. This can be seen across almost all plots which correspond to the complex queries. In these cases, our implementation works such that it will take the output of the base query and then uses it to compute the skyline (we exclude potential optimizations here). Still, while always yielding the same result, Spark may execute the queries in many different optimized ways heavily affecting the performance. Hence, the results obtained with the simpler queries such as those found in Section \ref{sect:experimental-results} seem to give a more precise picture of the performance of the various skyline algorithms. 

As already mentioned in Section \ref{sect:experimental-results}, there is a limit of how much distribution makes sense for the distributed algorithms. Since joins can also be computed in a distributed fashion, this effect may be amplified by more complex queries. This can, for example, be seen in Figure~\ref{fig:appendix_nodes_vs_time_musicbrainz} where $3$ executors is the optimum for which the query is executed the fastest. Adding further executors might slow down the query due to additional distribution and synchronization costs. In these benchmarks it is, however, impossible to exactly pinpoint which part of the query causes such loss of performance.

To conclude, for skyline queries on top of complex queries, it is not always clear if a particular increase of execution time or memory consumption is caused by the skyline computation or by peculiarities of the way how Spark processes the complex base query. Nevertheless, the results obtained with the complex queries are consistent with the ones obtained with the simple queries in Section \ref{sect:experimental-results}. Above all, it has again turned out that the reference queries are significantly slower in most cases and, in some cases, they also take significantly more memory than our specialized algorithms. As such, the specialized algorithms are potentially able to handle bigger data and also more complex queries than the ``plain'' SQL reference solution without running into serious problems.

\begin{figure*}[p]
    \begin{subfigure}{.5\linewidth}
      \centering
      \includegraphics[width=\linewidth]{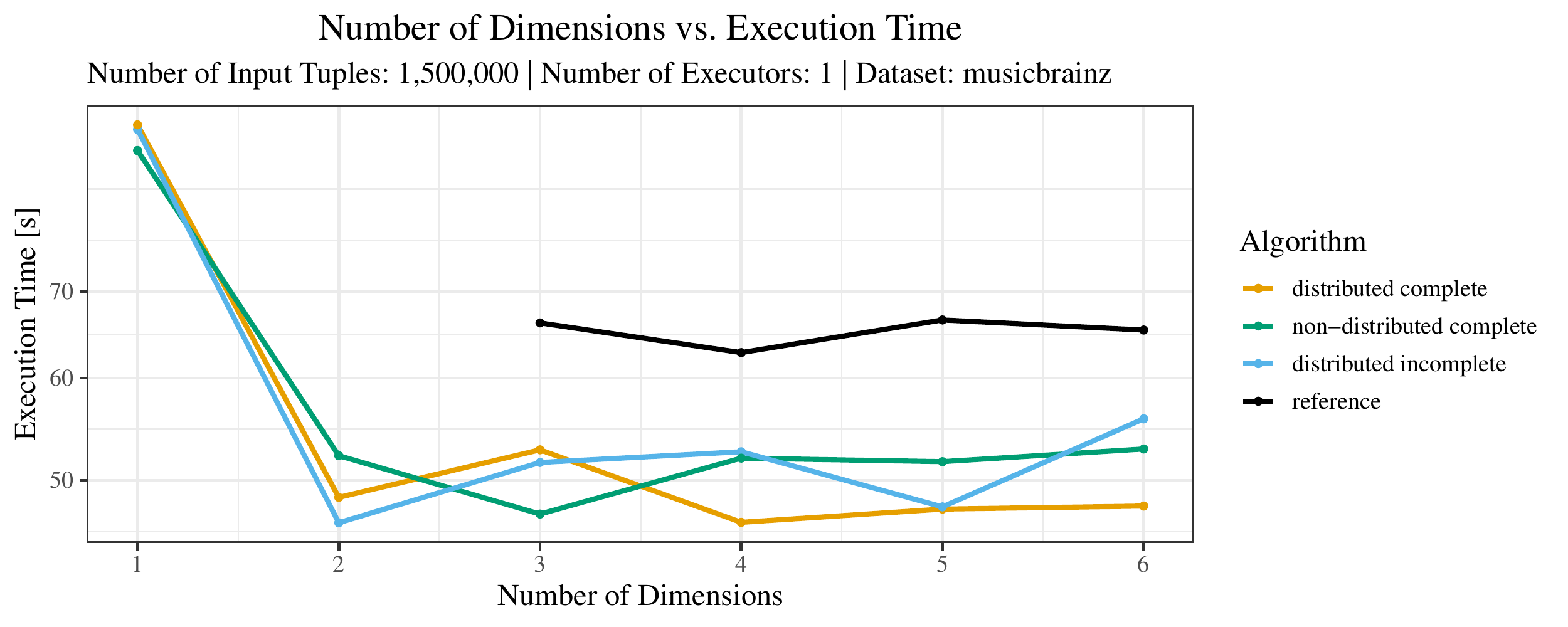}
    \end{subfigure}%
    \begin{subfigure}{.5\linewidth}
      \centering
      \includegraphics[width=\linewidth]{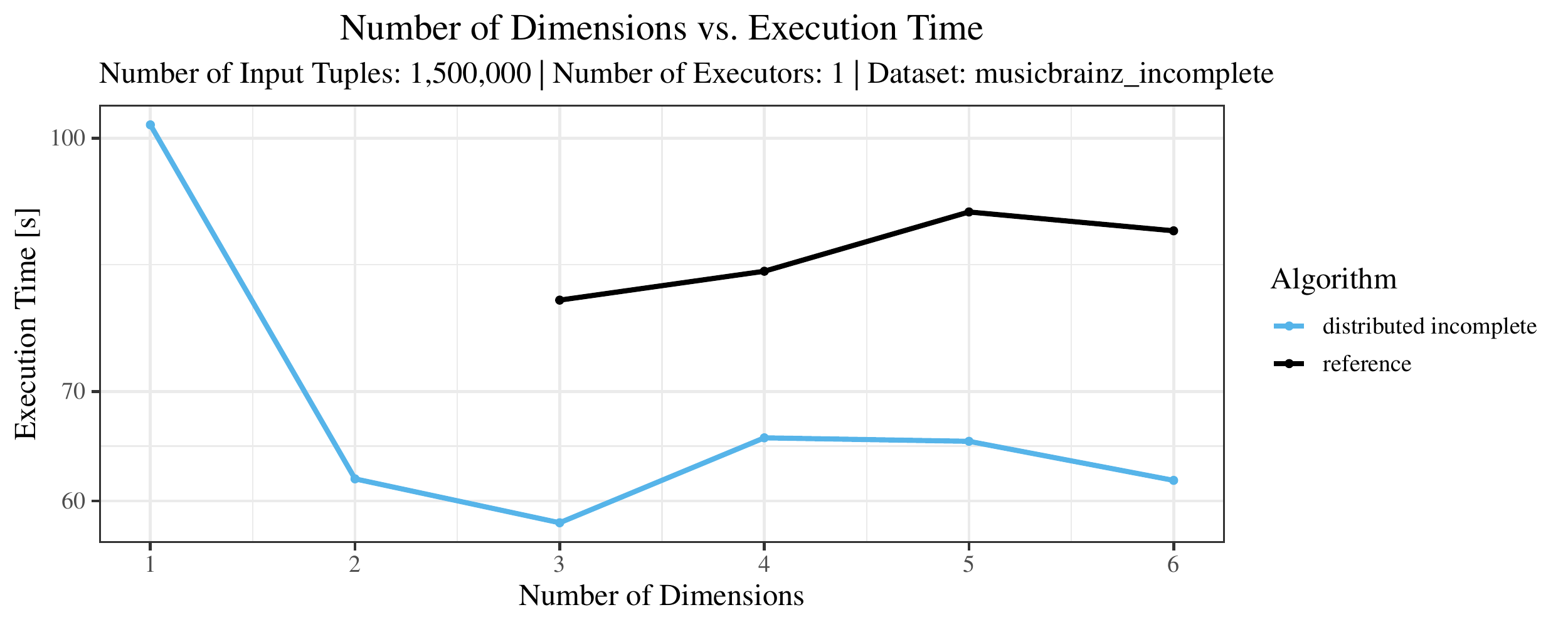}
    \end{subfigure}
    \begin{subfigure}{.5\linewidth}
      \centering
      \includegraphics[width=\linewidth]{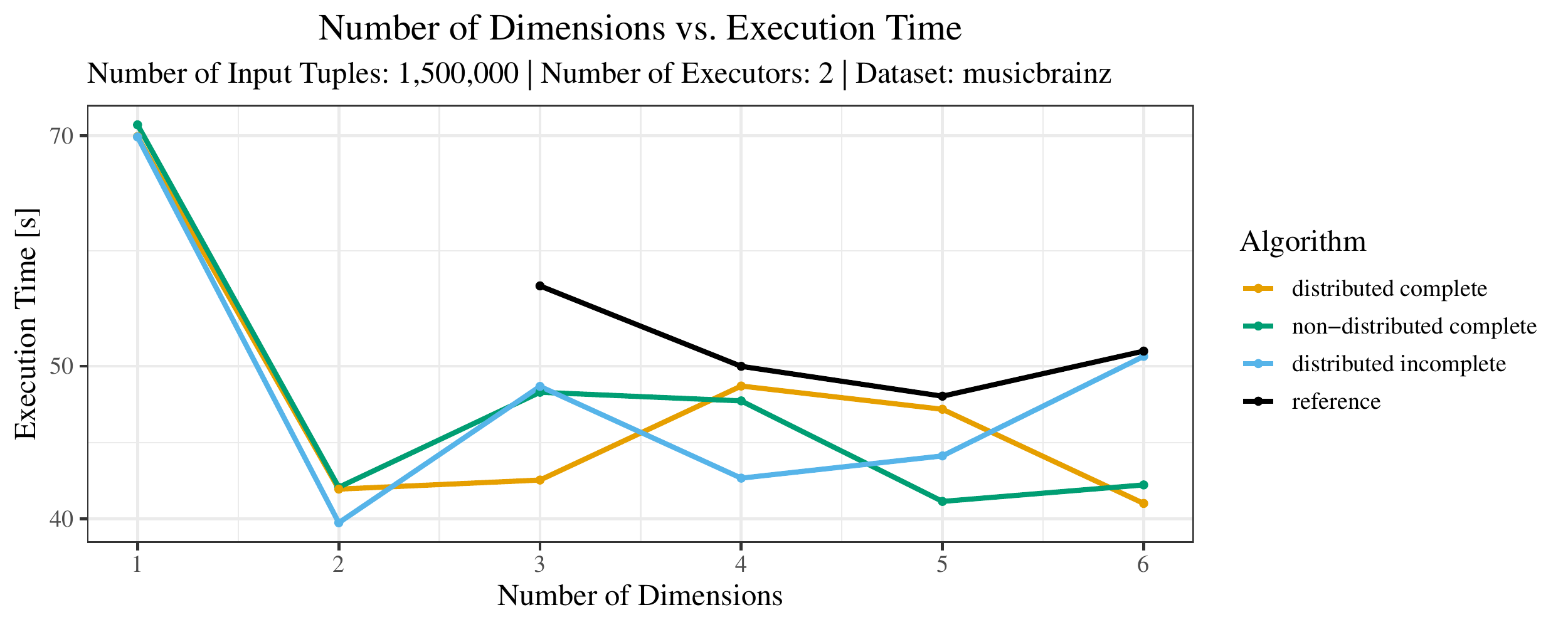}
    \end{subfigure}%
    \begin{subfigure}{.5\linewidth}
      \centering
      \includegraphics[width=\linewidth]{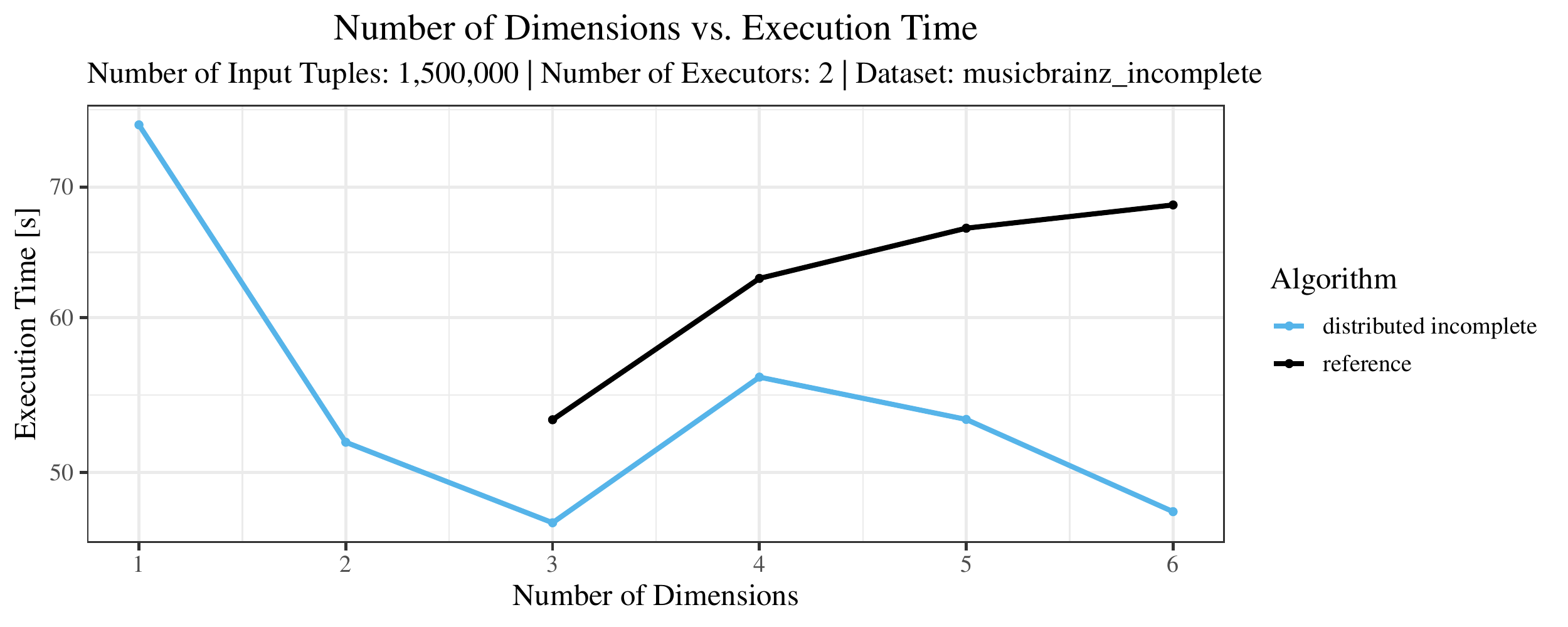}
    \end{subfigure}
    \begin{subfigure}{.5\linewidth}
      \centering
      \includegraphics[width=\linewidth]{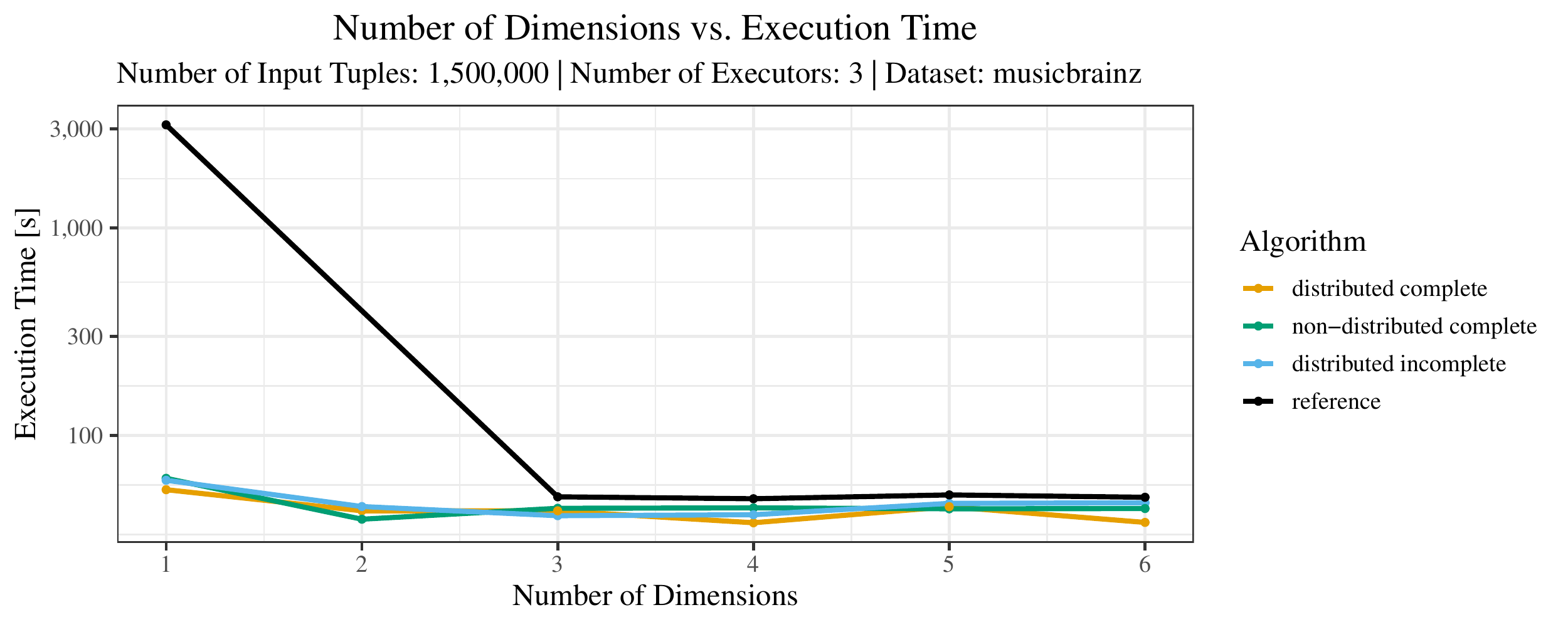}
    \end{subfigure}%
    \begin{subfigure}{.5\linewidth}
      \centering
      \includegraphics[width=\linewidth]{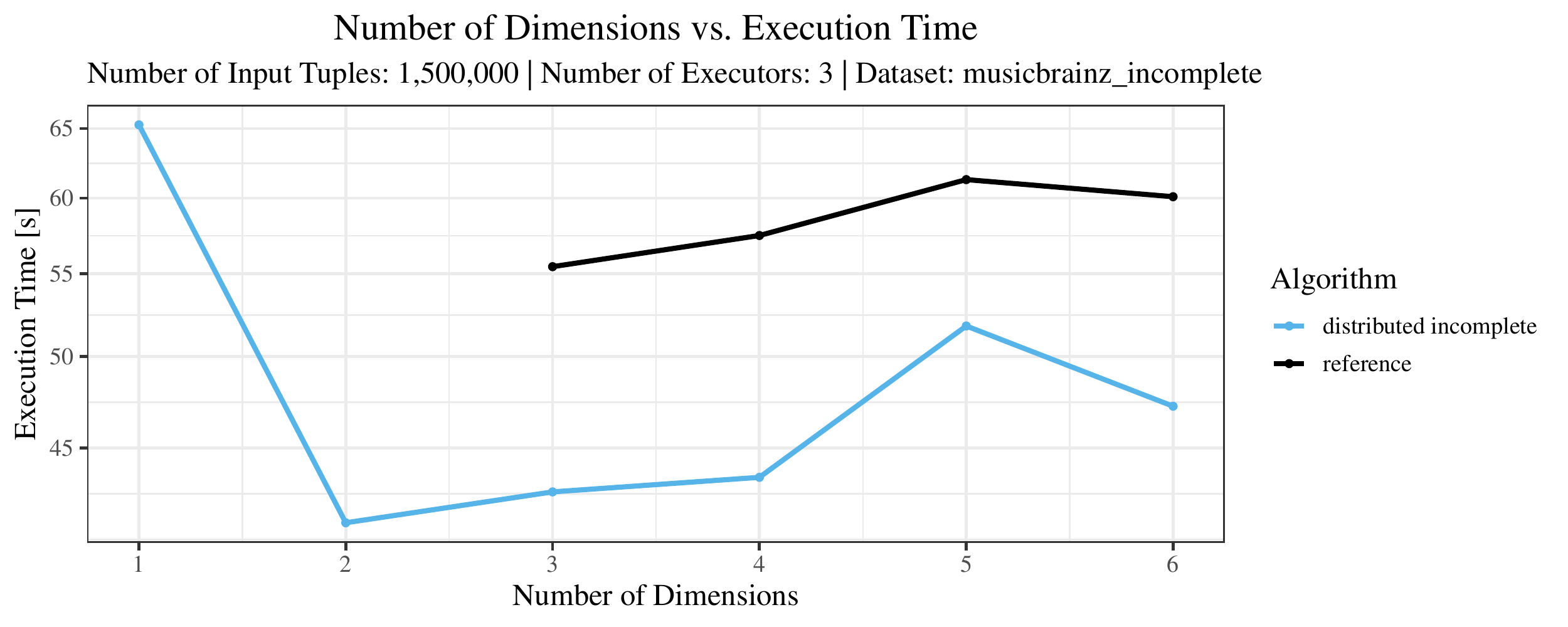}
    \end{subfigure}
    \begin{subfigure}{.5\linewidth}
      \centering
      \includegraphics[width=\linewidth]{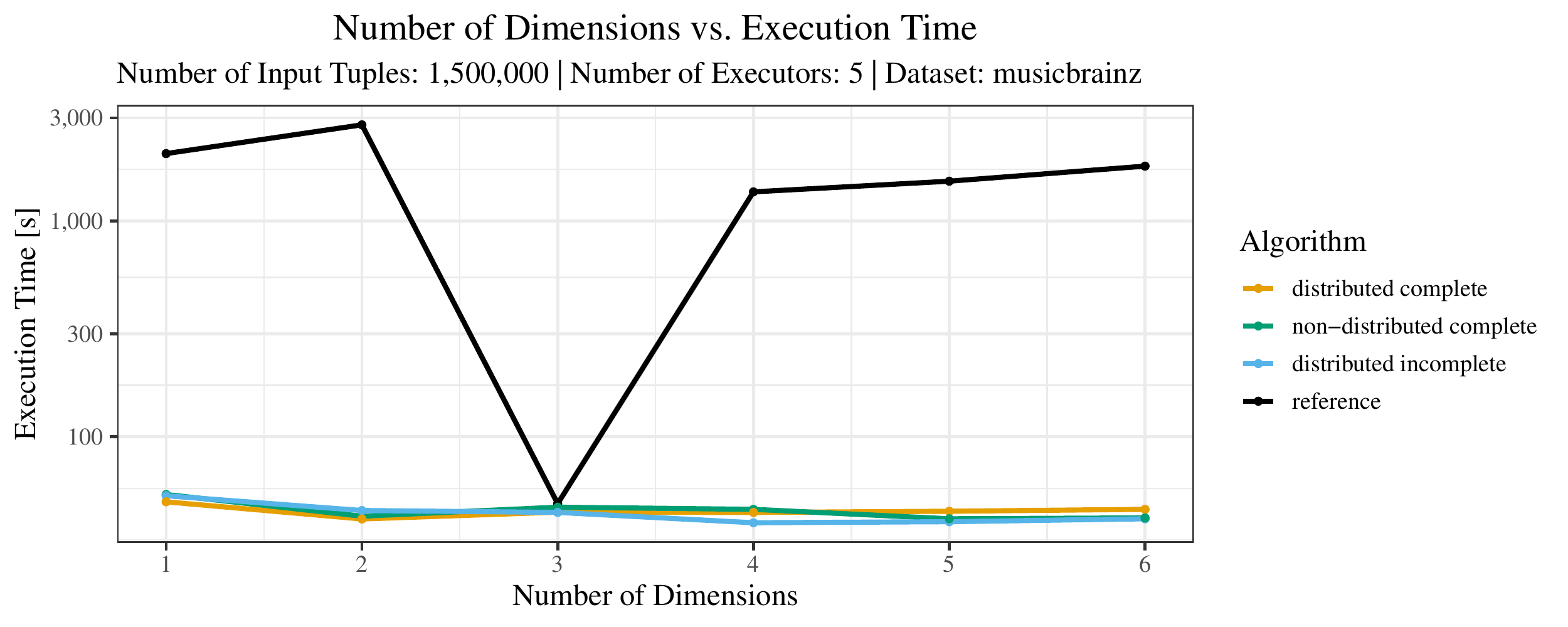}
    \end{subfigure}%
    \begin{subfigure}{.5\linewidth}
      \centering
      \includegraphics[width=\linewidth]{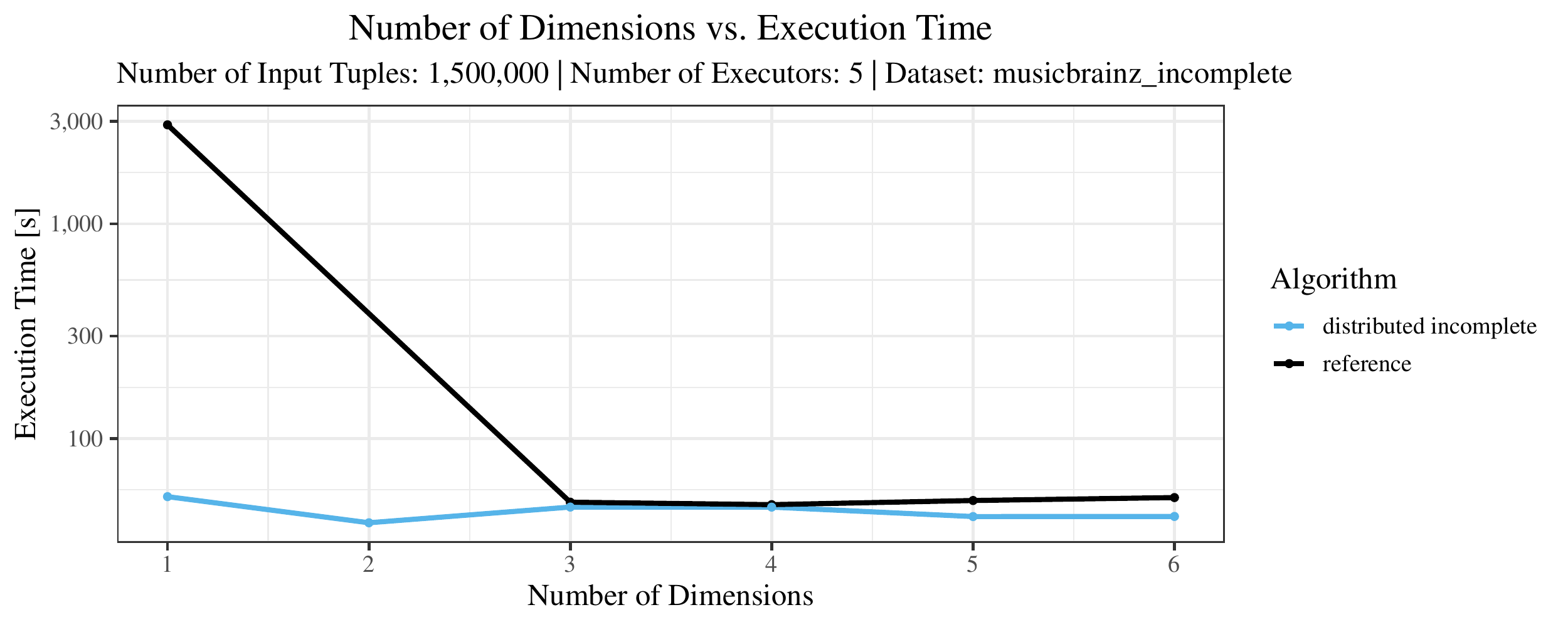}
    \end{subfigure}
    \begin{subfigure}{.5\linewidth}
      \centering
      \includegraphics[width=\linewidth]{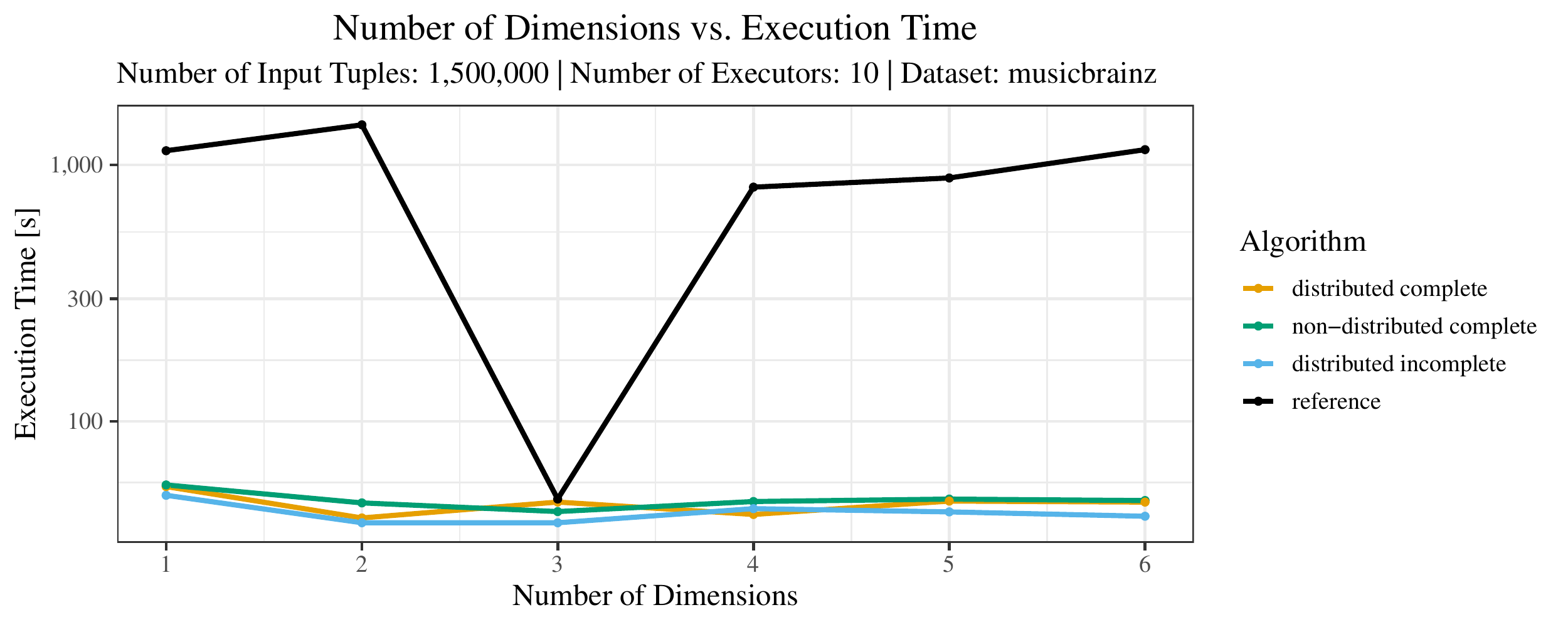}
    \end{subfigure}%
    \begin{subfigure}{.5\linewidth}
      \centering
      \includegraphics[width=\linewidth]{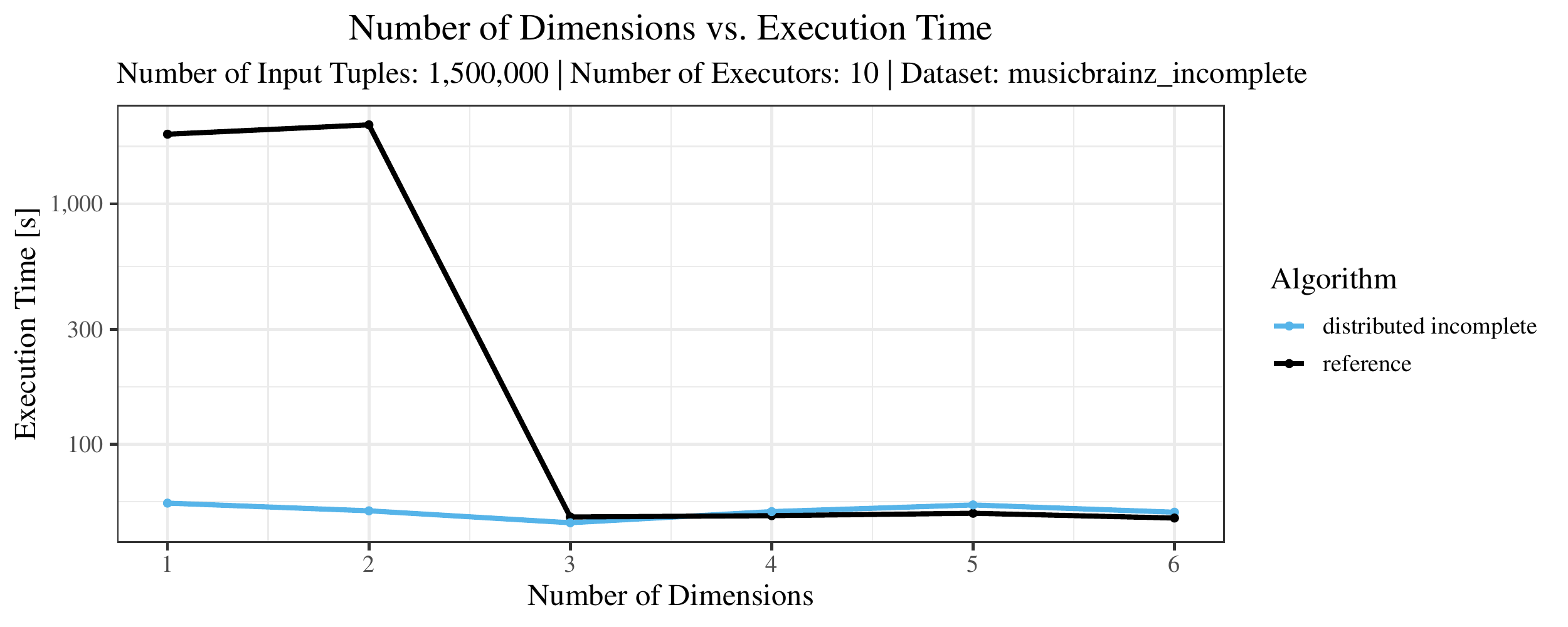}
    \end{subfigure}
    \caption{Number of dimensions vs. execution time using complex queries on the MusicBrainz dataset}
    \label{fig:appendix_dimensions_vs_time_musicbrainz}
\end{figure*}

\begin{figure*}[p]
    \begin{subfigure}{.5\linewidth}
      \centering
      \includegraphics[width=\linewidth]{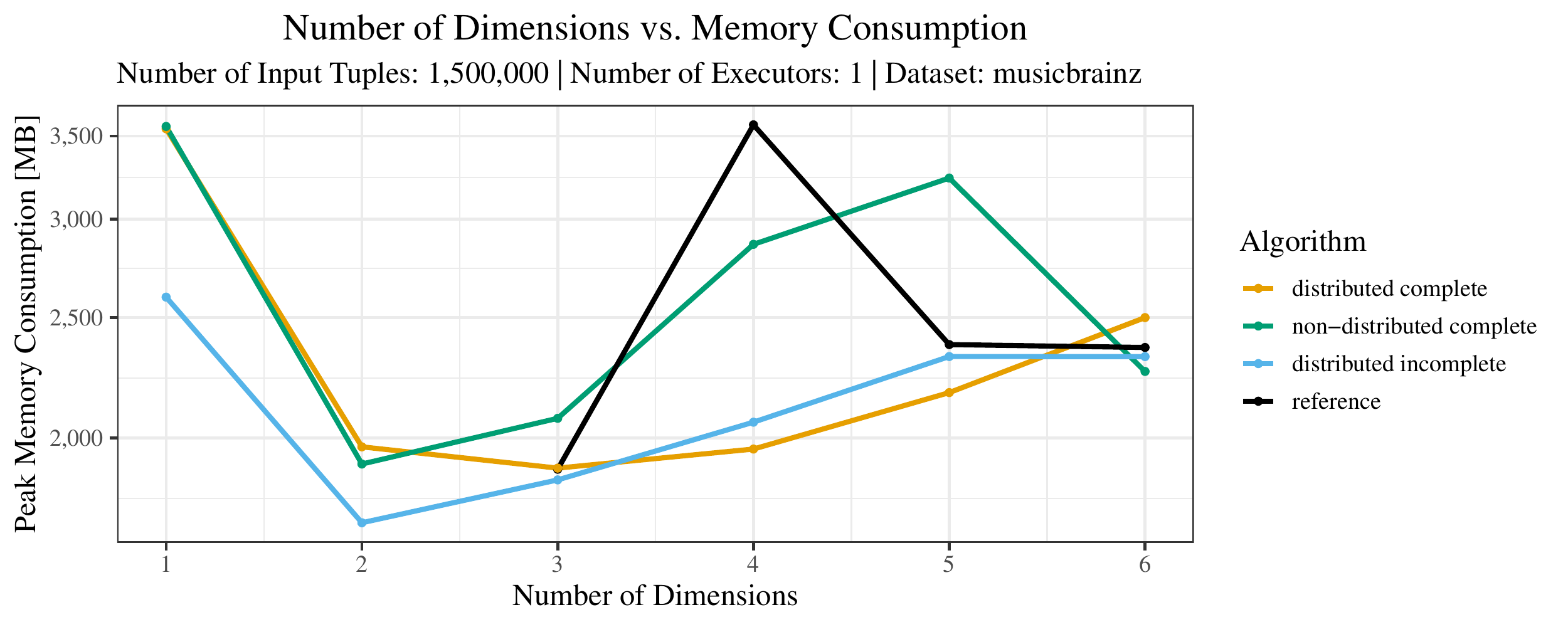}
    \end{subfigure}%
    \begin{subfigure}{.5\linewidth}
      \centering
      \includegraphics[width=\linewidth]{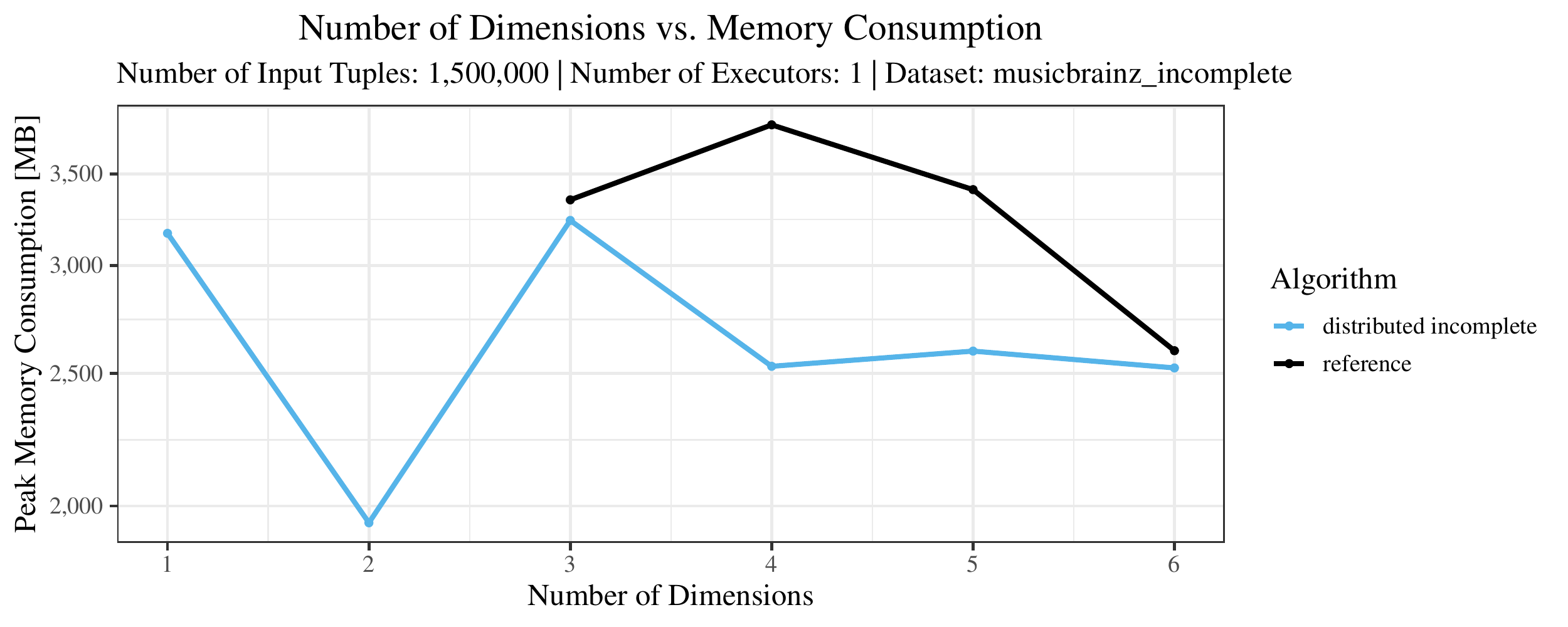}
    \end{subfigure}
    \begin{subfigure}{.5\linewidth}
      \centering
      \includegraphics[width=\linewidth]{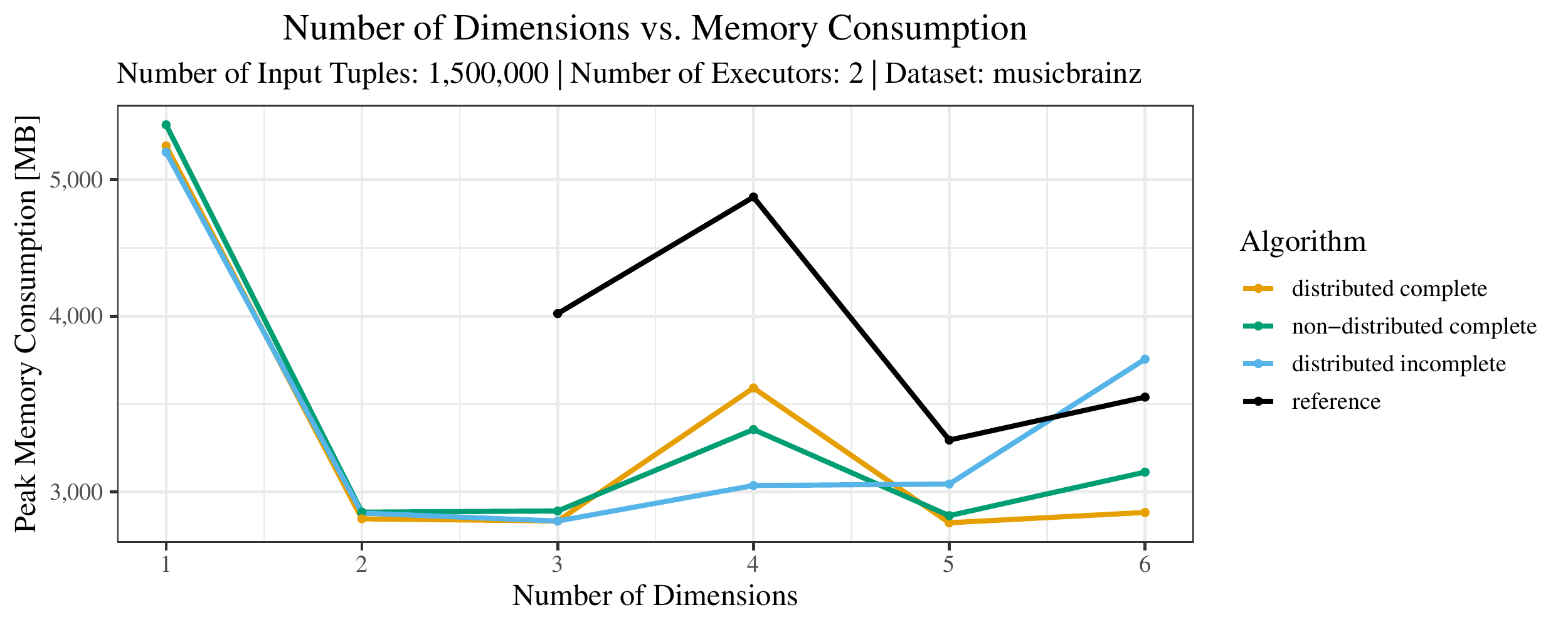}
    \end{subfigure}%
    \begin{subfigure}{.5\linewidth}
      \centering
      \includegraphics[width=\linewidth]{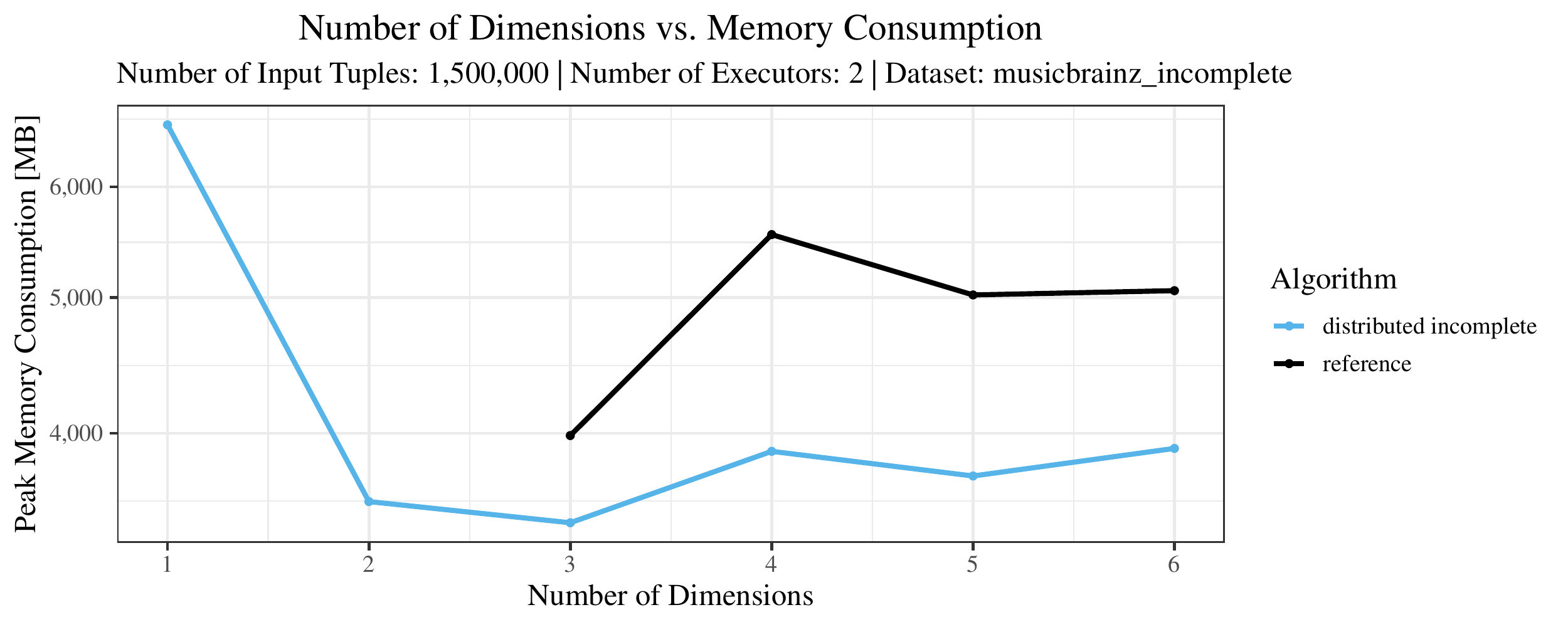}
    \end{subfigure}
    \begin{subfigure}{.5\linewidth}
      \centering
      \includegraphics[width=\linewidth]{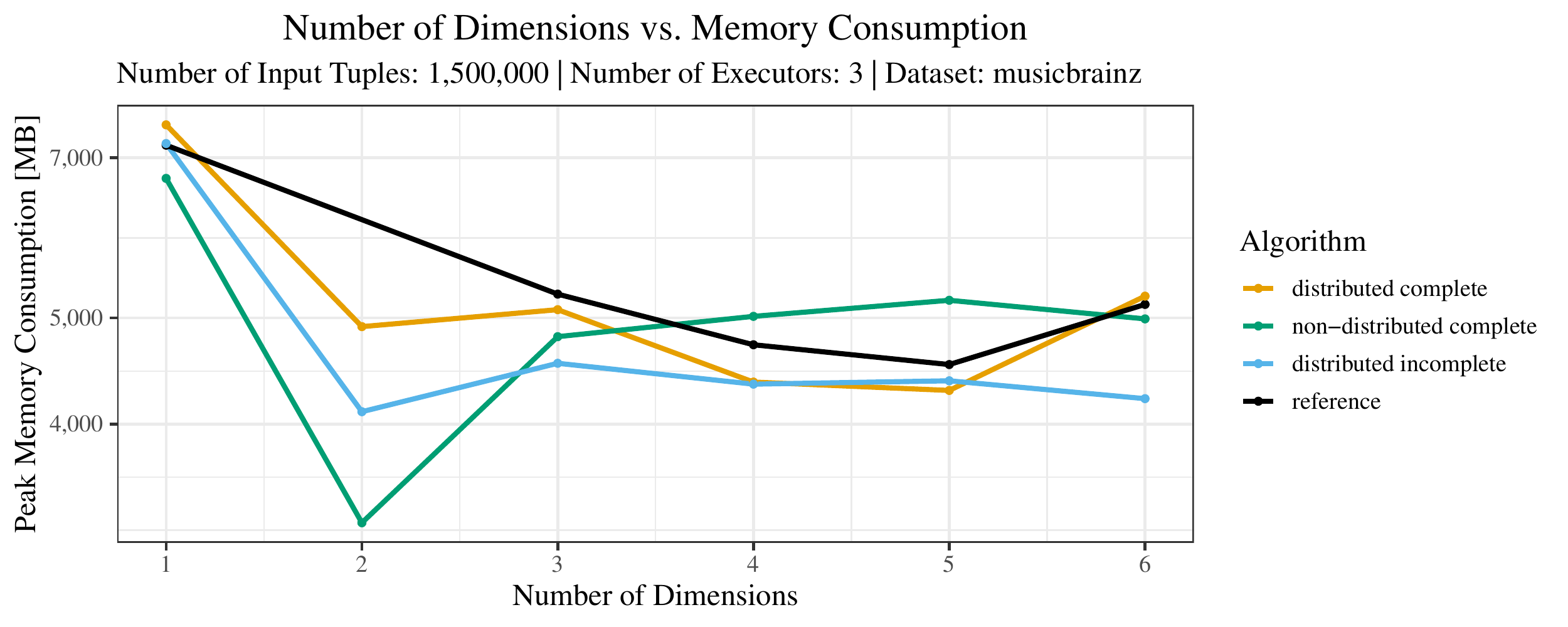}
    \end{subfigure}%
    \begin{subfigure}{.5\linewidth}
      \centering
      \includegraphics[width=\linewidth]{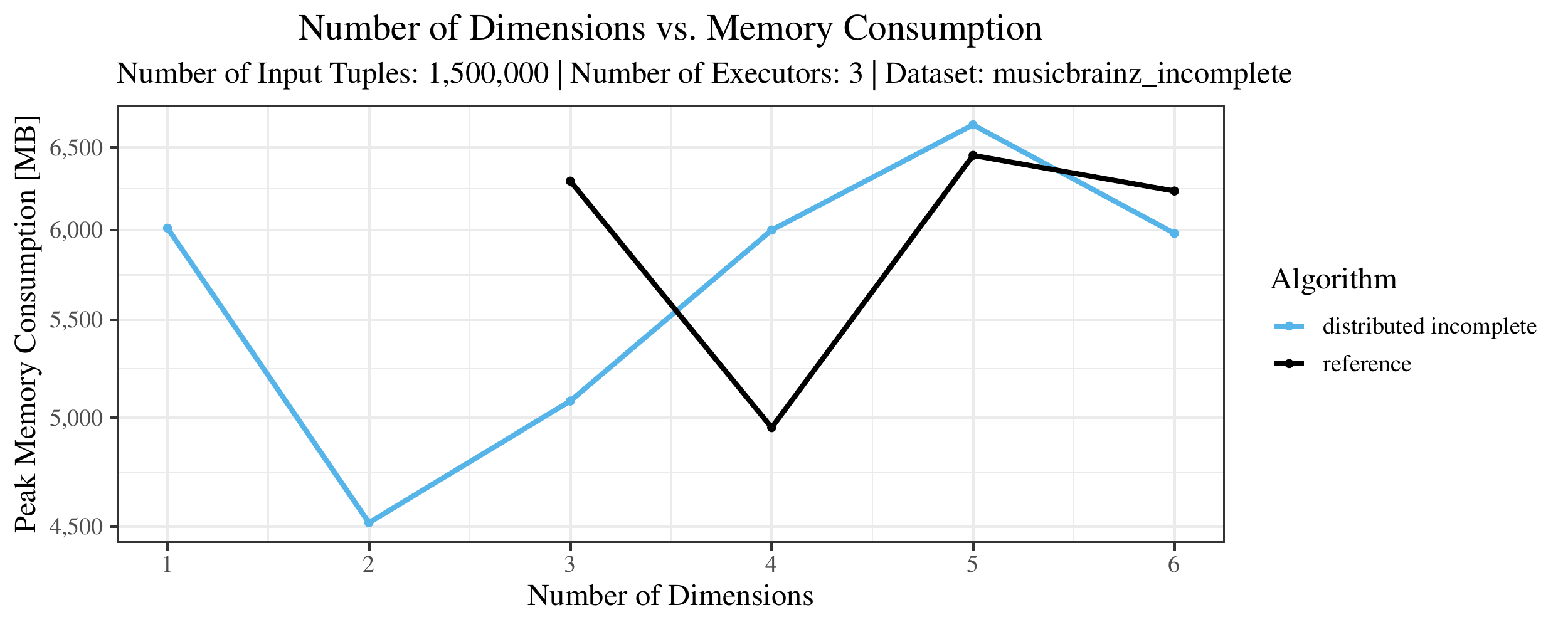}
    \end{subfigure}
    \begin{subfigure}{.5\linewidth}
      \centering
      \includegraphics[width=\linewidth]{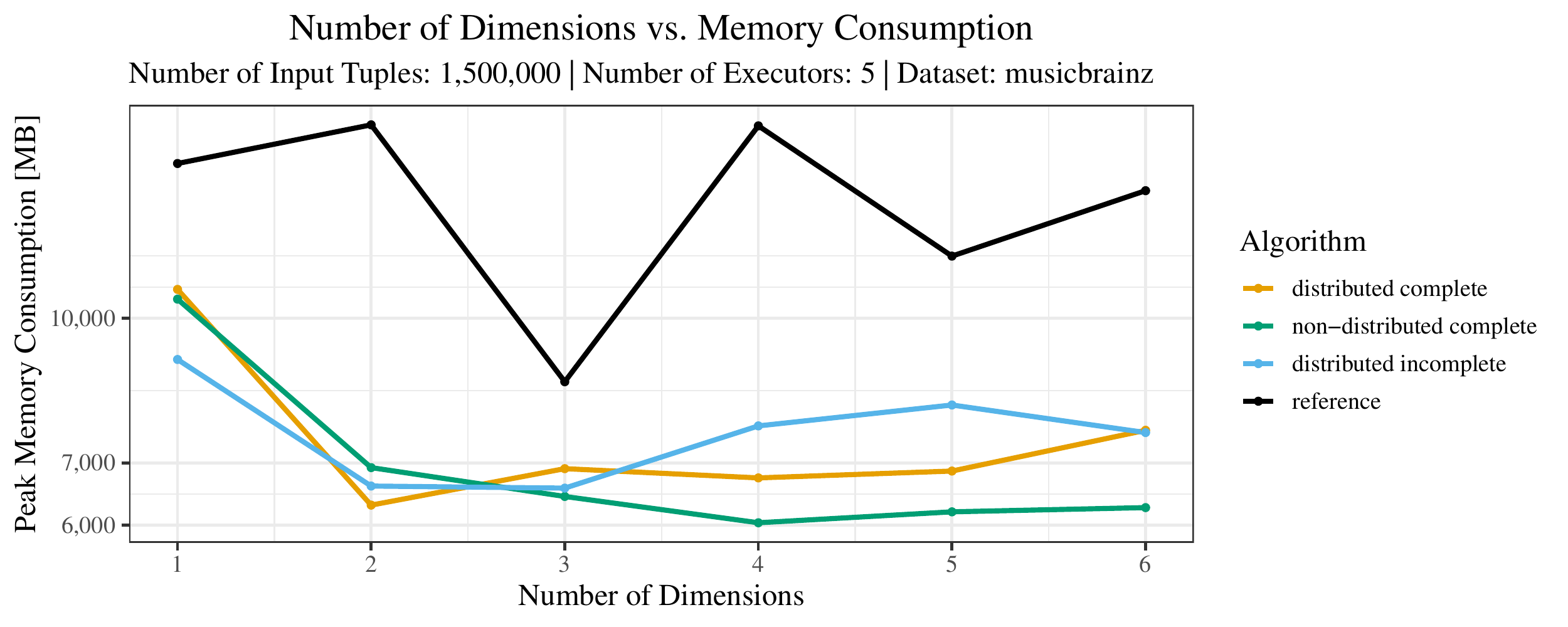}
    \end{subfigure}%
    \begin{subfigure}{.5\linewidth}
      \centering
      \includegraphics[width=\linewidth]{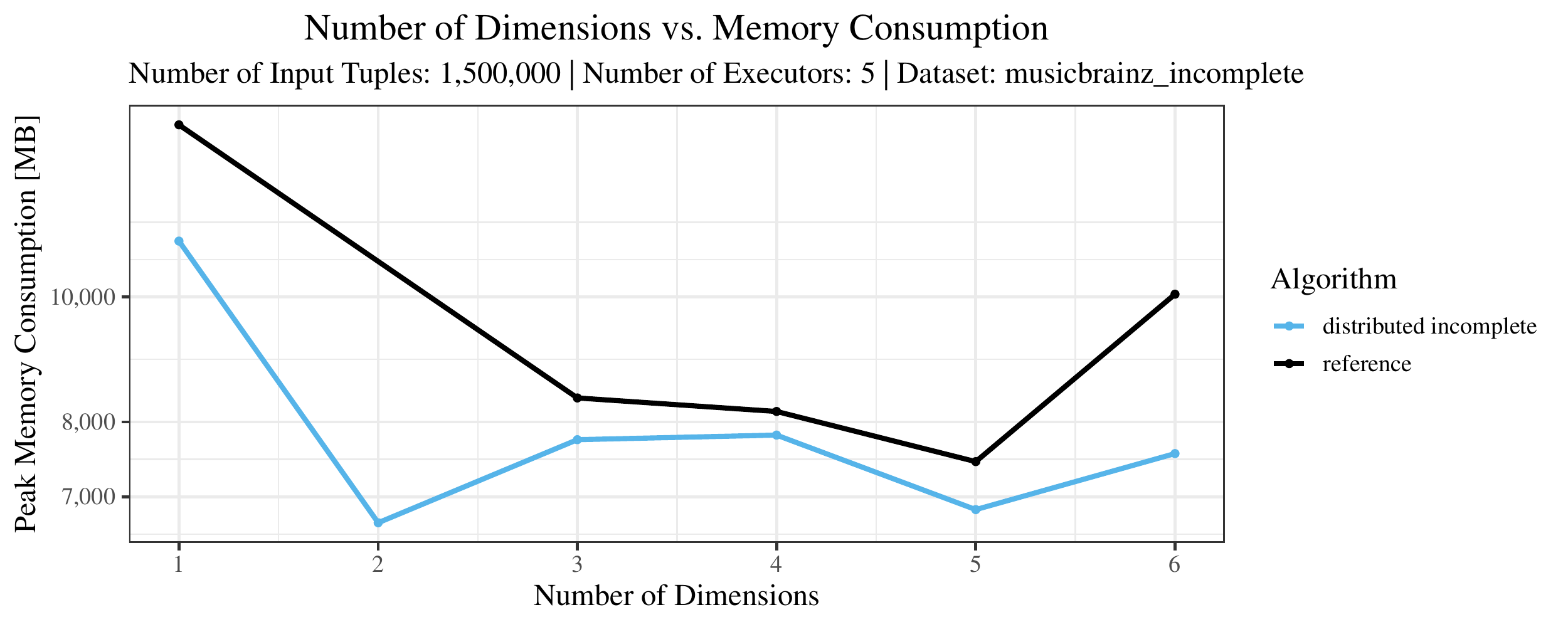}
    \end{subfigure}
    \begin{subfigure}{.5\linewidth}
      \centering
      \includegraphics[width=\linewidth]{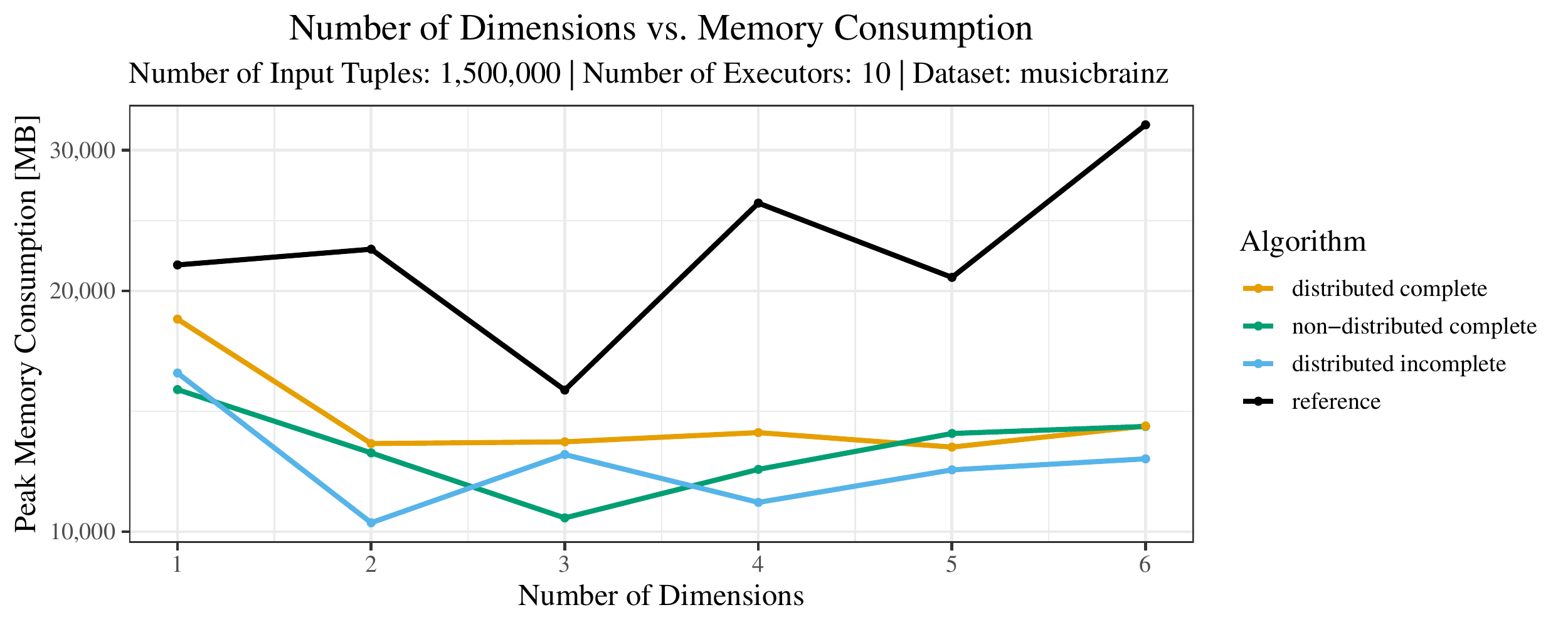}
    \end{subfigure}%
    \begin{subfigure}{.5\linewidth}
      \centering
      \includegraphics[width=\linewidth]{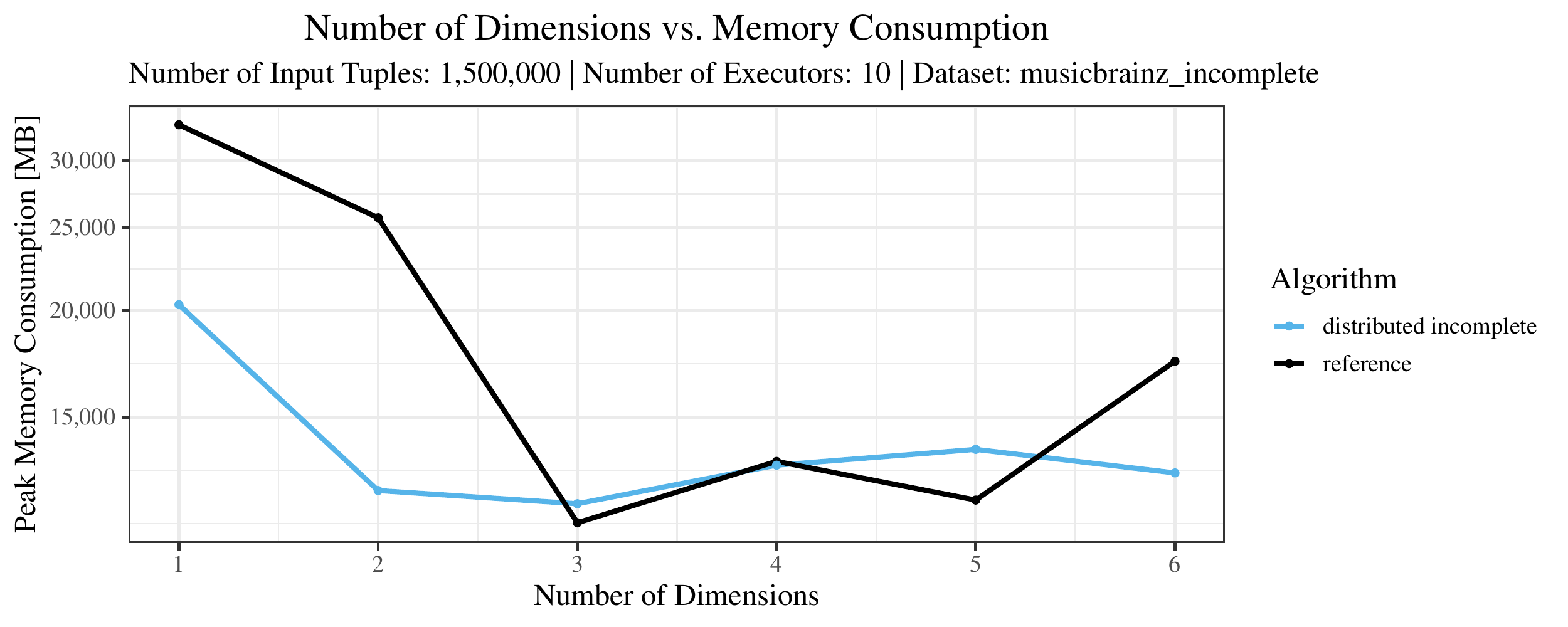}
    \end{subfigure}
    \caption{Number of dimensions vs. memory consumption using complex queries on the complete MusicBrainz dataset}
    \label{fig:appendix_dimensions_vs_memory_musicbrainz_complete}
\end{figure*}

\begin{figure*}[p]
    \begin{subfigure}{.5\linewidth}
      \centering
      \includegraphics[width=\linewidth]{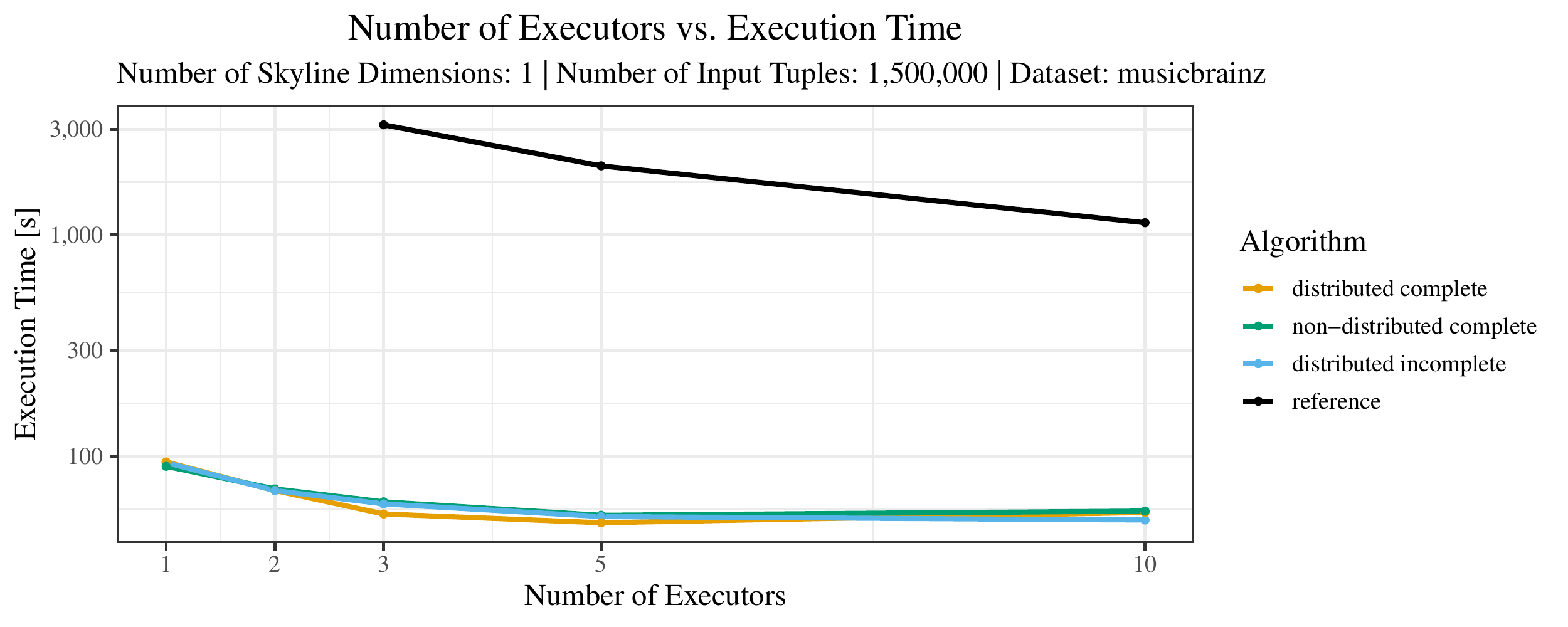}
    \end{subfigure}%
    \begin{subfigure}{.5\linewidth}
      \centering
      \includegraphics[width=\linewidth]{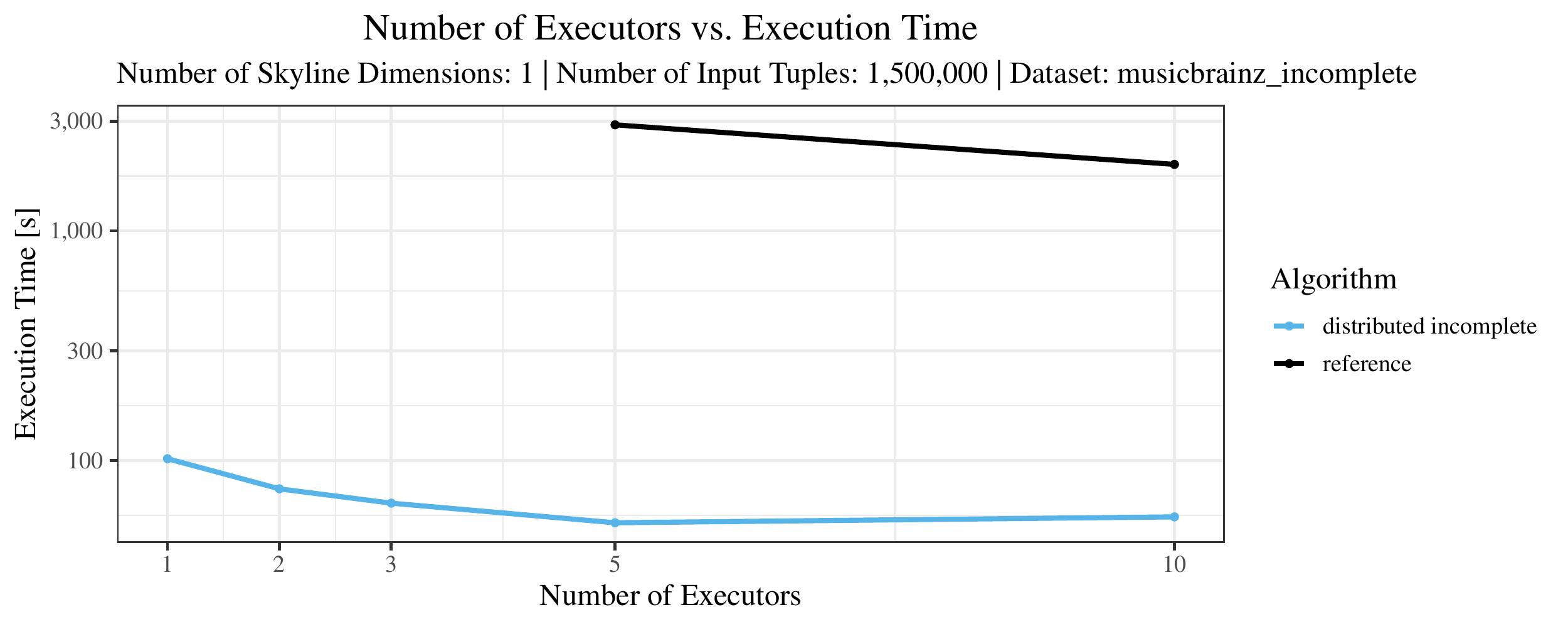}
    \end{subfigure}
    \begin{subfigure}{.5\linewidth}
      \centering
      \includegraphics[width=\linewidth]{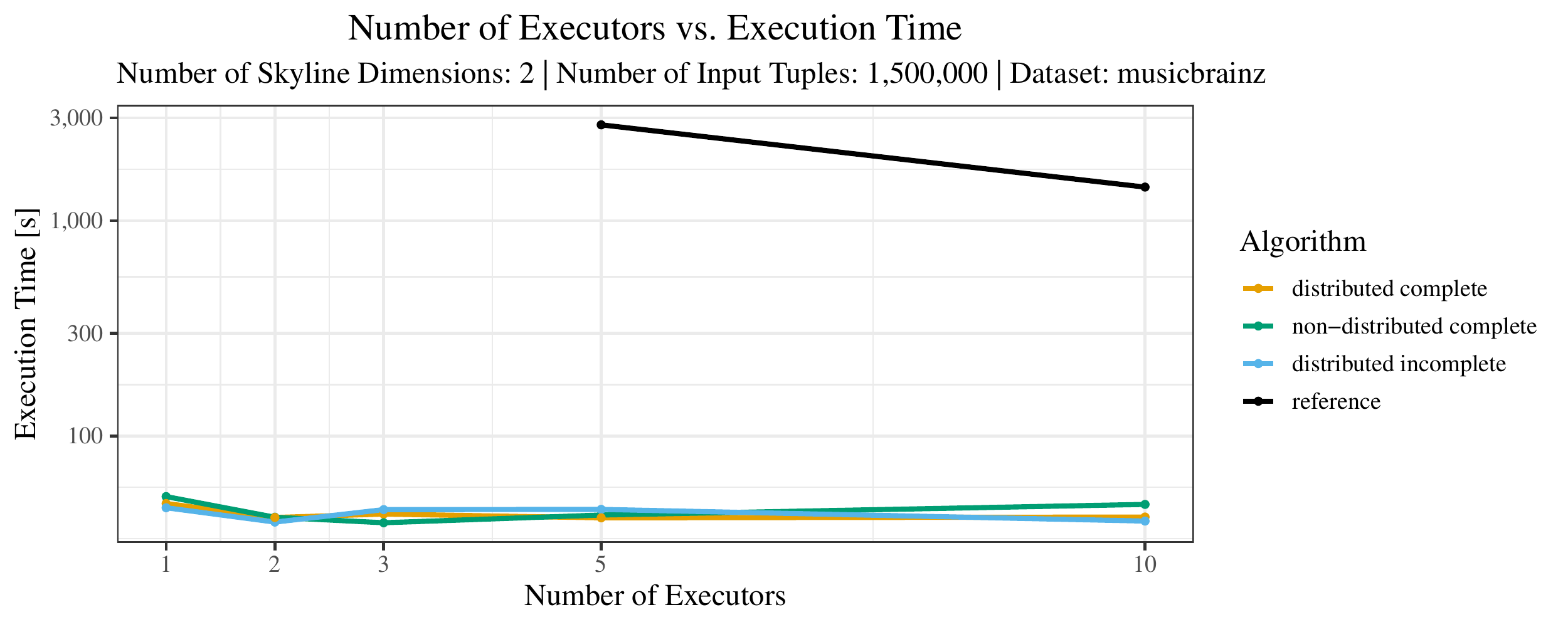}
    \end{subfigure}%
    \begin{subfigure}{.5\linewidth}
      \centering
      \includegraphics[width=\linewidth]{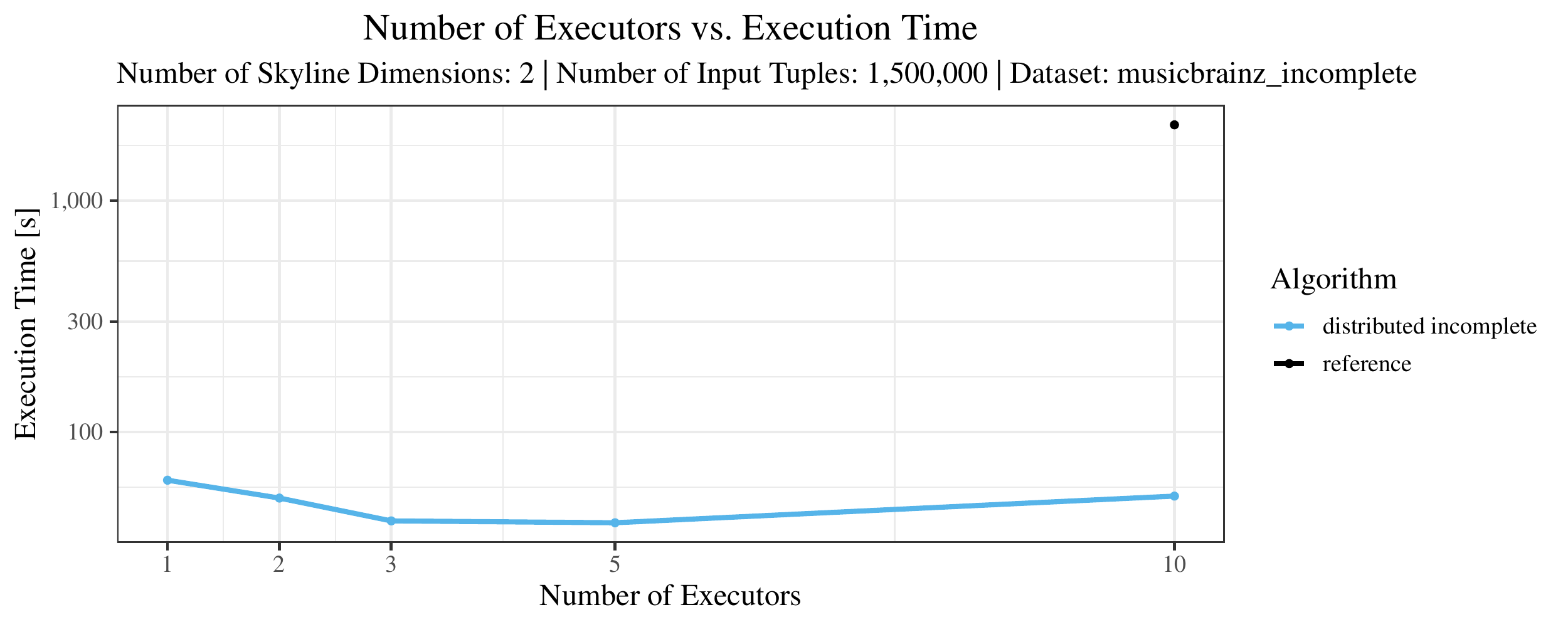}
    \end{subfigure}
    \begin{subfigure}{.5\linewidth}
      \centering
      \includegraphics[width=\linewidth]{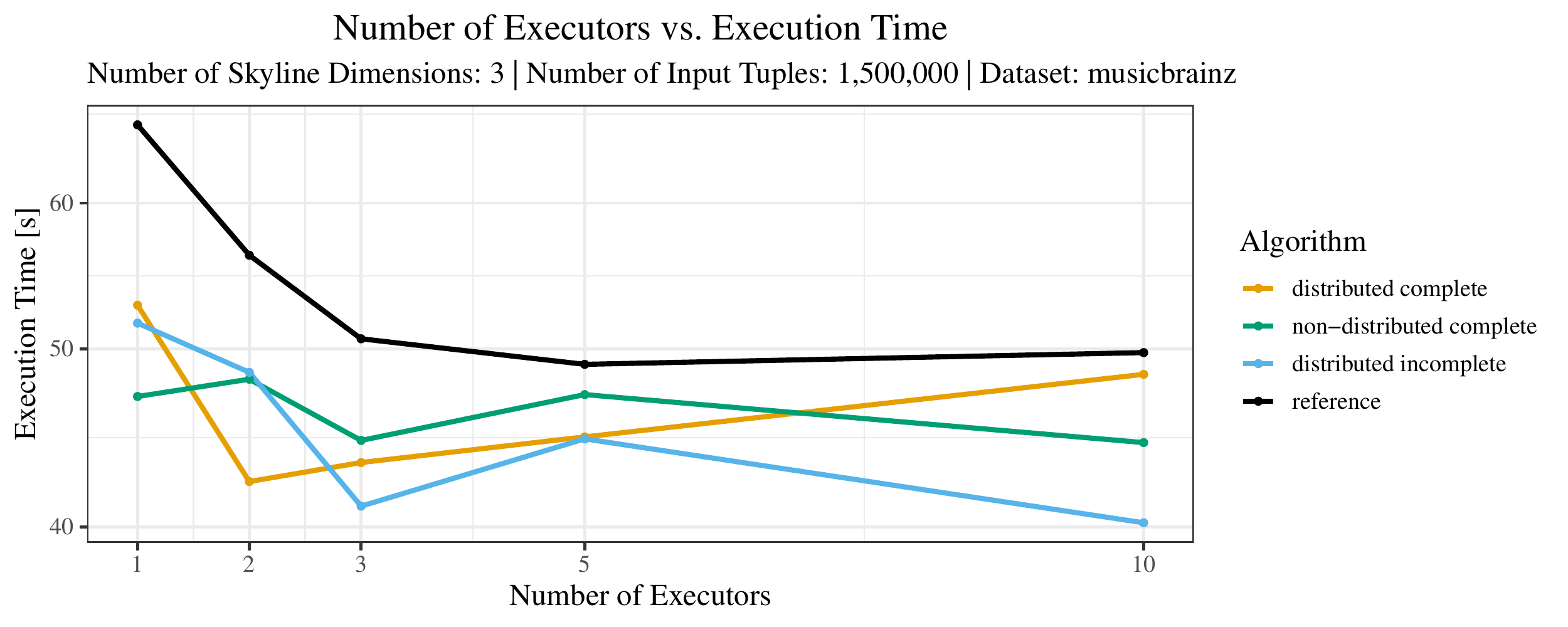}
    \end{subfigure}%
    \begin{subfigure}{.5\linewidth}
      \centering
      \includegraphics[width=\linewidth]{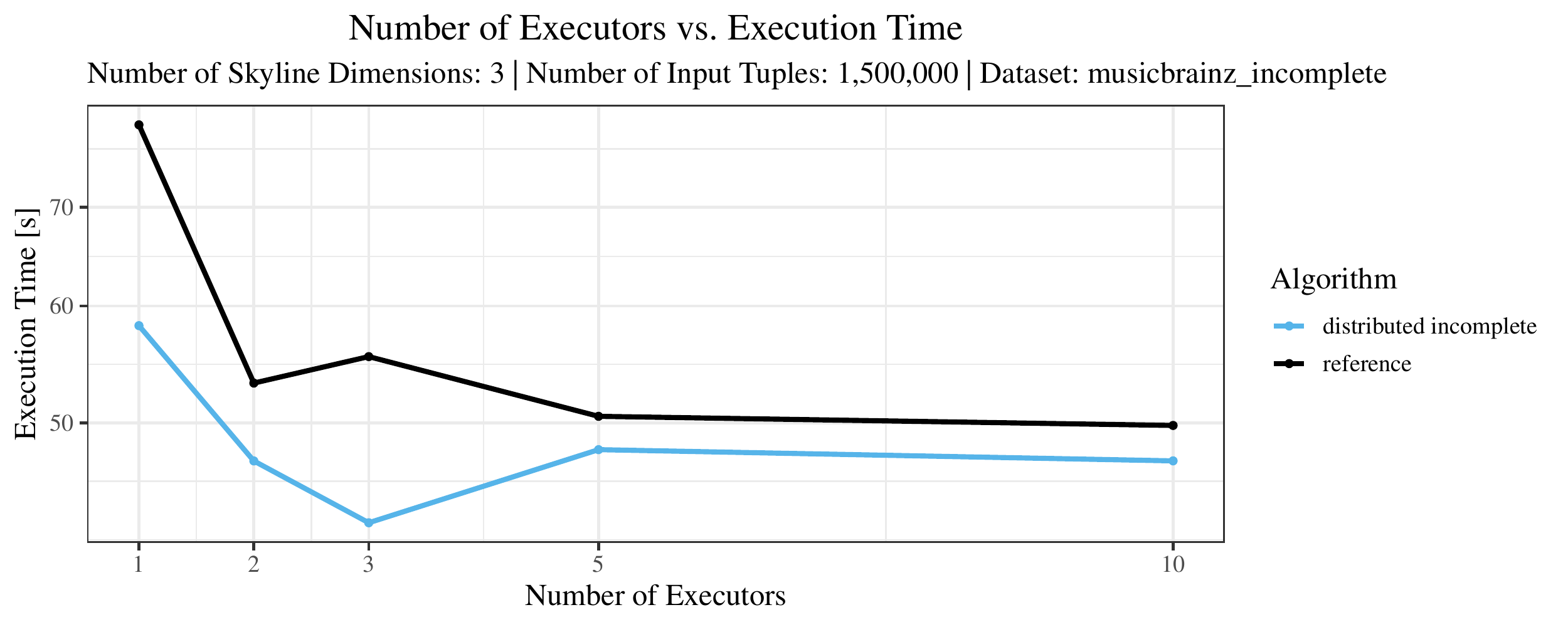}
    \end{subfigure}
    \begin{subfigure}{.5\linewidth}
      \centering
      \includegraphics[width=\linewidth]{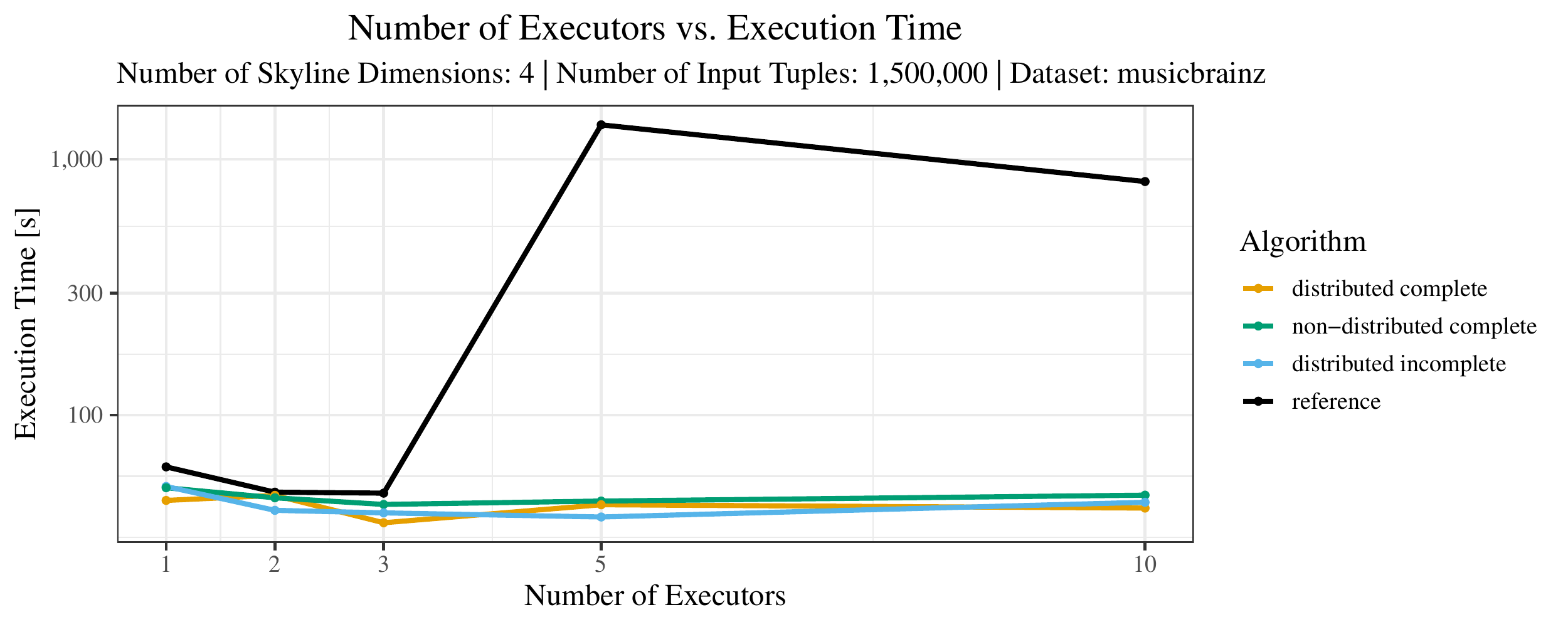}
    \end{subfigure}%
    \begin{subfigure}{.5\linewidth}
      \centering
      \includegraphics[width=\linewidth]{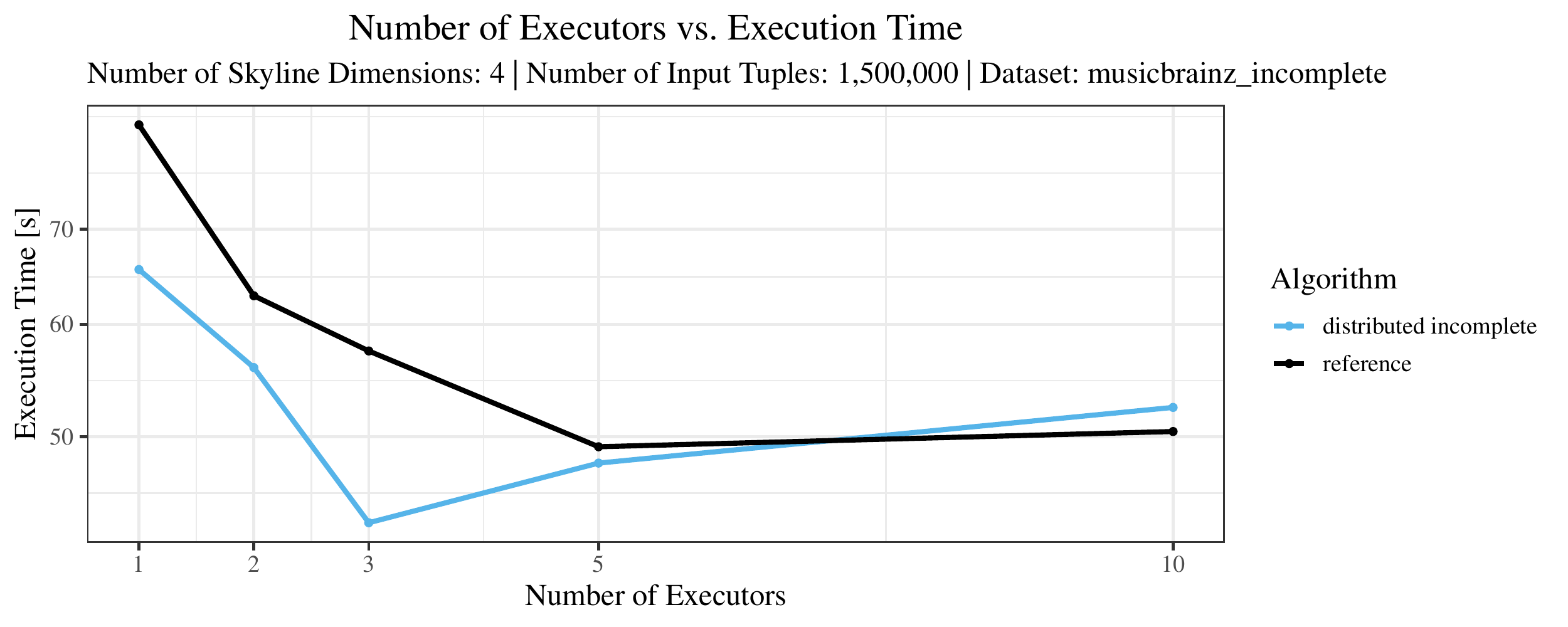}
    \end{subfigure}
    \begin{subfigure}{.5\linewidth}
      \centering
      \includegraphics[width=\linewidth]{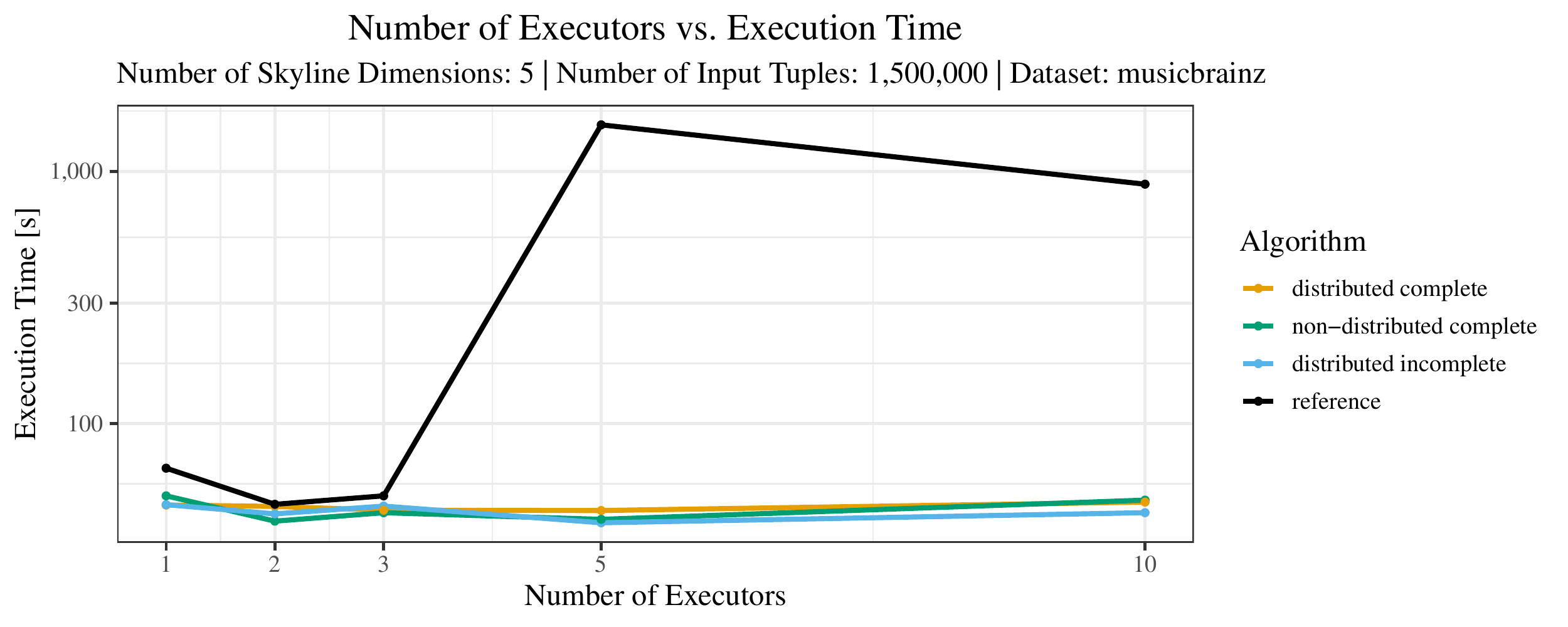}
    \end{subfigure}%
    \begin{subfigure}{.5\linewidth}
      \centering
      \includegraphics[width=\linewidth]{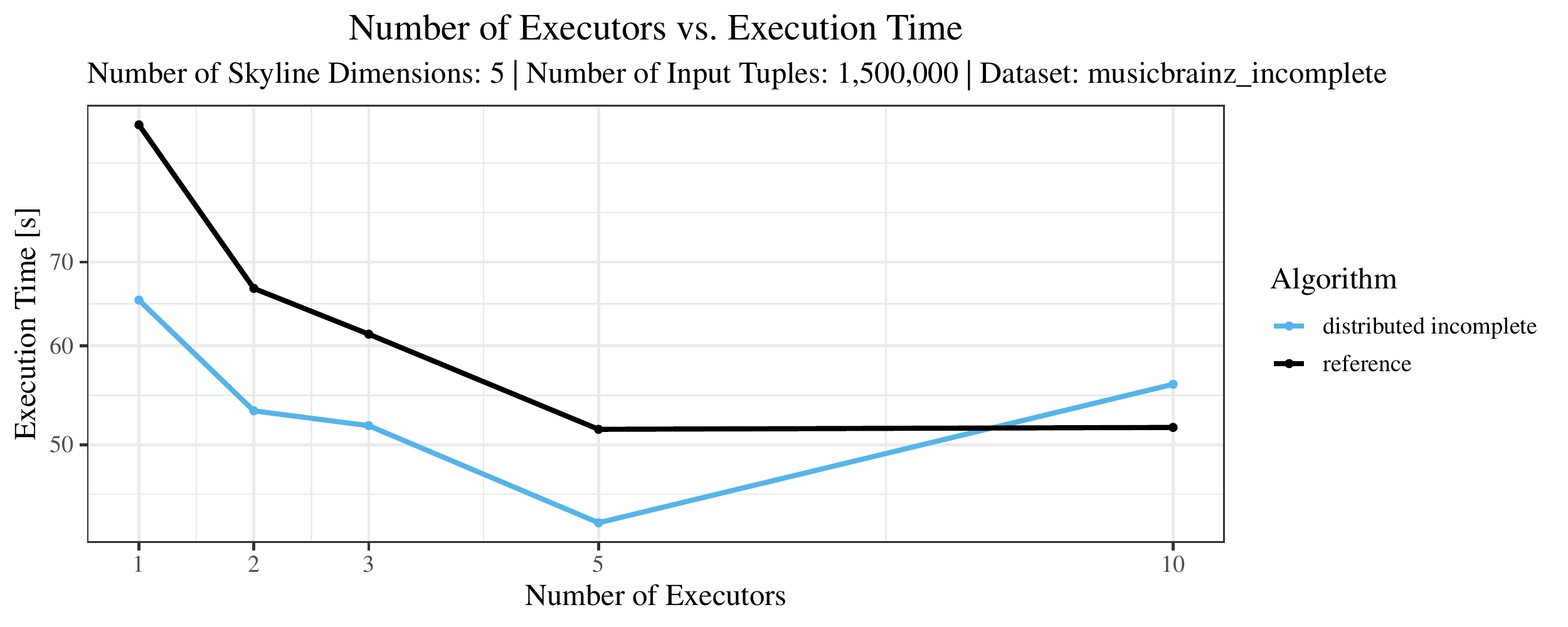}
    \end{subfigure}
    \begin{subfigure}{.5\linewidth}
      \centering
      \includegraphics[width=\linewidth]{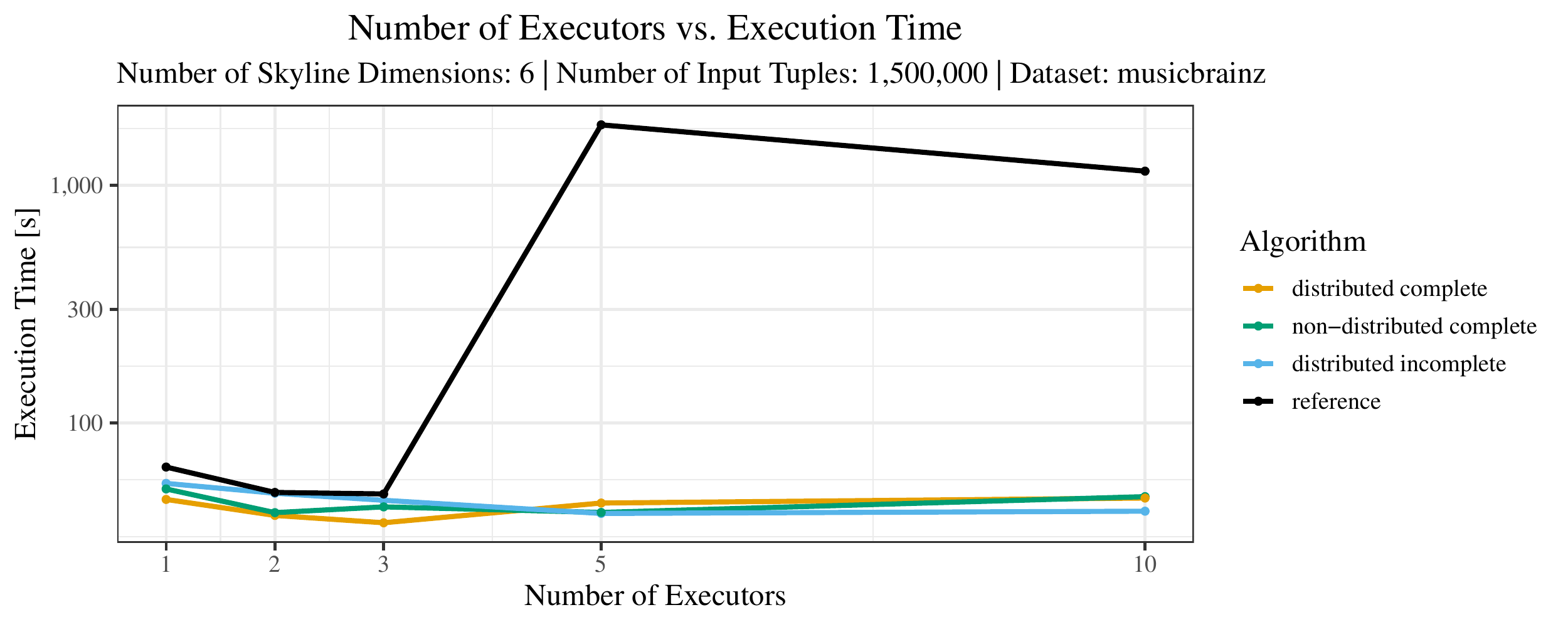}
    \end{subfigure}%
    \begin{subfigure}{.5\linewidth}
      \centering
      \includegraphics[width=\linewidth]{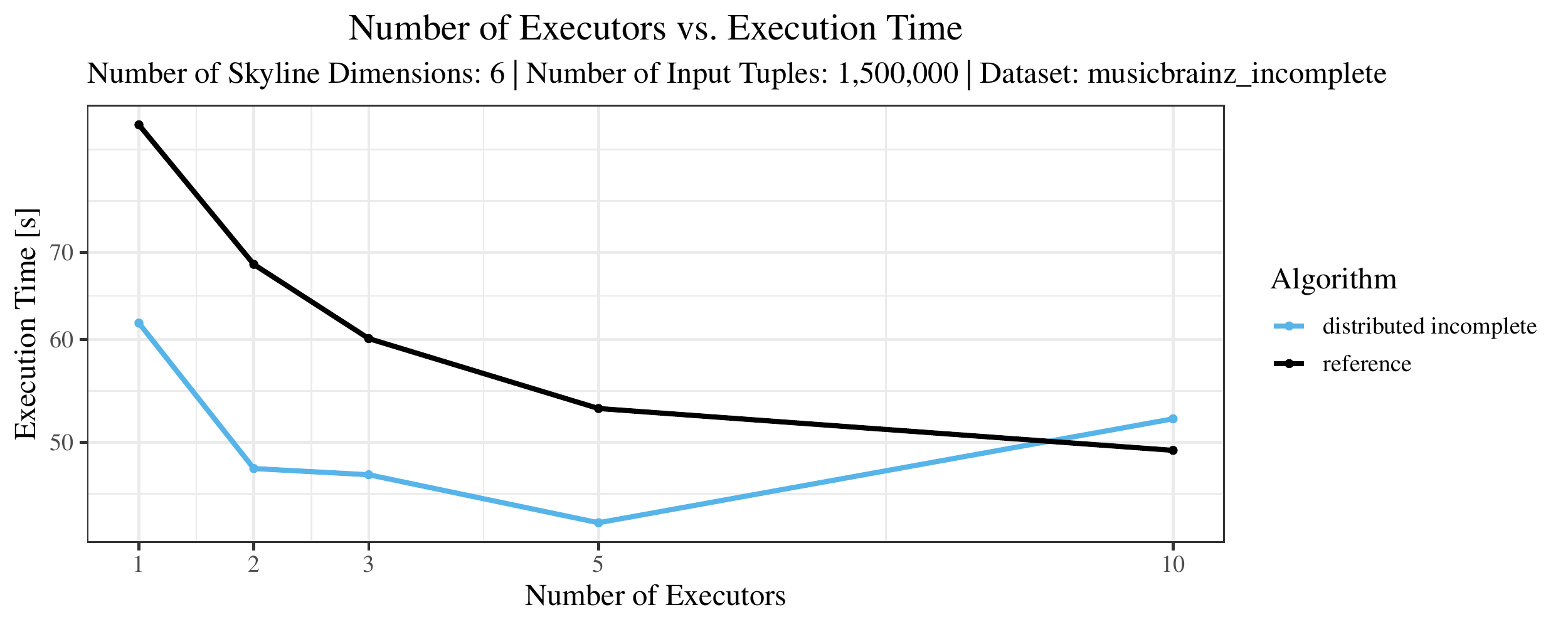}
    \end{subfigure}
    \caption{Number of executors vs. execution time using complex queries on the MusicBrainz dataset}
    \label{fig:appendix_nodes_vs_time_musicbrainz}
\end{figure*}

\begin{figure*}[p]
    \begin{subfigure}{.5\linewidth}
      \centering
      \includegraphics[width=\linewidth]{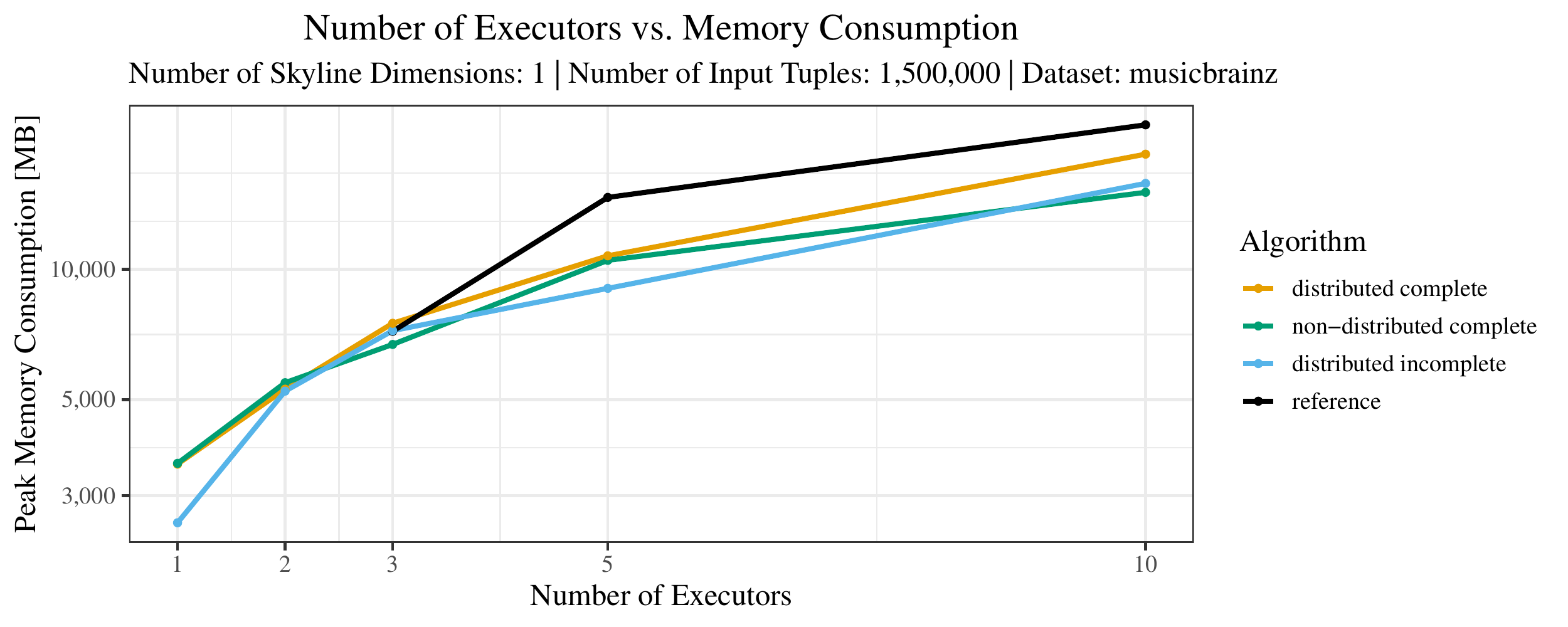}
    \end{subfigure}%
    \begin{subfigure}{.5\linewidth}
      \centering
      \includegraphics[width=\linewidth]{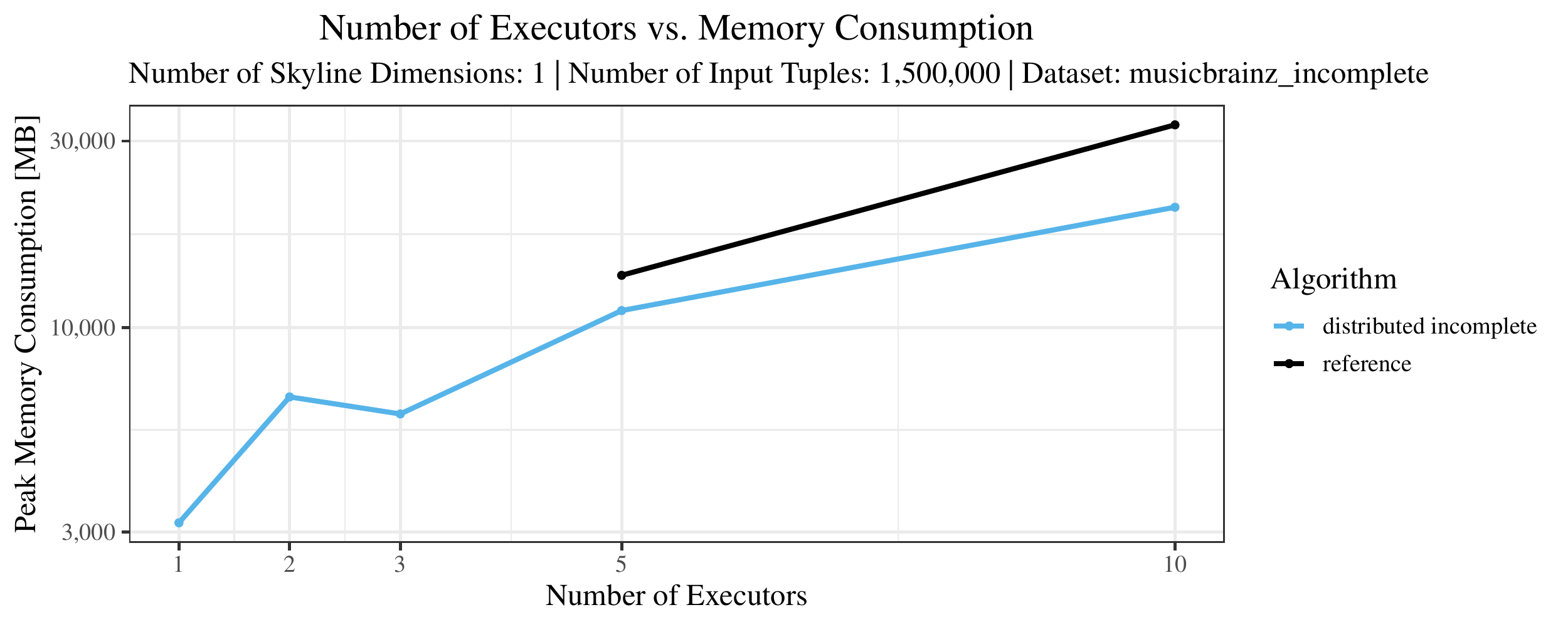}
    \end{subfigure}
    \begin{subfigure}{.5\linewidth}
      \centering
      \includegraphics[width=\linewidth]{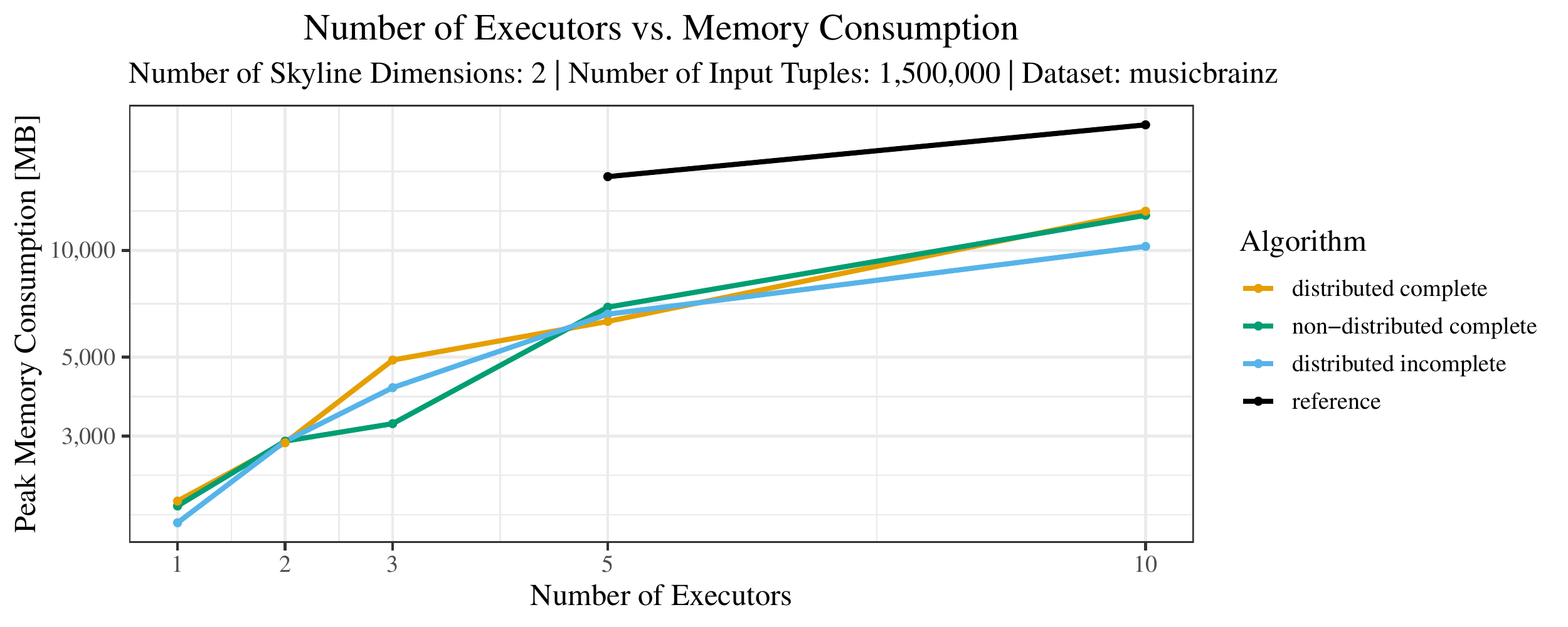}
    \end{subfigure}%
    \begin{subfigure}{.5\linewidth}
      \centering
      \includegraphics[width=\linewidth]{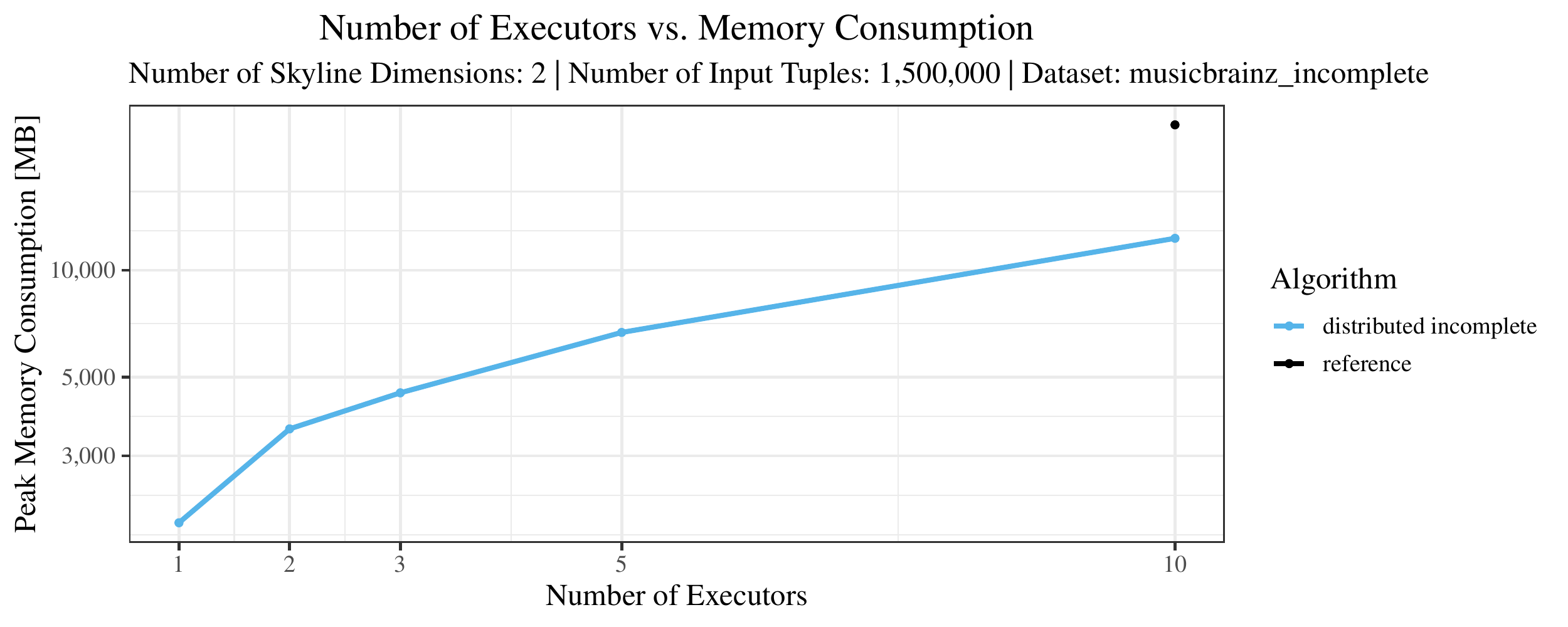}
    \end{subfigure}
    \begin{subfigure}{.5\linewidth}
      \centering
      \includegraphics[width=\linewidth]{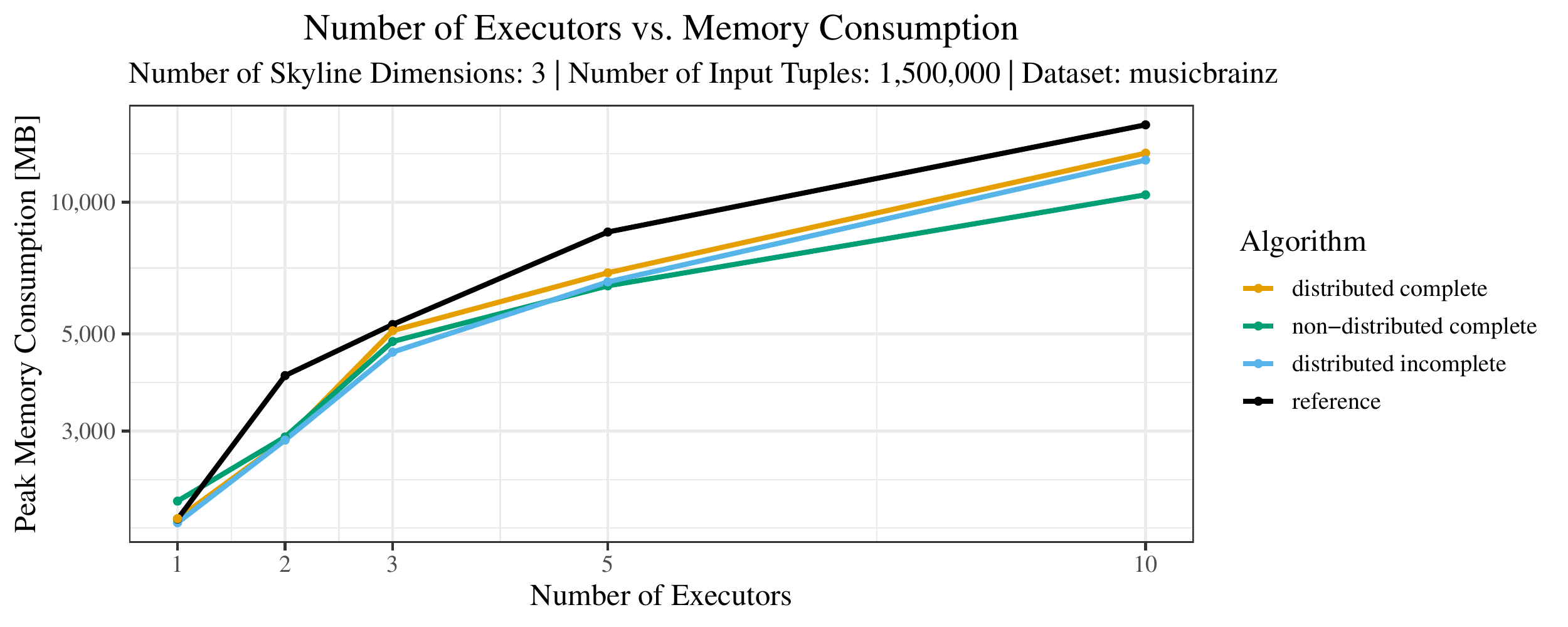}
    \end{subfigure}%
    \begin{subfigure}{.5\linewidth}
      \centering
      \includegraphics[width=\linewidth]{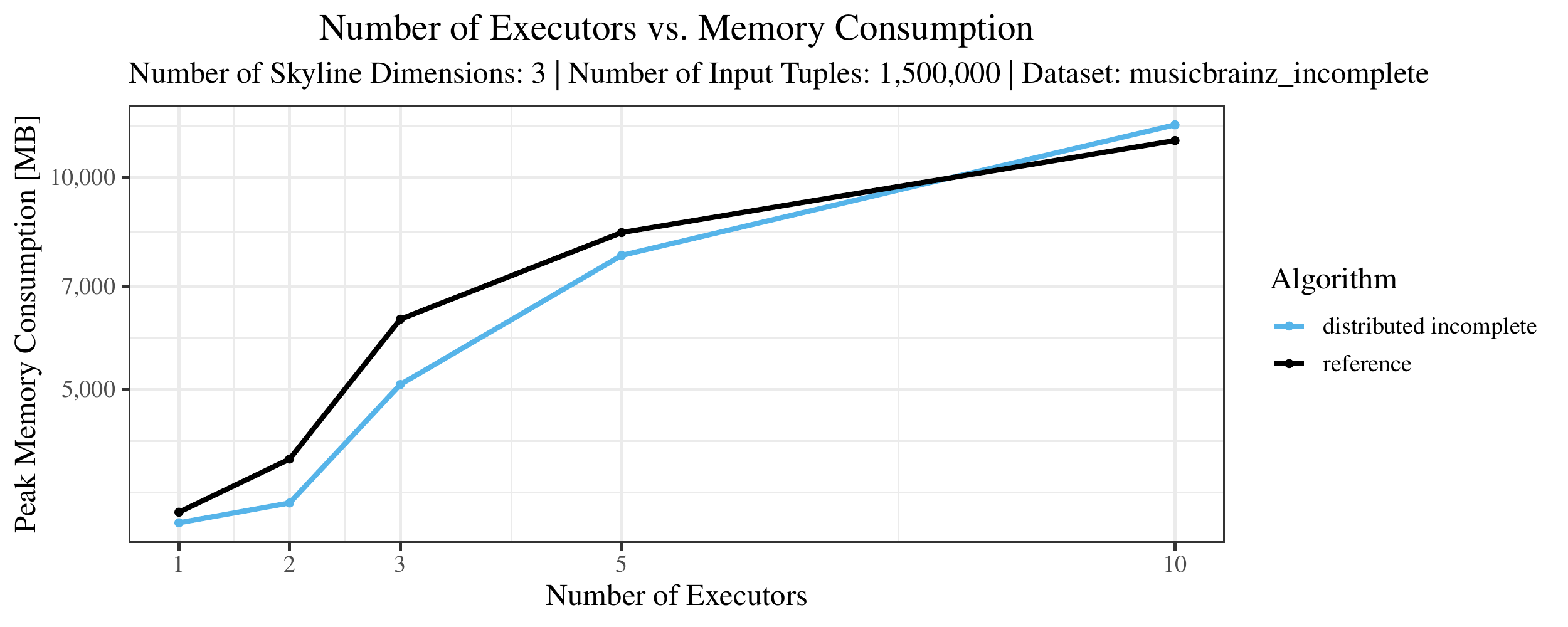}
    \end{subfigure}
    \begin{subfigure}{.5\linewidth}
      \centering
      \includegraphics[width=\linewidth]{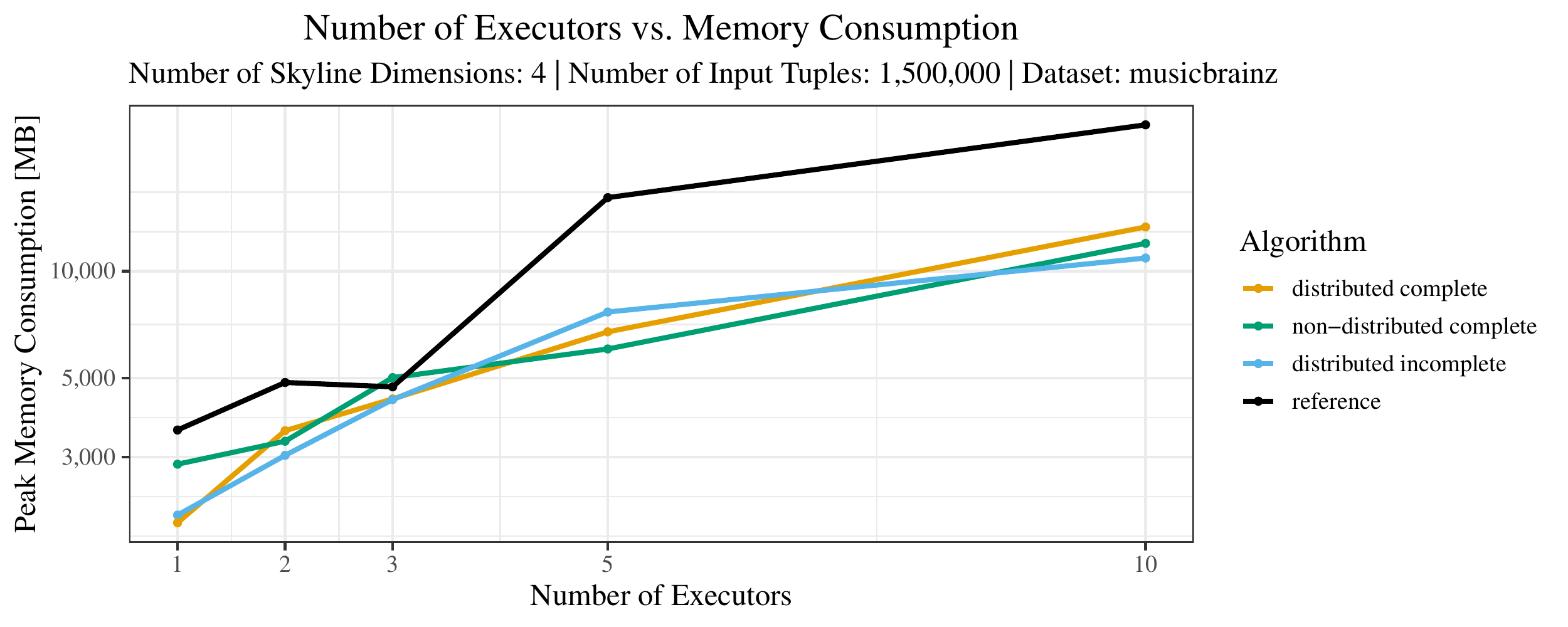}
    \end{subfigure}%
    \begin{subfigure}{.5\linewidth}
      \centering
      \includegraphics[width=\linewidth]{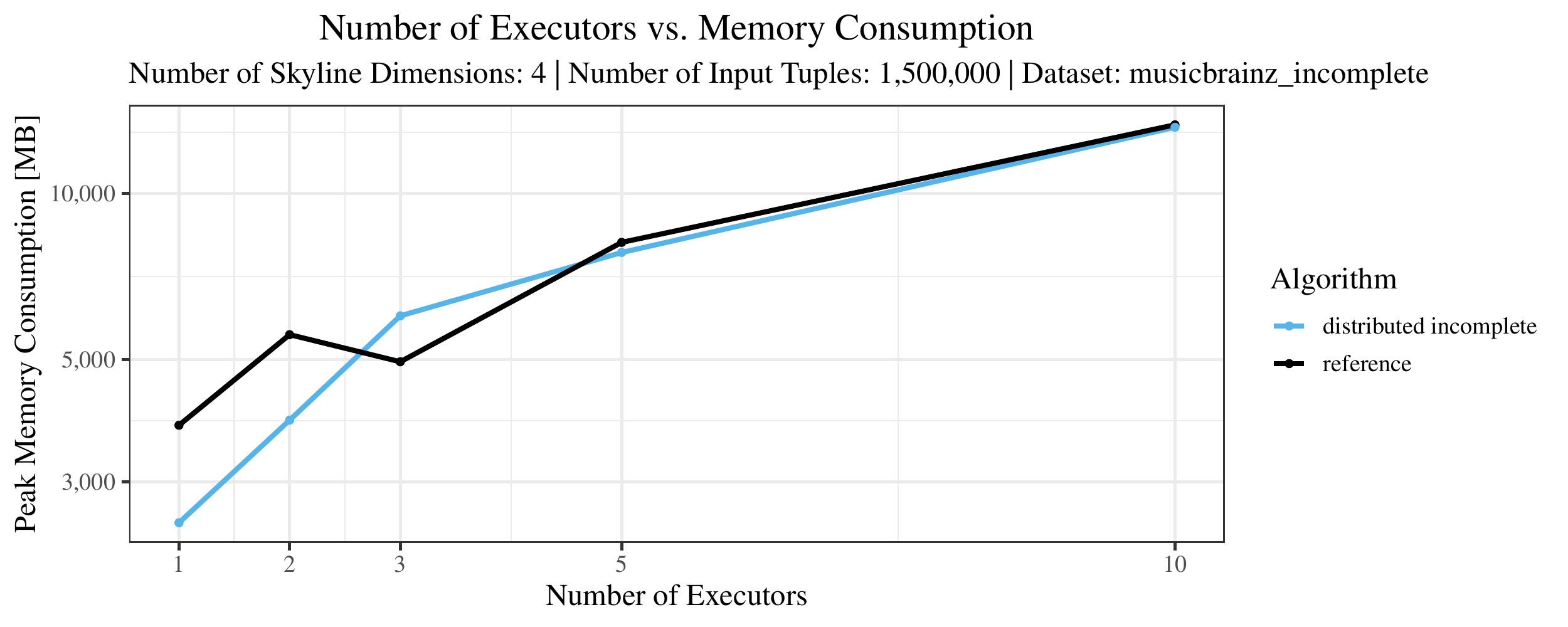}
    \end{subfigure}
    \begin{subfigure}{.5\linewidth}
      \centering
      \includegraphics[width=\linewidth]{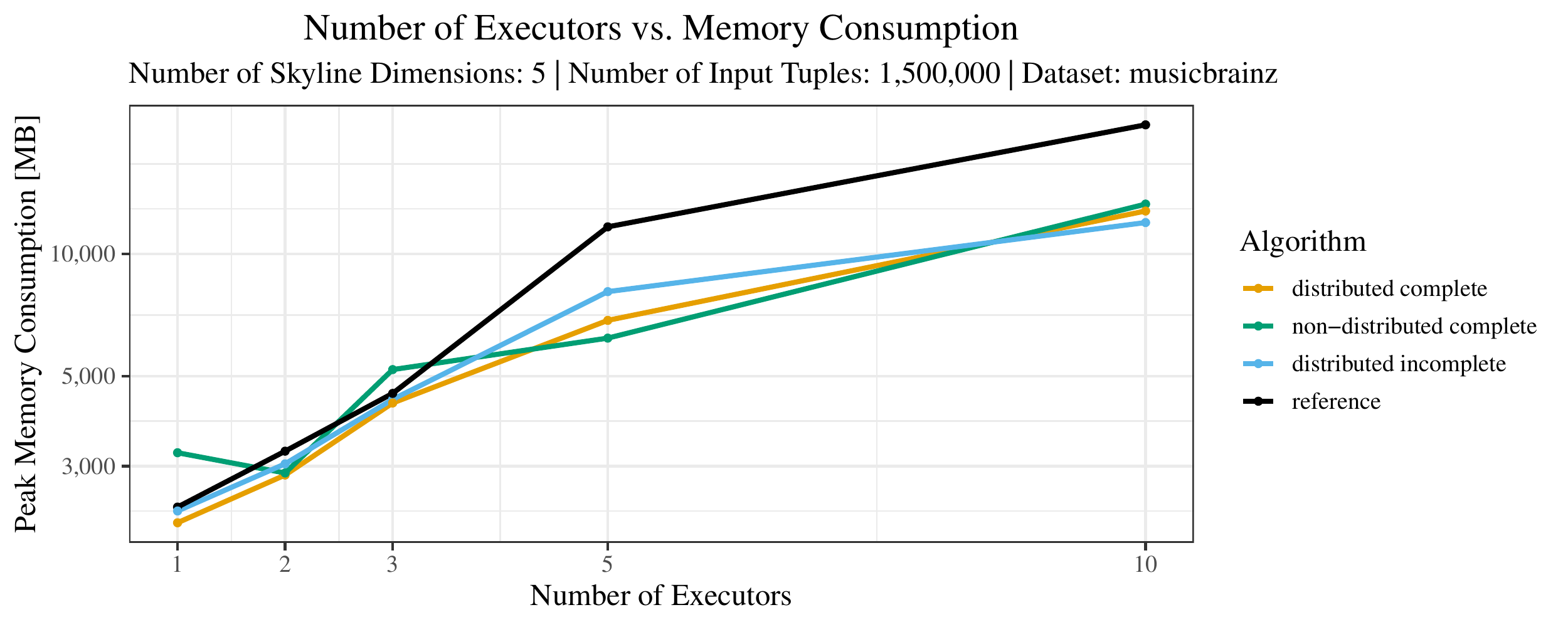}
    \end{subfigure}%
    \begin{subfigure}{.5\linewidth}
      \centering
      \includegraphics[width=\linewidth]{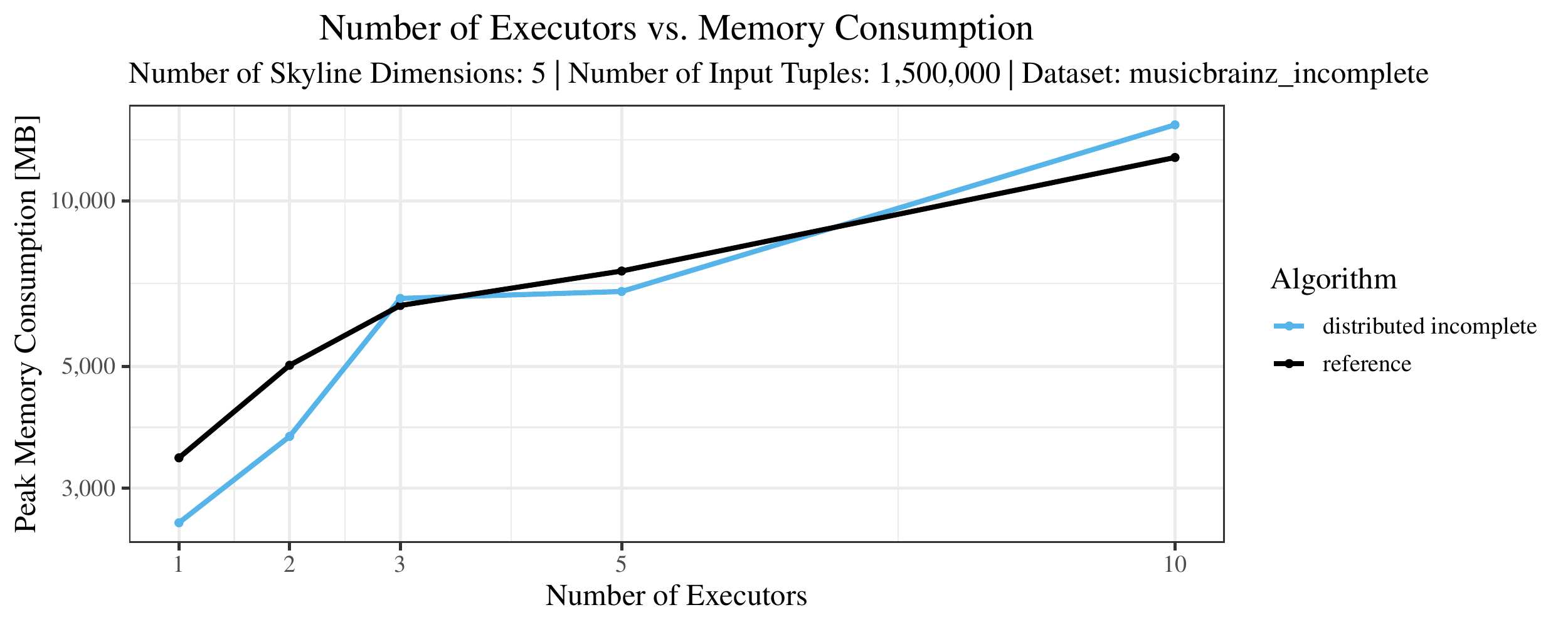}
    \end{subfigure}
    \begin{subfigure}{.5\linewidth}
      \centering
      \includegraphics[width=\linewidth]{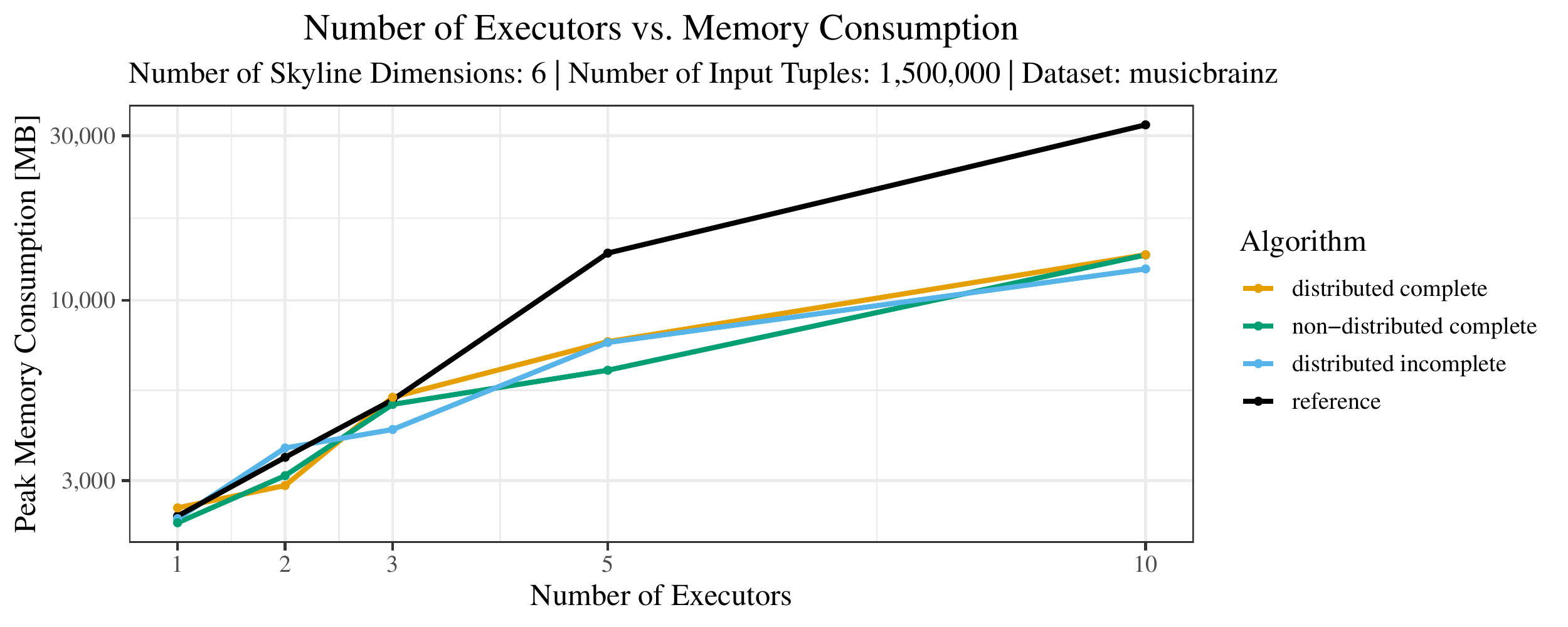}
    \end{subfigure}%
    \begin{subfigure}{.5\linewidth}
      \centering
      \includegraphics[width=\linewidth]{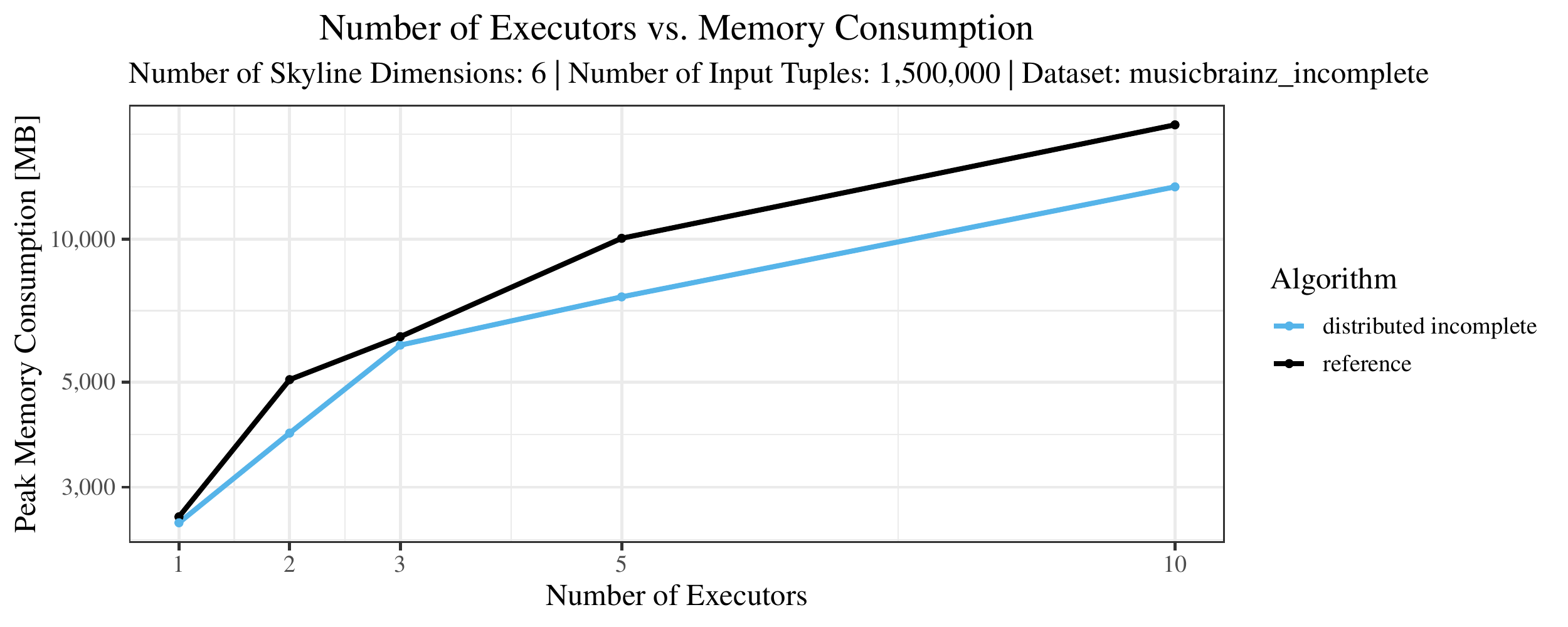}
    \end{subfigure}
    \caption{Number of executors vs. memory consumption using complex queries on the MusicBrainz dataset}
    \label{fig:appendix_nodes_vs_memory_musicbrainz}
\end{figure*}
\fi

\end{document}